\documentclass[twocolumn,numberedappendix]{aastex63}

\newcommand{\rev}{}

\newcommand{\hoststar}{TOI\,620}

\newcommand{\unit}{\mathrm}

\newcommand{\angd}{\ang[angle-symbol-over-decimal]}

\newcommand{\tess}{\emph{TESS}}

\newcommand{\gaia}{\emph{Gaia}}

\newcommand{\jwst}{\emph{JWST}}
\newcommand{\kepler}{\emph{Kepler}}

\newcommand{\wise}{\emph{WISE}}

\newcommand{\ms}{\mbox{m\,s$^{-1}~$}}
\newcommand{\kms}{\mbox{km\,s$^{-1}~$}}

\newcommand{\msun}{M$_{\odot}$}

\newcommand{\rsun}{R$_{\odot}$}
\newcommand{\lsun}{L$_{\odot}~$}

\newcommand{\mjup}{M$_{\rm J}~$}
\newcommand{\mjupe}{M$_{\rm J}$}

\newcommand{\mearth}{$M_{\oplus}~$}

\newcommand{\teff}{\ensuremath{T_{\rm eff}}}

\newcommand{\logg}{\ensuremath{\log{g}}}

\newcommand{\loggstar}{\ensuremath{\logg_\star}}

\newcommand{\rp}{\ensuremath{R_{\rm p}}}

\providecommand{\bjdtdb}{\ensuremath{\rm {BJD_{TDB}}}}

\providecommand{\mh}{\ensuremath{\left[{\rm m}/{\rm H}\right]}}
\providecommand{\teff}{\ensuremath{T_{\rm eff}}}

\providecommand{\msun}{\ensuremath{\,M_\Sun}}
\providecommand{\rsun}{\ensuremath{\,R_\Sun}}
\providecommand{\lsun}{\ensuremath{\,L_\Sun}}

\providecommand{\me}{\ensuremath{\,M_{\oplus}}}
\providecommand{\re}{\ensuremath{\,R_{\oplus}}}
\providecommand{\fave}{\langle F \rangle}
\providecommand{\fluxcgs}{10$^9$ erg s$^{-1}$ cm$^{-2}$}
\providecommand{\rhocgs}{g cm$^{-3}$}
\providecommand{\msday}{m s$^{-1}$ day$^{-1}~$}

\let\tablenum\relax
\usepackage{siunitx}
\usepackage{amssymb}
\usepackage{amsmath}
\usepackage{fontawesome}
\usepackage{gensymb}
\usepackage{mathrsfs}
\usepackage{upgreek}
\usepackage{fontenc}
\DeclareTextSymbolDefault{\dh}{T1}
\usepackage{hyperref}
\usepackage{float}
\usepackage{tabularx}

\shorttitle{TOI 620}
\shortauthors{Reefe et al.}

\graphicspath{{./}{figures/}}

\begin{document}

\title{A close-in puffy Neptune with hidden friends: The enigma of TOI 620}

\correspondingauthor{Michael A. Reefe}
\email{mreefe@gmu.edu}


\author[0000-0003-4701-8497]{Michael A. Reefe}
\affiliation{Department of Physics and Astronomy, George Mason University, 4400 University Drive, Fairfax, VA 22030, USA}

\author[0000-0002-4671-2957]{Rafael Luque}
\affiliation{Instituto de Astrof\'isica de Andaluc\'ia (IAA-CSIC), Glorieta de la Astronom\'ia s/n, 18008 Granada, Spain}

\author[0000-0002-5258-6846]{Eric Gaidos}
\affiliation{Department of Earth Sciences, University of Hawai`i at M\"{a}noa, Honolulu, HI 96822}

\author[0000-0001-7708-2364]{Corey Beard}
\affiliation{Department of Physics and Astronomy, University of California, Irvine, Irvine, CA 92697, USA}

\author[0000-0002-8864-1667]{Peter P. Plavchan}
\affiliation{Department of Physics and Astronomy, George Mason University, 4400 University Drive, Fairfax, VA 22030, USA}

\author{Marion Cointepas}
\affil{University of Grenoble Alpes, CNRS, IPAG, F-38000 Grenoble, France}
\affil{Observatoire de Gen\`eve, D\'epartement d’Astronomie, Universit\'e de Gen\`eve, Chemin Pegasi 51b, 1290 Versoix, Switzerland}

\author[0000-0002-2078-6536]{Bryson L. Cale}
\affiliation{NASA JPL, 4800 Oak Grove Drive, Pasadena, CA 91109}
\affiliation{IPAC, 770 South Wilson Avenue, Pasadena, CA 91125}

\author[0000-0003-0987-1593]{Enric Palle}
\affil{Instituto de Astrof\'isica de Canarias, C. Vía Láctea, s/n, 38205, Spain}

\author[0000-0001-5519-1391]{Hannu Parviainen}
\affiliation{Instituto de Astrofísica de Canarias (IAC), E-38200 La Laguna, Tenerife, Spain}
\affiliation{Dept. Astrofísica, Universidad de La Laguna (ULL), E-38206 La Laguna, Tenerife, Spain}

\author[0000-0002-2457-7889]{Dax L. Feliz}
\affiliation{Department of Physics and Astronomy, Vanderbilt University, 2201 West End Avenue, Nashville, TN 37235, USA}

\author[0000-0003-3773-5142]{Jason Eastman}
\affil{Center for Astrophysics, Harvard \& Smithsonian, 60 Garden Street, Cambridge, MA 02138, USA}

\author[0000-0002-3481-9052]{Keivan Stassun}
\affil{Department of Physics and Astronomy, Vanderbilt University, 2201 West End Avenue, Nashville, TN 37235, USA}

\author[0000-0002-2592-9612]{Jonathan Gagn\'e}
\affiliation{Université de Montréal, 2900 Edouard Montpetit Blvd, Montreal, Quebec H3T 1J4, Canada}

\author[0000-0002-4715-9460]{Jon M. Jenkins}
\affil{NASA Ames Research Center, Moffett Field, CA 94035, USA}

\author{Patricia~T.~Boyd}
\affiliation{NASA Goddard Space Flight Center, 8800 Greenbelt Rd, Greenbelt, MD 20771, USA}

\author{Richard~C.~Kidwell}
\affiliation{Space Telescope Science Institute, 3700 San Martin Dr, Baltimore, MD 21218, USA}

\author{Scott~McDermott}
\affiliation{Proto-Logic LLC, 1718 Euclid Street NW, Washington, DC 20009, USA}

\author[0000-0001-6588-9574]{Karen~A.~Collins}
\affiliation{Center for Astrophysics, Harvard \& Smithsonian, 60 Garden Street, Cambridge, MA 02138, USA}

\author[0000-0003-0241-2757]{William~Fong}
\affil{Kavli Institute for Astrophysics and Space Research, Massachusetts Institute of Technology, 77 Massachusetts Ave, Cambridge, MA 02139, USA}

\author[0000-0002-5169-9427]{Natalia~Guerrero}
\affil{Department of Astronomy, University of Florida, Gainesville, FL 32611, USA}
\affil{Kavli Institute for Astrophysics and Space Research, Massachusetts Institute of Technology, 77 Massachusetts Ave, Cambridge, MA 02139, USA}


\author{Jose-Manuel Almenara-Villa}
\affil{University of Grenoble Alpes, CNRS, IPAG, F-38000 Grenoble, France}

\author[0000-0003-4733-6532]{Jacob Bean}
\affil{Department of Astronomy \& Astrophysics, University of Chicago, 5640 South Ellis Avenue, Chicago, IL 60637, USA}

\author[0000-0002-5627-5471]{Charles A. Beichman}
\affil{NASA Exoplanet Science Institute, Caltech/IPAC, Mail Code 100-22, 1200 E. California Blvd., Pasadena, CA 91125, USA}

\author[0000-0003-1466-8389]{John Berberian}
\affiliation{Department of Physics and Astronomy, George Mason University, 4400 University Drive, Fairfax, VA 22030, USA}
\affiliation{Woodson High School, 9525 Main St, Fairfax, VA 22031, USA}

\author[0000-0001-6637-5401]{Allyson Bieryla}
\affiliation{Center for Astrophysics, Harvard \& Smithsonian, 60 Garden Street, Cambridge, MA 02138, USA}

\author[0000-0001-9003-8894]{Xavier Bonfils}
\affil{University of Grenoble Alpes, CNRS, IPAG, F-38000 Grenoble, France}

\author[0000-0002-7613-393X]{Fran\c cois Bouchy}
\affil{Observatoire de Gen\`eve, D\'epartement d’Astronomie, Universit\'e de Gen\`eve, Chemin Pegasi 51b, 1290 Versoix, Switzerland}

\author{Madison Brady}
\affil{Department of Astronomy \& Astrophysics, University of Chicago, 5640 South Ellis Avenue, Chicago, IL 60637, USA}

\author[0000-0001-7904-4441]{Edward M.~Bryant}
\affiliation{Dept.\ of Physics, University of Warwick, Gibbet Hill Road, Coventry CV4 7AL, UK}
\affiliation{Centre for Exoplanets and Habitability, University of Warwick, Coventry CV4 7AL, UK}

\author[0000-0001-8266-0894]{Luca Cacciapuoti}
\affiliation{University of Naples Federico II, Corso Umberto I, 40, 80138 Napoli NA, Italy}
\affiliation{NASA Goddard Space Flight Center, 8800 Greenbelt Rd, Greenbelt, MD 20771, USA}

\author[0000-0003-4835-0619]{Caleb I. Ca\~nas}
\altaffiliation{NASA Earth and Space Science Fellow}
\affiliation{Department of Astronomy \& Astrophysics, 525 Davey Laboratory, The Pennsylvania State University, University Park, PA, 16802, USA}
\affiliation{Center for Exoplanets and Habitable Worlds, 525 Davey Laboratory, The Pennsylvania State University, University Park, PA, 16802, USA}

\author[0000-0002-5741-3047]{David R. Ciardi}
\affil{NASA Exoplanet Science Institute, Caltech/IPAC, Mail Code 100-22, 1200 E. California Blvd., Pasadena, CA 91125, USA}

\author[0000-0003-2781-3207]{Kevin I. Collins}
\affiliation{Department of Physics and Astronomy, George Mason University, 4400 University Drive, Fairfax, VA 22030, USA}

\author[0000-0002-1835-1891]{Ian Crossfield}
\affil{University of Kansas, 1450 Jayhawk Blvd, Lawrence, KS 66045, USA}

\author[0000-0001-8189-0233]{Courtney D. Dressing}
\affil{Astronomy Department, University of California, Berkeley, CA 94720, USA}

\author{Philipp Eigmueller}
\affil{Institute of Planetary Research, German Aerospace Center, Rutherfordstrasse 2, 12489, Berlin, Germany}

\author[0000-0001-8364-2903]{Mohammed El Mufti}
\affiliation{Department of Physics and Astronomy, George Mason University, 4400 University Drive, Fairfax, VA 22030, USA}

\author[0000-0002-2341-3233]{Emma Esparza-Borges}
\affil{Instituto de Astrofísica de Canarias (IAC), E-38200 La Laguna, Tenerife, Spain}
\affil{Dept. Astrofísica, Universidad de La Laguna (ULL), E-38206 La Laguna, Tenerife, Spain}

\author[0000-0002-4909-5763]{Akihiko Fukui}
\affil{Komaba Institute for Science, The University of Tokyo, 3-8-1 Komaba, Meguro, Tokyo 153-8902, Japan}
\affil{Instituto de Astrof\'{i}sica de Canarias (IAC), 38205 La Laguna, Tenerife, Spain}


\author[0000-0002-8518-9601]{Peter Gao}
\affiliation{Earth and Planets Laboratory, Carnegie Institution for Science, 5241 Broad Branch Rd NW, Washington, DC 20015, USA}

\author[0000-0001-9596-8820]{Claire Geneser}
\affiliation{Department of Physics and Astronomy, Mississippi State University, 75 B. S. Hood Road, Mississippi State, MS 39762, USA}

\author[0000-0003-2519-6161]{Crystal~L.~Gnilka}
\affil{NASA Ames Research Center, Moffett Field, CA 94035, USA}

\author[0000-0002-9329-2190]{Erica Gonzales}
\affil{Department of Astronomy and Astrophysics, University of California, Santa Cruz, 1156 High St, Santa Cruz, CA 95064, USA}

\author[0000-0002-5463-9980]{Arvind F.\ Gupta}
\affiliation{Department of Astronomy \& Astrophysics, 525 Davey Laboratory, The Pennsylvania State University, University Park, PA, 16802, USA}
\affiliation{Center for Exoplanets and Habitable Worlds, 525 Davey Laboratory, The Pennsylvania State University, University Park, PA, 16802, USA}

\author[0000-0003-1312-9391]{Sam Halverson}
\affiliation{Department of Physics, Mathematics, and Astronomy, California Institute of Technology, 1200 E California Blvd, Pasadena, CA 91125, USA}

\author{Fred Hearty}
\affil{Department of Astronomy and Astrophysics, Pennsylvania State University, State College, PA 16801}

\author[0000-0002-2532-2853]{Steve~B.~Howell}
\affil{NASA Ames Research Center, Moffett Field, CA 94035, USA}

\author{Jonathan Irwin}
\affiliation{Center for Astrophysics, Harvard \& Smithsonian, 60 Garden Street, Cambridge, MA 02138, USA}

\author[0000-0001-8401-4300]{Shubham Kanodia}
\affil{Department of Astronomy and Astrophysics, Pennsylvania State University, State College, PA 16801}

\author[0000-0003-0534-6388]{David Kasper}
\affil{Department of Astronomy \& Astrophysics, University of Chicago, 5640 South Ellis Avenue, Chicago, IL 60637, USA}

\author[0000-0001-9032-5826]{Takanori Kodama}
\affil{Komaba Institute for Science, The University of Tokyo, 3-8-1 Komaba, Meguro, Tokyo 153-8902, Japan}

\author[0000-0001-9786-1031]{Veselin Kostov}
\affiliation{NASA Goddard Space Flight Center, 8800 Greenbelt Rd, Greenbelt, MD 20771, USA}

\author[0000-0001-9911-7388]{David W. Latham}
\affiliation{Center for Astrophysics, Harvard \& Smithsonian, 60 Garden Street, Cambridge, MA 02138, USA}

\author[0000-0001-9699-1459]{Monika Lendl}
\affil{Observatoire de Gen\`eve, D\'epartement d’Astronomie, Universit\'e de Gen\`eve, Chemin Pegasi 51b, 1290 Versoix, Switzerland}

\author[0000-0002-9082-6337]{Andrea Lin}
\affil{Department of Astronomy and Astrophysics, Pennsylvania State University, State College, PA 16801}

\author[0000-0002-4881-3620]{John H. Livingston}
\affil{Department of Astronomy, Graduate School of Science, The University of Tokyo, 7-3-1 Hongo, Bunkyo-ku, Tokyo 113-0033, Japan}

\author[0000-0001-8342-7736]{Jack Lubin}
\affiliation{Department of Physics and Astronomy, University of California, Irvine, Irvine, CA 92697, USA}

\author[0000-0001-9596-7983]{Suvrath Mahadevan}
\affil{Department of Astronomy and Astrophysics, Pennsylvania State University, State College, PA 16801}

\author[0000-0001-7233-7508]{Rachel Matson}
\affil{United States Naval Observatory, 3450 Massachusetts Ave NW, Washington, DC 20392, USA}

\author[0000-0003-0593-1560]{Elisabeth Matthews}
\affil{Observatoire de Gen\`eve, D\'epartement d’Astronomie, Universit\'e de Gen\`eve, Chemin Pegasi 51b, 1290 Versoix, Switzerland}

\author[0000-0001-9087-1245]{Felipe Murgas}
\affil{University of Grenoble Alpes, CNRS, IPAG, F-38000 Grenoble, France}
\affil{Instituto de Astrofísica de Canarias (IAC), E-38200 La Laguna, Tenerife, Spain}
\affil{Dept. Astrofísica, Universidad de La Laguna (ULL), E-38206 La Laguna, Tenerife, Spain}

\author[0000-0001-8511-2981]{Norio Narita}
\affil{Komaba Institute for Science, The University of Tokyo, 3-8-1 Komaba, Meguro, Tokyo 153-8902, Japan}
\affil{Astrobiology Center, 2-21-1 Osawa, Mitaka, Tokyo 181-8588, Japan}
\affil{Instituto de Astrof\'{i}sica de Canarias (IAC), 38205 La Laguna, Tenerife, Spain}

\author[0000-0003-3848-3418]{Patrick Newman}
\affiliation{Department of Physics and Astronomy, George Mason University, 4400 University Drive, Fairfax, VA 22030, USA}

\author[0000-0001-8720-5612]{Joe Ninan}
\affil{Department of Astronomy and Astrophysics, Pennsylvania State University, State College, PA 16801}

\author[0000-0002-5899-7750]{Ares Osborn}
\affiliation{Dept.\ of Physics, University of Warwick, Gibbet Hill Road, Coventry CV4 7AL, UK}
\affiliation{Centre for Exoplanets and Habitability, University of Warwick, Coventry CV4 7AL, UK}

\author[0000-0002-8964-8377]{Samuel N. Quinn}
\affiliation{Center for Astrophysics, Harvard \& Smithsonian, 60 Garden Street, Cambridge, MA 02138, USA}

\author[0000-0003-0149-9678]{Paul Robertson}
\affiliation{Department of Physics and Astronomy, University of California, Irvine, Irvine, CA 92697, USA}

\author[0000-0001-8127-5775]{Arpita Roy}
\affil{Space Telescope Science Institute, 3700 San Martin Dr, Baltimore, MD 21218, USA}

\author[0000-0001-5347-7062]{Joshua Schlieder}
\affil{NASA Goddard Space Flight Center, 8800 Greenbelt Rd, Greenbelt, MD 20771, USA}

\author[0000-0002-4046-987X]{Christian Schwab}
\affil{Department of Physics and Astronomy, University of Macquarie, Sydney, Australia}

\author[0000-0003-4526-3747]{Andreas Seifahrt}
\affil{Department of Astronomy \& Astrophysics, University of Chicago, 5640 South Ellis Avenue, Chicago, IL 60637, USA}

\author{Gareth D. Smith}
\affil{Astrophysics Group, Cavendish Laboratory, J.J. Thomson Avenue, Cambridge CB3 0HE, UK}

\author{Ahmad Sohani}
\affiliation{Department of Physics and Astronomy, Mississippi State University, 75 B. S. Hood Road, Mississippi State, MS 39762, USA}

\author[0000-0001-7409-5688]{Guðmundur Stef\'ansson}
\affil{Princeton University, Department of Astrophysical Sciences, 4 Ivy Lane, Princeton, NJ 08540, USA}

\author{Daniel Stevens}
\affil{Department of Astronomy and Astrophysics, Pennsylvania State University, State College, PA 16801}

\author[0000-0002-4410-4712]{Julian St\"urmer}
\affil{Landessternwarte, Zentrum f\"ur Astronomie der Universit\"at Heidelberg, K\"onigstuhl 12, 69117 Heidelberg, Germany}

\author[0000-0002-2903-2140]{Angelle Tanner}
\affiliation{Department of Physics and Astronomy, Mississippi State University, 75 B. S. Hood Road, Mississippi State, MS 39762, USA}

\author{Ryan Terrien}
\affil{Department of Physics and Astronomy, Carleton College, Sayles Hill Campus Center, North College Street, Northfield, MN 55057, USA}

\author{Johanna Teske}
\affiliation{Earth and Planets Laboratory, Carnegie Institution for Science, 5241 Broad Branch Rd NW, Washington, DC 20015, USA}

\author[0000-0002-4501-564X]{David Vermilion}
\affiliation{Department of Physics and Astronomy, George Mason University, 4400 University Drive, Fairfax, VA 22030, USA}
\affiliation{NASA Goddard Space Flight Center, 8800 Greenbelt Rd, Greenbelt, MD 20771, USA}

\author[0000-0002-6937-9034]{Sharon X. Wang}
\affiliation{Department of Astronomy, Tsinghua University, Beijing 100084, People's Republic of China}

\author[0000-0002-7424-9891]{Justin Wittrock}
\affiliation{Department of Physics and Astronomy, George Mason University, 4400 University Drive, Fairfax, VA 22030, USA}

\author[0000-0001-6160-5888]{Jason T.\ Wright}
\affiliation{Department of Astronomy \& Astrophysics, 525 Davey Laboratory, The Pennsylvania State University, University Park, PA, 16802, USA}
\affiliation{Center for Exoplanets and Habitable Worlds, 525 Davey Laboratory, The Pennsylvania State University, University Park, PA, 16802, USA}
\affiliation{Penn State Extraterrestrial Intelligence Center, 525 Davey Laboratory, The Pennsylvania State University, University Park, PA, 16802, USA}

\author[0000-0002-6532-4378]{Mathias Zechmeister}
\affil{Institut f\"ur Astrophysik, Georg-August-Universit\"at, Friedrich- Hund-Platz 1, D-37077 G\"ottingen, Germany}

\author[0000-0003-2872-9883]{Farzaneh Zohrabi}
\affiliation{Department of Physics and Astronomy, Louisiana State University, 202 Nicholson Hall, Baton Rouge, LA 70803, USA}

\begin{abstract}

We present the validation of a transiting low-density exoplanet orbiting the M2.5 dwarf TOI 620 discovered by the NASA \tess\ mission. We utilize photometric data from both \tess\ and ground-based follow-up observations to validate the ephemerides of the 5.09-day transiting signal and vet false positive scenarios. High-contrast imaging data are used to resolve the stellar host and exclude stellar companions at separations $\gtrsim \angd{;;0.2}$.  We \rev{obtain} follow-up spectroscopy and corresponding precise radial velocities (RVs) with multiple PRV spectrographs to confirm the planetary nature of the transiting exoplanet. We calculate a 5$\sigma$ upper limit \rev{of $M_P < 7.1$ \mearth and $\rho_P < 0.74$} \rhocgs, and we identify a non-transiting 17.7-day candidate. We also find evidence for a substellar (1--20 \mjupe) companion with a projected separation $\lesssim 20$ au from a combined analysis of \gaia, AO imaging, and RVs.  With the discovery of this outer companion, we carry out a detailed exploration of the possibilities that TOI 620 b might instead be a circum-secondary planet or a pair of eclipsing binary stars orbiting the host in a hierarchical triple system.  We find, under scrutiny, that we can exclude both of these scenarios from the multi-wavelength transit photometry, thus validating TOI 620 b as a low-density exoplanet transiting the central star in this system. The low density of TOI 620 b makes it one of the most amenable exoplanets for atmospheric characterization, \rev{such as with \jwst\ and \emph{Ariel},} validated or confirmed by the \tess\ mission to date.

\end{abstract}


\section{Introduction} 
\label{sect:intro}

The most successful method for discovering planets around other stars (exoplanets) is the photometric transit method, which measures the periodic dip in brightness from a star that is observed as a planet passes in front of it.  The orbital period and the size of the planet relative to the star can be readily derived from such observations \citep{Seager_2003}. After its launch in 2009, the \kepler\ mission \citep{Borucki2011, Howard2012} accelerated the discovery of Neptune- and terrestrial-size transiting exoplanets, while the \textit{Transiting Exoplanet Survey Satellite} \citep[\tess,][]{Ricker2015}, launched in 2018, has identified over 4000 candidate exoplanets orbiting relatively nearby, bright host stars suitable for further characterization.  However, the candidates discovered by the \tess\ mission need further supporting observations, such as archival photometry, ground-based light curves, high-contrast imaging, and reconnaissance spectroscopy, to validate and confirm that they are not false-positives. \rev{Out of these 4000 candidates}, 161 have been validated and/or confirmed to date. Due to the relatively large \tess\ pixels spanning $\angd{;;22}$ on the sky, fainter visual eclipsing binaries can blend with the nearby bright target stars and produce false-positives \rev{(barring instrumental artifacts).}  \rev{This is an important consideration particularly} when only a single transiting planet is found in the \tess\ 27-day time-baseline of sector observations, at lower ecliptic latitudes, and away from the ecliptic poles \citep{Lissauer_2012,vanderburg2019,rodriguez2020,hobson2021,addison2021,osborn2021,dreizler2020,brahm2020,nowak2020,teske2020,sha2021,gan2021,bluhm2020}.   

Complementary to exoplanet transit observations are radial velocity (RV) signals which undergo periodic variations from the stellar reflex motions of orbiting exoplanets, thereby inferring planet masses modulo an unknown inclination \citep{Fischer_2016, Mayor_1995}. We can leverage the strengths of both the RV and transit methods to provide independent confirmations on quantities that can be measured with both methods (such as the orbital period and ephemerides), constrain the orbital inclination, and determine mean densities.  Among sub-Jovian planets, mean density informs us about interior composition and the presence or absence of a thick atmosphere of H and He \citep[e.g.][]{Southworth_2010, Marcy_2014, Rogers_2015, Fulton_2017, Bitsch_2019, Zeng_2019}.

Direct exploration of exoplanet compositions and atmospheres can exploit differential observations during primary transits of the planet in front of the host star (i.e., spectroscopy of an atmosphere in transmission) or secondary eclipse of the planet by the star \citep[spectroscopy of an atmosphere in emission, e.g.,][]{2013ApJ...775..137L, Line2013ASR, Kreidberg2015ADO, Sing2016ACF, Deming2013INFRAREDTS, Greene2015CharacterizingTE}.  Although some limited observations can be done from the ground \citep[e.g.,][]{Nortman_2018,Allart_2018}, due to interference from Earth's atmosphere, most of these have been obtained by space telescopes such as \emph{HST} \citep[e.g.,][]{Ehrenreich_2015}.  The launch of \emph{JWST} \citep{Beichman2018} and \emph{Ariel} \citep{10.1117/12.2232370} will usher in a new era of spectral resolution, precision, and stability at the infrared wavelengths where many important atmospheric molecules have absorption features.  

Even with such advances in instrumentation, these demanding observations require planets transiting nearby bright but comparatively small stars for which the expected signal-to-noise (S/N) will be highest.  The primary mission of \emph{TESS} is to identify such systems: their suitability for transit and secondary eclipse spectroscopy can be quantified by two metrics related to the $S/N$ \citep{Kempton_2018}.  In addition, observations and models point to the planetary equilibrium temperature $T_{\rm eq}$ as a fundamental parameter in understanding exoplanet atmospheres: at $T_{\rm eq} > 2000$ K, atmospheres approach thermodynamic equilibrium, there are few or no condensates, and the observable role of photochemistry is minimal; below $T_{\rm eq} < 1300$ K, disequilibrium can readily occur, condensation and photochemistry can be important, and these atmospheres can be complex.  The coolest of the cool ($T_{\rm eq} \lesssim 500$ K) are those of interest to searches for biosignatures.  Finally, interpretation of observations requires an estimate of planet gravity and hence its mass \citep{Batalha2019,Madhusudhan2019,2010ApJ...712..974R}.  

\begin{figure}
    \centering
    \includegraphics[width=.49\textwidth]{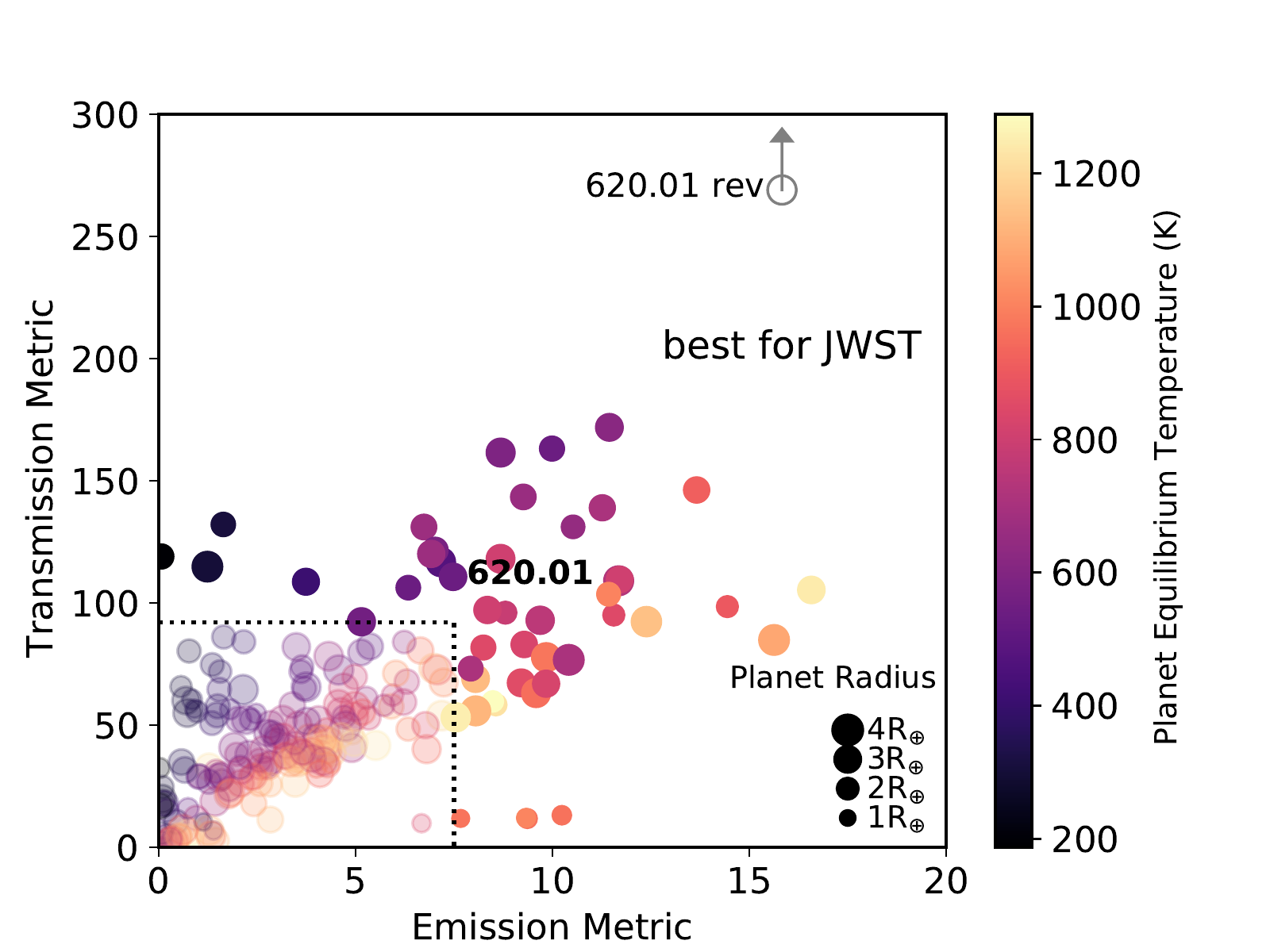}
    \caption{\citet{Kempton_2018} metrics of signal-to-noise for hypothetical observations of exoplanet atmospheres in transmission (during primary transit) and emission (during secondary eclipse) for a subset of \tess\ candidate or confirmed planets (TOIs) detected as of the end of September 2021 \citep[NASA Exoplanet Archive,][]{nasa_archive_planets,Akeson_2013}.  Only planets smaller than Neptune, with $T_{eq} <1300$K, and that are predicted to impart a Doppler RV signal $K>$3 \ms\ are shown.  Planet masses are estimated from the \citet{Chen_2016} mass-radius relation and are thus only preliminary.  The points' sizes are scaled with planet radius and the colors are keyed to $T_{eq}$.  Increased metric means higher S/N and the dashed lines indicate the boundary above and to the right of which systems are suitable for \emph{JWST} observations \citep{Kempton_2018}. Candidate planet TOI 620.01 is presented by its Exoplanet Archive-based value (filled point), and by a value/lower limit based on revised star and planet properties presented in this work (open grey point and arrow). }
    \label{fig:kempton}
\end{figure}

Figure \ref{fig:kempton} shows preliminary estimates for these two \rev{indices from \citet{Kempton_2018}---the transmission and emission spectroscopy metrics (TSM, ESM)---}for all \emph{TESS} candidate planets (\emph{TESS} Objects of Interest or TOIs), as of the end of September 2021, that have radii less than that of Neptune, and are predicted to impart Doppler RV signals with semi-amplitudes $K>3$ \ms, as a criterion for mass determination.  Objects outside the dashed zone are considered suitable targets for transit and/or secondary eclipse observations by \emph{JWST}.  Thus far, out of the thousands of \tess\ exoplanet candidates, less than 50 objects satisfy all these criteria, and thus these are some of the most important targets for validation and characterization.  Many of these are M dwarf systems where the small radius and low luminosity of the star mean relatively high transit S/N and low $T_{\rm eq}$.

One such system, TOI 620.01, is a candidate transiting sub-Neptune-size planet on a 5.09-day orbit around a nearby (33 pc), bright ($T=10$ mag) early M-type dwarf (Table \ref{tab:star}). We describe a multi-method, multi-wavelength, multi-instrument, and multi-team campaign to validate and characterize the planet and its host star, identify or rule out additional companions, and assess the suitability and value of the system for future atmospheric investigation by \emph{JWST} and other observatories. 

This paper is organized as follows: In ${\S}$\ref{sect:observations}, we present our baseline of two years of RV observations using the near-infrared (NIR) iSHELL spectrograph \citep{Cale2019}, along with a single season of RV data from CARMENES, MAROON-X and NEID. We supplement the \tess\ light curve with ground-based multi-wavelength follow-up observations from NGTS, LCO, MuSCAT2, TMMT, \rev{ExTrA}, and KeplerCam, and reconnaissance spectroscopy from TRES. ${\S}$\ref{sect:star} overviews the analysis and results of fitting of the host star properties, including multiplicity and age. We present analysis and results of the light curve transit fitting in ${\S}$\ref{sect:transit_analysis} and RV fitting in ${\S}$\ref{sect:rv_analysis}. In ${\S}$\ref{sect:discussion} we discuss the implications of our modeling, analyzing the effects of stellar activity and possible additional RV signals, and perform injection and recovery tests.  In ${\S}$\ref{sect:summary} we summarize our findings.  Finally, in the Appendices, we present more detailed explorations of alternative circum-secondary and hierarchical eclipsing binary scenario analyses that are motivated by the \gaia\ RUWE statistic, linear RV trend, and iSHELL SB2 analysis presented in the paper.

\begin{table}[]
    \centering
    \begin{tabularx}{.47\textwidth}{@{\extracolsep{\fill}}lcr}
        \hline
        Parameter & Value & Reference \\
        \hline
        & \textit{Identifiers} & \\
        TIC & 296739893 & S19 \\
        TOI & 620 & G21 \\
        G & 161-32 & G71 \\
        NLTT & 21863 & L79 \\
        Gaia DR2 \& EDR3 & 5738284016370287616 & G18 \\
        2MASS & J09284158-1209551 & S06 \\
        & \textit{Coordinates \& Velocities} & \\
        $\alpha$ & 09:48:41.59 & S19 \\
        $\delta$ & $-$12:09:55.75 & S19 \\
        Distance [pc] & $33.055 \pm 0.058$ & G18 \\ 
        Parallax ($\varpi$) [mas] & $30.283 \pm 0.061$ & G18, L18 \\
        $\mu_\alpha \cos\delta$ [mas yr$^{-1}$] & $35.87 \pm 0.11$ & G18 \\
        $\mu_\delta$ [mas yr$^{-1}$] & $-389.854 \pm 0.082$ & G18 \\
        X [pc] & $-12.521 \pm 0.012$ & this work \\
        Y [pc] & $-26.597 \pm 0.026$ & this work \\
        Z [pc] & $15.025 \pm 0.015$ & this work \\
        U [\kms] & $37.22 \pm 0.20$ & this work \\
        V [\kms] & $-40.13 \pm 0.42$ & this work \\
        W [\kms] & $-28.14 \pm 0.24$ & this work \\
        \hline
        & \textit{Physical Properties} & \\
        Spectral Type & M2.5V & S05\\
        $v\sin i$ [\kms] & $< 3$ & this work \\
        $P_{\rm rot}$ [days] & $8.99$ & this work\\
        & \textit{(see table \ref{tab:starparams})} & \\
        \hline
        & \textit{Magnitudes} & \\
        $B$ (APASS) & $13.58 \pm 0.24$ & H18 \\
        $V$ (APASS) & $12.265	\pm 0.019$ & H18 \\
        $g'$ (APASS) & $12.946 \pm 0.031$ & H18 \\
        $r'$ (APASS) & $11.678 \pm 0.018$ & H18 \\
        $i'$ (APASS) & $10.667 \pm 0.051$ & H18 \\
        $z'$ (APASS) & $10.064 \pm 0.079$ & H18 \\
        \gaia\ $G$ & $11.3104 \pm 0.0013$ & G18 \\
        \gaia\ $B_P$ & $12.4955 \pm 0.0022$ & G18 \\
        \gaia\ $R_P$ & $10.2525 \pm 0.0013$ & G18 \\
        $J$ (2MASS) & $8.837 \pm 0.030$ & S06 \\
        $H$ (2MASS) & $8.201 \pm 0.053$ & S06 \\
        $K$ (2MASS) & $7.954 \pm 0.027$ & S06 \\
        \wise\ 3.4 $\upmu$m & $7.839 \pm 0.024$ & W10 \\
        \wise\ 4.6 $\upmu$m & $7.809 \pm 0.019$ & W10 \\
        \wise\ 12 $\upmu$m & $7.733 \pm 0.020$ & W10 \\
        \wise\ 22 $\upmu$m & $7.51 \pm 0.14$ & W10 \\
        \hline
    \end{tabularx}
    \caption{Stellar parameters of TOI 620. The physical properties and distance are derived from an \texttt{EXOFASTv2} fit detailed in ${\S}$\ref{sect:exofast}.
    \textbf{References}:
    G18: \citet{Gaia_Collaboration_2018}, G21: \citet{Guerrero_2021}, H18: \citet{Henden_2018}, S19: \citet{Stassun_2019}, 
    S06: \citet{Skrutskie_2006}, S05: \citet{Scholz_2005}, W10: \citet{Wright_2010}, L18: \citet{2018A&A...616A...2L}, G71: \citet{1971lpms.book.....G}, L79: \citet{1979nltt.book.....L}
    \label{tab:star}}
\end{table}

\section{Observations}
\label{sect:observations}

In this section, we present an overview of all observational data used in our analysis.  All photometric light curve data is presented in ${\S}$\ref{sect:photometric_data}, all high contrast imaging observations are in ${\S}$\ref{sect:high_contrast_data}, and all RV observations are detailed in ${\S}$\ref{sect:rv_data}. A description of reconnaissance spectroscopy is also presented in ${\S}$\ref{sect:recon_spec_obs}. A summary of all space and ground-based transit data, high-contrast imaging data, and spectroscopic RV data is shown in Table \ref{tab:measurement_table}.  For more detailed in formation on the specific transit times or individual RV measurements, refer to Tables \ref{tab:transits} and \ref{tab:RVs} respectively, in the Appendix.

\begin{table*}[]
    \centering
    \begin{tabularx}{\textwidth}{@{\extracolsep{\fill}}XXXXX}
        \hline
        \hline
        Instrument / Facility & $N_{\rm transits}$ & Filter & Plate Scale & Precision \\
        \hline
        TESS & \rev{8} & \tess & $\angd{;;22}\, {\rm px}^{-1}$ & $1 \times 10^{-3}$\\
        NGTS / Paranal & \rev{2} & \emph{NGTS} & $\angd{;;5}\, {\rm px}^{-1}$ & $1 \times 10^{-2}$ \\
        CTIO 1 m / LCO & 1 & $z'$ & $\angd{;;0.389}\, {\rm px}^{-1}$ & $8 \times 10^{-4}$ \\
        TMMT / LCO & 1 & $I$ & $\angd{;;1.19}\, {\rm px}^{-1}$ & $2 \times 10^{-3}$ \\
        MuSCAT2 / TCS & 4 & $g'$, $i'$, $r'$, $z'$ & $\angd{;;0.44}\, {\rm px}^{-1}$ & $2 \times 10^{-3}$ \\
        KeplerCam / FLWO & 1 & $B$ & $\angd{;;0.672}\, {\rm px}^{-1}$ & $2 \times 10^{-3}$ \\
        LCRO / LCO & 1 & $i'$ & $\angd{;;0.773}\, {\rm px}^{-1}$ & $4 \times 10^{-3}$ \\
        \rev{ExTrA / La Silla} & \rev{4} & \rev{\emph{ExTrA}} & \rev{$\angd{;;0.870}\, {\rm px}^{-1}$} & \rev{$2 \times 10^{-3}$}
    \end{tabularx}
    
    \begin{tabularx}{\textwidth}{@{\extracolsep{\fill}}XXXX}
        \hline
        \hline
        Instrument / Facility & UT Observation Dates & Wavelength & Separation Range \\
        \hline
        \textit{Zorro} / Gemini South & 2020-03-16 & 562 nm, 832 nm & $\angd{;;0.02}$--$\angd{;;1.2}$ \\
        NIRC2 / Keck II & 2019-05-12 & Br$\gamma$ & $\angd{;;0.02}$--$\angd{;;4}$ \\
        NIRI / Gemini North & 2019-05-23 & Br$\gamma$ & $\angd{;;0.02}$--$\angd{;;7}$ \\
        NESSI / WIYN & 2019-11-09 & 562 nm, 832 nm & $\angd{;;0.04}$--$\angd{;;1.2}$\\
        ShaneAO / Lick & 2021-02-26, 2021-02-27 & $K_s$, $J$ & $\angd{;;0.6}$--$\angd{;;7}$ \\
    \end{tabularx}
    
    \begin{tabularx}{\textwidth}{@{\extracolsep{\fill}}llXXXXl}
        \hline
        \hline
        Instrument / Facility & $\lambda$ [\AA] & $\lambda / \Delta\lambda$ $[\times 10^{3}]$ & $N_{\rm nights}$ & $N_{\rm used}$ & $\sigma_{\rm RV}$ [\ms] & Pipeline \\
        \hline
        iSHELL / IRTF & 10600--53000 & 85 & 34 & 31 & 5.3 & \texttt{pychell} \citep{Cale2019} \\
        CARMENES-Vis / Calar Alto & 5200--\rev{9600} & 94.6 & 7 & 6 & 1.7 & \texttt{serval} \citep{zechmeister2018} \\
        CARMENES-NIR / Calar Alto & 9600--17100 & 80.4 & 7 & 7 & 7.2 & \texttt{serval} \citep{zechmeister2018} \\
        MAROON-X blue / Gemini North & 5000--6780 & 85 & 8 & 8 & 2.3 & \texttt{serval} \citep{zechmeister2018} \\
        MAROON-X red / Gemini North & 6540--9200 & 85 & 8 & 8 & 1.9 & \texttt{serval} \citep{zechmeister2018} \\
        NEID / WIYN & 4580--8920 & 120 & 8 & 8 & 1.1 & \texttt{serval} \citep{zechmeister2018} \\
        TRES / Tillinghast & 3850--9096 & 44 & 2 & 2 & -- & -- \\
        \hline
    \end{tabularx}
    \caption{Summary of all transit, high-contrast imaging, and radial velocity observations used in this work.  In the transit column headings, $N_{\rm transits}$ denotes the number of transits observed by that instrument and Precision denotes the order of magnitude of the normalized flux error for each instrument.  In the RV column headings, $N_{\rm nights}$ and $N_{\rm used}$ refer to the number of nights gathered and the number used, respectively.  The median intrinsic error bars $\sigma_{\rm RV}$ are calculated using only the nights used.}
    \label{tab:measurement_table}
\end{table*}

\subsection{Time-Series Photometry}
\label{sect:photometric_data}

\subsubsection{TESS Photometry}
TOI 620 (TIC 296739893; G 161-32; Gaia EDR3 5738284016370287616) was observed first in \tess\ sector 8 from UT 2019 February 2 to 2019 February 27, then in sector 35 during the \tess\ extended mission from UT 2021 February 9 to 2021 March 6.  The star is located at a distance of 33.06 pc and is relatively bright (e.g. $V = 12.265$, $J = 8.837$) making it an ideal candidate for study by \tess.  

\begin{figure*}
    \centering
    \includegraphics[width=.45\textwidth]{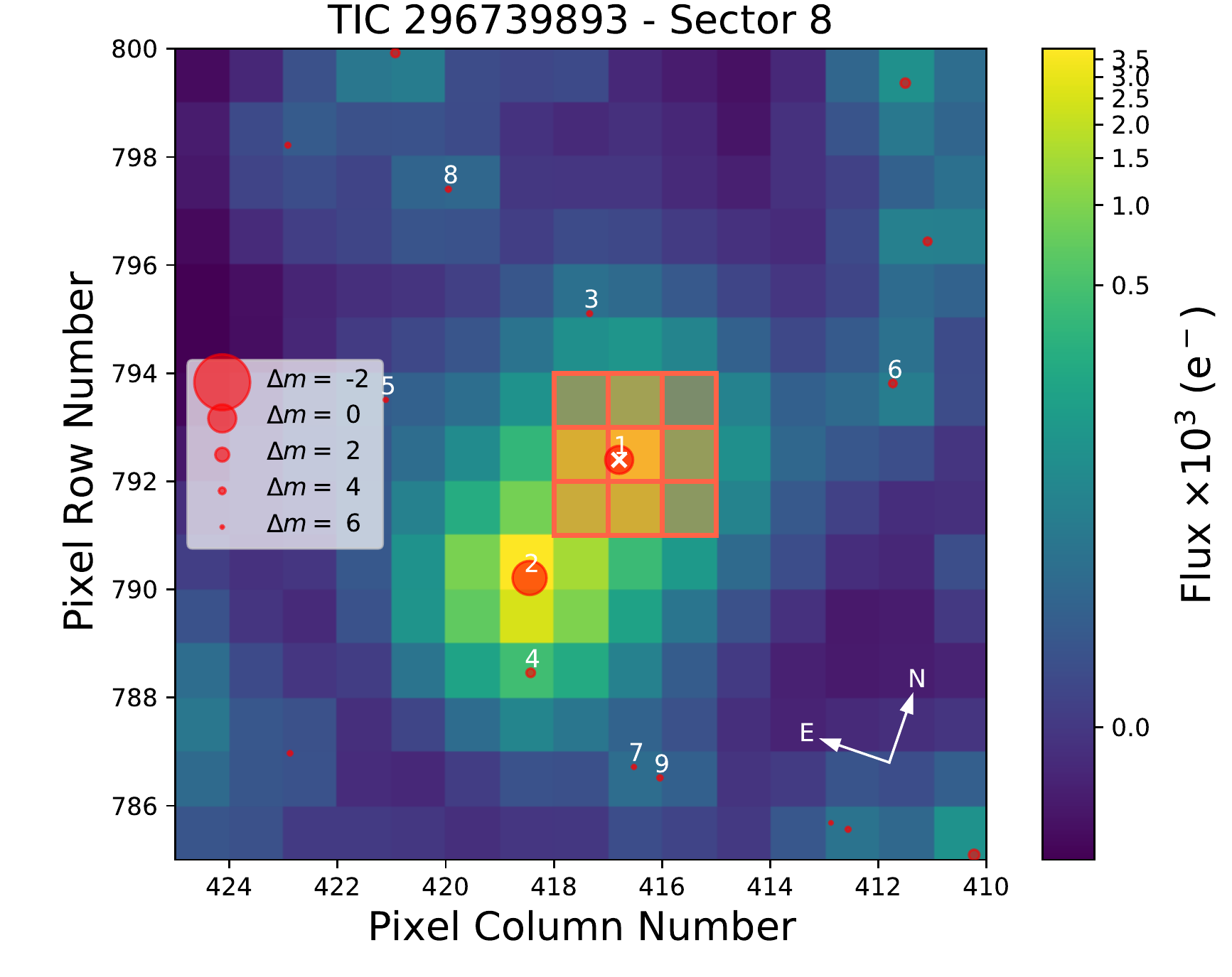}
    \includegraphics[width=.45\textwidth]{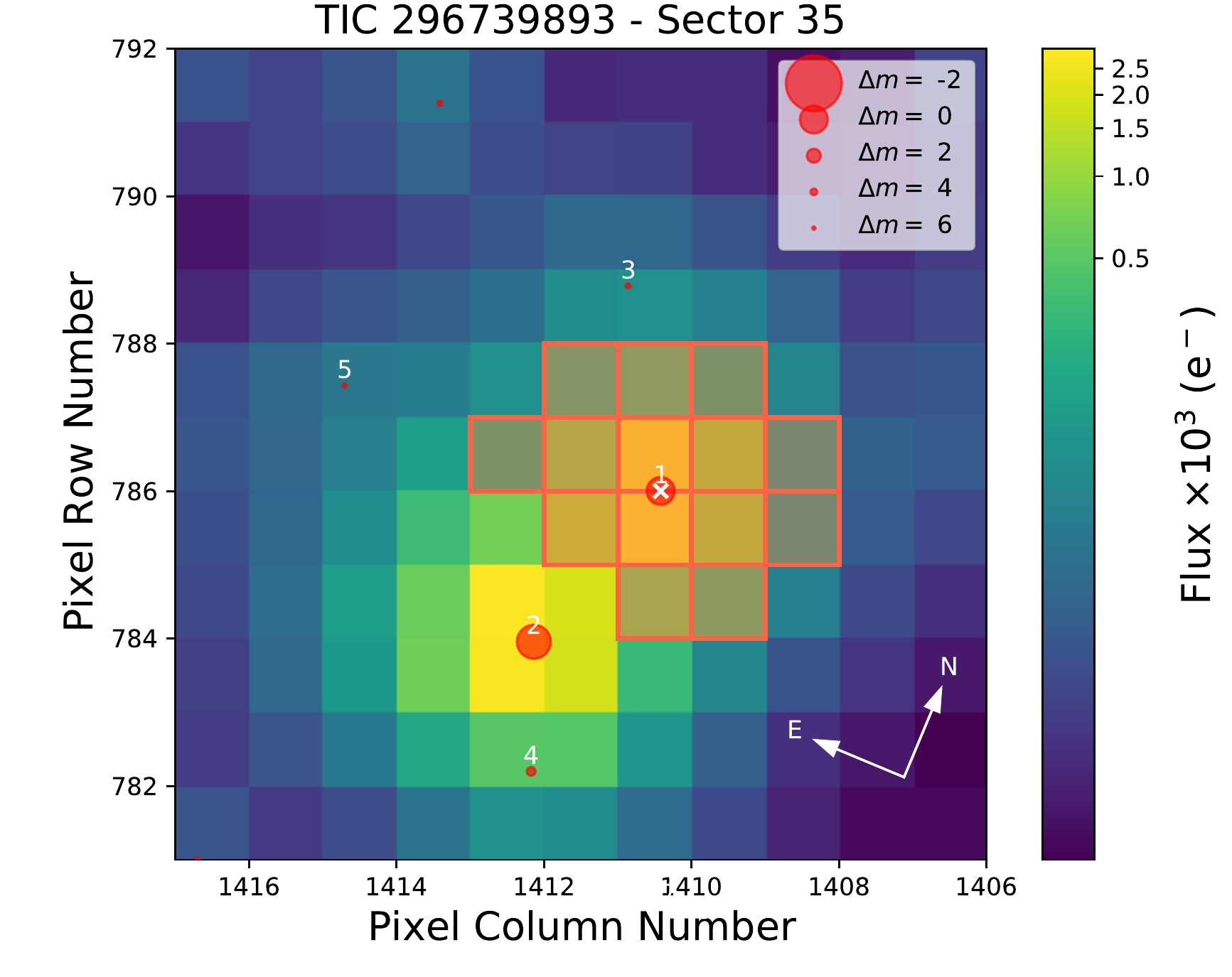}
    \caption{\tess\ target pixel file (TPF) data from sector 8 (left) and sector 35 (right) for TOI 620, created with \texttt{tpfplotter} \citep{Aller_2020}.  The pixels shown outlined in orange were the ones used to extract the light curve, while point sources from the Gaia DR2 catalog are labeled in red, with sizes in accordance to their relative magnitude from the target star.}
    \label{fig:tpf}
\end{figure*}

The data collection pipeline developed by the \tess\ Science Processing Operations Center \citep[SPOC,][]{Jenkins_2016} extracted the photometry for this target and performed a search for transiting planets using a wavelet-based matched filter \citep{2002ApJ...575..493J,2010SPIE.7740E..0DJ,2020TPSkdph} on 29 March 2019, detecting a strong transit signal. The data were fitted with a limb-darkened transit model \citep{Li:DVmodelFit2019} and subjected to a suite of diagnostic tests \citep{Twicken:DVdiagnostics2018} to distinguish between false positives and a planetary signal. The signature passed all the Data Validation tests, including the difference image centroiding test, which localized the souce of the transits to within $\angd{;;4.2980} \pm \angd{;;2.6862}$ of the target star. A search for additional planetary transit signatures failed to identify any. The TESS Science Office reviewed the vetting results and issued an alert for TOI-620.01 on 13 April 2019 \citep{Guerrero_2021}, which hereafter we also refer to as TOI-620 b. In Figure \ref{fig:tpf}, we show the \tess\ target pixel files (TPF) around the target star in sectors 8 and 35, where orange outlines show the aperture pixels used to extract the \tess\ light curve. A slightly brighter visual companion is located $\angd{;;55}$ to the S-SE, which does contribute (less than a few percent) to the \tess\ aperture for TOI 620; thus in the validation presented in this work, we do exclude this companion as the source of the transit and photometric variations. We specifically analyzed the detrended Presearch Data Conditioning Simple Aperture Photometry (PDC-SAP) light curve \citep{Smith_2012, Stumpe_2012, Stumpe_2014} obtained from the Mikulski Archive for Space Telescopes (MAST)\footnote{\href{https://mast.stsci.edu/portal/Mashup/Clients/Mast/Portal.html}{https://mast.stsci.edu/portal/Mashup/Clients/Mast/Portal.html}}. We normalize the light curves for each sector to unity. 

\subsubsection{NGTS / Paranal}
TOI 620 was observed by the Next Generation Transit Survey \citep[NGTS;][]{wheatley2018ngts} on the nights of 2019 April 20 and 2019 June 10. The NGTS photometric facility consists of twelve 0.2\,m diameter robotic telescopes, located at ESO's Paranal Observatory, Chile. On the night 2019 April 20, TOI 620 was observed using a single NGTS telescope and on 2019 June 10 two telescopes were used in the simultaneous multi-telescope observing mode to independently observe TOI 620 \citep[see][]{bryant2020multicam}. For both nights, TOI 620 was observed using the custom NGTS filter (520 - 890\,nm) and an exposure time of 10\,seconds. Across the two nights, a total of 2675 images were taken. The NGTS data were reduced using a custom aperture photometry pipeline, which performs source extraction and photometry using the \textsc{SEP} Python library \citep{bertin96sextractor, Barbary2016} and is detailed in \citet{bryant2020multicam}. The pipeline uses \gaia\ DR2 \citep{GAIA, GAIA_DR2} to automatically identify comparison stars which are similar in brightness, colour, and CCD position to TOI 620.

\subsubsection{CTIO 1 m / LCO}
The Las Cumbres Observatory at Cerro Tololo Inter-American Observatory in Chile \citep[LCO-CTIO;][]{Brown_2013} observed TOI 620 on the same night as NGTS using the 1-meter telescope, UT 2019 April 20, in the SDSS $z'$ filter.  The plate scale is $\angd{;;0.389}$ giving a full FOV of $\angd{;26.5;} \times \angd{;26.5;}$.  Exposure times were 30 seconds, and the sizes chosen for the aperture and sky annuli were 15 px ($\angd{;;5.835}$), 30 px ($\angd{;;11.67}$), and 45 px ($\angd{;;17.505}$), respectively.  The ingress was missed, but a full egress was captured.  Data was reduced using an AstroImageJ \citep[AIJ,][]{2017AJ....153...77C} pipeline.

\subsubsection{MuSCAT2 / TCS}
The MuSCAT2 camera at the Telescopio Carlos S\'anchez in Teide Observatory (Spain) \citep{Narita_2015, Narita_2018} has observed TOI 620 in its four simultaneous bands ($g'$, $i'$, $r'$, and $z'$) on four separate nights.  Different aperture sizes were used on each night ranging from $\angd{;;7.83}$--$\angd{;;13.92}$, with inner and outer sky annuli being an additional $\angd{;;10}$ and $\angd{;;18.7}$ out from the target aperture. Partial transits covering a full ingress and partial egress were observed on the nights of UT 2020 January 16, 2020 March 2, \rev{2020 April 16}, and 2021 January 7.  On the final night, January 7, the $g'$ band was unavailable for observations, so this night used only the other three filters.  The data were reduced using a custom Python pipeline developed specifically for MuSCAT2 \citep{Narita_2018}.

\subsubsection{KeplerCam / FLWO}
TOI 620 was also observed by KeplerCam \citep{Szentgyorgyi_2005} on UT 2020 January 26 in the B filter.  KeplerCam is a 4K $\times$ 4K Fairchild dectector on the 1.2m telescope at the Fred Lawrence Whipple Observatory (FLWO) atop Mt. Hopkins (Arizona, USA). The detector has a pixel scale of $\angd{;;0.672}\, \unit{pixel}^{-1}$ resulting in a field-of-view of $\angd{;23.1;} \times \angd{;23.1;}$. A full transit was observed using 60 second observations with $\sim \angd{;;1.9}$ FWHM and a $\angd{;;3.4}$ circular aperture. 

\subsubsection{TMMT / LCO}
We observed a transit of TOI 620 b on UT 2019 April 26 using the Three-hundred MilliMeter Telescope \citep[TMMT;][]{2017AJ....153...96M} at Las Campanas Observatory in Chile. TMMT is a f/7.8 FRC300 telescope from Takahashi on a German equatorial AP1600 GTO mount with an Apogee Alta U42-D09 CCD Camera, FLI ATLAS focuser, and Centerline filter wheel.

Observations were performed using the Bessell $I$ filter with exposure times of 70 $\unit{s}$. TMMT has a gain of $1.35 \unit{e/ADU}$ and a plate scale of $\angd{;;1.19}\, \unit{pixel}^{-1}$ for a field of view of $\angd{;40.8;}$. The target rose from an airmass of 1.06 at the start of the observations to a minimum airmass of 1.04 and then set to an airmass of 1.68 at the end of the observations. In addition to the standard bias, dark, and flat corrections, a fringe subtraction was also performed for the TMMT $I$-band images. 

We reduced this data using AIJ. The final light curve utilized a photometric aperture of 9 pixels ($\angd{;;5.97}$), and inner and outer sky annuli of 15 pixels ($\angd{;;23.9}$) and 25 pixels ($\angd{;;35.8}$), respectively.

\subsubsection{LCRO / LCO}
We also observed an ingress of TOI 620 b on UT 2020 November 27 using the $305 \unit{mm}$ Las Campanas Remote Observatory (LCRO) telescope at the Las Campanas Observatory in Chile. The LCRO telescope is an f/8 Maksutov-Cassegrain from Astro-Physics on a German Equatorial AP1600 GTO mount with an FLI Proline 16803 CCD Camera, FLI ATLAS focuser and Centerline filter wheel.

Observations were performed using the SDSS $i^\prime$ filter with exposure times of 120 $\unit{s}$. In this mode, LCRO has a gain of $1.52 \unit{e/ADU}$ and a plate scale of $\angd{;;0.773}\, \unit{pixel}^{-1}$ for a field of view of $\angd{;52.0;}$. The target rose from an airmass of 3.6 at the start of observations, to 1.07 at the end. 

We also reduced this data with AIJ, in the same fashion as the TMMT transit. For the final reduction, we selected a photometric aperture of 13 pixels ($\angd{;;10.0}$) with an inner sky annulus of 15 pixels ($\angd{;;11.6}$) and outer sky annulus of 20 pixels ($\angd{;;15.5}$).

\rev{
\subsubsection{ExTrA / La Silla}
The ExTrA facility \citep{2015SPIE.9605E..1LB} is composed of a near-infrared (0.85 to 1.55 $\upmu$m) multi-object spectrograph fed by three 60-cm telescopes located at La Silla observatory. We observed five full transits of TOI 620 b on UTC 2021 March 3, 2021 April 13 (with 2 telescopes), 2021 April 18 and 2021 June 3. We observed with one or two telescopes using the fibers with 8$''$ apertures. We used the low resolution mode of the spectrograph ($R \sim 20$) and 60-seconds exposures for all nights. At the focal plane of each telescope, five fiber positioners are used to pick the light from the target and four comparison stars. As comparison stars, we also observed 2MASS J09265392-1229161, 2MASS J09275007-1222230, 2MASS J09270219-1156332, and 2MASS J09261086-1200503, with $J$-magnitude \citep{Skrutskie_2006} and $T_{\rm eff}$ \citep{Gaia_Collaboration_2018} similar to TOI 620. The resulting ExTrA data were analysed using custom data reduction software.
}

\subsection{High Contrast Imaging}
\label{sect:high_contrast_data}

\subsubsection{Zorro / Gemini South}

TOI 620 was observed on 2020 March 16 UT using the \textit{Zorro} speckle instrument on Gemini South\footnote{\href{https://www.gemini.edu/sciops/instruments/alopeke-zorro/}{https://www.gemini.edu/sciops/instruments/alopeke-zorro/}} in Chile.  \textit{Zorro} provides simultaneous speckle imaging in two bands (562 nm and 832 nm) with output data products including a reconstructed image and robust contrast limits on companion detections \citep[e.g.][]{Howell_2016}. Five sets of $1000 \times 0.06$ sec exposures were collected and subjected to Fourier analysis in our standard reduction pipeline \citep[see][]{Howell_2011}.

\subsubsection{NIRI / Gemini North}

We collected observations of TOI 620 with the NIRI adaptive optics imager \citep{Hodapp_2003} at the Gemini North facility in Maunakea, Hawaii, USA on 2019-05-23. We collected nine frames, with individual exposure times of 1.8s, in the Br$\gamma$ filter, and dithered the telescope by $\sim \angd{;;3.3}$ between each frame in a grid pattern. A sky background was removed by median combining the individual science frames, thereby removing the signal from the star and any companions, and we also collected flat frames. For each image we first removed bad pixels, flat fielded, and subtracted the sky background. We then aligned the frames to the position of the star in each image, and co-added the sequence.

\subsubsection{NIRC2 / Keck II}

NIRC2 is designed for the Keck adaptive optics system in Maunakea, Hawaii, USA, as a near-infrared imager.  Observations of TOI 620 were performed with this instrument on UT 12 May 2019 so as to further constrain the parameter space of possibile companions in the TOI 620 system, as part of the standard process for doing so \citep{Ciardi_2015,2021FrASS...8...63S}.  A 3-point dither pattern is commonly used with NIRC2 to avoid using the noisier lower-left detector quadrant.  Observations were made in the Br$\gamma$ filter.

\subsubsection{NESSI / WIYN}

We observed TOI 620 with the NN-Explore Exoplanet Stellar Speckle Imager (NESSI; \citealt{Scott2018}) on the WIYN 3.5\,m telescope at Kitt Peak National Observatory, Arizona, USA on 2019 November 9. Sequences of 40 ms, diffraction-limited exposures were collected in the instrument's blue and red channels (with 562 nm and 832 nm filters, respectively). The data were reduced following \citet{Howell2011}.

\subsubsection{ShaneAO / Lick}

We obtained high-contrast AO images of TOI 620 from the 3m Shane Telescope at Lick Observatory, California, USA on the successive nights of 2021 Feb 26 and 27. The AO imaging was carried out in the $K_s$ and $J$ bandpasses using the ShARCS camera \citep{Srinath2014}. We observed both bandpasses with a five-point dither pattern (see, e.g., \citet{Furlan2017}), imaging the star at four quadrants of the detector as well as the center. We used custom Python software to perform standard image processing, including flat-fielding, sky subtraction, and subpixel image alignment. 

\subsection{Radial Velocities}
\label{sect:rv_data}

In this section we present the RV data collected for TOI 620 from four different PRV spectrographs spanning the visible through NIR wavelengths. Taken in isolation, each spectrograph did not obtain a substantial number of RV epochs (with the exception of the lower precision iSHELL).  However, collectively the RVs are sufficient in number \citep{Plavchan_2015} to permit a robust search for TOI 620 b.

\subsubsection{iSHELL / IRTF}

We have gathered a total of 379 observations of TOI 620 over 34 nights using the iSHELL instrument at NASA IRTF in Maunakea, Hawaii, USA from UT 26 January 2020 to UT 4 June 2021.  iSHELL observes in a range of wavelengths around 2350 nm.  Exposure times were 300 seconds, and were repeated anywhere from 9-17 times consecutively per night to obtain a signal-to-noise ratio (SNR) of 87-155 per spectral pixel. A methane isotopologue ($^{13}$CH$_4$) gas cell is used to provide a common optical path wavelength reference and to constrain the variable line spread function (LSF) of the spectrograph \citep{2012PASP..124..586A,2013DPS....4520402P}. Raw iSHELL data are processed in pychell with updated methods to those described in \citet{Cale2019}.  For a detailed description of these updated methods, refer to Appendix \ref{sect:forward_modeling}.

\subsubsection{CARMENES / Calar Alto}
The CARMENES (Calar Alto high-Resolution search for M dwarfs with Exo-earths with Near-infrared and optical Echelle Spectrographs) instrument located at Calar Alto Observatory in Spain \citep{Quirrenbach_2018} consists of visual and near-infrared arms covering a wavelength range of 520--960 nm and 960--1710 nm, respectively.  We obtained 7 measurements with exposure times of $\sim 1800$ s of TOI 620 from UT 3 February 2021 to 28 March 2021 in both the visual and NIR arms, but we were not able to use the first visual arm measurement from UT 2021-02-03 due to drift in the Fabry-P\'erot wavelength calibration device.  The CARMENES RVs were processed using the \texttt{SERVAL} pipeline \citep{zechmeister2018}.

\subsubsection{MAROON-X / Gemini North}
\label{sec:maroonx}
The MAROON-X (M dwarf Advanced Radial velocity Observer Of Neighboring eXoplanets) instrument \citep{Seifahrt_2018} is mounted at the Gemini North facility at Maunakea, Hawaii, USA, and like CARMENES it consists of two arms of differing wavelength ranges.  The blue arm covers 500--678 nm, while the red arm covers 654--920 nm, both with a resolving power of $R \approx 85,000$.  We observed TOI 620 with this instrument from UT 24 February 2021 to 3 June 2021, gathering a total of 8 measurements in both arms, with exposure times of 300 seconds.  The RVs are processed using a dedicated version of the \texttt{SERVAL} pipeline.

In the middle of the timespan that RVs were collected with the MAROON-X instrument, the observatory cooling system failed, causing a significant state change in the instrument's calibration, affecting the absolute RV offsets.  To correct for the relative errors introduced by this state change, we applied offset terms for each time range between the dates which the instrument was affected.  This occurred once between 2021 February 24 and April 17, and once between 2021 April 30 and May 7.  These offsets are applied in addition to the standard $\gamma$ offsets applied to each instrument in the main MCMC analysis presented later herein.  We subtracted a random sample from a normal distribution with a prior center and standard deviation summarized in Table \ref{tab:RVs} for each timespan.  The values of the offsets and errors were estimated using data from stars of a similar type that were observed during the same observation runs.

\subsubsection{NEID / WIYN}

We obtained precise broadband-optical, fiber-fed radial velocities of TOI 620 using the newly-commissioned NEID spectrometer \citep{Schwab_2016} on the 3.5\,m WIYN Telescope at Kitt Peak National Observatory, Arizona, USA. All NEID nights on WIYN are queue scheduled, and we obtained 8 queue-scheduled observations of TOI 620 between January and May of 2021.

Each NEID observation consisted of 2$\times$ 900-second exposures in the instrument's High Resolution (HR) mode, which yields a resolving power of $R \sim 120,000$. The exposures were taken without a simultaneous source on the calibration fiber in order to avoid cross-contamination with the relatively faint target. The exposures have a median S/N of 11.8 per 1D extracted pixel evaluated at $\lambda = 550$ nm. 

Basic data reduction and spectral extraction were performed by the automated NEID data pipeline.  The barycentric corrections were performed using the algorithms from \citet{2014PASP..126..838W} implemented in \texttt{barycorrpy} \citep{Kanodia_2018}.  We extracted precise RVs from the extracted spectra using a modified version of the \texttt{SERVAL} pipeline \citep{zechmeister2018}, which we describe further in a forthcoming publication (Stefansson et al. 2021, in prep). \texttt{SERVAL} uses the template-matching technique \citep{anglada2012} that is particularly effective for cool stars. For the RV reduction, we used NEID order indices 40 to 104, spanning wavelengths from $4580$ \AA\, to $8920$ \AA. We note that the \texttt{SERVAL} RVs are consistent with the RVs computed by the automated pipeline, which uses the CCF mask technique, but yields significantly higher RV precision for M-dwarf stars as it is capable of using a higher fraction of the RV information content inherent in M-dwarf spectra.

\subsection{Recon Spectroscopy: TRES}
\label{sect:recon_spec_obs}
The Tillinghast Reflector Echelle Spectrograph \citep[TRES;][]{Furesz_2008, Szentgyorgyi_2005} obtained two reconnaissance spectra of TOI 620 on UT 2019 April 22 and 2019 April 25, covering a wavelength range of 385--909.6 nm.  Spectra were processed using methods outlined in \cite{Buchhave_2010} and \cite{Quinn_2014}, with the exception of the cross-correlation template, for which the high-S/N median observed spectrum is used instead.  The extracted spectra are available at the NASA Exoplanet Archive EXO-FOP data repository \citep{exofop,Akeson_2013}.

\section{System Characterization}
\label{sect:star}

In this section we examine properties of the TOI 620 stellar system itself and model the luminous bodies of the system under various assumptions.  In ${\S}$\ref{sect:recon_spec} we present analysis of the reconnaissance spectroscopy measurements of TOI 620 from TRES.  Then in ${\S}$\ref{sect:bulk_star_props} we model the TOI 620 star, under the assumption that it is a single star, using MIST isochrones and an SED.  In ${\S}$\ref{sect:high_contrast_imaging} we present the results of analyses of the high contrast imaging data, followed by historical imaging data in ${\S}$\ref{sect:historical_imaging}.  Then in ${\S}$\ref{sec:companion} we explore the possibility of stellar multiplicity, and ${\S}$\ref{sect:sb2_analysis} presents a two-star model of the iSHELL spectra.

\subsection{TRES Spectroscopy Results}
\label{sect:recon_spec}

From examining the TRES spectra's NIR TiO lines, we find absolute velocities of 6.25 and 6.23 \kms.  The corresponding absolute velocities from the Mg b-containing order are 6.48 and 6.17 \kms, both with errors of $\sim 0.25$ \kms, meaning we see no significant RV variation between quadratures of TOI 620 b's orbit (assuming it to be circular). The best fit is achieved with no rotational broadening, so we can confidently place an upper limit on the rotational velocity of $v\sin i < 3$ \kms.  Estimates for the stellar parameters of TOI 620 can be made from the TRES observations, where we obtain \teff\  $\sim$3750--4000 K, \loggstar\ $\sim$ 4.0, and \mh\ $\sim$ 0.  However, since the TRES modeling pipeline uses ATLAS model atmospheres, which are known to not provide accurate stellar spectra for \teff\ $< 4500$ K, and TOI 620 is an M dwarf with a \teff\ in this range, the results for these stellar parameters are approximate.  Looking at the TRES activity spectroscopic features, we don't find any significant emission in H$\alpha$.  We do identify line core flux emission in the Sodium doublet, but these are relatively narrow emission features and associated with telluric contamination. We also do not identify any Lithium absorption consistent with ages of $<$50 Myr.

\subsection{Fitting bulk stellar properties}
\label{sect:bulk_star_props}

\begin{table}
    \centering
    \begin{tabularx}{\columnwidth}{@{\extracolsep{\fill}}XXXX}
        \hline
        Parameter [units] & Initial Value $(P_0)$ & Priors & Prior Citation \\
        \hline
        \hline
        $A_V$ [mag] & 0 & $\mathcal{U}(P_0, 0.12)$ & S11 \\
        $\varpi$ [mas] & 30.283 & $\mathcal{N}(P_0, 0.061)$ & G18 \\
        $[$Fe/H$]$ & 0 & $\mathcal{N}(0, 1)$ & this work \\
        \hline
        \gaia\ $G$ & 11.31 & $\mathcal{N}(P_0, 0.02)$ & G18 \\
       \gaia\ $B_P$ & 12.50 & $\mathcal{N}(P_0, 0.02)$ & G18\\
       \gaia\ $R_P$ & 10.25 & $\mathcal{N}(P_0, 0.02)$ & G18 \\
          $J$ 2MASS & 8.837 & $\mathcal{N}(P_0, 0.010)$ & S06 \\
          $H$ 2MASS & 8.201 & $\mathcal{N}(P_0, 0.053)$ & S06 \\
          $K$ 2MASS & 7.954 & $\mathcal{N}(P_0, 0.027)$ & S06 \\
        \wise 1 & 7.839 & $\mathcal{N}(P_0, 0.030)$ & W10 \\
        \wise 2 & 7.809 & $\mathcal{N}(P_0, 0.030)$ & W10 \\
        \wise 3 & 7.733 & $\mathcal{N}(P_0, 0.030)$ & W10 \\
        \wise 4 & 7.51 & $\mathcal{N}(P_0, 0.14)$ & W10 \\
        \hline
    \end{tabularx}
    \caption{Prior probability distributions for our \texttt{EXOFASTv2} MCMC simulations.  $\mathcal{N}(\mu, \sigma)$ signifies a Gaussian prior with mean $\mu$ and standard deviation $\sigma$.  $\mathcal{U}(\ell, r)$ signifies a uniform prior with left bound $\ell$ and right bound $r$.  $A_V$ is the extinction in the $V$ band, and $\varpi$ is the parallax. Parameters that are not included here, including stellar $M_*$, $R_*$, $T_{\rm eff}$, were not constrained by any priors, and were given an initial MCMC starting value of Sun-like to assess the robustness of the MCMC posterior convergence on an M dwarf host star.
    \emph{References}: S11: \citet{Schlafly_2011}, G18: \citet{Gaia_Collaboration_2018}, S06: \citet{Skrutskie_2006}, W10: \citet{Wright_2010}.}
    \label{tab:exofast_priors}
\end{table}

\begin{figure}
    \centering
    \includegraphics[width=\columnwidth]{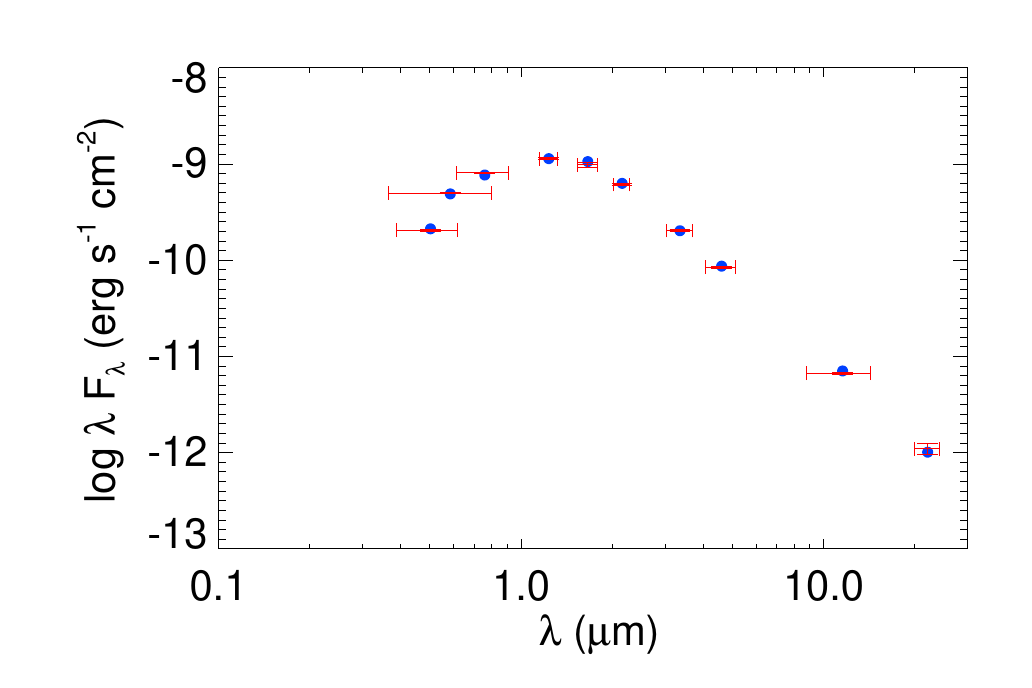}
    \caption{\rev{SED fit of flux as a function of wavelength for TOI 620.  Blue points are the best-fit values, and red points are the corresponding model values and errors.}}
    \label{fig:exofast_star}
\end{figure}

We next look at all of our stellar magnitudes and parallax data to more accurately determine characteristics of the host star, such as effective temperature, gravity, metallicity, etc.  We perform a joint amoeba fit followed by a Markov-Chain Monte Carlo (MCMC) simulation fitting both stellar properties and planet properties from the transit data of TOI 620 b assuming a single-planet, single-star \rev{scenario} simultaneously \rev{with \texttt{EXOFASTv2} \citep{Eastman_2013, Eastman_2019}.}  Details on the single-planet transit analysis are in the next section (\ref{sect:transit_analysis}), but here we present the results of the stellar modeling.  We start the MCMC with as few assumptions as possible--namely, we place no priors on the spectral type, and we employ parallel tempering with 8 parallel threads, following \citet{Eastman_2019}.  We place priors on V-band extinction, parallax (corrected as prescribed by \citet{2018A&A...616A...2L}), and metallicity summarized in Table \ref{tab:exofast_priors}.  We simultaneously fit with Mesa Isochrones and Stellar Tracks \citep[MIST,][]{Dotter_2016,Choi_2016,Paxton_2011,Paxton_2013,Paxton_2015} and a spectral energy distribution (SED) function.  For the SED, we include magnitudes from \gaia\ DR2, \emph{2MASS}, and \wise\ which are both precise and span a large wavelength range for broadband M dwarf SED characterization \citep{2015ApJ...804...64M}.  The results of our stellar MCMC modeling are shown in Figure \ref{fig:exofast_star} and Table \ref{tab:starparams}.  

\startlongtable
\begin{deluxetable*}{lcchhhhhhhhhhhhhhhhhhhhhhhhhhh}
\tablecaption{\centering \rev{Stellar parameters: Median values and 68\% confidence interval for TOI620, created using EXOFASTv2 commit number 7971a947}}
\label{tab:starparams}
\tablehead{\colhead{~~~Parameter\hspace{4cm}} & \colhead{\hspace{4.0cm}Units}\hspace{4.0cm} & \multicolumn{28}{c}{Values\hspace{4.0cm}}\hspace{4.0cm}}
\startdata
\smallskip\\\multicolumn{2}{l}{Stellar Parameters:}&\smallskip\\
~~~~$M_*$\dotfill &Mass (\msun)\dotfill &$0.577^{+0.024}_{-0.023}$\\
~~~~$R_*$\dotfill &Radius (\rsun)\dotfill &$0.550\pm0.017$\\
~~~~$L_*$\dotfill &Luminosity (\lsun)\dotfill &$0.0515\pm0.0015$\\
~~~~$F_{Bol}$\dotfill &Bolometric Flux (cgs)\dotfill &$1.511^{+0.045}_{-0.044} \times 10^{-9}$\\
~~~~$\rho_*$\dotfill &Density (cgs)\dotfill &$4.89^{+0.41}_{-0.39}$\\
~~~~$\log{g}$\dotfill &Surface gravity (cgs)\dotfill &$4.718^{+0.024}_{-0.025}$\\
~~~~$T_{\rm eff}$\dotfill &Effective Temperature (K)\dotfill &$3708^{+57}_{-56}$\\
~~~~$[{\rm Fe/H}]$\dotfill &Metallicity (dex)\dotfill &$0.35^{+0.11}_{-0.12}$\\
~~~~$[{\rm Fe/H}]_{0}$\dotfill &Initial Metallicity$^{2}$ \dotfill &$0.31^{+0.10}_{-0.11}$\\
~~~~$Age$\dotfill &Age (Gyr)\dotfill &$7.2^{+4.6}_{-4.7}$\\
~~~~$EEP$\dotfill &Equal Evolutionary Phase$^{3}$ \dotfill &$312^{+14}_{-31}$\\
~~~~$A_V$\dotfill &V-band extinction (mag)\dotfill &$0.048^{+0.043}_{-0.034}$\\
~~~~$\sigma_{SED}$\dotfill &SED photometry error scaling \dotfill &$1.73^{+0.63}_{-0.40}$\\
~~~~$\varpi$\dotfill &Parallax (mas)\dotfill &$30.283\pm0.061$\\
~~~~$d$\dotfill &Distance (pc)\dotfill &$33.022^{+0.067}_{-0.066}$\\
\enddata
\tablenotetext{}{See Table 3 in \citet{Eastman_2019} for a detailed description of all parameters}
\tablenotetext{2}{\rev{The metallicity of the star at birth}}
\tablenotetext{3}{\rev{Corresponds to static points in a star's evolutionary history. See \S2 in \citet{Dotter_2016}.}}
\end{deluxetable*}

\subsection{Stellar Rotation}
\label{sect:stellar_rotation_period}

\begin{figure}
    \centering
    \includegraphics[width=\columnwidth]{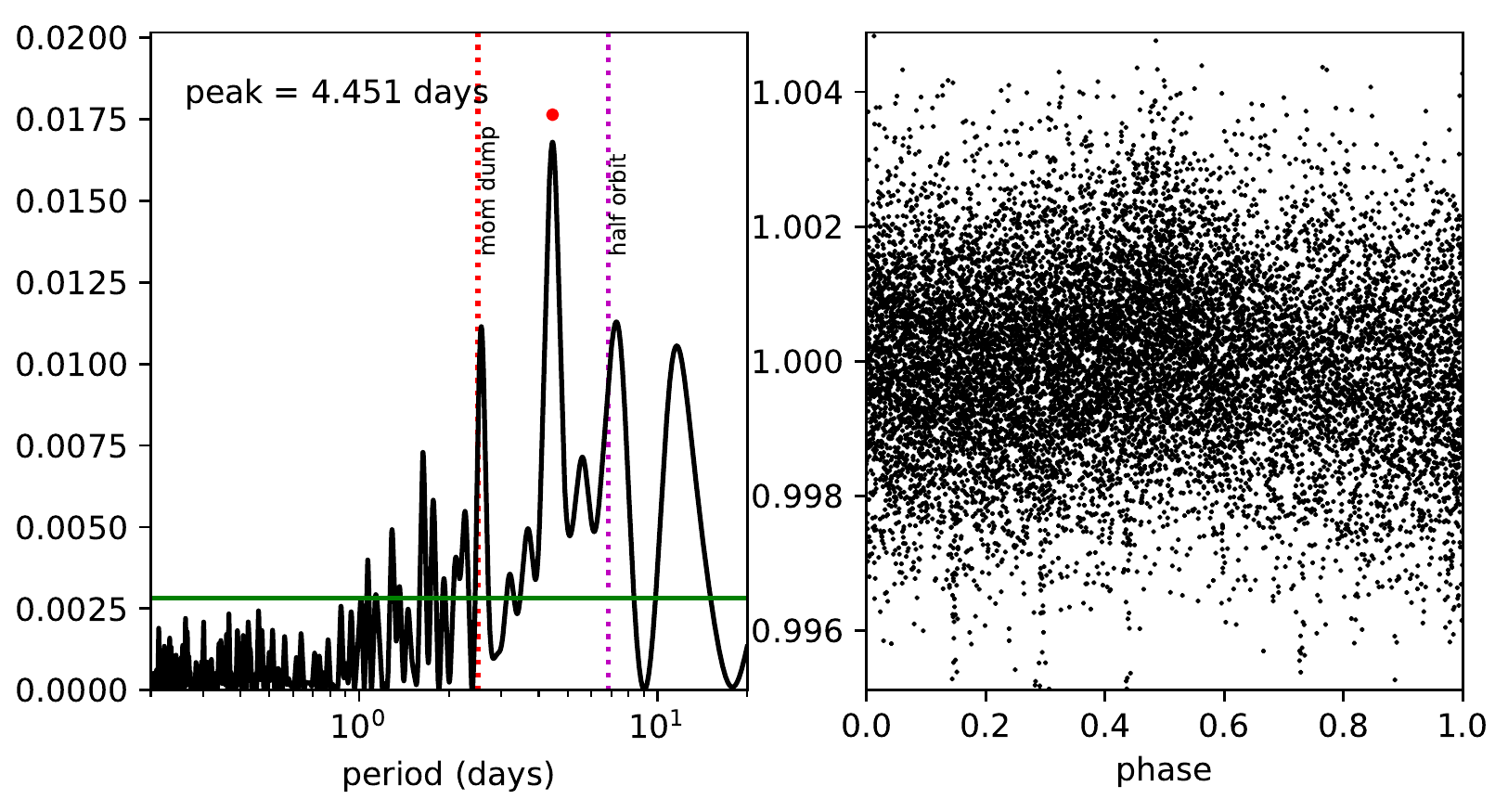}
    \includegraphics[width=\columnwidth]{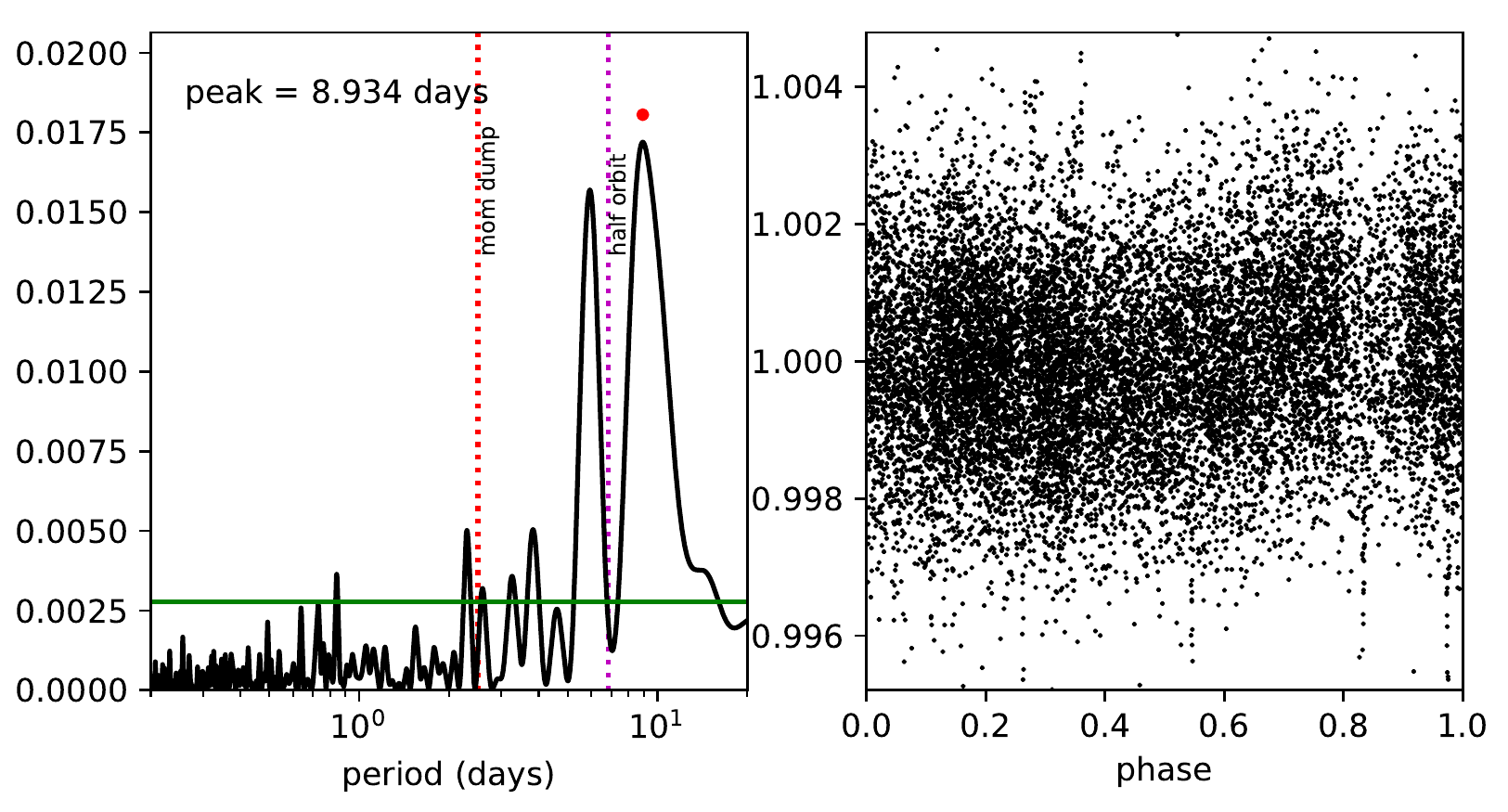}
    \caption{Lomb-Scargle periodograms (left panels) of \tess\ Sectors 8 (top) and 35 (bottom) photometry, showing peaks corresponding to plausible stellar rotation at 4.45 and 8.93 days, respectively.  The horizontal green line is the power corresponding to a false alarm probability of 0.1\%, and the vertical dotted lines mark the half-orbit and thruster firing  (momentum dump) intervals of \tess\ where systematics are expected.  The normalized light curves phased using these peak periods are shown in the right panels.}
    \label{fig:ls_periodograms_rot}
\end{figure}

Stellar rotation can manifest itself as periodic variation in a light curve (due to spots) or in time-series of an activity indicator (due to active regions) or by rotational broadening of lines in a stellar spectrum.  Lomb-Scargle (LS) periodograms \citep{1976Ap&SS..39..447L,1982ApJ...263..835S} of the \tess\ Sector 8 and 35 light curves (Fig \ref{fig:ls_periodograms_rot}) contain significant peaks at 4.45 days and 8.93 days, respectively. These signals could in principle come from the primary star at one-half and one times the stellar rotation period respectively \citep[e.g.,][]{2013A&A...557A..11R}.  These photometric variations could also originate from an unrelated neighboring star that contributes signal within the photometry aperture, or they could be artifacts, \rev{which} we now address.  The two nearest resolved stars in the \gaia\ EDR3 catalog are TIC\,296739889 ($\angd{;;9.2}$) and TIC\,296739884 ($\angd{;;34.3}$), \rev{which} have estimated $T$-mag \rev{contrasts} of 7.0 and 6.8, respectively.  The amplitude of the detected periodic signals ($\sim10^{-3}$; Fig \ref{fig:ls_periodograms_rot}) is comparable to their brightness, making them implausible sources.  The third-nearest star, TIC\,296739875, is comparable in brightness to TOI 620 ($T=10.12$), and is the aforementioned (${\S}$\ref{sect:photometric_data}) companion $\angd{;;55}$ to the S-SE. The light curve of the latter undoubtedly includes scattered light from the former.  However, light curves constructed from aperture photometry performed on this star for either sector do not contain a 4.45- or 8.93-day signal (nor any 5.09 day eclipsing signal).  The sinsusoidal periodic signals are not recovered from the uncorrected simple aperture photometry (SAP) of TOI 620 b, raising the possibility that they are a processing product.  But such signals do not systematically appear in the light curves of all 19 other stars with 2-min cadence photometry falling within 1 deg of TOI 620 b, ruling out a common processing origin. 

Ground-based light curves of TOI 620 from transient and transit searches do not have sufficient photometric precision to detect the $\sim$0.2\% quasi-sinusoidal photometric variations seen in the \tess\ light curve \rev{(${\S}$\ref{sect:ffp_analysis})}.  Time-series of the activity indicators (H$\alpha$, Na D, and CRX) extracted from the CARMENES, MAROON-X, and NEID spectra do not contain significant \rev{periodicity.}


An analysis of the TiO lines in the TRES spectra described in ${\S}$\ref{sect:recon_spec} limits the rotational broadening $v \sin i$ to $<$3 \kms. With the assumption that the rotation axis is close to the plane of the sky (which holds if the orbit of the transiting planet is aligned with the stellar rotation), this is marginally consistent with a rotation period of 8.93 days but not 4.45 days. The signal could conceivably arise from an unresolved low-luminosity companion suggested by the \gaia\ astrometric error (see Sec. \ref{sec:companion}); both ultra-cool dwarfs and white dwarfs typically rotate much faster than $\sim$9 days \citep{2018ApJ...859..153S,2015ASPC..493...65K}.  On the other hand, late-type ($>$M4) field M dwarfs exhibit a wide range of rotation periods of $\sim$0.1-100 day \citep{2016ApJ...821...93N}. \rev{The most harmonious explanation is that the 8.9-day signal is either the rotation period of TOI 620 and that this reflects the influence of an (undetected) companion, or that it is the rotation of a late-type M dwarf companion itself.}

\subsection{Stellar Activity and Age }
\label{sect:stellar_age}

\rev{Our stellar age posterior from \texttt{EXOFASTv2}, while appearing to imply an older star at $7.2^{+4.6}_{-4.7}$ Gyr, is not constraining, nor does it take into account the stellar rotation period analysis from the previous section. Since the star is on the main sequence, the  broadband magnitudes, parallax, and galactic extinction values alone are not enough for us to provide a constrained age estimate.} A rotation period of 8.93 days is intermediate between the \teff-rotation sequences of \emph{single} M dwarfs in the \rev{120 Myr-old Pleiades and} 670 Myr-old Praesepe clusters \citep[see Fig 7 in ][]{2020ApJ...904..140C}.  Thus the TOI 620 system could be a mere few hundred Myrs old.  However, binary stars tend to be more rapidly rotating than their single counterparts \citep[e.g.,][]{2018AJ....156..275S,2019ApJ...871..174S} due to tides (for systems with separations $\ll 1$ au) or the rapid dissipation of primordial disks that would otherwise be a sink for angular momentum (for systems with separations $\lesssim$100 au).  Importantly, we find no corroborating evidence for a young age: an exhaustive comparison of the $UVW$ space motions of the star with that of nearby open moving groups and clusters reveals no matches \citep[][Gagn\'e, priv. comm.]{2018ApJ...862..138G}, nor are the values ($+8.5$, $+13.4$, $+6.5$) \kms close to the Local Standard of Rest \citep[e.g, ($-8.6$, $-4.8$,$-7.3$) \kms,][]{2019RAA....19...68D}.  The star shows no emission in H$\alpha$ \citep{2014MNRAS.443.2561G} nor is there emission in the core of the Na I D lines \rev{(${\S}$\ref{sect:recon_spec})}. The star was not detected by \emph{ROSAT} in 0.1-2.4 keV X-rays \citep[Second ROSAT all-sky survey source catalog,][]{2016A&A...588A.103B}, \emph{GALEX} in the FUV (1340-1806\AA) or NUV (1693-3006\AA) pass-bands, nor by APASS in the Sloan $u'$ (3000-4000\AA) pass-band, as might be expected for a nearby rapidly-rotating, magnetically active star with a bright chromosphere.  \rev{The relatively short rotation period can then be interpreted as the result of potential binary interactions, or potentially the nature of an unseen companion, rather than youth of TOI 620 itself.}

\subsection{High Contrast Imaging}
\label{sect:high_contrast_imaging}

\begin{figure*}
    \centering
    \quad
    \includegraphics[width=.6\textwidth]{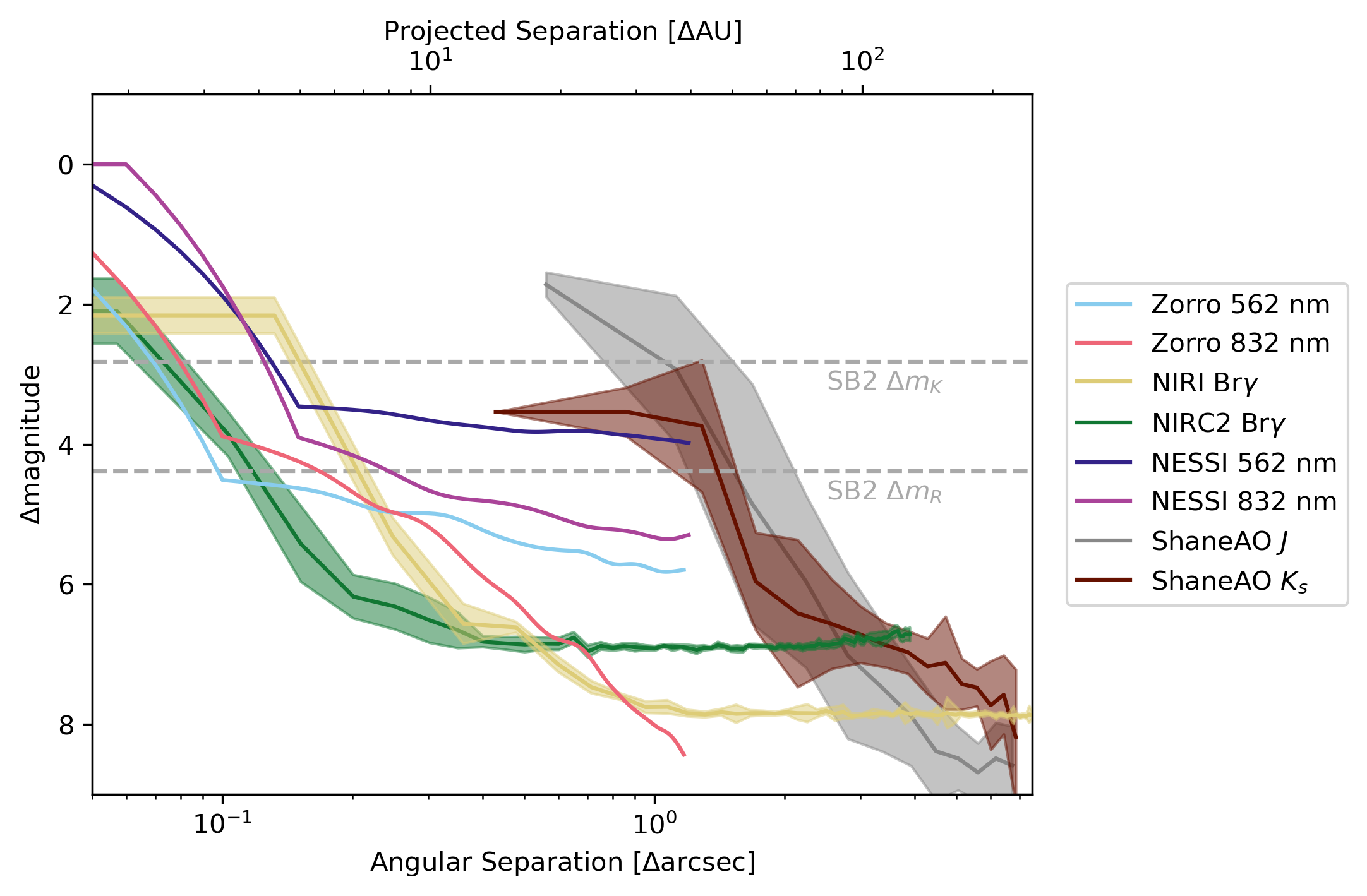}
    \caption{Plots of angular separation vs. the change in magnitude $\Delta m$ from each of our high contrast imaging observations.  The width of the NIRI, NIRC2, and ShaneAO curves indicates the range of uncertainty for each curve.  The 5$\sigma$ sensitivity curves for NIRI and NIRC2 achieve excellent sensitivity to resolve stellar companions, and do not identify any such companions in the field of view.}
    \label{fig:high_contrast}
\end{figure*}

\begin{figure*}
    \centering
    \quad
    \includegraphics[width=.75\textwidth]{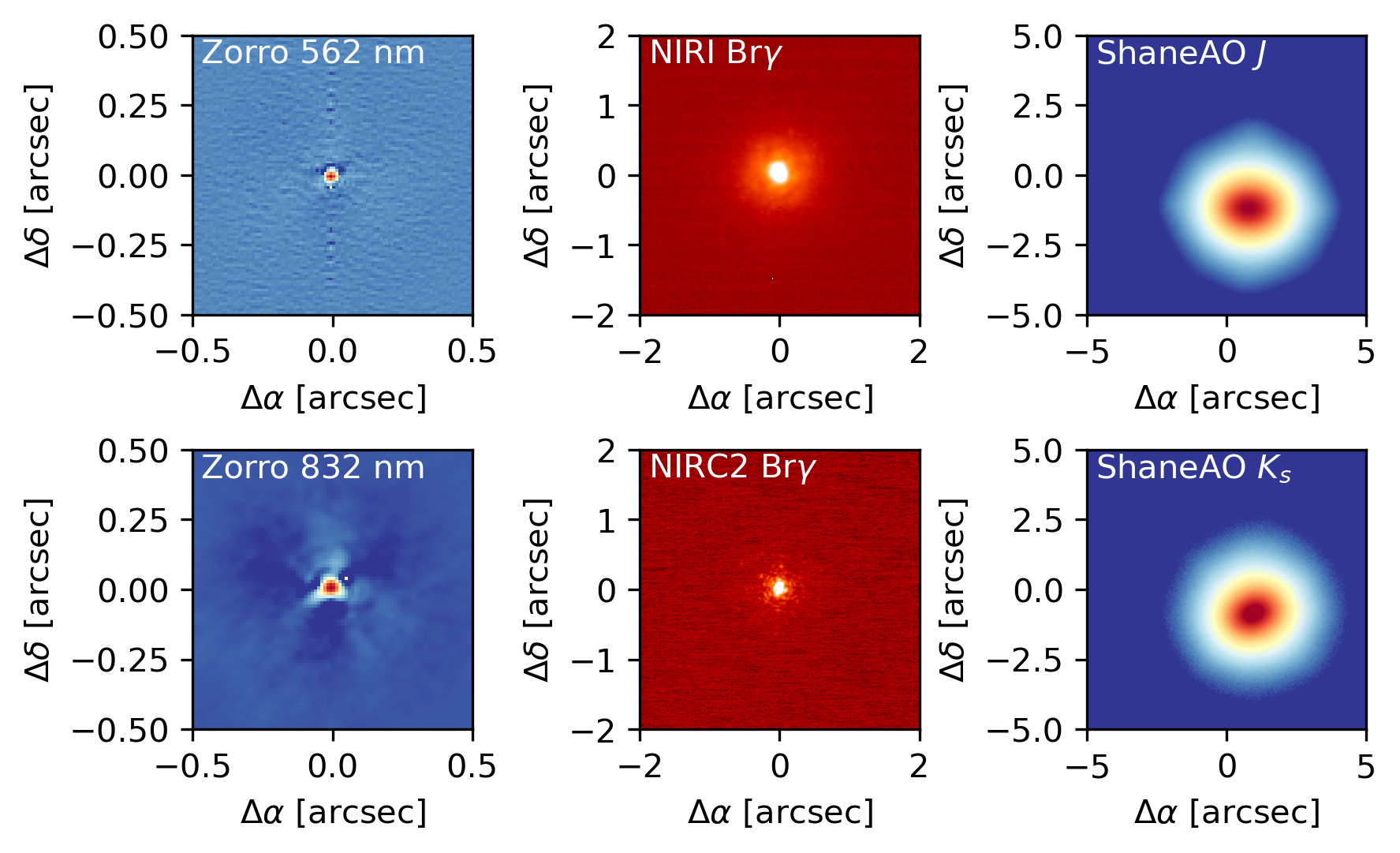}
    \caption{Reconstructed images from \textit{Zorro} (top-left: 562 nm, bottom-left: 832 nm), NIRI (top-middle), NIRC2 (bottom-middle), and ShaneAO (top-right: J, bottom-right: Ks) which appear single to the limit of each instrument's resolution.}
    \label{fig:speckle_images}
\end{figure*}

High contrast imaging observations are key in allowing us to constrain the properties of the host star, particularly in studying whether or not there are any bound companion stars within $\sim \angd{;;0.2}$--$\angd{;;5}$ projected separation of the primary star.

Figure \ref{fig:high_contrast} shows the final contrast curves and Figure \ref{fig:speckle_images} shows the images for all of our high-contrast imaging observations included in this work, with each instrument and filter labeled appropriately.  For the \textit{Zorro} 562 nm and 832 nm data, we find that TOI 620 is a single star with no companion brighter than 4.5 magnitudes at the diffraction limit (20 mas) and no companion brighter than 8.5 magnitudes at $\angd{;;1.2}$. At the distance of TOI 620 ($d = 33$ pc) these angular limits correspond to spatial limits of 0.7 to 40 au.  

In the \rev{NIRC2} analysis, we searched for companions visually, and did not detect point sources anywhere in the field of view, which extends to $\sim\angd{;;4}$ from the host star in all directions.  To test the sensitivity of our observations, we injected fake companions throughout the image, and tested the flux at which these companions could be redetected at 5$\sigma$. We averaged the sensitivity over position angle to create the \rev{NIRC2} sensitivity curve included in Figure \ref{fig:high_contrast}.  We achieve excellent sensitivity to stellar companions even with this very short observing sequence, due to the good weather conditions and the brightness of the host star. Our \rev{NIRC2} observations are sensitive to companions 6.2 mag fainter than the host beyond 200 mas, and are sensitive to companions 6.8 mag fainter than the host in the background limited regime, beyond $\angd{;;0.5}$.  

\rev{Analysis of the NIRI contrast curve was performed in a similar fashion to the NIRC2 analysis described above.} Analyzing the NESSI data, we detect no companions down to a magnitude difference $\Delta m \approx 4$ at $\angd{;;0.2}$ and $\Delta m \approx 5$ at $\angd{;;1}$.  For the ShaneAO data, we computed the variance in flux in a series of concentric annuli centered on the target star in the combined image. The resulting 5$\sigma$ contrast curves are shown in Figure \ref{fig:high_contrast}.

\subsection{Background Stars from Historical Imaging}
\label{sect:historical_imaging}

\begin{figure}
    \centering
    \includegraphics[width=.45\textwidth]{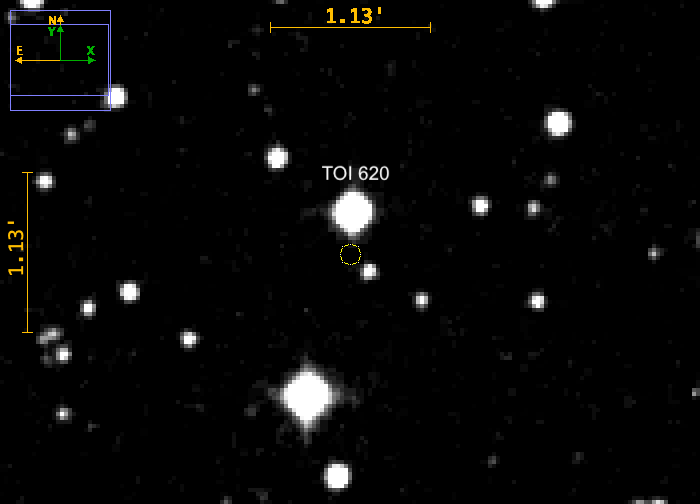}
    \caption{A historical DSS image of the TOI 620 system in 1954, in the DSS red band, where the current location of TOI 620 is circled in yellow. The diameter of the circle is $\approx \angd{;;11}$, or $\sim$ half the size of a \tess\ pixel.  It is visibly clear that no stars would be in the background of TOI 620's current position.  The star to the lower-right of its current position is the aforementioned TIC 296739889 at a separation of $\angd{;;9.2}$, with a $T$ mag contrast of 7.  Its close separation indicates its light would be blended with TOI 620's in the \tess\ aperture.}
    \label{fig:1954}
\end{figure}

TOI 620 is classified as a high proper motion star, with a $|\mu_\delta| > 350$ mas yr$^{-1}$.  This means we can look back at historical images of the night sky around the TOI 620 system and see how the star has moved, so we can see whether there are any background stars that TOI 620 has approached and thus would have its light diluted with at the current epoch.  A historical image of the TOI 620 system from the Digitized Sky Survey (DSS) in 1954 is shown in Figure \ref{fig:1954}, which shows that there are no background stars at the current coordinates of TOI 620, with the closest source being TIC 296739889 at $\angd{;;9.2}$, as mentioned in ${\S}$\ref{sect:stellar_age}.  There are also no \gaia\ EDR3 sources within $\angd{3;;}$ that have common proper motions to TOI 620, indicating that it does not have a wide binary companion down to the sensitivity of \gaia.

\subsection{An Unresolved Companion?}
\label{sec:companion}
Neither our AO imaging nor spectroscopy contain unambiguous evidence for a (sub)stellar companion; nevertheless, \gaia\ astrometry points to the existence of such an object.  The Reduced Unit Weighted Error (RUWE), a measure of the goodness of fit of the astrometry to a single-star solution corrected for chromatic effects, is 1.395.  RUWE values approaching 1.4 (where the average deviation squared is twice the error squared) have been empirically found to be highly correlated with stellar multiplicity \citep[][Kraus et al., in prep.]{Belokurov2020}.  TOI\,620 is free of the effects (extreme color or high variability) that might make such a RUWE value suspect, although some anomalous RUWE values can be due to instrumental effects.  Instead, the astrometric error could also be produced by (i) the presence of a second unresolved source causing a shift in the apparent photocenter location along the \gaia\ scan track that depend on the angle between the binary axis and the scan direction \citep[i.e.][]{Ziegler_2019}; and/or (ii) motion of the system photocenter on the sky due to orbital motion.  The former effect requires that the companion be luminous but does not require orbital motion and will increase with angular separation up to a point ($\gtrsim \angd{;;0.7}$) where the binary is resolved by \gaia.  The latter effect also increases with angular separation but requires significant orbital motion (which decreases with semi-major axis) and will be most prominent at intermediate separations.  

\begin{figure}
\centering
\includegraphics[width=\columnwidth]{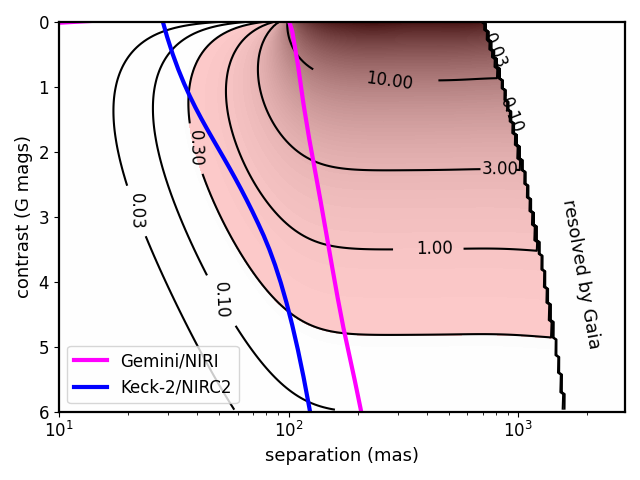}
\caption{RMS \gaia\ photocenter shift in units of mas expected for an unresolved binary with a given separation and \gaia\ contrast ratio that produce an asymmetric point spread function.  The per-observation centroid measurement error for \hoststar\ is 0.3 mas and the photocenter RMS would exceed this (and RUWE would be $\gtrsim$1.4) in the red shaded region.  The blue and magenta curves are the 5$\sigma$ detection contrast ratio limits on any companion from our Keck-2/NIRC2 and Gemini-N/NIRI AO imaging.}
\label{fig:photocenter_shift}
\end{figure}

There will be a limited range of scenarios (i.e. companion mass or luminosity and semi-major axis or separation) that can produce the astrometric error but are compatible with our AO and RV observations \citep[e.g.,][]{Wood2021}.   We performed Monte Carlo simulations combined with analytical predictions of \gaia\ astrometric deviation (Gaidos et al., in prep.).  For the asymmetry effect we assumed that the 43 scans used in EDR3 for TOI\,620.01 were distributed uniformly with angle on the sky, and axisymmetric Gaussian PSFs with a FWHM of $\angd{;;0.1074}$ \citep{Rowell2021}. \gaia\ is assumed to resolve sources with separations greater than $0.7 + 0.15\Delta G$ arcsec \citep{Brandecker2019}.  Figure \ref{fig:photocenter_shift} shows the mean photocenter deviation in mas as a function of $\rho$ and $\Delta G$.  The mean centroid error we approximate as $0.53 \sigma_{\varpi} \sqrt{N}$, where $\sigma_{\varpi}$ is the error in parallax and $N$ is the number of scans used in the astrometric fit \citep{Belokurov2020}.  For TOI 620 b the mean centroid error is 0.297 mas, and the region of parameter space where the expected deviation exceeds this value (and hence RUWE is $\gtrsim 1.4$) is shaded in red.   The 5$\sigma$ contrast ratio detection limits for our NIRC2 and NIRI AO imaging, converted from $\Delta K$ (or Br $\gamma$) to $\Delta G$ using absolute photometry of a set of M dwarfs \citep{Mann2015,Mann2019}, are also plotted.  Most of the region that would explain a high RUWE in this manner is ruled out by these data. (The remaining sliver at small separation is also ruled out by our RV measurements, see below).  Thus, apparent photocenter motion due to an asymmetric PSF is unlikely to explain the high RUWE value.

This leaves actual photocenter shift due to Keplerian orbital motion (in a binary system) as the explanation for the high RUWE.  We again calculated the permitted range of parameters for this scenario, assuming a ``sub-thermal" eccentricity distribution (uniform with the square-root of the eccentricity), and a log-normal distribution of semi-major axis with a mean of $\log 5.3$\,au and a standard deviation of 0.87 dex \citep{Duchene2013}. This is based on the observed distribution of stellar companions but the distribution of giant planets also has a peak between 1 and 10 au \citep{Nielsen2019,2019ApJ...874...81F,2018A&A...612L...3M}.   We used isotropic distributions for inclination, mean anomaly, and argument of periapsis, and two mass distributions: a ``brown-dwarf-rich" uniform distribution,  and ``brown-dwarf-poor" log-normal, the latter centered at 1 \mjup with a standard deviation of 0.6 dex to reflect the mass distribution of giant planets found in RV surveys of exoplanets \citep{Malhotra2015} and an apparent ``desert" in brown dwarfs close to stars, especially low-mass stars such as TOI\,620 \citep{Nielsen2019}.  We required that the mean RMS photocenter motion equal or exceed centroid error, we imposed constraints from our AO imaging, and required the absolute radial acceleration to be $<0.089$ m~sec$^{-1}$~day$^{-1}$ (96\% upper limit; ${\S}$\ref{sec:primary_analysis}).  

\begin{figure*}
\centering
\includegraphics[width=0.49\textwidth]{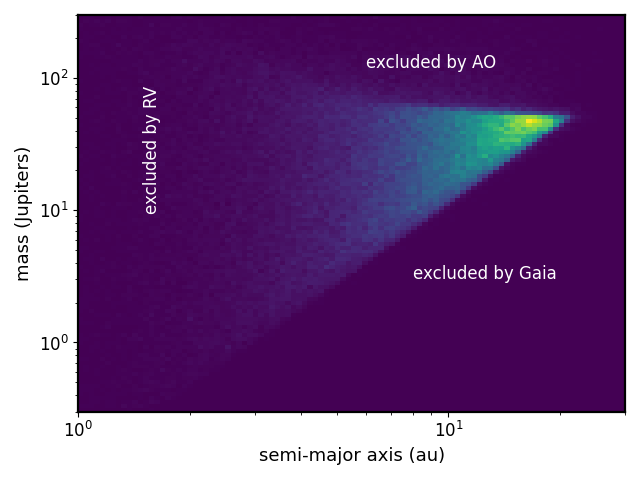}
\includegraphics[width=0.49\textwidth]{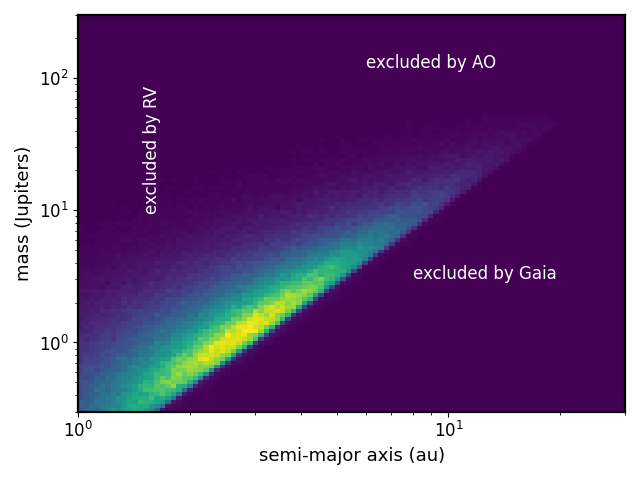}
\caption{Limits on the separation and contrast ratio of a potential companion detected by  \emph{Gaia} (and RVs).  The colormap linearly represents the posterior probability of a companion in the separation -- companion mass plane, as ascertained by Monte Carlo calculations.  The left panel assumes a uniform prior on companion mass, while the right panel uses a log normal prior centered on 1 Jupiter mass and with a standard deviation of 0.6 dex.}
\label{fig:companion}
\end{figure*}

The range of mass and semi-major axis that are permitted by these constraints occupy a narrow band running from a fractional mass of Jupiter at a few au, to tens of Jupiter masses at 30\,au,   Figures \ref{fig:companion} a and b show the results for the uniform and log-normal mass priors (left and right, respectively).  A range of scenarios is clearly possible but either a Jupiter-mass companion at $\sim$3\,au, or an ultracool dwarf at 20-30\,au are favored. More massive companions are permitted, but only under the unlikely scenario where the projected separation is much less than the true orbital separation, \rev{which we explore in the next section}.

\subsection{Spectroscopic Binary Analysis \& Results}
\label{sect:sb2_analysis}
We next explore the possibility that TOI 620 possesses a low-mass spectroscopic stellar companion hidden at a small projected separation, much smaller than the true orbital separation, a relatively low-probability occurrence.  TOI 620 shows no evidence for an equal-mass SB2 binary in any of our spectroscopic data, and our broadband SED analysis shows no indication for an over-luminosity as is common for binaries; this limits any stellar companions to TOI 620 to mid and late M dwarfs with an unequal mass ratio (much lower than the primary mass).  

The most favorable flux contrast ratio for a hypothetical mid- to late-M dwarf companion would be in the near-infrared.  Consequently, we model our 10-times iterated iSHELL stellar template (see appendix \ref{sect:forward_modeling} on how this iterated stellar template is generated) with a two-component spectroscopic binary model consisting of two BT-Settl models generated from the Spanish Virtual Observatory (SVO) website\footnote{\href{http://svo2.cab.inta-csic.es/theory/newov2/index.php}{http://svo2.cab.inta-csic.es/theory/newov2/index.php}} \citep[e.g.][]{2020ApJ...899...29K}, with an arbitrary radial velocity offset between them. Our iSHELL stellar template is effectively an iterated and empirically deconvolved cumulative high SNR ($>$500) spectrum of our target star, comprised of the summation of all of the iSHELL observations in the stellar rest frame, after modeling out tellurics, the gas cell absorption, the blaze function, and other instrumental effects that are incoherent in the stellar rest frame when sampled across many Solar System barycenter velocities.  \rev{The iSHELL observations themselves cover a time baseline of $<$2 years--shorter than putative orbital periods of any potential stellar companions (${\S}$\ref{sect:rv_analysis}). As such, time-averaging of relative RV shifts between the stars is not an issue.}

We correct the SVO wavelengths (which have a resolution of $10^{-3}$ \AA) for the index of refraction of the atmosphere ($n_{\rm air} = 1.000293$) and use a piecewise cubic Hermite interpolating polynomial (PCHIP) to interpolate the data with \texttt{scipy}.  We assume the rotational velocities of each star are small enough to have minimal Doppler broadening effects and do not fit for them.  We fit the temperature of each star and the radial velocity of each star (for a total of 4 free parameters), and we use the Tables from \citet{Pecaut_2013} to determine the difference in magnitudes at K band from the best-fit temperatures and apply the flux ratios to our model. We linearly interpolate values inbetween entries from \citet{Pecaut_2013}.  Our binary flux model can be summarized as 

\begin{equation}
    F(\lambda) = (1-D)F_1(\lambda) + DF_2(\lambda)
\end{equation}

\noindent
where $F(\lambda)$ is the total flux at wavelength $\lambda$, \rev{$F_1(\lambda)$ and $F_2(\lambda)$ are the flux from the primary and secondary, respectively}, with arbitrary radial velocity offsets. $D$ is the dilution, or the fraction of light from the secondary divided by the total light from the system. 

We first perform a maximum likelihood fit by minimizing the negative log of the likelihood, where our log-likelihood function is defined as

\begin{equation}
    \ln \mathcal{L} = -\frac{1}{2}\sum_{i=1}^{N}\bigg(\frac{d(\lambda_i)-F(\lambda_i)}{\sigma_i}\bigg)^2 + \ln(2\pi\sigma_i^2)
\end{equation}

\noindent
where the subscript $i$ enumerates each wavelength datapoint up to $N$ total, $d(\lambda_i)$ is the value of our iteration 10 deconvolved stellar template from \texttt{pychell}, and $\sigma_i$ is the error in our observed and deconvolved stellar spectrum $d(\lambda_i)$ at wavelength $\lambda_i$ for all iSHELL orders considered. We assume that $\sigma_i=0.01$ is a constant across our spectrum, a conservative assumption given the cumulative SNR of our observations.

The maximum a posteriori values are then used as starting points for an MCMC simulation, where we impose a Gaussian prior on the \teff~of the primary corresponding to our posterior from \texttt{EXOFASTv2}, and we impose hard boundaries on the radial velocities between $\pm 200$ \kms.  We run a series of MCMC simulations with differing upper boundaries on the \teff~of the secondary corresponding to flux ratios of $<$50\%, $<$20\%, $<$15\%, $<$10\%, and $<$5\%.  We find in the $<$50\% and $<$20\% limiting flux ratio cases that the \teff\ of the secondary hits the upper boundary in \teff\ with no relative radial velocity offset, indicating that our spectra are best described by an equal-temperature binary, a scenario that is excluded by our SED and high-contrast imaging analysis. Since TOI 620 is not over-luminous, and since we do not impose this constraint in this spectroscopic analysis, this effectively implies that the single-star solution is preferred for flux ratios of $>$20\%.  

\begin{figure*}
    \centering
    \quad
    \includegraphics[width=.47\textwidth]{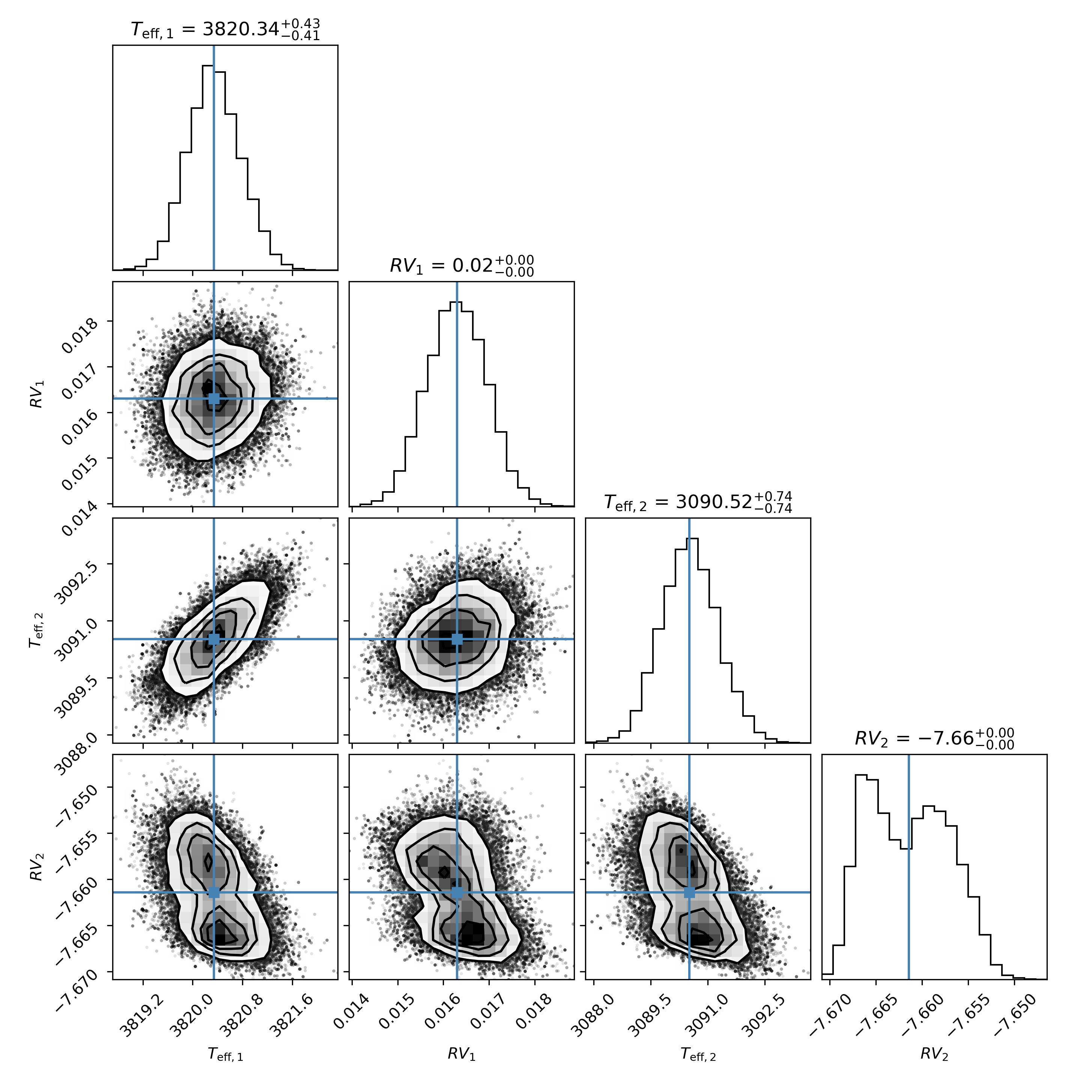}
    \includegraphics[width=.47\textwidth]{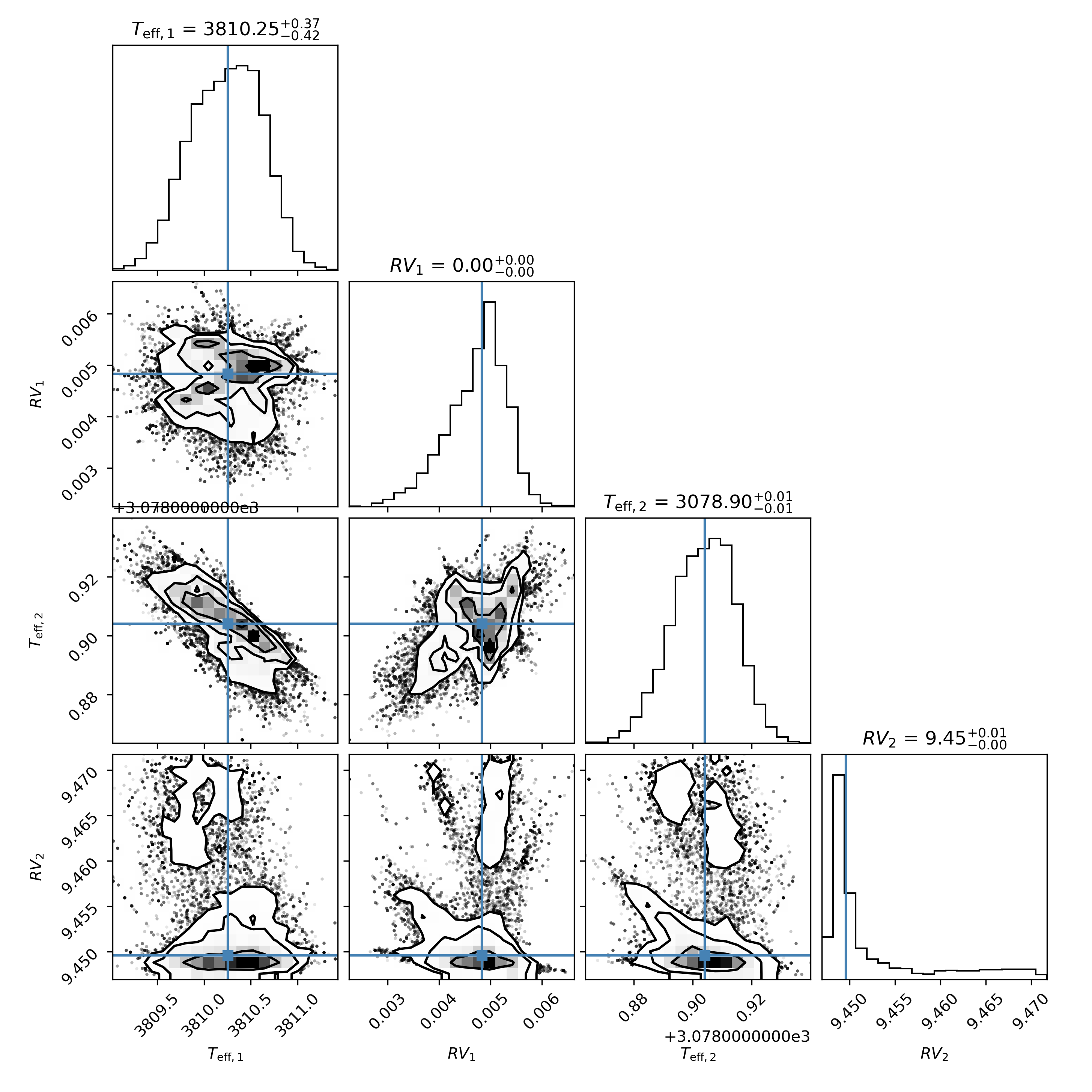}
    \caption{MCMC cornerplot of our spectral binary analysis of the iSHELL iteration 10 stellar template. Plots along the diagonal show 1-dimensional histograms of the posterior distributions of each parameter.  Off-diagonal plots show the covariance between each model parameter. 
    \rev{The posterior distributions are bisected to show a zoom of each of the two individual posterior peaks on the left and right cornerplots}, showing that they are centralized maxima and are not edge solutions in our model parameter space.}
    \label{fig:sb2_cornerplot}
\end{figure*}

However, in the $<$10\% and $<$15\% flux ratio limited MCMC cases we find consistent and robust doubly peaked solutions that suggest the possibility of a hierarchical eclipsing binary (HEB) system scenario for TOI 620 in which two smaller stars are both orbiting each other and then the pair is orbiting the more massive primary star, in contrast to the companion analysis in ${\S}$\ref{sec:companion}. The cornerplot of our $<$10\% flux-ratio scenario is shown in Figure \ref{fig:sb2_cornerplot} \rev{with} each peak in the posteriors separated and plotted individually. The doubly-peaked posteriors are not confined by the prior bounds on the flux ratio. Finally, for the $<$5\% flux ratio limited MCMC scenario, we again recover a maximum posterior probability at the upper-limit to the flux-ratio range explored, indicating that a robust two-star model is only favored for flux-ratios in the 10--20\% range.  We do not explore a three-star model.

\begin{figure*}
    \centering
    \includegraphics[width=\textwidth]{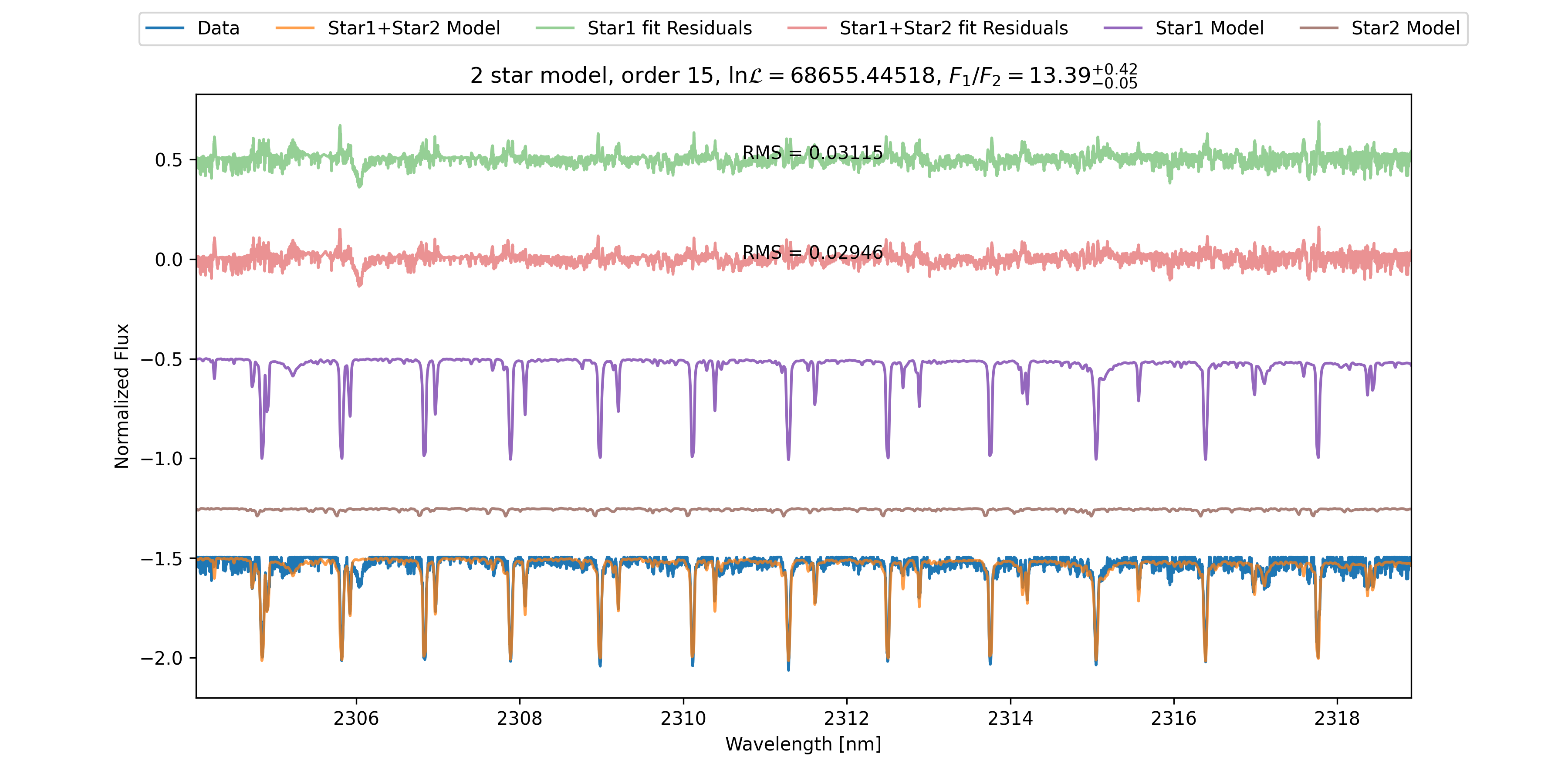}
    \caption{Fits of our spectral binary model of the iSHELL iteration 10 stellar template in the 10\% flux ratio limited case. The blue line is the iteration 10 stellar template (see ${\S}$\ref{sect:forward_modeling}), the orange line is the combined final flux model, the purple and brown lines are the individual fluxes from the primary and secondary, respectively, and the green and red lines are the residuals from the single-star and double-star fits, respectively, with RMS displayed. The wavelength range shown corresponds to echelle order 15 on the iSHELL spectrograph. An example stellar spectral absorption line at 2306 nm that is missing from the BT-Settl models is readily apparent.  Note, in the stellar rest frame, the telluric absorption feature residuals from all of our 379 iSHELL spectra sampled at a range of barycenter velocities are de-coherent and not contributing significantly to the observed residuals.}
    \label{fig:sb2_spectral_plot}
\end{figure*}

We show the two-star spectral fit of iSHELL's echelle order 15 ($\lambda\ 2304$--$2320$ nm; $m=226$), which is relatively free of macro-tellurics compared to other K-band orders, in Figure \ref{fig:sb2_spectral_plot} \citep{2013SPIE.8864E..1JP}, with an RMS$\sim$0.03. Given the SNR $>$ 500 of our empirical, deconvolved stellar spectrum, a residual RMS $>2\times10^{-3}$ is significant. However, the BT-Settl synthetic stellar models are incomplete in NIR opacity sources for M dwarf atmospheres, resulting in missing stellar absorption features and other systematics. Consequently, the residual RMS of our best fitting model is greater than the expected RMS from the cumulative SNR of our observations. Our results are nonetheless compelling despite this model incompleteness, as systematics would not produce the isolated local maxima in the likelihood function that we observe, particularly when averaging over 13 iSHELL orders. While we show order 15 as a representative example, the MCMC modeling was performed jointly across orders 5-17 ($m=216$--$228$) from our iSHELL data spanning a significant fraction of K-band. Performing a model comparison between this double-star model in the $<$20\% flux-ratio regime, and the best-fit single star model, we find a $\Delta \ln \mathcal{L} = 68655.45$. From this we compute the corresponding difference in the small-sample Akaike Information Criterion \citep[AICc;][]{Akaike_1974, Burnham_2002}, $\Delta {\rm AICc} = -137302.89$.  Since the value is negative, this indicates that the 2 star model is favored over the one star model, though the magnitude of how much it is favored is dependent on the spectroscopic flux error of 0.01 that we imposed earlier. 

From this analysis, we conclude that the iSHELL data indicates the possible presence of one or two low-mass stellar companions with a $K$-band flux ratio of 13.39$^{+0.42}_{-0.05}$, \teff$_1 = 3090$ K and \teff$_2 = 3079$ K and RVs relative to the primary of $-7.66$ and $+9.45$ \kms respectively. The RV separation of $\sim$ 17 \kms is approximately consistent with a Keplerian orbital velocity for the orbital period of the candidate exoplanet (if it were instead an HEB), and slightly offset from the velocity of the primary by a reasonable $\sim$1 \kms.  

Alternatively, we do not exclude and do not explore in this analysis that this favored SB2 solution could be an artifact of not rotationally broadening our stellar models, which could also potentially yield a false symmetric set of binary companions. Nonetheless, motivated by the Gaia RUWE statistic and this SB2 analysis of our iSHELL spectra, we must carefully consider and explore the possibility that TOI 620 is a circum-secondary planet or HEB false-positive in more detail.

For this $K$-band flux ratio, and the corresponding $R$-band flux ratio for M dwarfs of these temperatures of $\sim$57, they would have been detected by the high-contrast imaging in ${\S}$\ref{sect:high_contrast_imaging} for projected separations $>\angd{;;0.2}$; additionally, given the iSHELL slit width of $\angd{;;0.375}$ and typical seeing conditions of less than one arc-second, the projected separation must also be less than one arc-second. However, no such companions are detected. Thus, if these stellar companions exist, they must possess a projected separation of $<\angd{;;0.2}$ or $<6.6$ au, and deeper high-contrast near-infrared imaging or aperture mask photometry will be required in the future to exclude this possibility. Given the visible flux-contrast ratio, any such companions would easily be hidden in the high-resolution spectroscopic data. However, such a false-positive circum-secondary or HEB scenario could be uncovered from chromatic transit photometry. We next turn to our transit analysis to explore the analysis of the primary star transit and these possible false-positive scenarios.

\section{Transit Analysis \& Results}
\label{sect:transit_analysis}

It is clear from the \gaia\ RUWE statistic and spectroscopic binary analysis, in conflict with the high contrast imaging, that there are three distinct possibilities for the source of the transit signal, but from these results alone it is difficult to determine whether we have a single system with a circum-primary planet, a double with a circum-secondary planet, or even triple hierarchical system with an eclipsing binary pair.  Motivated by this quandary, we present results from three separate analyses of the TOI 620 light curves in \texttt{EXOFASTv2} under the assumptions of a circum-primary planet (${\S}$\ref{sect:exofast}), circum-secondary planet (${\S}$\ref{sect:csec_transit_main}), and HEB star system (${\S}$\ref{sect:heb_transit_main}) to determine which, if any, of these scenarios is the most plausible given the data. To gain additional insight, we first perform traditional vetting analysis in ${\S}$\ref{sect:false_positives} to see if any of the usual oddities that would indicate an EB or other false positive appear in the light curve. 

\subsection{Vetting against False Positives}
\label{sect:false_positives}
\rev{The first of our vetting tests was performed with the DAVE vetting pipeline \citep{Kostov_2019}, which shows no significant odd-even differences between consecutive transits (confirming the measured period is not an integer multiple of the true period), no significant photocenter motion during the transits (confirming the target is the source of the transits), and no significant secondary eclipses. Phased transit data and photocenter plots are provided in appendix section \ref{sect:app_primary}.  We confirm these results with the EDI-Vetter Unplugged tool \citep{Zink_2020}\footnote{\href{https://github.com/jonzink/EDI_Vetter_unplugged}{https://github.com/jonzink/EDI\_Vetter\_unplugged}}, which checks for a similar suite of eclipsing binary indicators, and found no evidence pointing to a false-positive scenario.}

\rev{We additionally perform a false positive probability analysis with \texttt{vespa} \citep{Morton_2012}, which uses galactic population statistics for stellar multiplicity, transit depth, duration, and ingress/egress duration to calculate the probability that the target is an EB, BEB, HEB, or planet. Using the \tess\ transit data, this gives a FPP of 1 in 22179.  The priors, likelihoods, and probabilities are shown in Figure \ref{fig:vespa} (in the appendix) along with TOI 620's location in $\log\delta$-$T$-$T/\tau$ space compared to typical planet populations.}

\subsection{True Positive Scenario}
\label{sect:exofast}

\begin{table}
    \centering
    \begin{tabularx}{\columnwidth}{@{\extracolsep{\fill}}XXlX}
        \hline
        Parameter [units] & Initial Value $(P_0)$ & Priors & Prior Citation \\
        \hline
        \hline
        $P$ [days] & 5.098831 & $\mathcal{U}(P_0 \pm 10\%)$ & E19 \\
        $T_C$ [days] & 8518.005713 & $\mathcal{U}(P_0 \pm P/3)$ & E19 \\
        $R_p/R_*$ & 0.053 & -- & this work \\
        \hline
    \end{tabularx}
    \caption{Prior probability distributions for our \texttt{EXOFASTv2} MCMC simulations.  $\mathcal{N}(\mu, \sigma)$ signifies a Gaussian prior with mean $\mu$ and standard deviation $\sigma$.  $\mathcal{U}(\ell, r)$ signifies a uniform prior with left bound $\ell$ and right bound $r$.  Parameters that are missing, including $e$, $\omega$, etc. are \rev{initialized to circular and edge-on values, with no imposed priors}. Also note that the time of conjunction $T_C$ has been subtracted by 2450000. \emph{References}: E19: \citet{Eastman_2019}}
    \label{tab:exofast_priors_transit}
\end{table}

\rev{We} next performed an analysis with the signal generated by a planet orbiting the known star. After normalizing the \tess\ PDC-SAP data as described in ${\S}$\ref{sect:observations}, we jointly model the \tess\ and all ground-based light curve follow-up observations with \texttt{EXOFASTv2}. Our minimal priors are detailed in Table \ref{tab:exofast_priors_transit}, including the period $P$, time of conjunction $T_C$, and \rev{radius ratio} $R_p/R_*$.

\begin{figure*}
    \centering
    \includegraphics[width=.75\columnwidth]{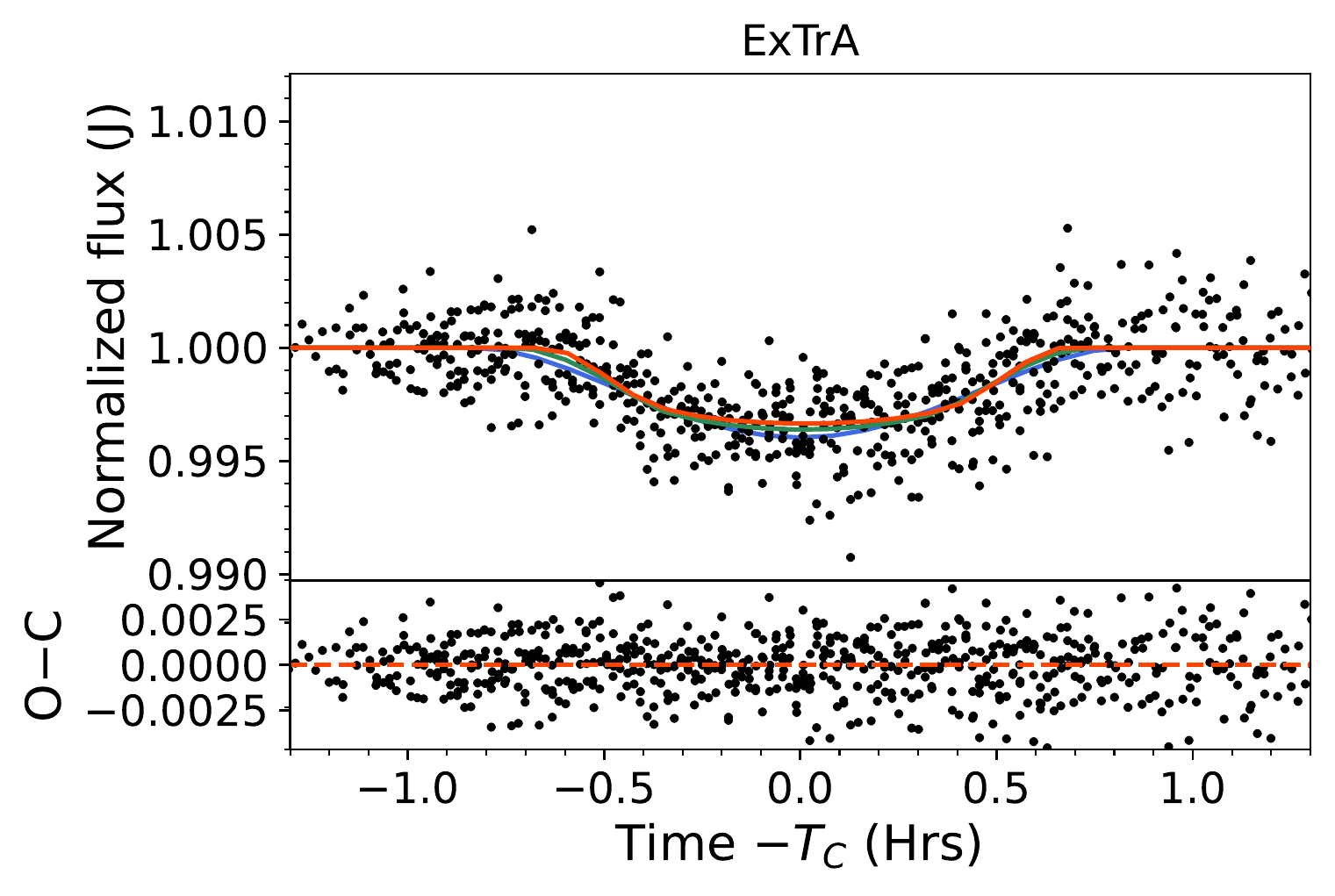}
    \includegraphics[width=.73\columnwidth]{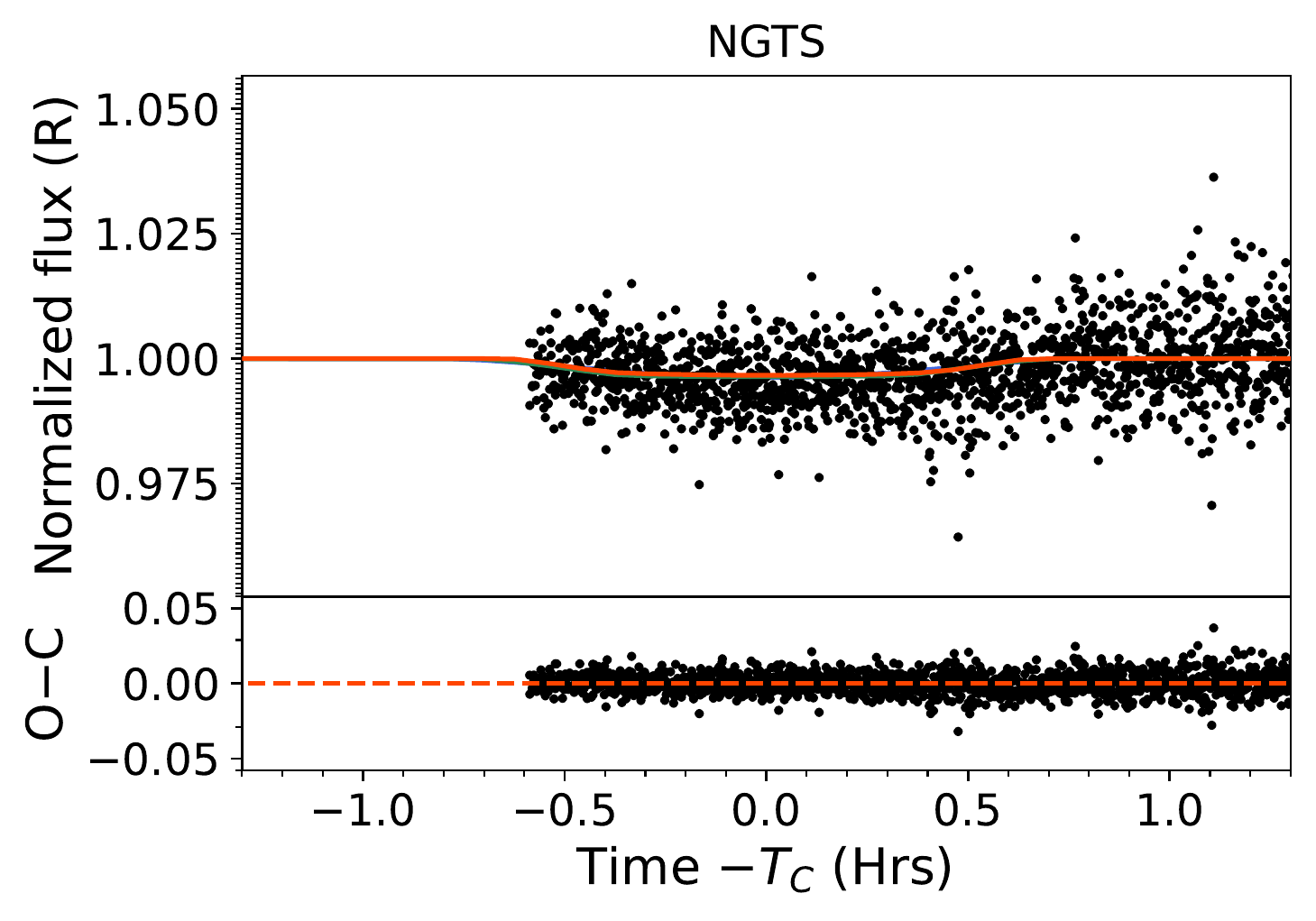}
    \includegraphics[width=.75\columnwidth]{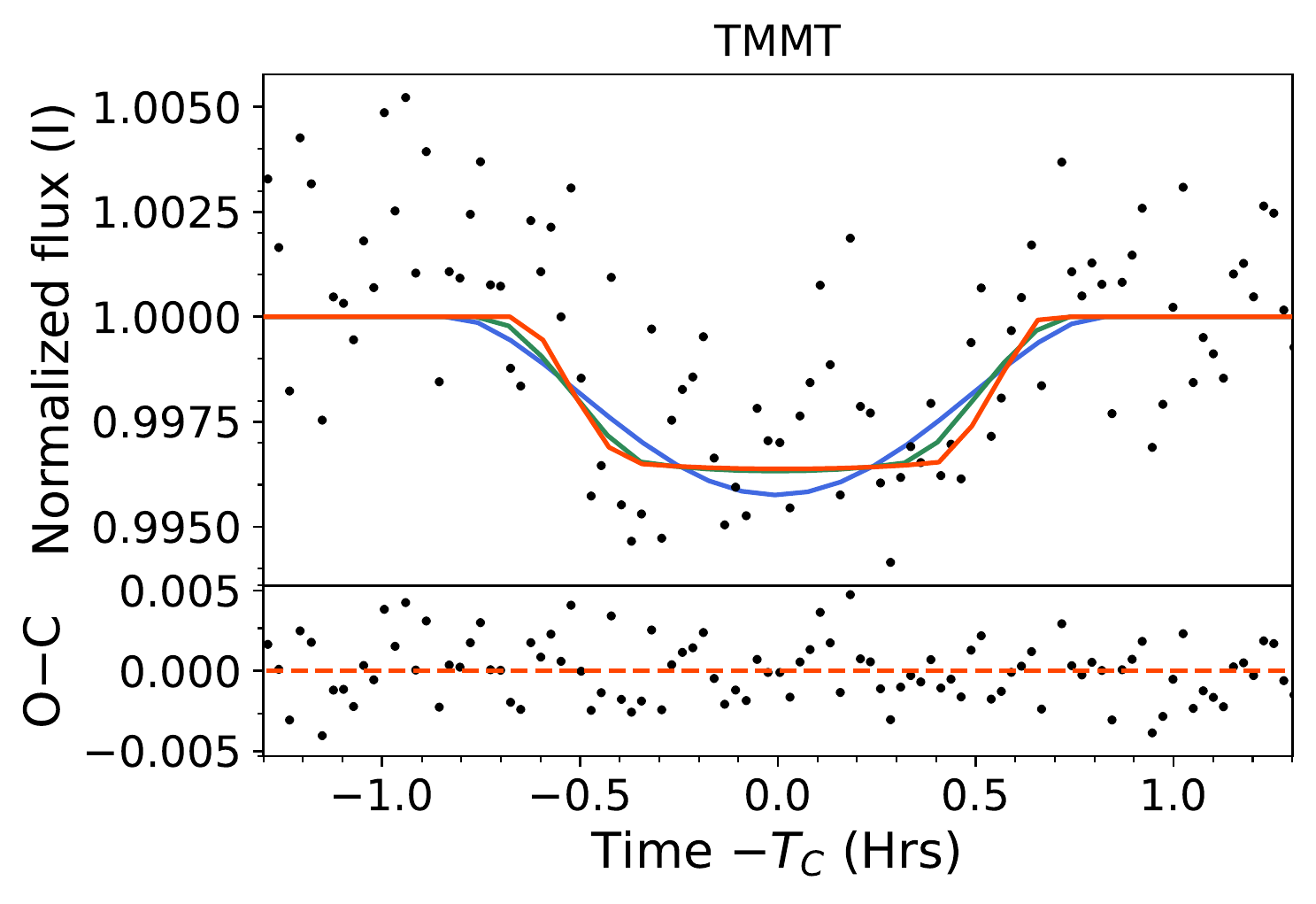}
    \includegraphics[width=.73\columnwidth]{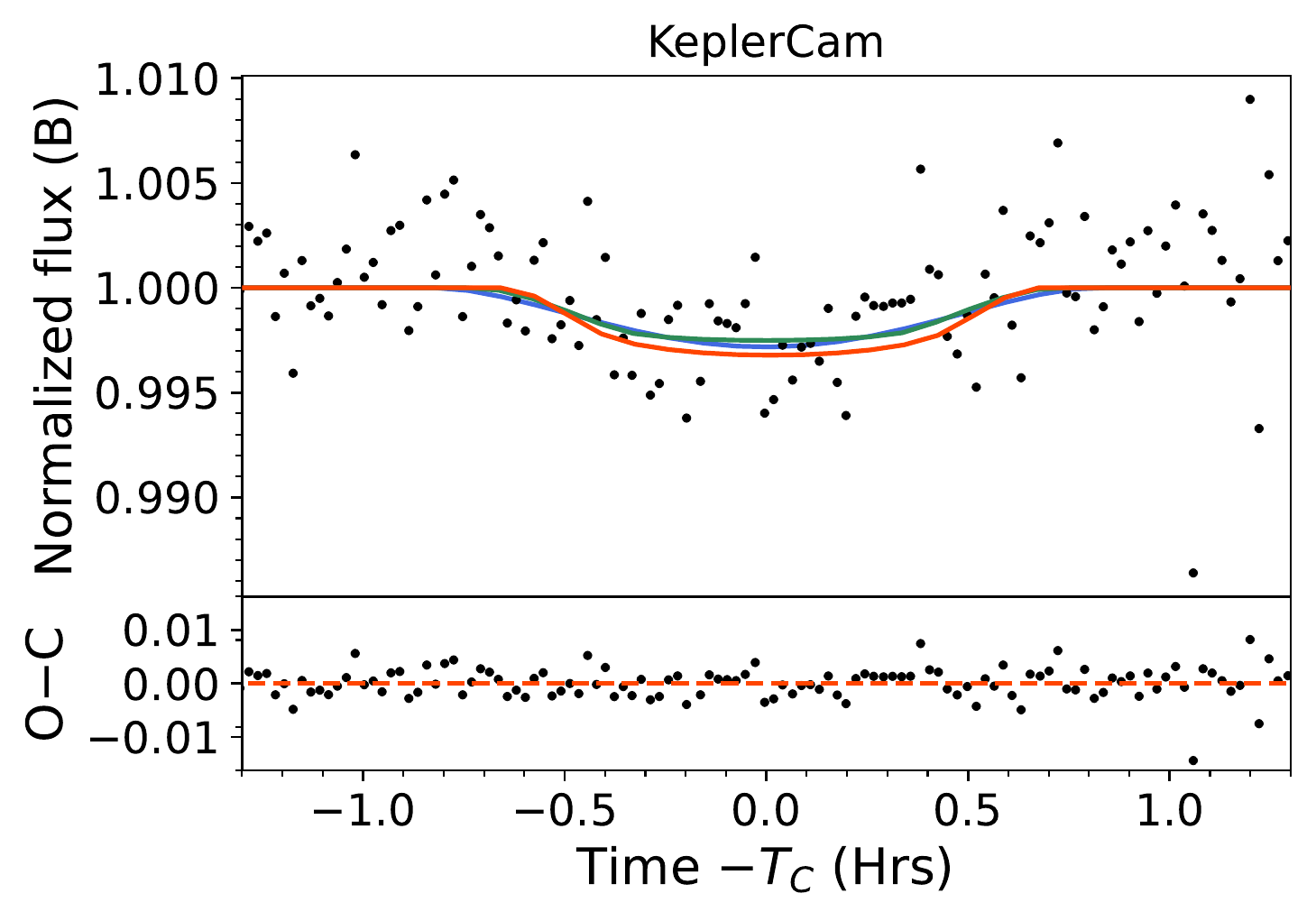}
    \includegraphics[width=.75\columnwidth]{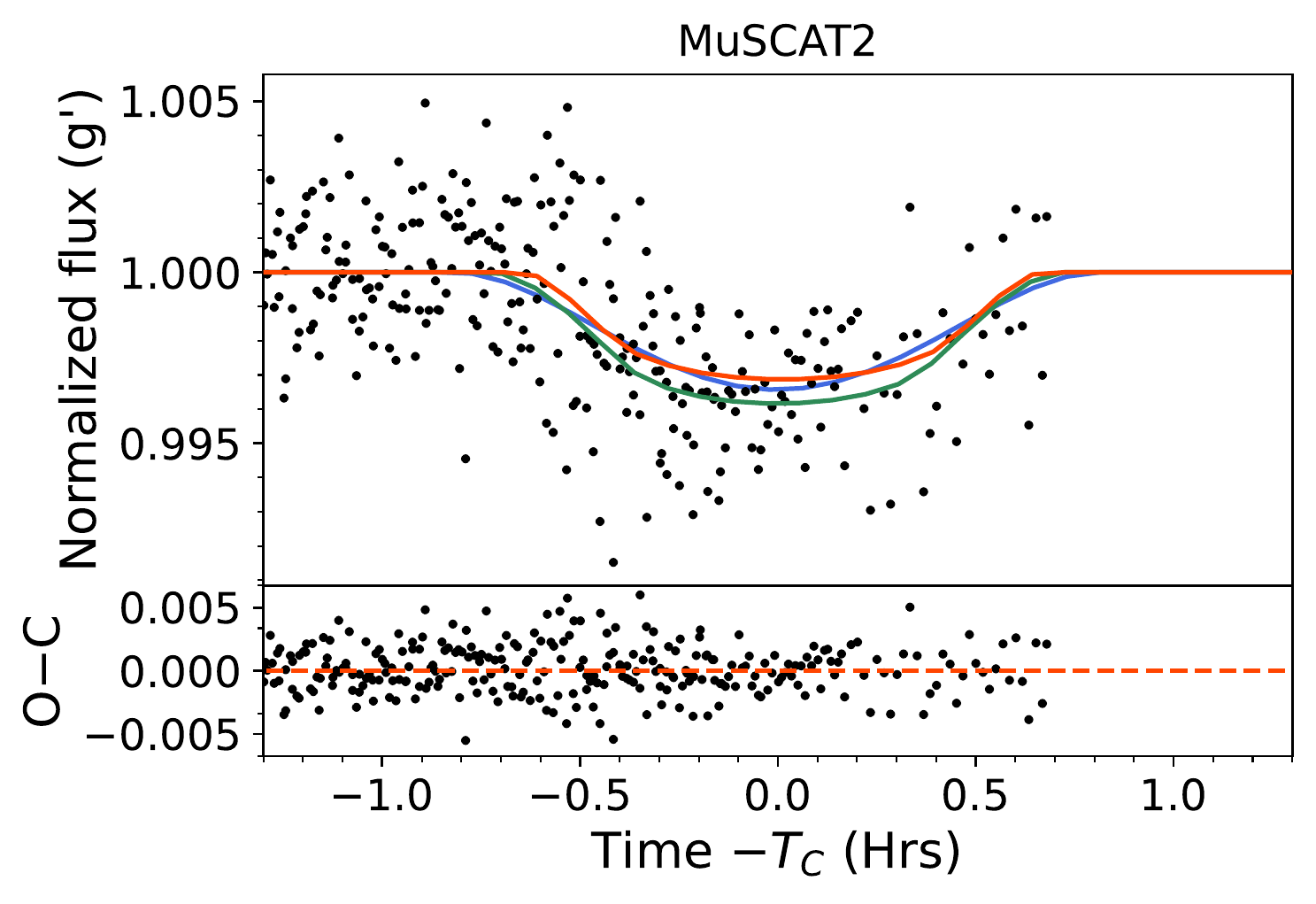}
    \includegraphics[width=.75\columnwidth]{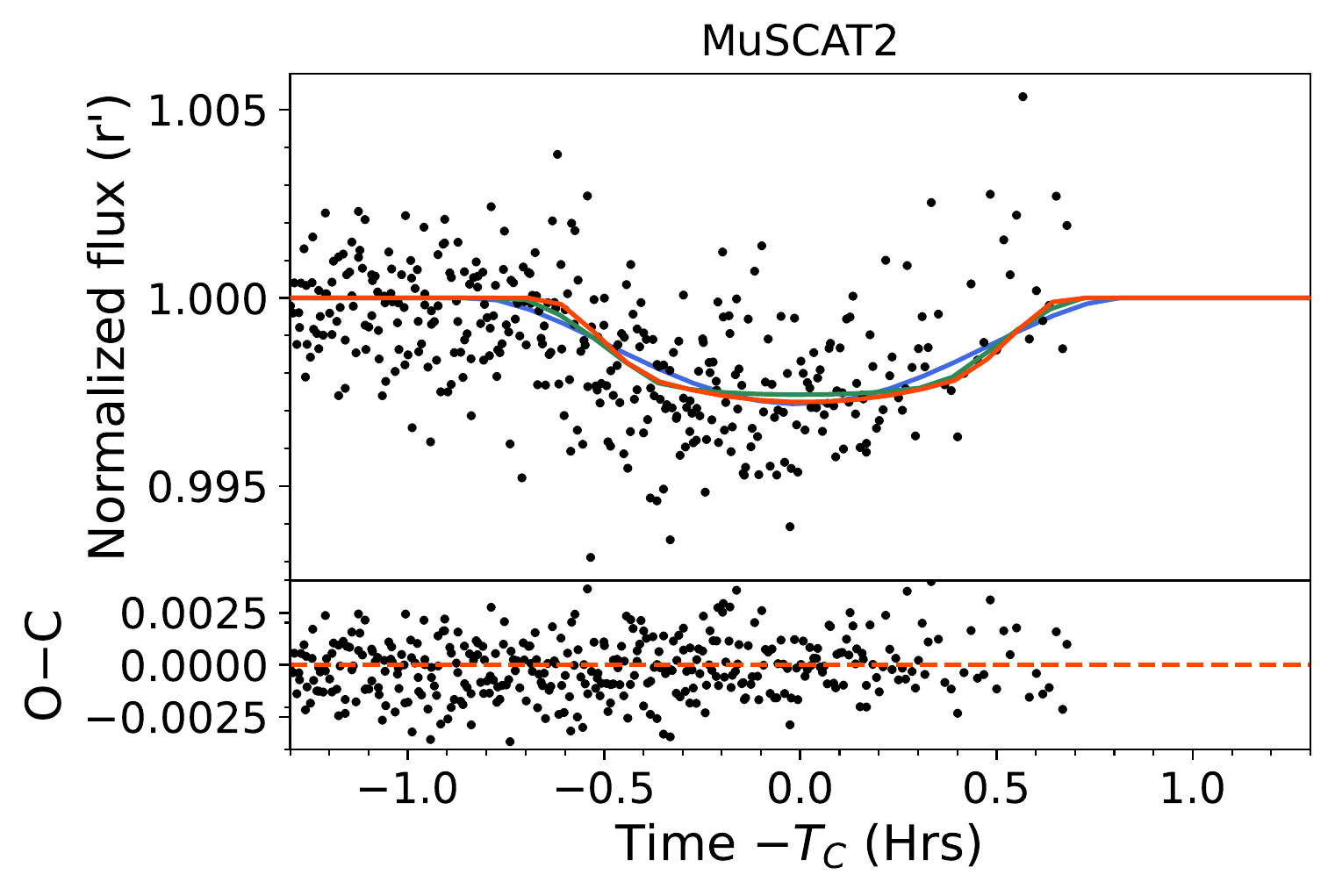}
    \includegraphics[width=.75\columnwidth]{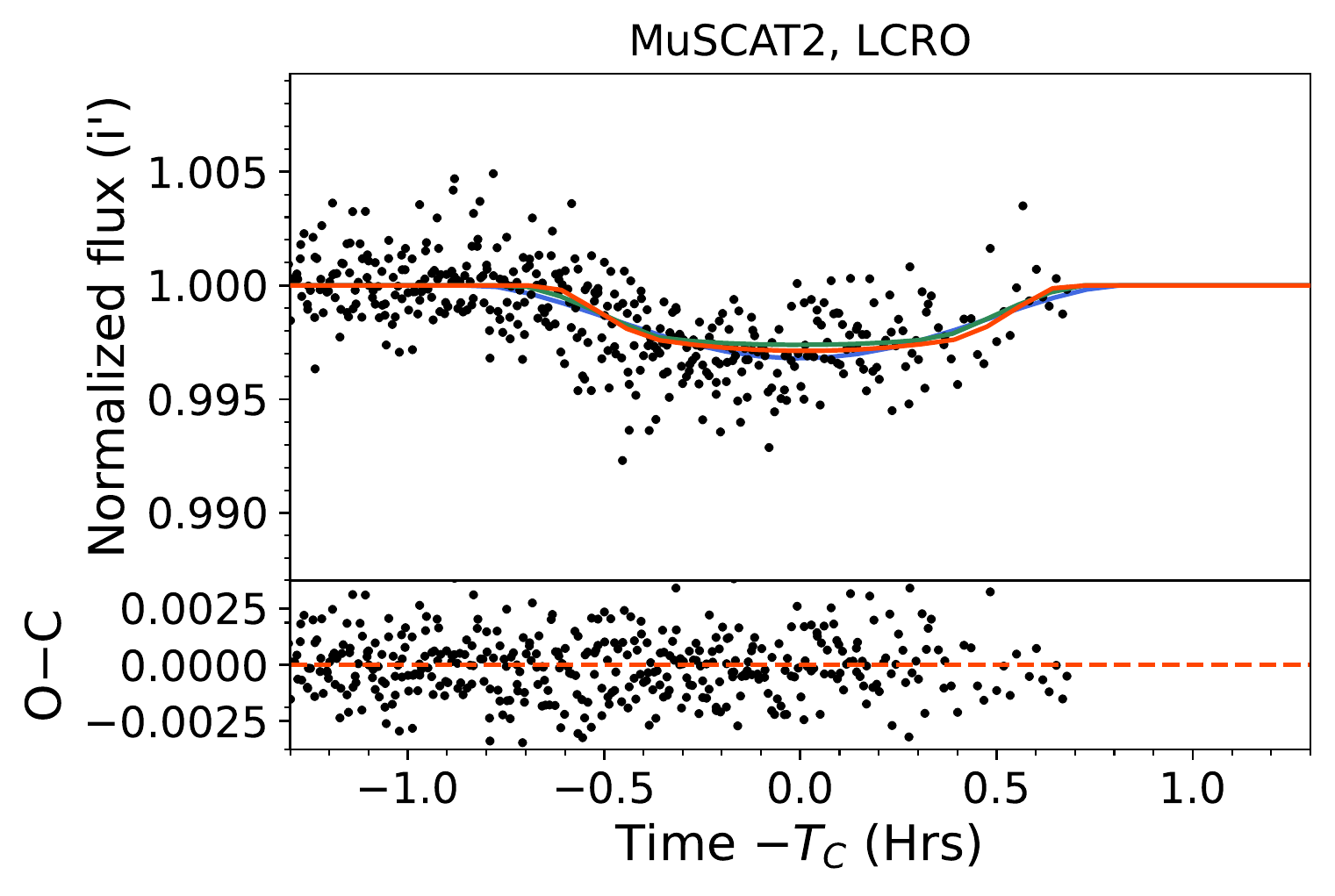}
    \includegraphics[width=.75\columnwidth]{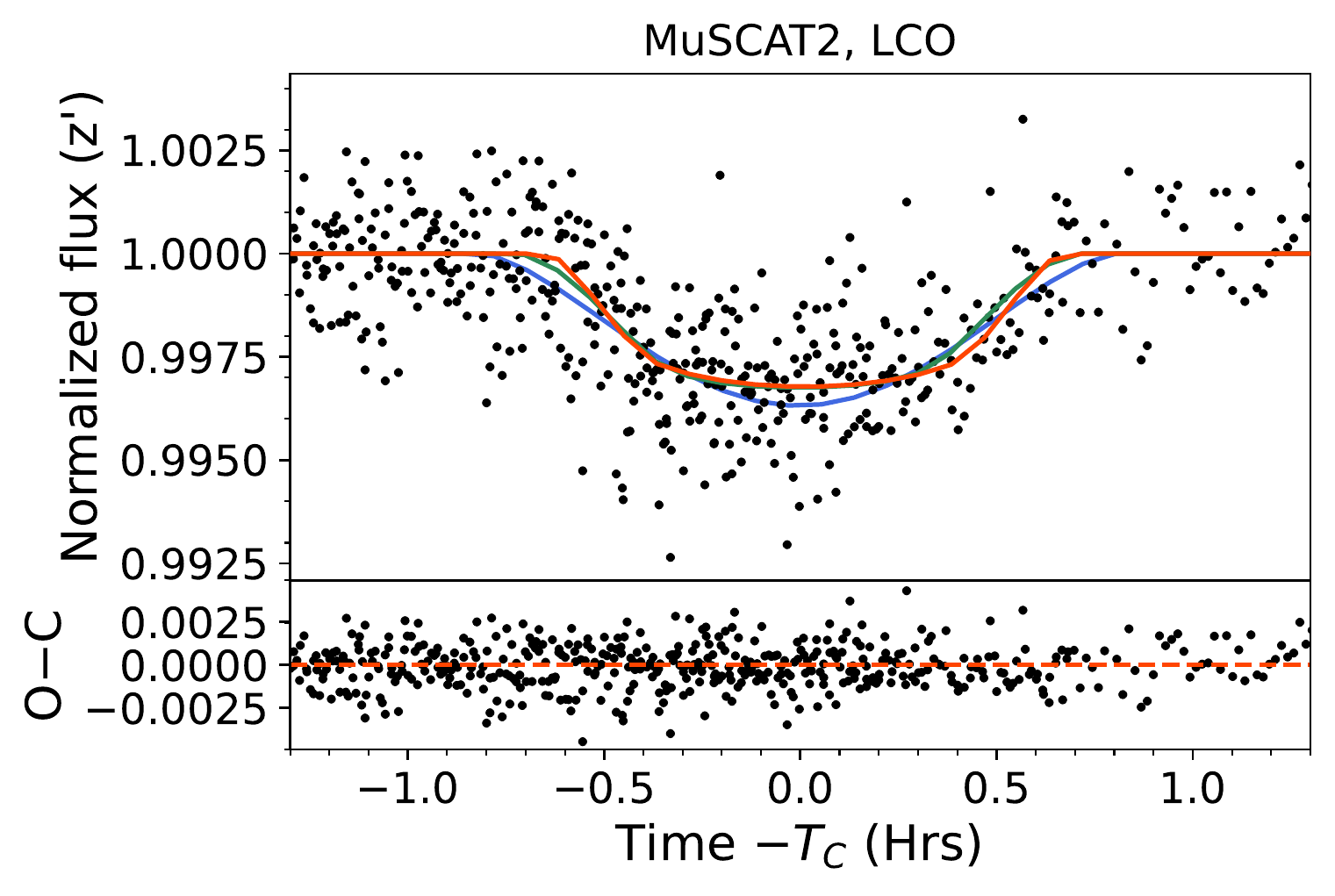}
    \includegraphics[width=.75\columnwidth]{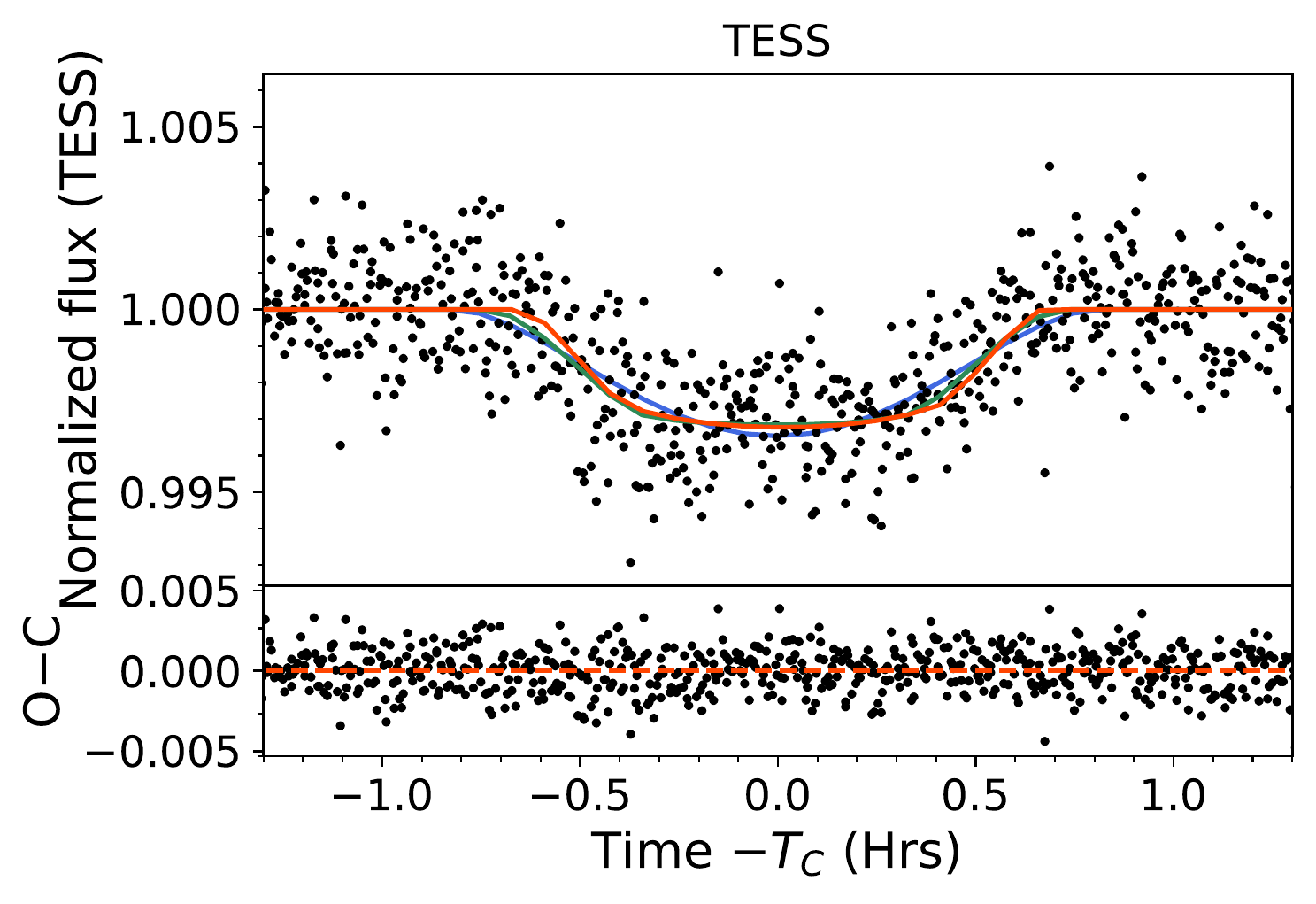}
    \caption{\rev{Transit data and models for each filter used in the \texttt{EXOFASTv2} analysis of TOI 620.  Filters shown, from top-left to bottom-right, are $J$, $R$, $I$, $B$, $g'$, $r'$, $i'$, $z'$, and \tess.  In each filter's plot, the top plot shows the combined transit data from all observations in that filter, phased to the period of the planet (5.09887 days), with each line showing the median transit model for the circum-primary (red), circum-secondary (green), and HEB (blue) cases. The dilution in the circum-secondary and HEB scenarios were allowed to independently vary in each bandpass and converge on unphysical values, as shown in Figure \ref{fig:dilution_p}.  The bottom plots show the residuals for the circum-primary model only.}}
    \label{fig:exofast_transits}
\end{figure*}

\begin{figure*}
    \centering
    \includegraphics[width=\textwidth]{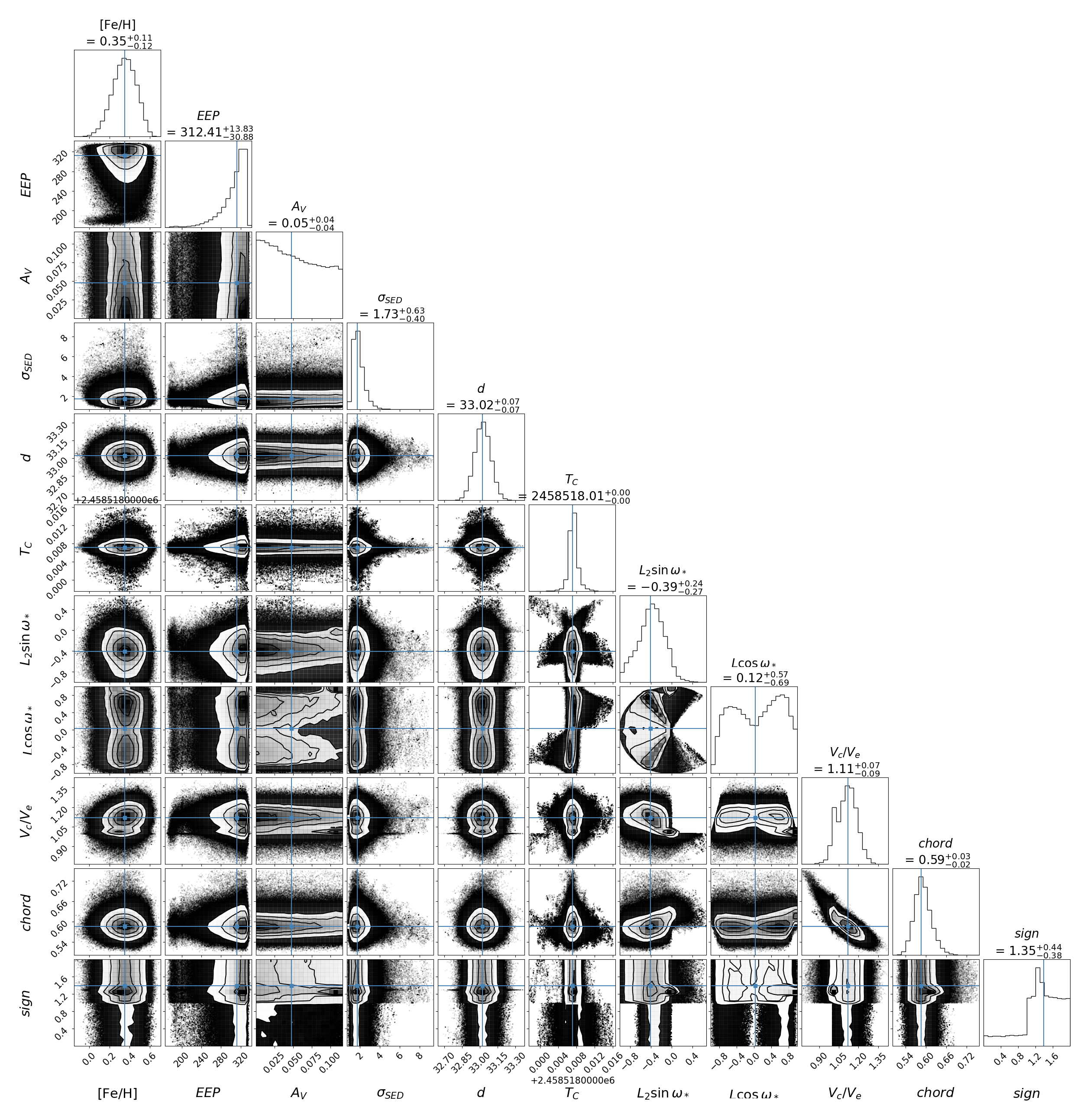}
    \caption{MCMC cornerplot of our joint ground and space-based transit model MCMC for TOI 620.  Plots along the diagonal show 1-dimensional histograms of the posterior distributions of each parameter. Off-diagonal plots show the covariance between each model parameter. \rev{Note that ${\rm chord} = \sqrt{(1-R_P/R_*)^2-b^2}$ and $L_2 = {\rm sign}(V_c/V_e-1)L$ \citep[see ][]{Eastman_2019}.} Not all model parameters that were varied are shown in this cornerplot.  Namely, we do not show the detrending parameters of each dataset, linear and quadratic limb darkening coefficients from each wavelength band, $R_*$, $R_{*,\rm SED}$, $T_{\rm eff}$, $T_{\rm eff, SED}$, $[{\rm Fe/H}]_0$, $\log M_*/M_\odot$, $M_P$, $R_P/R_*$, and $\log P$.  \rev{The posteriors not shown are well behaved, and there is no strong covariance between any of them except the linear and quadratic limb-darkening coefficients, which show anti-correlation, and the metallicity and initial metallicity, which show strong correlation.} All MCMC cornerplots presented in this paper are generated using \texttt{corner.py} \citep{corner}.}
    \label{fig:exofast_cornerplot}
\end{figure*}

The posterior values for this initial MCMC run are then used as the initial values for a second iteration run (though we keep the same uniform and Gaussian priors as the initial run) that we allow to run longer and we confirm the second MCMC converges on the same results within 1$\sigma$ to check for the robustness of the MCMC posteriors. \rev{Each MCMC is run for 225,000 steps, and we measure convergence by ensuring the maximum Gelman-Rubin statistic of the chains is $\lesssim$1.1 \citep{Gelman_1992} at the end of the simulation.} The transit models of the MCMC simulation are shown in Figure \ref{fig:exofast_transits}, \rev{while the SED model is in Figure \ref{fig:exofast_star}}. A cornerplot showing a subset of the most interesting posteriors is presented in Figure \ref{fig:exofast_cornerplot}. The median posterior values and 68\% confidence interval 1$\sigma$ Gaussian equivalent uncertainties are shown in Table \ref{tab:TOI620}. 

\begin{figure*}
    \centering
    \quad
    \includegraphics[width=\textwidth]{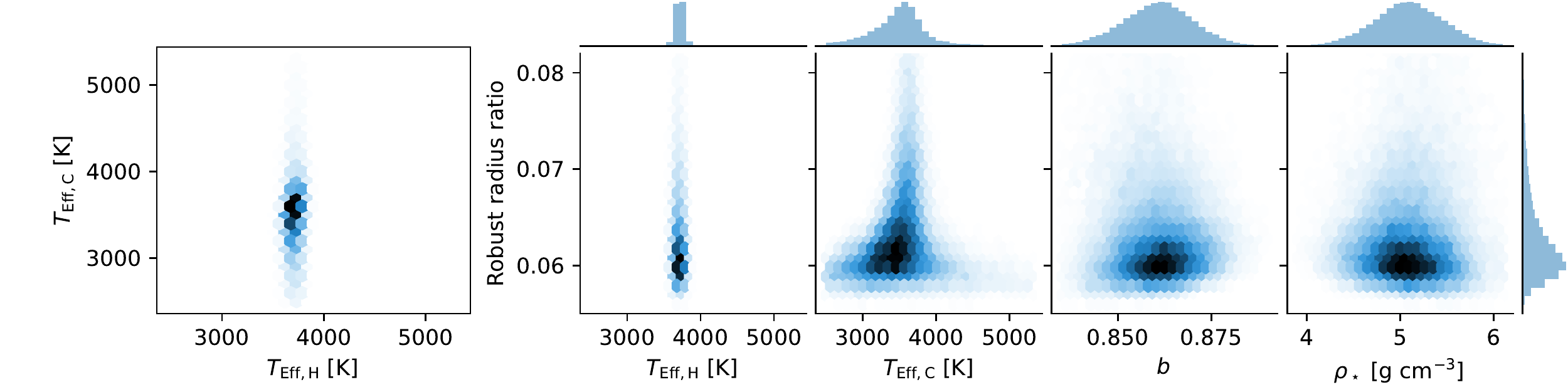}
    \includegraphics[width=\textwidth]{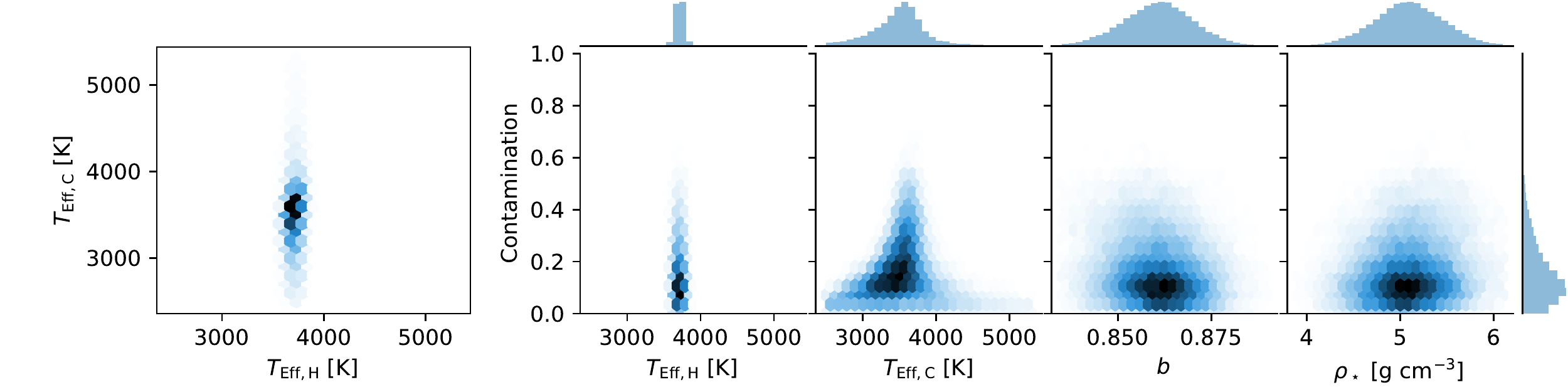}
    \includegraphics[width=.5\textwidth]{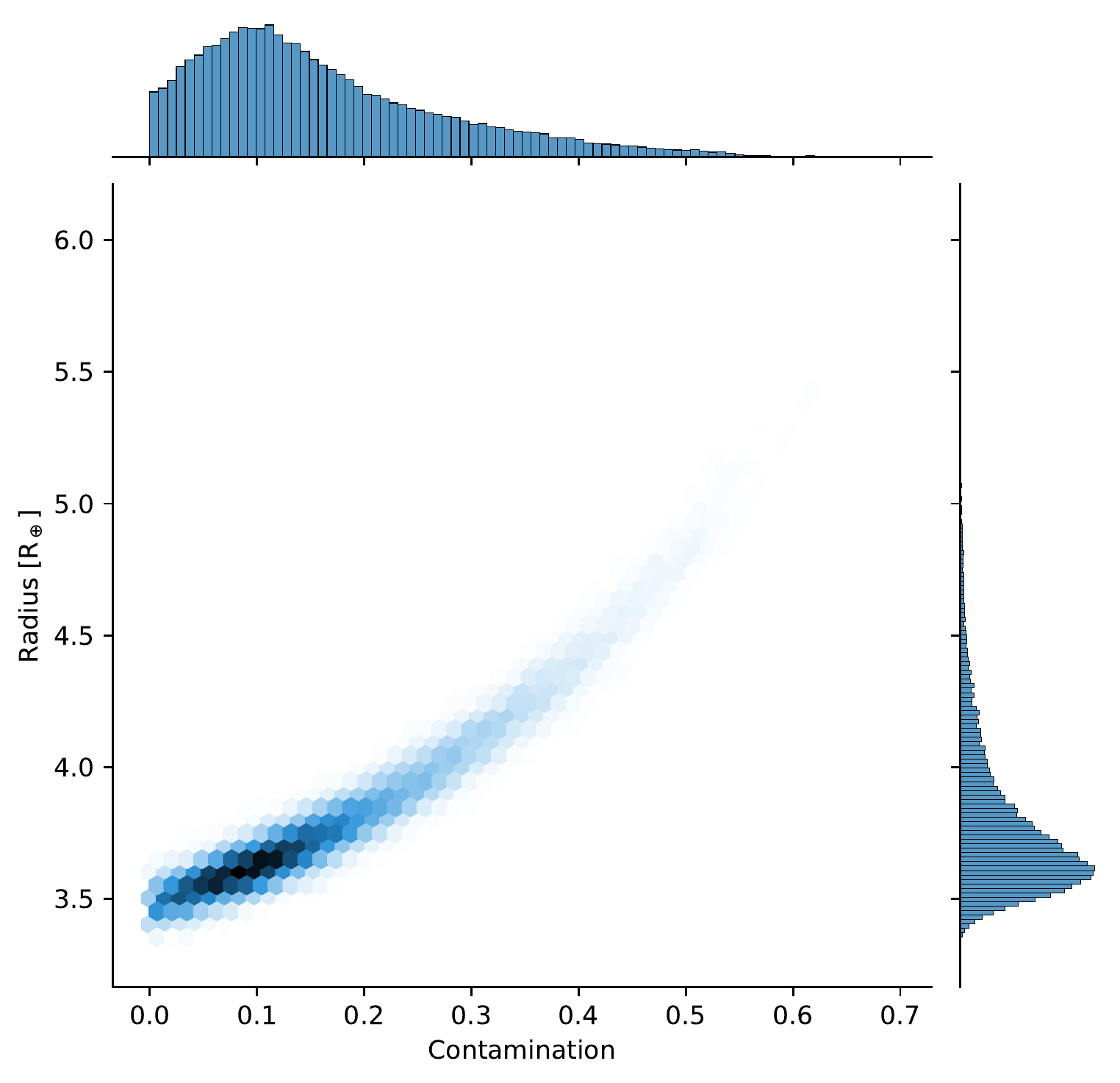}
    \caption{One-dimensional model parameter posterior histograms and model parameter covariances for \teff~of the host [H], \teff~ of the contaminant [C], impact parameter, and host stellar density against the planet-to-star radius ratio (\textit{top}) and flux contamination (\textit{middle}) from a secondary star.\\\textit{Bottom}: The corrected or ``true'' exoplanet radius as a function of the flux contamination from a secondary.}
    \label{fig:muscat_ror_cont}
\end{figure*}

We also perform a separate study of the MuSCAT2 data specifically, due to its simultaneous observations in the $g'$, $i'$, $r'$, and $z'$ bands to search for a possible chromatic variation of transit depth, and put constraints on any resulting contamination from a companion in the circum-primary scenario. The study is done using \texttt{PyTransit} \citep{Parviainen_2015} and follows the multicolor candidate validation approach described in  \citet{Parviainen_2019} and \citet{Parviainen_2020}.  With the assumption that TOI 620 b is a circum-primary planet, we look at what constraints can be placed on a secondary star present in the system. We are able to rule out any significant contamination of $>20\%$ in flux from companions of different spectral types, whereas stars of similar in spectral type to the host star are limited to brightness ratios $< 40\%$ relative to the host star.  The former is consistent with our SB2 analysis in ${\S}$\ref{sect:sb2_analysis} (and does not exclude the potential companions identified therein), and the latter is further constrained by our SED analysis in ${\S}$\ref{sect:bulk_star_props} since there is no significant over-luminosity of the primary.  In Figure \ref{fig:muscat_ror_cont}, we show posteriors and covariances for a set of model parameters (effective temperature of the host and contaminant stars, impact parameter, and host stellar density) against the ``true'' planet-to-star radius ratio, and the flux contamination ratio from the secondary star. We also show the effective planet radius -- after correcting for the flux contamination from a secondary -- as a function of the flux contamination; the planet remains roughly Neptune-sized, even in the presence of \rev{up} to 40\% flux contamination from a secondary. \rev{As a caveat to our analysis, the radius value and uncertainty presented in Table \ref{tab:TOI620} do not account for any inflation from a potential flux contaminant, since we only considered a single-star host in this model.}

\clearpage
\startlongtable
\begin{deluxetable*}{lcchhhhhhhhhhhhhhhhhhhhhhhhhhhhhhhh}
\tablecaption{Planetary parameters: Median values and 68\% confidence interval for TOI620, created using EXOFASTv2 commit number 7971a947}
\label{tab:TOI620}
\tablehead{\colhead{~~~Parameter} & \colhead{Units} & \multicolumn{28}{c}{Values}}
\startdata
\smallskip\\\multicolumn{2}{l}{Planetary Parameters:}&b\smallskip\\
~~~~$P$\dotfill &Period (days)\dotfill &$5.0988179^{+0.0000045}_{-0.0000046}$\\
~~~~$R_P$\dotfill &Radius (\re)\dotfill &$3.76\pm0.15$\\
~~~~$M_P$\dotfill &Predicted Mass$^{4}$ (\me)\dotfill &$15.4^{+5.5}_{-3.6}$\\
~~~~$T_C$\dotfill &Time of conjunction$^{5}$ (\bjdtdb)\dotfill &$2458518.00718^{+0.00093}_{-0.00083}$\\
~~~~$T_T$\dotfill &Time of minimum projected separation$^{6}$ (\bjdtdb)\dotfill &$2458518.00717^{+0.00050}_{-0.00051}$\\
~~~~$T_0$\dotfill &Optimal conjunction Time$^{7}$ (\bjdtdb)\dotfill &$2458992.19724^{+0.00078}_{-0.00069}$\\
~~~~$a$\dotfill &Semi-major axis (AU)\dotfill &$0.04825^{+0.00065}_{-0.00066}$\\
~~~~$i$\dotfill &Inclination (Degrees)\dotfill &$87.47^{+0.18}_{-0.30}$\\
~~~~$e$\dotfill &Eccentricity \dotfill &$0.22^{+0.28}_{-0.12}$\\
~~~~$\omega_*$\dotfill &Argument of Periastron (Degrees)\dotfill &$-84^{+54}_{-64}$\\
~~~~$T_{eq}$\dotfill &Equilibrium temperature$^{8}$ (K)\dotfill &$603.6^{+5.5}_{-5.3}$\\
~~~~$\tau_{\rm circ}$\dotfill &Tidal circularization timescale (Gyr)\dotfill &$14^{+12}_{-13}$\\
~~~~$K$\dotfill &RV semi-amplitude$^{4}$ (m/s)\dotfill &$8.9^{+3.6}_{-2.2}$\\
~~~~$R_P/R_*$\dotfill &Radius of planet in stellar radii \dotfill &$0.0627\pm0.0017$\\
~~~~$a/R_*$\dotfill &Semi-major axis in stellar radii \dotfill &$18.87^{+0.52}_{-0.51}$\\
~~~~$\delta$\dotfill &$\left(R_P/R_*\right)^2$ \dotfill &$0.00393^{+0.00021}_{-0.00020}$\\
~~~~$\delta_{\rm B}$\dotfill &Transit depth in B (fraction)\dotfill &$0.00257^{+0.00096}_{-0.0022}$\\
~~~~$\delta_{\rm I}$\dotfill &Transit depth in I (fraction)\dotfill &$0.00346^{+0.00036}_{-0.00065}$\\
~~~~$\delta_{\rm J}$\dotfill &Transit depth in J (fraction)\dotfill &$0.00357^{+0.00029}_{-0.00043}$\\
~~~~$\delta_{\rm R}$\dotfill &Transit depth in R (fraction)\dotfill &$0.00328^{+0.00047}_{-0.00098}$\\
~~~~$\delta_{\rm g'}$\dotfill &Transit depth in g' (fraction)\dotfill &$0.00293^{+0.00073}_{-0.0017}$\\
~~~~$\delta_{\rm i'}$\dotfill &Transit depth in i' (fraction)\dotfill &$0.0010^{+0.0015}_{-0.0022}$\\
~~~~$\delta_{\rm r'}$\dotfill &Transit depth in r' (fraction)\dotfill &$-0.0003^{+0.0019}_{-0.0027}$\\
~~~~$\delta_{\rm z'}$\dotfill &Transit depth in z' (fraction)\dotfill &$0.00295^{+0.00060}_{-0.0010}$\\
~~~~$\delta_{\rm TESS}$\dotfill &Transit depth in TESS (fraction)\dotfill &$0.00255^{+0.00075}_{-0.0011}$\\
~~~~$\tau$\dotfill &Ingress/egress transit duration (days)\dotfill &$0.0137\pm0.0019$\\
~~~~$T_{14}$\dotfill &Total transit duration (days)\dotfill &$0.0565^{+0.0013}_{-0.0015}$\\
~~~~$T_{FWHM}$\dotfill &FWHM transit duration (days)\dotfill &$0.0428^{+0.0013}_{-0.0015}$\\
~~~~$b$\dotfill &Transit Impact parameter \dotfill &$0.887^{+0.014}_{-0.017}$\\
~~~~$b_S$\dotfill &Eclipse impact parameter \dotfill &$0.66^{+0.11}_{-0.13}$\\
~~~~$\tau_S$\dotfill &Ingress/egress eclipse duration (days)\dotfill &$0.0061^{+0.0018}_{-0.0017}$\\
~~~~$T_{S,14}$\dotfill &Total eclipse duration (days)\dotfill &$0.0590^{+0.0021}_{-0.0062}$\\
~~~~$T_{S,FWHM}$\dotfill &FWHM eclipse duration (days)\dotfill &$0.0527^{+0.0021}_{-0.0066}$\\
~~~~$\delta_{S,2.5\mu m}$\dotfill &Blackbody eclipse depth at 2.5$\mu$m (ppm)\dotfill &$1.058^{+0.10}_{-0.095}$\\
~~~~$\delta_{S,5.0\mu m}$\dotfill &Blackbody eclipse depth at 5.0$\mu$m (ppm)\dotfill &$39.6^{+2.7}_{-2.5}$\\
~~~~$\delta_{S,7.5\mu m}$\dotfill &Blackbody eclipse depth at 7.5$\mu$m (ppm)\dotfill &$115.8^{+7.1}_{-6.7}$\\
~~~~$\rho_P$\dotfill &Density$^{4}$ (cgs)\dotfill &$1.59^{+0.57}_{-0.36}$\\
~~~~$logg_P$\dotfill &Surface gravity$^{4}$ \dotfill &$3.03^{+0.13}_{-0.11}$\\
~~~~$\Theta$\dotfill &Safronov Number \dotfill &$0.0242^{+0.0085}_{-0.0054}$\\
~~~~$\fave$\dotfill &Incident Flux (\fluxcgs)\dotfill &$0.0283^{+0.0018}_{-0.0047}$\\
~~~~$T_P$\dotfill &Time of Periastron (\bjdtdb)\dotfill &$2458516.5^{+2.2}_{-1.6}$\\
~~~~$T_S$\dotfill &Time of eclipse (\bjdtdb)\dotfill &$2458515.47^{+0.98}_{-0.87}$\\
~~~~$T_A$\dotfill &Time of Ascending Node (\bjdtdb)\dotfill &$2458516.62^{+0.51}_{-0.62}$\\
~~~~$T_D$\dotfill &Time of Descending Node (\bjdtdb)\dotfill &$2458519.41^{+0.68}_{-0.47}$\\
~~~~$V_c/V_e$\dotfill & \dotfill &$1.120^{+0.066}_{-0.073}$\\
~~~~$((1-R_P/R_*)^2-b^2)^{1/2}$\dotfill & \dotfill &$0.586^{+0.023}_{-0.021}$\\
~~~~$sign$\dotfill & \dotfill &$1.39^{+0.42}_{-0.51}$\\
~~~~$e\cos{\omega_*}$\dotfill & \dotfill &$0.00^{+0.30}_{-0.27}$\\
~~~~$e\sin{\omega_*}$\dotfill & \dotfill &$-0.147^{+0.081}_{-0.11}$\\
~~~~$M_P\sin i$\dotfill &Minimum mass$^{4}$ (\me)\dotfill &$15.4^{+5.5}_{-3.6}$\\
~~~~$M_P/M_*$\dotfill &Mass ratio$^{4}$ \dotfill &$0.000080^{+0.000029}_{-0.000019}$\\
~~~~$d/R_*$\dotfill &Separation at mid transit \dotfill &$20.0^{+1.7}_{-2.3}$\\
~~~~$P_T$\dotfill &A priori non-grazing transit prob \dotfill &$0.0468^{+0.0060}_{-0.0037}$\\
~~~~$P_{T,G}$\dotfill &A priori transit prob \dotfill &$0.0530^{+0.0068}_{-0.0041}$\\
~~~~$P_S$\dotfill &A priori non-grazing eclipse prob \dotfill &$0.0599^{+0.022}_{-0.0056}$\\
~~~~$P_{S,G}$\dotfill &A priori eclipse prob \dotfill &$0.0680^{+0.025}_{-0.0065}$\\
\smallskip\\\multicolumn{2}{l}{Wavelength Parameters:}&B&I&J&R&g'&i'&r'&z'&TESS\smallskip\\
~~~~$u_{1}$\dotfill &linear limb-darkening coeff \dotfill &$0.68^{+0.47}_{-0.44}$&$0.28^{+0.32}_{-0.20}$&$0.21^{+0.26}_{-0.15}$&$0.38^{+0.38}_{-0.27}$&$0.55^{+0.49}_{-0.37}$&$1.07^{+0.26}_{-0.37}$&$1.24^{+0.21}_{-0.30}$&$0.54\pm0.33$&$0.69^{+0.29}_{-0.35}$\\
~~~~$u_{2}$\dotfill &quadratic limb-darkening coeff \dotfill &$0.04^{+0.48}_{-0.44}$&$0.05^{+0.29}_{-0.23}$&$0.56^{+0.23}_{-0.32}$&$0.06^{+0.36}_{-0.30}$&$0.28^{+0.40}_{-0.50}$&$-0.30^{+0.43}_{-0.27}$&$-0.48^{+0.34}_{-0.19}$&$0.11^{+0.40}_{-0.38}$&$-0.04^{+0.42}_{-0.33}$\\
\smallskip\\\multicolumn{2}{l}{Transit Parameters:}&TESS UT 2019-02-03 (TESS)&TESS UT 2019-02-08 (TESS)&TESS UT 2019-02-13 (TESS)&TESS UT 2019-02-23 (TESS)&NGTS UT 2019-04-20 (R)&LCO UT 2019-04-20 (z')&TMMT UT 2019-04-25 (I)&NGTS UT 2019-06-10 (R)&MuSCAT2 UT 2020-01-16 (g')&MuSCAT2 UT 2020-01-16 (i')&MuSCAT2 UT 2020-01-16 (r')&MuSCAT2 UT 2020-01-16 (z')&KeplerCam UT 2020-01-26 (B)&MuSCAT2 UT 2020-03-02 (g')&MuSCAT2 UT 2020-03-02 (i')&MuSCAT2 UT 2020-03-02 (r')&MuSCAT2 UT 2020-03-02 (z')&MuSCAT2 UT 2020-04-16 (g')&MuSCAT2 UT 2020-04-16 (i')&MuSCAT2 UT 2020-04-16 (r')&MuSCAT2 UT 2020-04-16 (z')&LCRO UT 2020-11-27 (i')&MuSCAT2 UT 2021-01-07 (i')&MuSCAT2 UT 2021-01-07 (r')&MuSCAT2 UT 2021-01-07 (z')&TESS UT 2021-02-11 (TESS)&TESS UT 2021-02-16 (TESS)&TESS UT 2021-02-27 (TESS)&ExTrA UT 2021-03-04 (J)&TESS UT 2021-03-04 (TESS)&ExTrA UT 2021-04-13 (J)&ExTrA UT 2021-04-19 (J)&ExTrA UT 2021-06-03 (J)\smallskip\\
~~~~$\sigma^{2}$\dotfill &Added Variance \dotfill &$-0.00000039^{+0.00000012}_{-0.00000011}$&$-0.00000018^{+0.00000012}_{-0.00000011}$&$-0.00000032^{+0.00000011}_{-0.00000010}$&$-0.00000010^{+0.00000013}_{-0.00000012}$&$-0.0000276^{+0.0000011}_{-0.0000010}$&$0.00000115^{+0.00000024}_{-0.00000021}$&$-0.00000225^{+0.00000044}_{-0.00000037}$&$-0.0000571^{+0.0000016}_{-0.0000014}$&$-0.00000169^{+0.00000035}_{-0.00000030}$&$-0.00000153^{+0.00000015}_{-0.00000013}$&$-0.00000039^{+0.00000020}_{-0.00000017}$&$-0.000001023^{+0.00000010}_{-0.000000091}$&$0.00000367^{+0.00000080}_{-0.00000069}$&$-0.00000147^{+0.00000039}_{-0.00000033}$&$-0.00000091^{+0.00000023}_{-0.00000019}$&$-0.00000162^{+0.00000026}_{-0.00000022}$&$-0.00000121^{+0.00000017}_{-0.00000015}$&$-0.00000192^{+0.00000055}_{-0.00000047}$&$-0.00000223^{+0.00000042}_{-0.00000035}$&$-0.00000094^{+0.00000037}_{-0.00000032}$&$-0.00000139^{+0.00000043}_{-0.00000035}$&$-0.0000018^{+0.0000019}_{-0.0000015}$&$-0.00000102^{+0.00000017}_{-0.00000014}$&$-0.00000141^{+0.00000020}_{-0.00000017}$&$-0.00000151^{+0.00000019}_{-0.00000017}$&$0.00000057^{+0.00000014}_{-0.00000013}$&$0.00000037\pm0.00000012$&$0.00000034^{+0.00000012}_{-0.00000011}$&$-0.00000024^{+0.00000033}_{-0.00000029}$&$0.00000021^{+0.00000011}_{-0.00000010}$&$-0.00000053^{+0.00000029}_{-0.00000025}$&$0.00000002^{+0.00000033}_{-0.00000028}$&$-0.00000002^{+0.00000023}_{-0.00000019}$\\
~~~~$F_0$\dotfill &Baseline flux \dotfill &$0.999887\pm0.000063$&$0.999887^{+0.000063}_{-0.000064}$&$1.000085^{+0.000061}_{-0.000062}$&$0.999681\pm0.000065$&$0.99843\pm0.00020$&$0.99969^{+0.00013}_{-0.00014}$&$1.00074\pm0.00017$&$0.99998\pm0.00038$&$0.99991\pm0.00014$&$1.000006\pm0.000093$&$1.00030\pm0.00011$&$1.000212\pm0.000078$&$1.00063^{+0.00021}_{-0.00020}$&$1.00013\pm0.00017$&$1.00015\pm0.00013$&$0.99986\pm0.00014$&$1.00016\pm0.00011$&$0.99966\pm0.00018$&$0.99963\pm0.00016$&$0.99982\pm0.00015$&$0.99983\pm0.00016$&$1.00071\pm0.00035$&$0.99999\pm0.00011$&$0.99986\pm0.00012$&$1.00000\pm0.00011$&$1.000102\pm0.000068$&$1.000281\pm0.000064$&$1.000454\pm0.000063$&$1.00000\pm0.00013$&$0.999968\pm0.000061$&$0.99999\pm0.00015$&$0.99999\pm0.00015$&$0.99999\pm0.00012$\\
~~~~$C_{0}$\dotfill &Additive detrending coeff \dotfill &--&--&--&--&$0.0001^{+0.0020}_{-0.0021}$&--&$0.00061\pm0.00059$&$-0.0048\pm0.0015$&$0.00064\pm0.00077$&$0.00076\pm0.00049$&$0.00032^{+0.00056}_{-0.00055}$&$0.00037\pm0.00045$&--&$-0.00002^{+0.00060}_{-0.00061}$&$0.00052\pm0.00039$&$-0.00039\pm0.00039$&$0.00062\pm0.00035$&$-0.00169^{+0.00069}_{-0.00070}$&$-0.00203\pm0.00055$&$-0.00080^{+0.00053}_{-0.00054}$&$-0.00064\pm0.00056$&$-0.0006\pm0.0010$&$-0.00002\pm0.00033$&$0.00027\pm0.00033$&$-0.00004\pm0.00027$&--&--&--&--&--&--&--&--\\
~~~~$M_{0}$\dotfill &Multiplicative detrending coeff \dotfill &--&--&--&--&$0.00010^{+0.00065}_{-0.00067}$&$-0.00037\pm0.00026$&--&$0.0036\pm0.0010$&$0.00004^{+0.00042}_{-0.00043}$&$0.00032^{+0.00027}_{-0.00026}$&$0.00016\pm0.00026$&$0.00023\pm0.00020$&$0.00037\pm0.00060$&$-0.00006\pm0.00068$&$0.00021\pm0.00051$&$0.00014^{+0.00039}_{-0.00040}$&$0.00005\pm0.00029$&$-0.00027\pm0.00057$&$-0.00035\pm0.00050$&$-0.00006\pm0.00044$&$-0.00015^{+0.00046}_{-0.00047}$&--&$-0.00011^{+0.00037}_{-0.00038}$&$0.00009^{+0.00045}_{-0.00044}$&$-0.00010\pm0.00033$&--&--&--&--&--&--&--&--\\
~~~~$M_{1}$\dotfill &Multiplicative detrending coeff \dotfill &--&--&--&--&--&--&--&--&$-0.00100\pm0.00078$&$-0.00095^{+0.00052}_{-0.00053}$&$0.00084\pm0.00061$&$0.00030\pm0.00044$&--&$0.00008^{+0.00069}_{-0.00068}$&$-0.00020\pm0.00046$&$0.00005\pm0.00049$&$-0.00012^{+0.00045}_{-0.00046}$&$0.00082^{+0.00084}_{-0.00083}$&$0.00109\pm0.00068$&$0.00039^{+0.00066}_{-0.00065}$&$0.00017^{+0.00064}_{-0.00065}$&--&$-0.00010\pm0.00027$&$-0.00017\pm0.00038$&$-0.00008\pm0.00036$&--&--&--&--&--&--&--&--\\
\enddata
\tablenotetext{}{See Table 3 in \citet{Eastman_2019} for a detailed description of all parameters}
\tablenotetext{4}{Uses measured radius and estimated mass from \citet{Chen_2016}}
\tablenotetext{5}{Time of conjunction is commonly reported as the "transit time"}
\tablenotetext{6}{Time of minimum projected separation is a more correct "transit time"}
\tablenotetext{7}{Optimal time of conjunction minimizes the covariance between $T_C$ and Period}
\tablenotetext{8}{Assumes no albedo and perfect redistribution}
\end{deluxetable*}

\subsection{\rev{Circum-Secondary Scenario}}
\label{sect:csec_transit_main}

\begin{figure}
    \centering
    \includegraphics[width=\columnwidth]{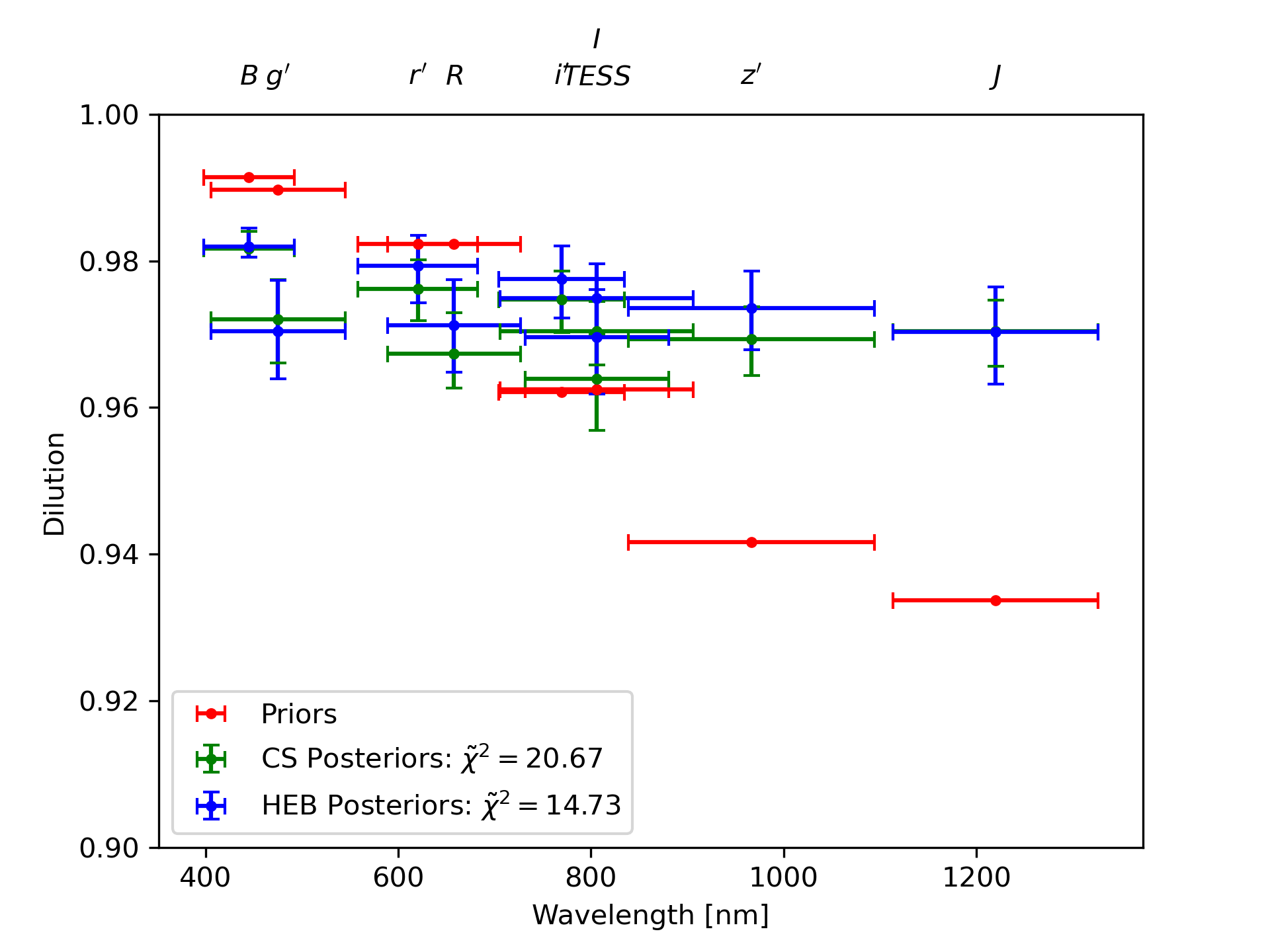}
    \caption{\rev{Our MCMC dilution priors (red) vs. posteriors of the secondary for the circum-secondary model (green) and the HEB model (blue). We see a prior distribution that decreases with wavelength as expected for flux contamination from a hotter primary star, whereas the posteriors are mostly flat and become inconsistent at short and long wavelengths.}}
    \label{fig:dilution_p}
\end{figure}

\rev{We now model the transit data under the assumption of a circum-secondary transiting planet, using posteriors from our iSHELL SB2 analysis as priors for the star and dilution terms in \texttt{EXOFASTv2}.  The dilution priors for each filter are made under the approximations that \tess\ $\approx I$ $\approx$ $i^\prime$, $R \approx r^\prime$, and $B \approx g^\prime$, and we set lower bounds on the dilution to prevent them from being driven to 0 and recovering the circum-primary solution.  We exclude SED modeling with \texttt{EXOFASTv2} since the broadband apparent magnitudes will be dominated by the flux from the primary.  Qualitatively, the dilution of the secondary by the primary is greater at bluer wavelengths due to their relative effective temperatures and photometric colors expected from \citet{Pecaut_2013}; the net result would be a shallower transit in the blue than in the red. However, we find our converged median MCMC dilution posteriors to be relatively flat, and thus inconsistent with the expected dilution curve for a cooler mid-M dwarf secondary, as shown in Figure \ref{fig:dilution_p}.  By rerunning the MCMC with the dilution parameters fixed at their priors, we recover transit depths that are inconsistent with the data, most notably in the B band (Figure \ref{fig:csec_b_transit}). The priors, posteriors, and transit times are listed in Appendix section \ref{sect:csec_transit}, along with an additional analysis of the MuSCAT data.}

\subsection{\rev{HEB Scenario}}
\label{sect:heb_transit_main}

\begin{table}[]
    \centering
    \begin{tabular}{lcccc}
    \hline
    Model & $\ln\mathcal{L}$ & $\Delta$AICc & $\Delta$BIC & $N$ free \\
    \hline
    \hline
    Circum-primary & 45015.895 & 0.000 & 0.000 & 157 \\
    CS & 45001.005 & 29.781 & 29.700 & 157 \\
    CS (fixed) & 44888.868 & 235.801 & 164.067 & 148 \\
    HEB & 44961.968 & 107.856 & 107.782 & 157 \\
    HEB (fixed) & 44854.627 & 304.282 & 232.555 & 148 \\
    \hline
    \end{tabular}
    \caption{\rev{A model comparison test between the transiting circum-primary, circum-secondary (CS) and HEB models.  $N$ free is the number of free model parameters, the vast majority of which are light curve and associated detrending parameter normalization constants for each ground-based data set.  The models marked as ``fixed'' are the ones where the dilution terms are held fixed to their prior values.}}
    \label{tab:modelcomp_exofast}
\end{table}

\rev{We next model the transit data under the assumption of an HEB scenario, deriving our priors in the same way as the circum-secondary analysis. This scenario comes with the added complication of not knowing whether we observe both primary and secondary eclipses, meaning the true period could be 5.09881 days or twice that, at 10.19762 days.  As such, we create three models corresponding to 1) the 5.09 day period, seeing only even or odd transits, 2) the 10.20 day period's even transits, and 3) the 10.20 day period's odd transits. Our posteriors only converge in case 1, indicating we can rule out the 10.20 day period.  We still expect the dilution posteriors to be shallower in the blue compared to the red, but again we find that our median MCMC dilution posteriors are flat and inconsistent with our expectations, as in Figure \ref{fig:dilution_p}.  By rerunning the MCMC with the dilution parameters fixed at their priors, we again recover transit depths that are inconsistent with the data (Figure \ref{fig:heb_b_transit}). The priors, posteriors, and transit times are listed in Appendix section \ref{sect:heb_transit}.  We also perform a model comparison test by calculating the $\ln \mathcal{L}$, AICc, and Bayesian Information Criterion (BIC) for the circum-primary, circum-secondary, and HEB models in Table \ref{tab:modelcomp_exofast}, which shows all models other than the circum-primary are ruled out to relative probabilities (compared to the circum-primary model) of $P_{\rm rel} = \exp(-\Delta {\rm AICc}/2) \leqslant 3.41 \times 10^{-7}$.}

\section{RV Analysis \& Results}
\label{sect:rv_analysis}

In this section, we present RV data analysis and results. Similarly to the transit analysis, we present results from three separate analyses of the TOI 620 RV data assuming the planet is circum-primary (${\S}$\ref{sec:primary_analysis}), circum-secondary (${\S}$\ref{sect:csec_rv}), or an HEB (${\S}$\ref{sect:heb_rv}), the latter two of which are detailed in the Appendix and summarized in ${\S}$\ref{sect:discussion}.

\subsection{Keplerian RV Analyses \& Results}
\label{sec:primary_analysis}

\begin{table*}
\centering
\begin{tabular}{ccccc}
    \hline
    Parameter [units] & Initial Value ($P_{0}$) & Priors & MAP Value & MCMC Posterior \\
    \hline
    \hline
    $P_b$ [days] & 5.09881 & \faLock & -- & -- \\
    $T_{C,b}$ [days] & 2458518.007 & \faLock & -- & -- \\
    $e_b$ & $10^{-5}$ & $\mathcal{U}(0, 0.5)$ & $10^{-5}$ & $0.26^{+0.18}_{-0.20}$ \\
    $\omega_b$ & $\pi/2$ & $\mathcal{U}(-\pi, \pi)$ & $\pi/2$ & $-0.10^{+1.91}_{-1.63}$ \\
    $K_b$ [m s$^{-1}$] & 5 & $\mathcal{U}(0, \infty)$ & $8.97 \times 10^{-6}$ & $0.33^{+0.50}_{-0.25}$ \\
    \hline
    $\gamma_{\rm iSHELL}$ [m s$^{-1}$] & $-8.448$  & $\mathcal{U}(P_0 \pm 100)$ & -1.95 & $3.83^{+3.08}_{-3.05}$ \\
    $\gamma_{\rm CARMENES-Vis}$ [m s$^{-1}$] & $-0.783$  & $\mathcal{U}(P_0 \pm 100)$ & -0.78 & $-1.60^{1.96}_{-2.00}$ \\
    $\gamma_{\rm CARMENES-NIR}$ [m s$^{-1}$] & $-1.510$  & $\mathcal{U}(P_0 \pm 100)$ & -1.51 & $-2.25^{+3.88}_{-3.76}$ \\
    $\gamma_{\rm NEID}$ [m s$^{-1}$] & $-1.525$ & $\mathcal{U}(P_0 \pm 100)$ & -1.53 & $-0.25^{+1.46}_{-1.50}$ \\
    $\gamma_{\rm MAROON-X-blue}$ [m s$^{-1}$] & $0.392$  & $\mathcal{U}(P_0 \pm 100)$ & -3.89 & $-5.89^{+1.33}_{-1.27}$ \\
    $\gamma_{\rm MAROON-X-red}$ [m s$^{-1}$] & $-1.769$  & $\mathcal{U}(P_0 \pm 100)$ & -6.72 & $-7.95^{+1.50}_{-1.53}$ \\
    \hline
    $\sigma_{\rm iSHELL}$ [m s$^{-1}$] & 5 & $\mathcal{N}(P_0, 2)$; $\mathcal{U}(10^{-5}, 100)$ & 13.42  & $13.44^{+1.11}_{-1.07}$ \\
    $\sigma_{\rm CARMENES-Vis}$ [m s$^{-1}$] & 5 & $\mathcal{N}(P_0, 2)$; $\mathcal{U}(10^{-5}, 100)$ & 3.89  & $4.34^{+1.57}_{-1.28}$ \\
    $\sigma_{\rm CARMENES-NIR}$ [m s$^{-1}$] & 5 & $\mathcal{N}(P_0, 2)$; $\mathcal{U}(10^{-5}, 100)$ & 5.00 & $5.35^{+1.88}_{-1.97}$ \\
    $\sigma_{\rm NEID}$ [m s$^{-1}$] & $5$ & $\mathcal{N}(P_0, 2)$; $\mathcal{U}(10^{-5}, 100)$ & 5.00 & $5.39^{+1.08}_{-0.95}$\\
    $\sigma_{\rm MAROON-X-blue}$ [m s$^{-1}$] & 1 & $\mathcal{N}(P_0, 2)$; $\mathcal{U}(10^{-5}, 100)$ & 1.00  & $1.61^{+1.04}_{-0.95}$ \\
    $\sigma_{\rm MAROON-X-red}$ [m s$^{-1}$] & 1 & $\mathcal{N}(P_0, 2)$; $\mathcal{U}(10^{-5}, 100)$ & 2.92  & $3.32^{+0.89}_{-0.68}$ \\
    \hline
    $\dot{\gamma}$ [\msday] & $10^{-5}$ & $\mathcal{U}(-50, 50)$ & 0.06 & $0.08^{+0.01}_{-0.01}$ \\  
    \hline
\end{tabular}
\caption{\rev{The model parameters and prior distributions used in our RV model that considers the transiting b planet and the linear $\dot{\gamma}$ trend, as well as the recovered MAP fit and MCMC posteriors. \faLock\ indicates the parameter is fixed.  $\mathcal{N}(\mu, \sigma)$ signifies a Gaussian prior with mean $\mu$ and standard deviation $\sigma$.  $\mathcal{U}(\ell, r)$ signifies a uniform prior with left bound $\ell$ and right bound $r$.}}
\label{tab:bc_priors}
\end{table*}

\begin{table}[]
    \centering
    \begin{tabular}{cccccc}
        \hline
        Planets & $\ln \mathcal{L}$ & $\Delta$ AICc & $\Delta$ BIC & N free & $\chi^2_{\mathrm{red}}$ \\
        \hline
        \hline
        None & -278.95 & 0.00 & 0.00 & 13 & 1.79 \\
        b & -282.22 & 15.89 & 19.54 & 16 & 1.97 \\
        \hline
    \end{tabular}
    \caption{\rev{A model comparison test for planet b, showing that the most favorable model includes no planets.}}
    \label{tab:bc_modelcomp}
\end{table}

Each planet is modeled in \texttt{pychell} \citep{Cale2019} with a standard basis set of five orbital parameters: the period $P$, time of conjunction $T_C$, eccentricity $e$, argument of periastron $\omega$, and RV semi-amplitude $K$, with subscripts denoting the planet each parameter is associated with (in this case, planet b and a candidate c).  We also include for each instrument an absolute RV offset term ($\gamma$), and a jitter term ($\sigma$) which quantifies RV white noise not accounted for by any modeled planet(s), stellar activity, or under-estimated RV precision systematics. The $\dot{\gamma}$ term models a linear trend in the RVs over the entire baseline of observations and is indicative of a very long period companion for which we have only captured a small portion of the orbit's phase, so it can be approximated as a line; we do not find evidence for a jerk $\ddot{\gamma}$ and exclude it from our model.  From our \texttt{EXOFASTv2} analysis, we \rev{found $P_b = 5.0988179^{+0.0000045}_{-0.0000046}$ and $T_C = 2458518.00718^{+0.00093}_{-0.00083}$,} both of which have an uncertainty which are orders of magnitude finer than we can hope to resolve with the precision, sampling and time baseline of our RV measurements.  We thus fix both RV model parameters at these transit posterior values for all RV analyses in this paper to reduce the model parameter space. We exclude consideration of high-eccentricity $e>0.5$ orbital solutions, and we find no compelling evidence in our RV measurements obtained to date to support such a high eccentricity. We first present the analysis assuming a circum-primary planet, mirroring our analysis of the light curves in section \ref{sect:transit_analysis}.

\begin{figure}
    \centering
    \includegraphics[width=.47\textwidth]{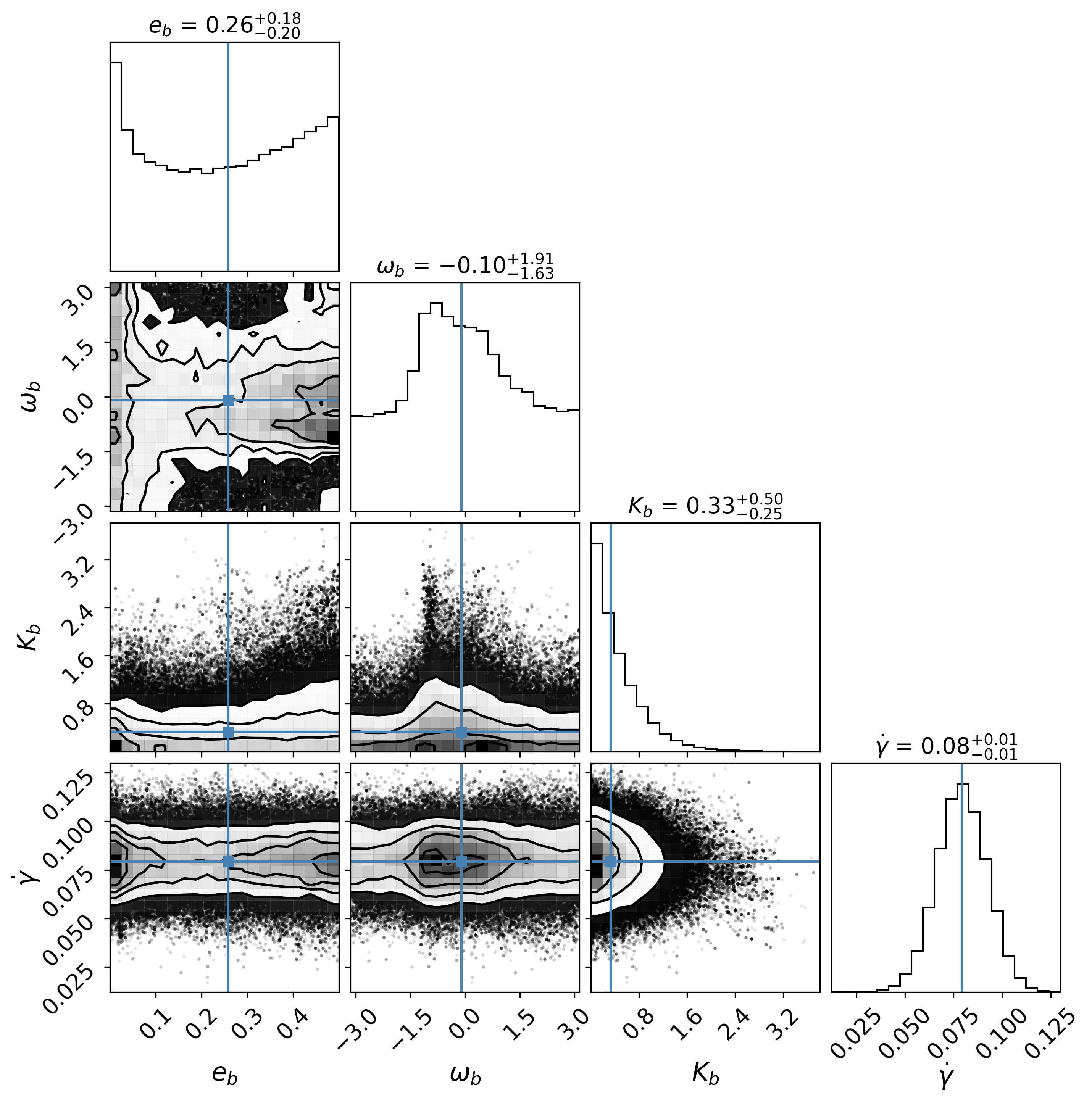}
    \caption{MCMC cornerplot of our eccentric ($e<0.5$) TOI 620 b circum-primary model with all RV datapoints included, showing the posterior distributions of each model parameter that we allowed to vary. The gamma offsets and jitter terms are not shown, as they are all uncorrelated and to a good approximation are ideal Gaussian distributions. We obtain similar posteriors for $K_b$ and $\dot{\gamma}$ for the assumption of a circular orbit.}
    \label{fig:bc_cornerplot}
\end{figure}

\begin{figure}
    \centering
    \includegraphics[width=.45\textwidth]{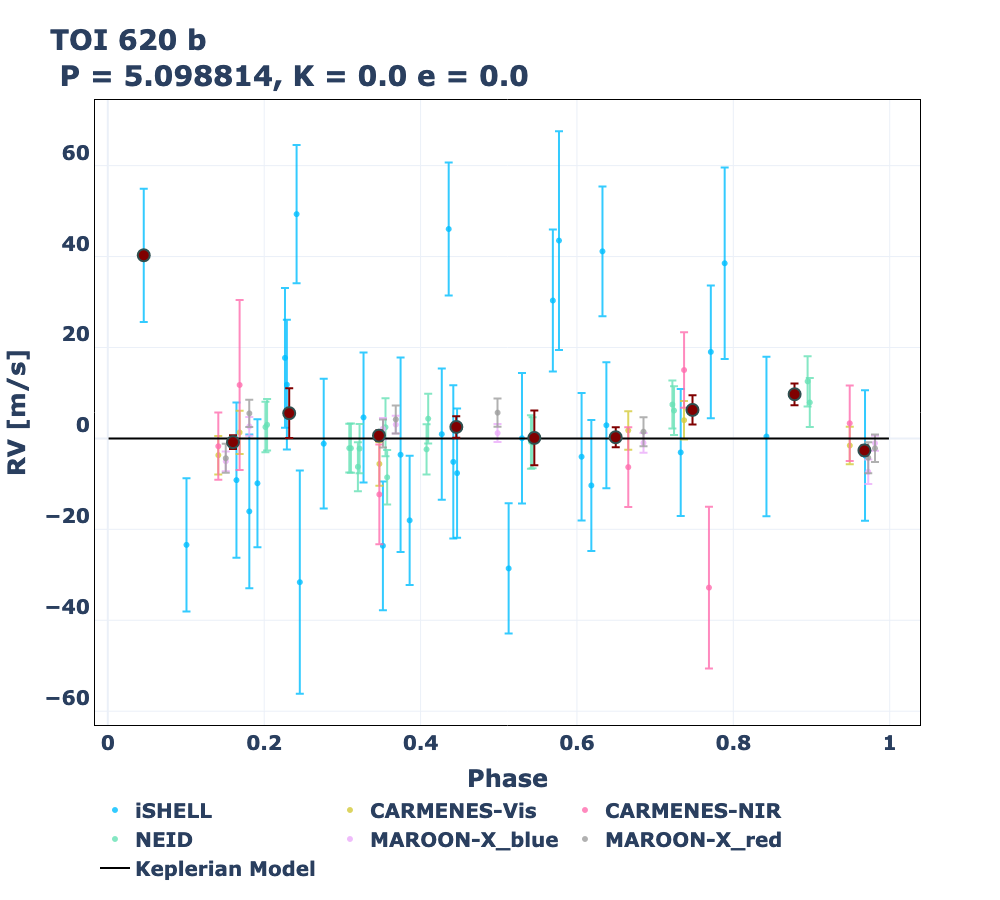}
    \caption{RV time-series phased to the period of b, with the black model representing the planet's MAP model fit. The maroon data points are binned RVs every 0.1 in phase. A careful inspection reveals that the RVs are marginally more consistent with a negative $K_b$, although in this circum-primary scenario we assume a prior of $K_b>0$.}
    \label{fig:bc_phasedrvs}
\end{figure}

\begin{figure*}
    \centering
    \includegraphics[width=\textwidth]{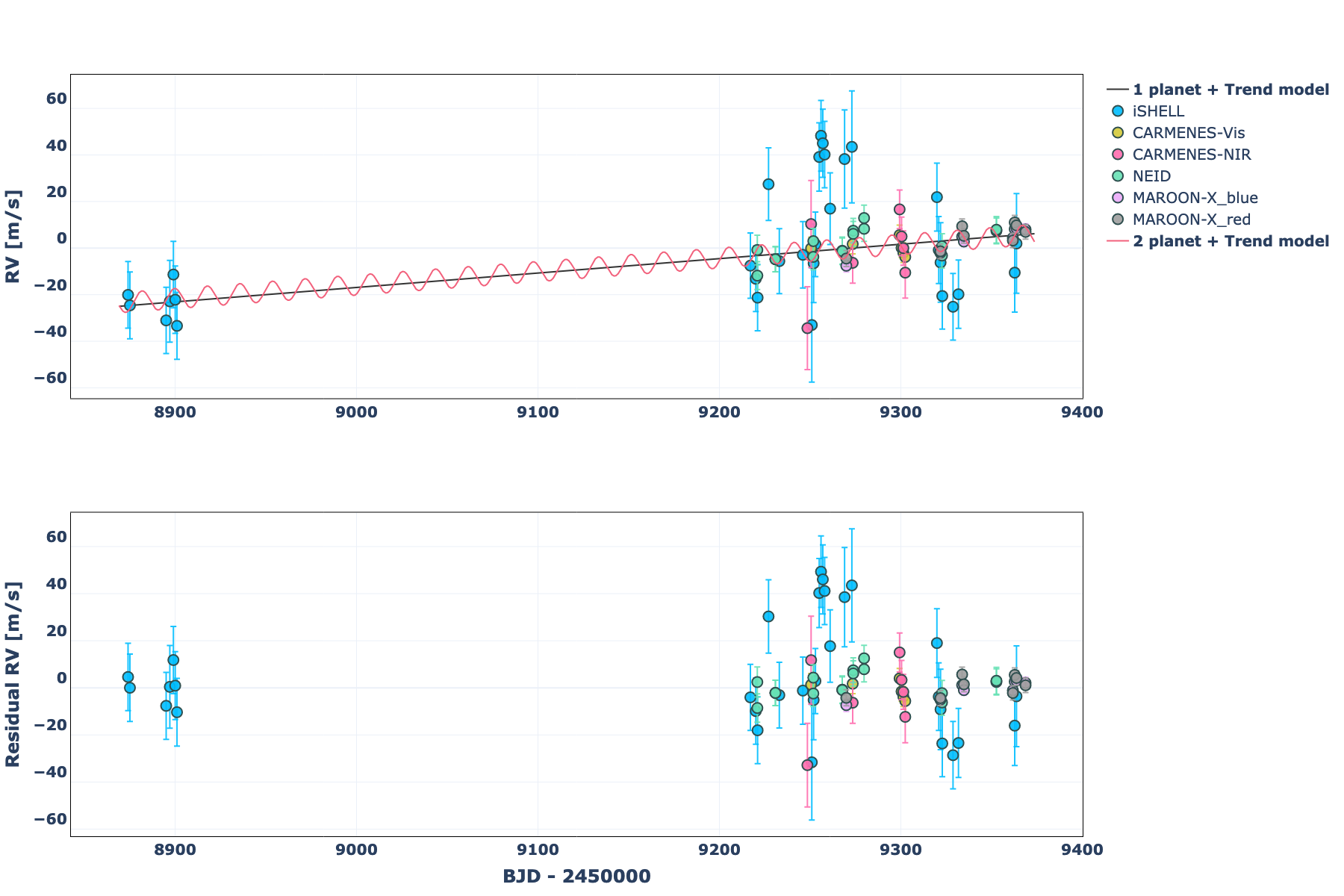}
    \caption{\rev{Full unphased RV time-series as a function of time, with the black dashed lines representing our 1-planet and 2-planet (${\S}$\ref{sect:planetc}) MAP models with the linear trend. The top plot shows the RVs for each instrument and errorbars over the full time baseline of observations, while the bottom plot shows the residuals (data $-$ model). The 1 and 2 planet models have slightly different ($<2 \ms$) RV offsets ($\gamma$) for each data set as enumerated in Tables \ref{tab:bc_priors} and \ref{tab:2p_priors} respectively, but only the 1 planet RV offsets are applied to the data as shown for clarity.}}
    \label{fig:bc_fullrvs}
\end{figure*}


The single-planet and linear trend RV model yields no recovery of the RV semi-amplitude for the transiting b planet, and the MAP value drops to 0.  The expected mass and RV semi-amplitude provided by the \texttt{EXOFASTv2} fit of the transit data (using the \citet{Chen_2016} mass-radius relationship) were \rev{15.4 \mearth and 8.9 \ms,} respectively, which we can robustly exclude.\footnote{\rev{This is supported by checking with the predicted values from MRExo \citep{2019ascl.soft12020K}.}} The RV models with MCMC median values are shown with our RV measurements in Figure \ref{fig:bc_fullrvs}, while in Figure \ref{fig:bc_phasedrvs} they are phased to the period of b. The MCMC chains (excluding $\gamma$ offsets and $\sigma$ jitter model parameter terms) are shown in Figure \ref{fig:bc_cornerplot}. \rev{The cornerplot shows in particular the well-behaved posterior of $\dot{\gamma}$ at $0.08 \pm 0.01$ \msday} and the nondetection on $K_b$. We get very similar results by assuming circular orbits. 

\rev{A model comparison with and without planet b, is shown} in Table \ref{tab:bc_modelcomp}  and show that our most favored model is that with no planets (e.g. a non-detection of the transiting planet). To confirm that there is not a single dataset in use that is diminishing our recovery of TOI 620 b, we run 5 separate fits, ignoring the $\dot{\gamma}$ term and the iSHELL data, and in each fit we remove a single dataset (other than iSHELL).  The MAP and MCMC recovered $K$ values are listed in Table \ref{tab:k_recovery} and show that there is not a single dataset that is significantly diminishing the recovery compared to the others.

Finally, we calculate a 5-$\sigma$ upper limit to the mass of TOI 620 b of $M_P \leqslant 7.1$ \mearth.
Using the radius of planet b from the \texttt{EXOFASTv2} analysis in ${\S}$\ref{sect:exofast}, \rev{$\rp = 3.76 \pm 0.15$ \re, we derive a 5-$\sigma$ upper limit to the density of TOI 620 b of $\rho_P \leqslant 0.74$ \rhocgs.} \rev{Of the 149 confirmed exoplanets from the NASA exoplanet archive with radii between 3--5 \re\ and with measurements for both a mass and radius, the average mass is $\approx 18.7$ \me\ and the minimum is 2.07 \me, putting our upper limit for TOI 620 b well below average and closer to the minimum.} Thus we can conclude \rev{that} TOI 620 b is a very low-density planet \rev{(see Figure \ref{fig:mass_radius})}.

\begin{table*}[]
    \centering
    \begin{tabular}{c|ccccc}
        Removed Dataset & CARMENES-Vis & CARMENES-NIR & MAROON-X blue & MAROON-X red & NEID \\
        \hline
        \hline
        MAP $K$ [\ms] & $3.07 \times 10^{-10}$ & $2.24 \times 10^{-9}$ & $1.91 \times 10^{-9}$ & $4.95 \times 10^{-9}$ & $3.36 \times 10^{-7}$ \\
        MCMC $K$ [\ms] & $0.85^{+1.14}_{-0.63}$ & $0.56^{+0.81}_{-0.42}$ & $0.41^{+0.60}_{-0.30}$ & $0.47^{+0.73}_{-0.35}$ & $2.06^{+1.48}_{-1.35}$ \\
        \rev{ $5\sigma$ $M$ [\me]} & \rev{11.88} & \rev{9.88} & \rev{11.58} & \rev{10.69} & \rev{15.74} \\
    \end{tabular}
    \caption{The MAP and MCMC recovered semi-amplitudes, 68\% confidence intervals\rev{, and 5$\sigma$ upper mass limits} of TOI 620 b in cases where we remove a single dataset (and iSHELL), fixing $\dot{\gamma}=0$, and allow for eccentricities up to 0.5.  In no combination of remaining data sets do we recover a detection for TOI 620 b, and in all cases we \rev{still} arrive at much lower velocity semi-amplitude upper-limits that expected from a mass-radius relation.}
    \label{tab:k_recovery}
\end{table*}

\section{Discussion}
\label{sect:discussion}

In this section, we first cover whether we can be confident that we are constraining the Doppler amplitude of TOI 620 b in ${\S}$\ref{sect:amplitude_constraint} by considering the origin of the transit signal, our RV noise, and systematics (including stellar activity and additional Keplerian RV signals).  Then in ${\S}$\ref{sect:massive_companions} we consider the evidence for and against the existence of additional massive companion(s) (whether stellar or not) to the primary TOI 620 host star. We summarize our findings on the age of TOI 620 in ${\S}$\ref{sect:discussion_age}.  Finally, in ${\S}$\ref{sect:density_implications} we discuss the nature and implications of an abnormally low-density TOI 620 b on its composition, evolution, and potential for future observations.

\subsection{Are we actually constraining the Doppler amplitude of TOI 620 b?}
\label{sect:amplitude_constraint}

The lack of recovery of a Doppler signal from TOI 620 b, despite seeing a clear transit signal, could be explained by one or more of the following:  First, the transit signal could be associated with an unresolved fainter companion to TOI 620 with a flux too low to impart a significant Doppler signal on the combined flux from the system, e.g. a circum-secondary transiting object, or a hierarchical eclipsing binary. Second, the Keplerian signal could be masked by an additional source of noise, e.g., either stellar ``jitter'' or apparent Doppler shifts produced by the combination of spectra from two dissimilar stars.  Third, TOI 620 could host multiple planets that each impart significant Keplerian signals with different periods and which cannot be disentangled with the limited number of RVs obtained so far.  Fourth, the planet could have a mass below our detection threshold.  We discuss these possibilities in turn.

\subsubsection{Excluding the Circum-Secondary and Hierarchical Eclipsing Binary scenarios with the chromatic transit light curves}

We rule out any nearby resolved stars as the source of the transit signal (${\S}$\ref{sect:false_positives}), and due to the high proper motion of the system, we can exclude blended background eclipsing binaries. From our \texttt{EXOFASTv2} analysis of all the ground and \tess\ transits, and also from an independent analysis considering only the MuSCAT2 transits, we can decisively rule out an unresolved companion with a circum-secondary (CS) transiting object (${\S}$\ref{sect:csec_transit_main}), and we can also rule out a hierarchical eclipsing binary (HEB) with a pair of low-mass eclipsing stellar companions (${\S}$\ref{sect:heb_transit_main}). \rev{Under the CS and HEB scenarios, we would observe a variable transit depth as a function of wavelength, and this is definitively excluded. The same result is obtained by analyzing only the simultaneous quad-band MuSCAT2 transit light curves.} 

\rev{This effect is readily observable in the $B$ band transits, which in these scenarios would produce shallower transits due to an increasing flux contrast between the primary and secondary at shorter wavelengths, but this is not observed. Similarly, the HEB model for the \tess\ LC produces an eclipse (``V'') shape that is inconsistent with the data, which shows steeper ingresses and egresses due to a smaller planet-to-star radius ratio.} While we do not explore the possibility that the secondary could be a hotter white dwarf, we do not see any evidence for UV excess or discrepant broadband photometric colors.



\subsubsection{Assessing the impact of stellar activity on our ability to recover a mass for TOI 620 b from our RVs}
\label{sect:ffp_analysis}

Stellar oscillations, photospheric convection, magnetic activity, and the rotation of starspots can impart net shifts in spectral lines which appear as systematic noise in RV time-series \citep{2020AJ....159..235L}.  There is no evidence that TOI 620 is a particularly active star, but even the moderate activity characteristic of middle-aged stars can obfuscate the Keplerian signals of low-mass planets, and require careful analysis to mitigate \citep[e.g.][]{Cale_2021,Plavchan_2015,Dumusque_2010,Dumusque_2011,Rajpaul_2015,Vanderburg_2016}.  One approach is to use photometry of the star (reflecting the extent, rotation, and migration of starspots) as a proxy for activity and its affect on the RV signals, and essentially regress and subtract the photometry-estimated RV variation.  In an attempt to recover the Keplerian signal of TOI 620 b, we perform a Gaussian Process (GP) regression of the RV time series using a quasi-periodic kernel (or covariance matrix) with hyper-parameters determined by an $FF^\prime$ analysis \citep{2012MNRAS.419.3147A} of the \tess\ light curve:

\begin{equation}
    \mathbf{K}(t_i,t_j) = \eta_\sigma^2 \exp \bigg[-\frac{\Delta t^2}{2\eta_\tau^2} - \frac{1}{2\eta_\ell^2}\sin^2\bigg(\pi \frac{\Delta t}{\eta_P} \bigg) \bigg]
\end{equation}

\noindent (where $\Delta t \equiv |t_j - t_i|$).  In this definition, $\eta_\sigma$ is the amplitude of the auto-correlation in the signal, $\eta_\tau$ is the mean star spot lifetime, $\eta_\ell$ is a smoothness parameter, and $\eta_P$ is the stellar rotation period.  We use the $FF^\prime$ analysis technique to estimate $\eta_\tau$, $\eta_\ell$, and $\eta_P$, but since the \tess\ photometry and RV observations were not contemporaneous, we must fit for the wavelength-dependent amplitudes $\eta_\sigma$ using the RV data alone.  That being said, the variation in the \tess\ light curve is around 0.2\%, and if we assume a $v\sin i \sim 2$ \kms, this leads to an expected RV semi-amplitude from stellar rotation of $\sim 4$ \ms.

\begin{figure*}
    \centering
    \includegraphics[width=\textwidth]{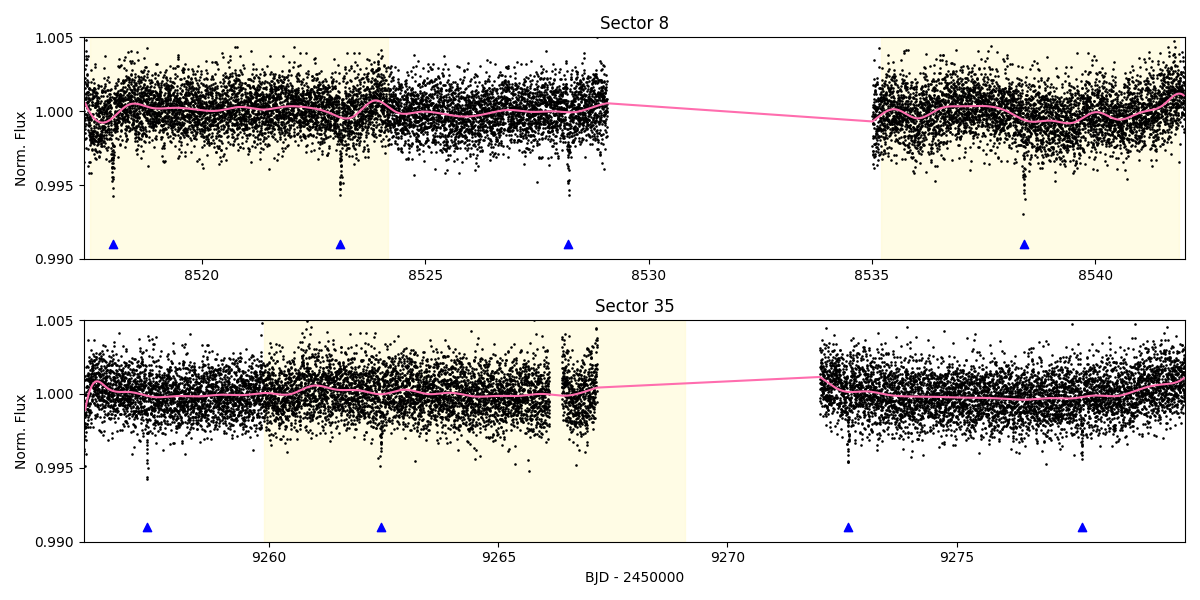}
    \caption{\rev{The \tess\ PDCSAP light curve of TOI 620 from sector 8 (top) and sector 35 (bottom) plotted as a function of time.  The times of transit for TOI 620 b are indicated with blue triangles.  Data gaps during sectors are due to data downlinks, momentum dumps, and other data artifacts and systematics such as scattered light.  The best-fit cubic spline polynomial is shown in pink, and is fitted to the data with the transits of TOI 620 b masked out.  Any interpolation in the data dump regions are thrown out. The yellow shaded regions show the predicted transit windows for TOI 620 c from our 2-planet MCMC analysis, showing no obvious presence of a more massive transiting companion to TOI 620 b (see ${\S}$\ref{sect:planetc})}} 
    \label{fig:lightcurve_ffp}
\end{figure*}

To derive an $FF^\prime$ RV stellar activity model, we first median-normalize the \tess\ PDC-SAP light curve and mask out the transits and the edges of the light curve data.  We then choose knots at evenly spaced intervals of 0.5 days (excluding any that happen to fall within the \tess\ data dump regions) and use \texttt{scipy} to fit a cubic spline to the data which is used to compute both $F$ and its first derivative, $F^\prime$, and multiply them together.  We also then arbitrarily divide the $FF^\prime$ curve by its standard deviation for normalization, since the amplitude will be fit for separately with the RV data.  The light curves with the best fit cubic splines in each sectors are shown in Figure \ref{fig:lightcurve_ffp}. We perform an MCMC fitting analysis with wide uniform priors on $\eta_\sigma \sim \mathcal{U}(0.03, 6)$, $\eta_\tau \sim \mathcal{U}(5, 2000)$, and $\eta_\ell \sim \mathcal{U}(0.1, 0.6)$, while we use a Gaussian prior for $\eta_P$ based on periodogram analysis at $\sim$8.9 days.  Instead of accounting for intrinsic error bars in the data, we also fit a jitter term $\sigma_{LC} \sim \mathcal{U}(10^{-5}, 0.2)$.  We find a strong doubly peaked posterior on $\eta_P$ \rev{at $8.99^{+0.03}_{-0.04}$ and $9.94^{+0.03}_{-0.04}$ days, the former of which is close to the Lomb-Scargle periodogram peak.}

Taking the results of the $FF^\prime$ analysis and using them as priors for GP hyperparameters, we incorporate separate quasi-periodic GPs for each RV instrument in our dataset, fixing the model values for $\eta_\tau$ and $\eta_P$, while allowing $\eta_\ell$ and $\eta_\sigma$ to vary.  Each instrument has an independent amplitude, $\eta_{\sigma, i}$, but the other three hyperparameters are shared between all instruments. A full summary of the priors used in these runs is shown in the Appendix in Table \ref{tab:gp_rv_priors}. The GP analysis is performed using the methods outlined in \citet{Cale_2021}. \rev{Using the 8.99-day $\eta_P$, we find well behaved posteriors, but all GP amplitudes are consistent with 0. We encounter similar results with the 9.94-day $\eta_P$. Thus, accounting for stellar activity} does not significantly improve our recovery of $K_b$. In other words, stellar activity is not degrading our recovery of a mass for TOI 620 b.

Another potential source of systematic error would be the superposition of two or more dis-similar spectra as a consequence of unresolved stellar companions; this is separate from the Keplerian signal imparted by any such companion. Contamination of the light from the primary by the secondary could result in blended spectral lines and  distorted line shapes. The changing relative Doppler shift of the spectra along the orbit could lead to time variation in the line shape and spurious variation in the measured RV.  A standard diagnosis of this effect is the line bisector \citep{2002A&A...392..215S}; but there is no evidence for such a systematic in our PRV spectra.  Additionally, this effect would exhibit a strongly chromatic effect on our RVs, more pronounced in red and NIR wavelengths than the blue visible, due to the  lower companion flux contrast towards the red. However, we do not observe any significant difference in the recovery of $K_b$ when excluding the red and NIR RVs in Table \ref{tab:k_recovery}.

\subsubsection{Are there additional short-orbital period Keplerian companions masking our RV recovery of TOI 620 b?}
\label{sect:planetc}

\begin{figure*}
    \centering
    \includegraphics[width=\textwidth]{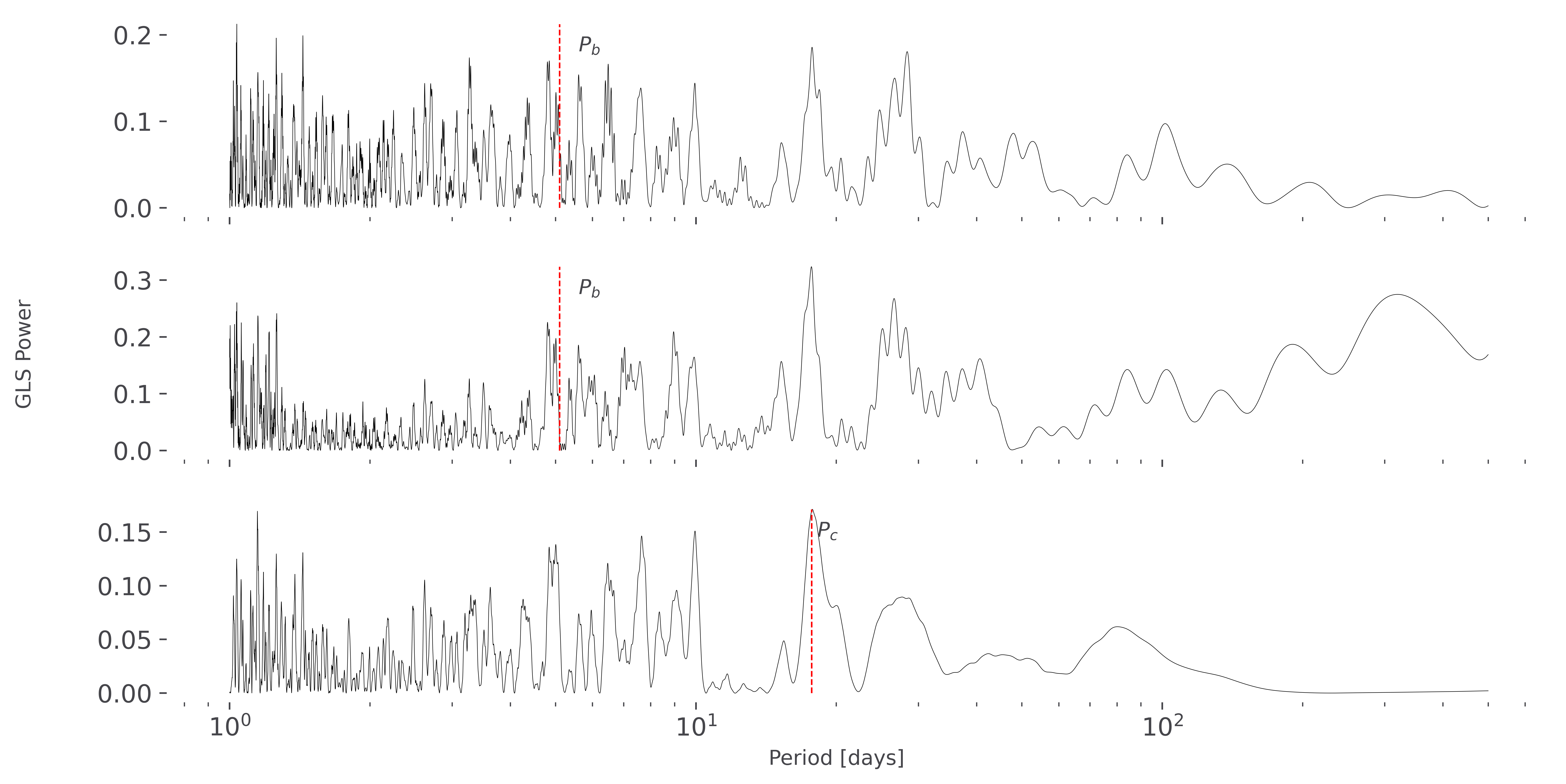}
    \caption{A series of GLS periodograms examining the signals in our TOI 620 RV time-series.  The horizontal-axes are all logarithmic.  All three panels depict a single planet search with a floating $T_C$. The top panel includes no extra planets, the middle panel models out TOI 620 b to search for a second planet, and the bottom panel models out both TOI 620 b and the linear RV trend to search for a second planet. The periods of b and the tentative c are marked with dashed red vertical lines in each panel.}
    \label{fig:gls_iterations}
\end{figure*}

Additional planets in the system might manifest themselves as low-significance signals in Fourier analysis of the RV time-series.  \rev{We search for such signals with a Generalized Lomb-Scargle (GLS) periodogram,} iteratively removing each identified signal in our model, and examining the residuals, \rev{shown in Figure \ref{fig:gls_iterations}. We confirm this analysis by creating a corresponding $\ln \mathcal{L}$ periodogram, which shows consistent results.}

\begin{figure*}
    \centering
    \includegraphics[width=.45\textwidth]{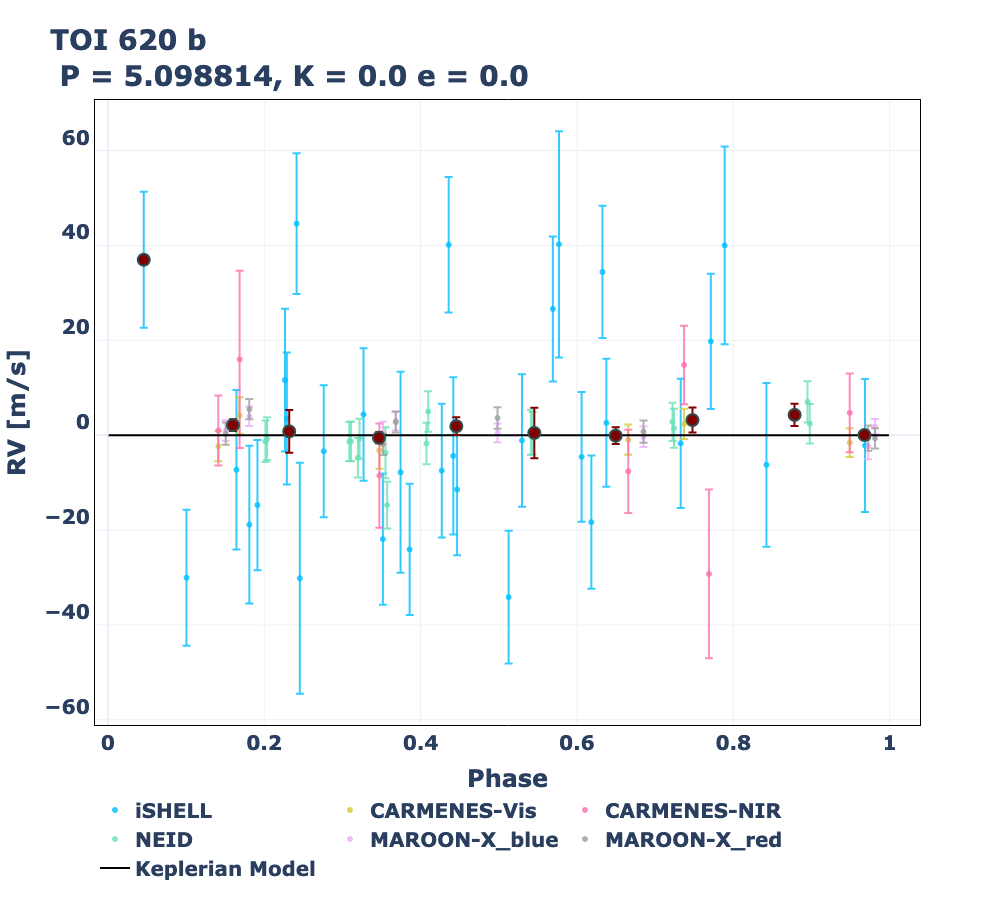}
    \includegraphics[width=.45\textwidth]{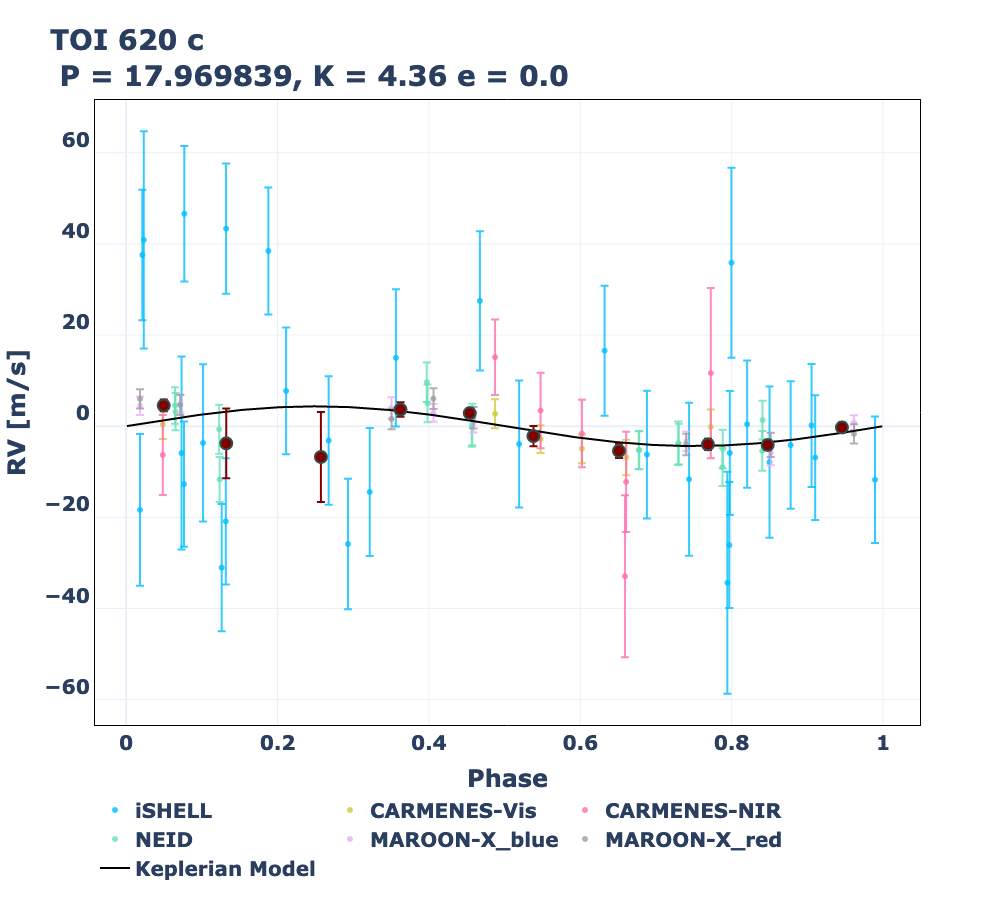}
    \caption{RV time-series plot for our two-planet model phased to the period of b (left) and c (right), with the black model representing the planet's MAP model fit. The maroon points are binned RVs every 0.1 in orbital phase.}
    \label{fig:2p_phasedrvs}
\end{figure*}

\begin{table*}
\centering
\begin{tabular}{ccccc}
    \hline
    Parameter [units] & Initial Value ($P_{0}$) & Priors & MAP Value & MCMC Posterior \\
    \hline
    \hline
    $P_b$ [days] & 5.09881 & \faLock &  -- & -- \\
    $T_{C,b}$ [days] & 2458518.007 & \faLock & -- & -- \\
    $e_b$ & $10^{-5}$ & $\mathcal{U}(0, 0.5)$ & $10^{-5}$ & $0.19^{+0.22}_{-0.18}$ \\
    $\omega_b$ & $\pi/2$ & $\mathcal{U}(-\pi, \pi)$ & 3.14 & $-0.41^{+2.45}_{-1.74}$ \\
    $K_b$ [m s$^{-1}$] & 5 & $\mathcal{U}(0, \infty)$ & $4.46 \times 10^{-3}$ & $0.32^{+0.48}_{-0.24}$ \\
    \hline
    $P_c$ [days] & 17.7 & $\mathcal{N}(P_0, 0.1)$ & 17.97 & $17.72^{+0.08}_{-0.07}$ \\
    $T_{C,c}$ [days] & 2458518.007 & $\mathcal{U}(P_0 \pm P_c/2)$ & 2458526.86 & $2458520.49^{+3.19}_{-3.47}$ \\
    $e_c$ & $10^{-5}$ & $\mathcal{U}(0, 1)$ & $1.02 \times 10^{-5}$ & $0.20^{+0.30}_{-0.18}$\\
    $\omega_c$ & $\pi/2$ & $\mathcal{U}(-\pi, \pi)$ & 3.14 & $1.45^{+0.76}_{-2.48}$ \\
    $K_c$ & 10 & $\mathcal{U}(0, \infty)$ & 4.36 & $4.74^{+1.14}_{-1.06}$ \\
    \hline
    $\gamma_{\rm iSHELL}$ [m s$^{-1}$] & $-8.448$  & $\mathcal{U}(P_0 \pm 100)$ & 0.70 & $2.48^{+2.98}_{-3.05}$ \\
    $\gamma_{\rm CARMENES-Vis}$ [m s$^{-1}$] & $-0.783$  & $\mathcal{U}(P_0 \pm 100)$ & 0.60 & $0.35^{+1.76}_{-1.79}$ \\
    $\gamma_{\rm CARMENES-NIR}$ [m s$^{-1}$] & $-1.510$  & $\mathcal{U}(P_0 \pm 100)$ & -1.51 & $-0.28^{+3.83}_{-3.87}$ \\
    $\gamma_{\rm NEID}$ [m s$^{-1}$] & $-1.525$ & $\mathcal{U}(P_0 \pm 100)$ & 1.40 & $0.99^{+1.25}_{-1.30}$ \\
    $\gamma_{\rm MAROON-X-blue}$ [m s$^{-1}$] & $-0.033$  & $\mathcal{U}(P_0 \pm 100)$ & -5.40 & $-5.92^{+1.30}_{-1.31}$ \\
    $\gamma_{\rm MAROON-X-red}$ [m s$^{-1}$] & $-1.770$  & $\mathcal{U}(P_0 \pm 100)$ & -6.87 & $-7.02^{+1.46}_{-1.54}$ \\
    \hline
    $\sigma_{\rm iSHELL}$ [m s$^{-1}$] & 5 & $\mathcal{N}(P_0, 2)$; $\mathcal{U}(10^{-5}, 100)$ & 13.04 & $13.32^{+1.16}_{-1.10}$ \\
    $\sigma_{\rm CARMENES-Vis}$ [m s$^{-1}$] & 5 & $\mathcal{N}(P_0, 2)$; $\mathcal{U}(10^{-5}, 100)$ & 2.69 & $3.20^{+1.78}_{-1.49}$ \\
    $\sigma_{\rm CARMENES-NIR}$ [m s$^{-1}$] & 5 & $\mathcal{N}(P_0, 2)$; $\mathcal{U}(10^{-5}, 100)$ & 5.00 & $5.08^{+1.94}_{-1.92}$ \\
    $\sigma_{\rm NEID}$ [m s$^{-1}$] & $5$ & $\mathcal{N}(P_0, 2)$; $\mathcal{U}(10^{-5}, 100)$ & 3.64 & $4.17^{+1.11}_{-0.95}$ \\
    $\sigma_{\rm MAROON-X-blue}$ [m s$^{-1}$] & 1 & $\mathcal{N}(P_0, 2)$; $\mathcal{U}(10^{-5}, 100)$ & 0.97 & $0.79^{+0.86}_{-0.55}$\\
    $\sigma_{\rm MAROON-X-red}$ [m s$^{-1}$] & 1 & $\mathcal{N}(P_0, 2)$; $\mathcal{U}(10^{-5}, 100)$ & 1.98 & $2.68^{+0.82}_{-0.65}$ \\
    \hline
    $\dot{\gamma}$ [\msday] & $10^{-5}$ & $\mathcal{U}(-50, 50)$ & 0.06 & $0.06^{+0.01}_{-0.01}$ \\  
    \hline
\end{tabular}
\caption{\rev{The model parameters and prior distributions used in our RV model that considers the transiting b planet, an additional c planet at 17.7 days, and the linear $\dot{\gamma}$ trend, as well as the recovered MAP fit and MCMC posteriors.  \faLock\ indicates the parameter is fixed. $\mathcal{N}(\mu, \sigma)$ signifies a Gaussian prior with mean $\mu$ and standard deviation $\sigma$.  $\mathcal{U}(\ell, r)$ signifies a uniform prior with left bound $\ell$ and right bound $r$.}}
\label{tab:2p_priors}
\end{table*}

\begin{figure*}
    \centering
    \includegraphics[width=.8\textwidth]{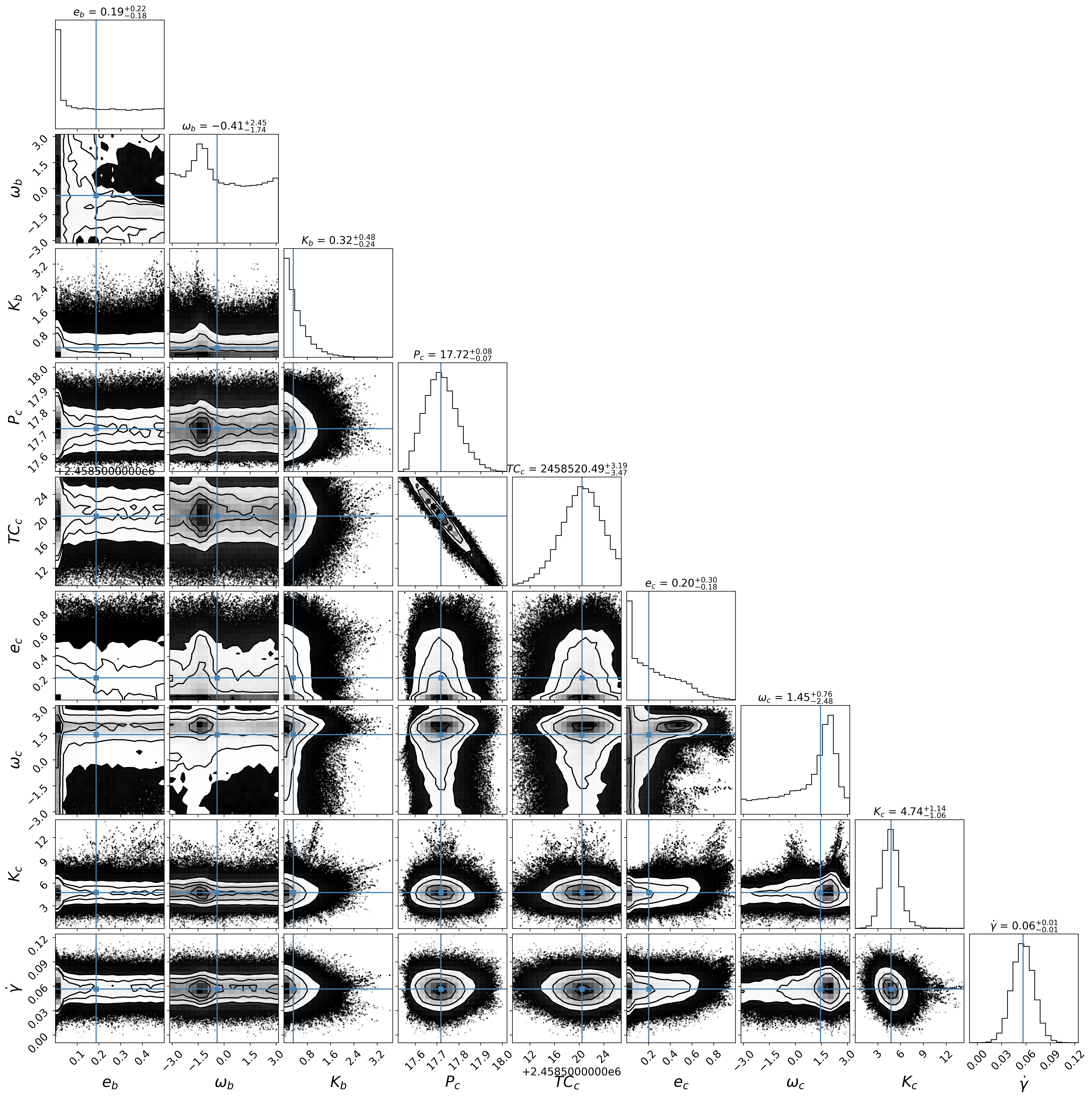}
    \caption{\rev{MCMC cornerplot of our b \& c model with all RV datapoints included, showing the posterior distributions of each model parameter that we allowed to vary.  The gamma offsets and jitter terms are not shown, as they are all uncorrelated and to a good approximation are ideal Gaussian distributions.}}
    \label{fig:2p_cornerplot}
\end{figure*}

We recover a 17.7 day signal at $>4\sigma$ significance and with a semi-amplitude corresponding to a planet mass of $M_P\sin i = 13.1^{+2.8}_{-2.9}$ \mearth and a semi-major axis $a_{\rm p} = 0.111^{+0.002}_{-0.002}$ au using the mass of the primary host star and corresponding uncertainty from our \texttt{EXOFASTv2} analysis in ${\S}$\ref{sect:bulk_star_props}. The MAP and MCMC posteriors are listed in Table \ref{tab:2p_priors} along with the model's priors, and we present the full and phased RV plots in \rev{Figures \ref{fig:2p_phasedrvs} and \ref{fig:bc_fullrvs}}, and an abbreviated MCMC cornerplot in Figure \ref{fig:2p_cornerplot}. Our model comparison (see Table \ref{tab:2p_modelcomp}) shows that the models that includes a second (``c'') planet candidate is the most favored, but any model including the b planet is disfavored and models without the ``c'' planet are not ruled out. In other words, the inclusion of a c planet in our RV model does not improve our recovery of a mass for TOI 620 b, nor are we statistically confident in the recovery of the c planet. Analysis of the predicted $T_C$ also shows that this planet is likely not transiting. The uncertainty in the $T_C$ is high, and the predicted transit windows are marked in \rev{Figure \ref{fig:lightcurve_ffp}}. While the TOI 620 b transits are readily apparent by eye, no transits for TOI 620 c are seen.

Taken as a whole, we cannot confirm the c planet as statistically significant ($>5\sigma$) with only the RV data and model comparison presented herein.  We also note that 17.7 days is approximately twice the $\sim 8.9$-day signal seen in the \tess\ light curves \rev{(Fig. \ref{fig:lightcurve_ffp})}, and could instead be potentially related to stellar activity. $K_c$ is very close to the expected RV semi-amplitude from stellar rotation of $\sim 4$ \ms. \rev{Assuming 17.7 days is the true stellar rotation period, we would have identified the 8.9-day harmonic as the most prominent signal instead since our analysis of a single \tess\ sector light curve would be insensitive to the true period.} If this is the case, then our RV data contain no direct evidence for additional planets in the system, and additional RV follow-up will be needed to rule in or out a c planet and/or stellar activity. 

\begin{table}[]
    \centering
    \begin{tabular}{cccccc}
        \hline
        Planets & $\ln \mathcal{L}$ & $\Delta$ AICc & $\Delta$ BIC & N free & $\chi^2_{\mathrm{red}}$ \\
        \hline
        \hline
        c & -263.22 & 0.00 & 0.00 & 18 & 1.77 \\
        None & -278.27 & 13.98 & 8.46 & 13 & 1.77 \\
        b, c & -267.39 & 19.44 & 21.32 & 21 & 1.92 \\
        b & -280.69 & 28.15 & 26.27 & 16 & 1.89 \\
        \hline
    \end{tabular}
    \caption{\rev{A model comparison test for planets b and c showing that the most favorable model includes only the c planet.}}
    \label{tab:2p_modelcomp}
\end{table}

\subsubsection{The detection threshold mass of TOI 620 b}
\label{sect:injection_recovery}

To assess our detection efficiency vs. planet mass, we carry out injection and recovery tests of the RV data. In these tests, we inject simulated Keplerian signals with known orbital parameters ($\{P,e,\omega,T_P,K\}$) into our combined RV data.  The time of injected periastron ($T_P$) is arbitrarily set as 2459273.623416268, which is close to the median of the data, and $e_i$ and $\omega_i$ are set to 0.  We consider 20 values of period from 1.12345--10.12345 days, and 30 values of semi-amplitude from 0.1--100 \ms, both evenly spaced in log space to broadly sample our RV sensitivity as a function of orbital period.  We model each data set assuming a Keplerian signal with a circular orbit, and with no Keplerian signal but including a linear trend.  We also model the injected planet with the period $P_i$ and ephemerides $T_{C,i}$ fixed at the injected values. The only parameter of the injected planet that we do allow to vary is $K_i$, which we start at 5 \ms and invoke $\mathcal{U}(0, \infty)$.

\begin{figure*}
    \centering
    \includegraphics[width=0.8\textwidth]{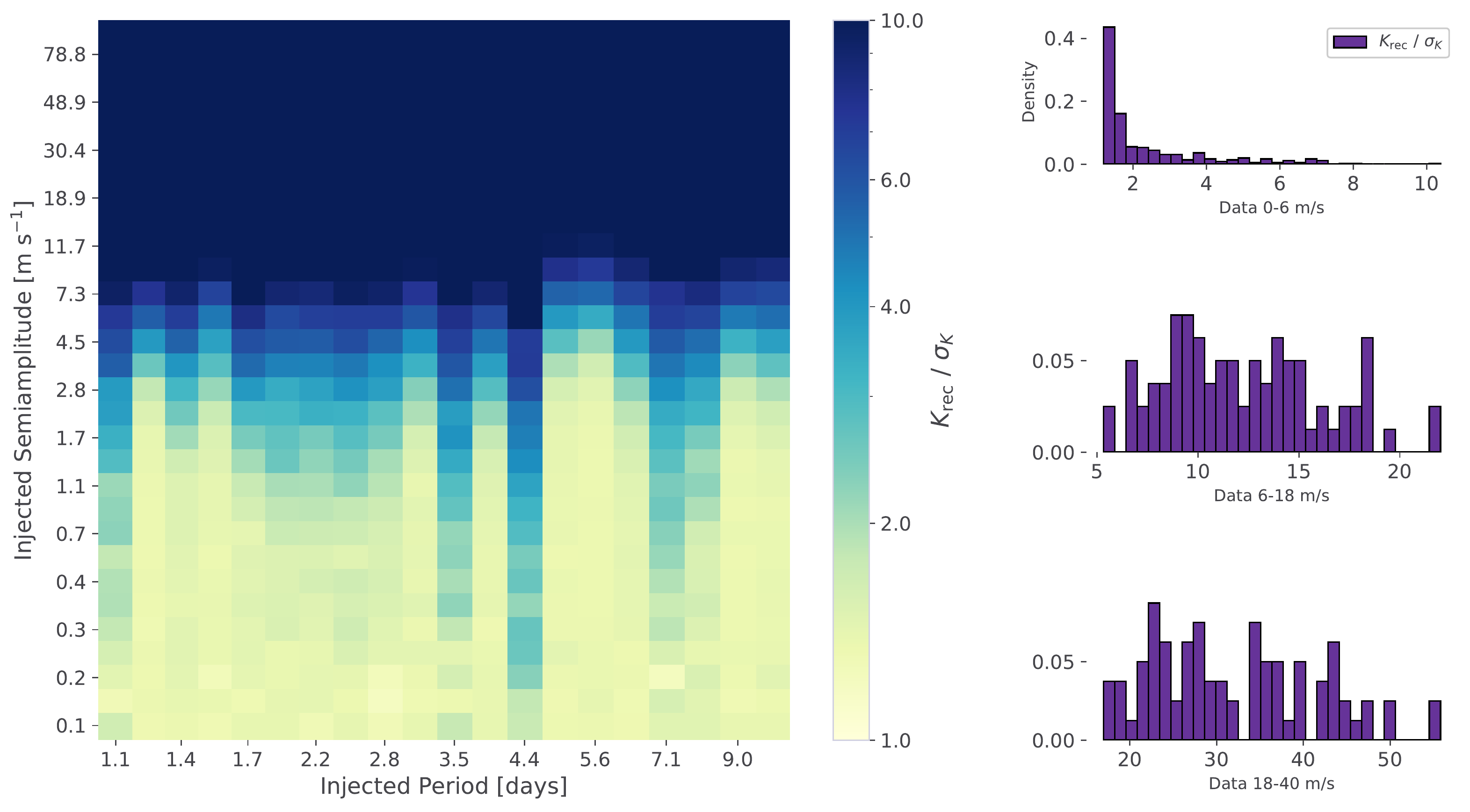}
    \caption{\rev{Left: a 2D histogram depicting the recovered semi-amplitude of the injected planet in units of the recovered uncertainty, where values above $\sim$3 represent recoveries.  Each tile represents an MCMC analysis performed where the injected planet has a period and semi-amplitude determined by the horizontal and vertical axis position of the tile.  Both axes scale logarithmically, but the color bars do not. Right: the right three panels are 1D histograms of the same data depicted in the left histogram, but binned into different groups based on the value of the injected semi-amplitude. The top histogram covers 0--6 \ms, the middle 6--18 \ms, and the bottom 18--40 \ms.  The vertical axes of these histograms are normalized probability densities rather than bin counts.}}
    \label{fig:injection_recovery}
\end{figure*}

For each planet that is injected, we run an MCMC to determine how well we recover $K_i$ and its corresponding uncertainty, and two MAP fits: one with a model that includes the injected planet, and one without. A two-dimensional histogram of this recovery data is shown in Figure \ref{fig:injection_recovery}, \rev{where the left panel presents the recovered semi-amplitude as a fraction of its uncertainty.  The right 3 charts are one-dimensional histograms depicting the same data as the left panel, for specific ranges of $P_i$}.  We can therefore conclude that significant recoveries are possible for our model for a range of orbital periods for any semi-amplitudes above $\sim 3$ m s$^{-1}$. This is the sensitivity limit above which  our model begins to recover semi-amplitudes of the same order as the injected values, with a confidence of $\geqslant 3\sigma$. \rev{This strengthens our confidence that our upper limit on the Doppler amplitude of TOI 620 b is not limited by RV noise}, as a 3 \ms sensitivity should be well within reasonable expectations to recover a signal at 6.5--8.4 \ms as anticipated from the \citet{Chen_2016} mass-radius relation.

\subsection{Are there additional massive companions in the system?}
\label{sect:massive_companions}

\begin{figure}
    \centering
    \includegraphics[width=.47\textwidth]{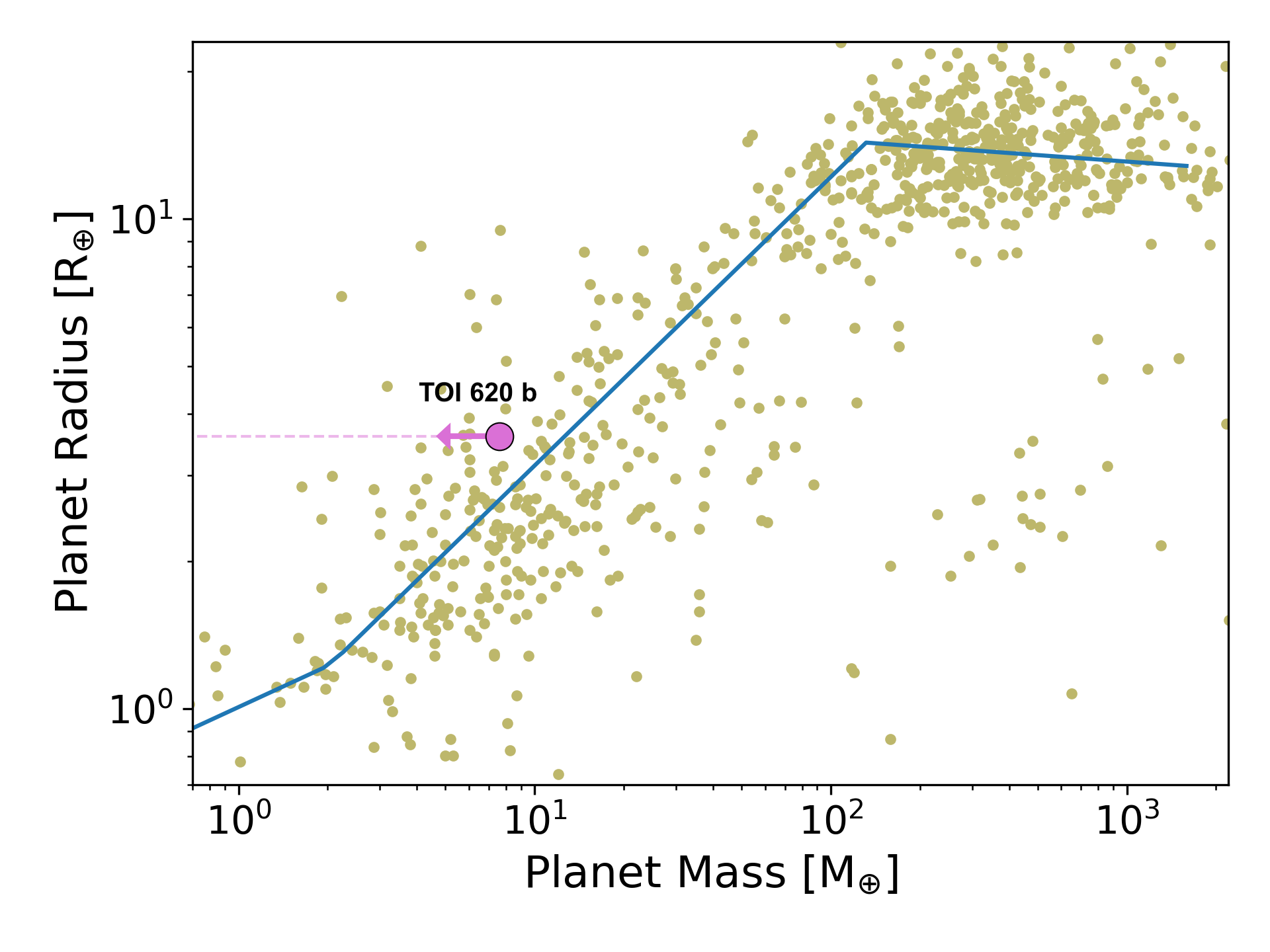}
    \caption{\rev{The mass-radius diagram for all exoplanets with provided radii and masses from the NASA Exoplanet Archive. TOI 620 b's $5\sigma$ upper limit is plotted in pink with an arrow and dashed line.  The arrow extends back down to the median mass from our MCMC chains. The blue line traces the \citet{Chen_2016} relation.  This plot demonstrates that TOI 620 is among some of the lowest-density Neptune-sized planets known.}}
    \label{fig:mass_radius}
\end{figure}

The evidence for a second, massive companion in the system consists of the elevated value of the RUWE statistic (${\S}$\ref{sec:companion}), which describes the error in the fit of the \gaia\ astrometry to a single-star solution, the non-zero linear trend ($0.08 \pm 0.01$ \msday) in our RV data (${\S}$\ref{sec:primary_analysis}), and the double-star or hierarchical binary solutions to our spectroscopic modeling of the high-resolution iSHELL NIR spectra (${\S}$\ref{sect:sb2_analysis}).  However, the historical imaging and high contrast imaging rule out any background star contamination, or any bound stellar companions with angular separations $\gtrsim \angd{;;0.2}$. Therefore any luminous companions would have to be in a pathological situation where the projected angular separation is much less than the true separation; otherwise, the companion must be a less luminous brown dwarf or massive Jupiter as constrained in ${\S}$\ref{sec:companion}. Further long-term RV monitoring, deeper high-contrast imaging, and \rev{the upcoming \gaia\ DR3 scheduled to be released in April--June, which presumably will include a revised and more accurate RUWE value for TOI 620,} may unveil this hidden companion.

\subsection{How old is the TOI 620 system?}
\label{sect:discussion_age}

A young age for TOI 620 could help explain the anomolously low-density for TOI 620 b.  However, considering all the data relevant to the age of TOI 620, the short rotation period of $\sim$ 9 days is the only indication we have that TOI 620 is potentially a young system.  As covered in section \ref{sect:stellar_age}, the $UVW$ space motion of the star does not implicate it as a member of any cluster, and we see no emission in H$\alpha$, Na I D lines, X-rays, FUV, NUV, or APASS $u'$. Additionally, the 17.7 day RV signal in ${\S}$\ref{sect:planetc} may instead indicate that the rotation period is twice as long and TOI 620 relatively older.  Thus, we cannot at this time put a constraint on the age of the system.

\subsection{The implications of a low-density TOI 620 b}
\label{sect:density_implications}

Exhausting all other explanations for our non-recovery of a Doppler RV signal, we derive a 5-$\sigma$ upper-limit in the circum-primary favored scenario of $M_P \leqslant 7.1$ \mearth. When combined with radius estimates from the circum-primary transit data analysis, \rev{$\rp = 3.76 \pm 0.15 \re$, this gives us a 5$\sigma$ upper limit on the density of $\rho_P \leqslant 0.74$ g cm$^{-3}$,} making TOI 620 b one of the puffiest Neptunes ever discovered (see Figure \ref{fig:mass_radius}).  



\rev{We calculate updated transmission and emission spectroscopy metrics for TOI 620 b as described in \citet{Kempton_2018}. Using the 5$\sigma$ upper limit on the mass, this gives a \textit{lower} limit on the TSM to be 327, which already places it within the first (most valuable) TSM quartile defined in \citet{Kempton_2018}}, and higher than any other Neptune-sized TOI using the Chen-Kipping mass-radius relation, as in Figure \ref{fig:kempton}. \rev{The new ESM estimate is also significantly larger, at 17.}

As mentioned in ${\S}$\ref{sect:stellar_age}, \rev{there is no evidence (including excess UV/X-ray brightness) that TOI 620 is a young ($< 1$ Gyr) star}.  Thus, we can rule youth out as a possible mechanism for the low density of TOI 620 b.  \rev{There have been a few proposed alternative mechanisms for super-puff formations and observations, including an augmentation of the observed planetary radius due to the presence of photochemical hazes in the upper atmosphere \citep{2020DPS....5221303G}.  \citet{2016ApJ...817...90L} instead propose that super-puffs form in the outer disk regions ($> 1$ au) with lower opacities and are able to accrete more H/He via rapid cooling.  They then migrate inwards to their present locations.  In contrast, \citet{2019ApJ...886...72M} proposes that obliquity tides inflate the radii of super-puffs.  This explanation does not require a dominantly H/He atmosphere, but does require planets to be close to the host star, which may be plausible with TOI 620's semi-major axis of $< 0.05$ au.}  Due to its abnormally low density, this planet would be a good candidate for He $\lambda$108030 nm transmission spectroscopy and low resolution spectrophotometry in order to detect broad spectral features such as Rayleigh scattering or the broad wings of Na and K \citep[e.g. WASP-127 b and WASP-21 b,][]{2018A&A...616A.145C,2020A&A...642A..54C} in future work.  Although WASP-127 b and WASP-21 b are both hotter than TOI 620, TOI 620 would allow for the exploration of the broad continuum of much cooler atmospheres dominated by other species.  


\section{Conclusions \& Future Work}
\label{sect:summary}
In this work, we have presented a rigorous exploration and validation of the TOI 620 system based on two seasons of RV measurements from iSHELL, CARMENES, MAROON-X, and NEID, and photometric data from \tess\ and ground-based follow-up observations from NGTS, LCO, MuSCAT2, TMMT, LCRO, \rev{ExTrA} and KeplerCam, and high-resolution images from Gemini South, NIRI, NIRC2, NESSI, and ShaneAO.  

Taking all of the transit and RV analysis results into account, we can conclude that TOI 620 b is a highly under-dense transiting exoplanet orbiting the primary M2.5 star, \rev{with $P = 5.0988179 \pm 0.0000045$ days, $T_C = 2458518.00718 \pm 0.00093$, $\rp = 3.76 \pm 0.15$ \re, and 5$\sigma$ upper limits of $M_P \leqslant 7.1$ \mearth and $\rho_P \leqslant 0.74$ \rhocgs.} From the RV trend, \gaia\ RUWE statistic and high-contrast imaging, we also find a possible additional hidden Jupiter-mass companion planet at $\sim 3$ au \textit{or} an ultracool dwarf at 20--30 au.  We additionally present a candidate periodic signal in the data at 17.7 days that shows up prominently in the residuals of both our GLS and $\ln \mathcal{L}$ periodograms. Injection and recovery analyses show that we can reliably recover planets in the RVs down to $\sim 3$ \ms. We are also able to robustly exclude circum-secondary and hierarchical eclipsing binary scenarios from the chromatic transit light curves.

More RV data is needed to further constrain the mass of TOI 620 b, rule in or out the candidate RV signal at 17.7 days, and to continue to monitor the linear RV trend for a turnover. Deeper high contrast imaging and aperture masking interferometry with instruments like \textit{Keck} could resolve a possible bright companion at smaller angular separations than have been analyzed in this paper, and further constrain the mass -- semi-major axis parameter space allowed for the hidden massive outer companion. Finally, the nearby TOI 620 b with its NIR-bright host star is among the best targets for atmospheric characterization with \jwst, given its abnormally low density and large atmospheric scale-height with a \rev{TSM $\geqslant 327$.}

\begin{acknowledgements}
MAR and PPP acknowledge support from NASA (Exoplanet Research Program Award \#80NSSC20K0251, TESS Cycle 3 Guest Investigator Program Award \#80NSSC21K0349, JPL Research and Technology Development, and Keck Observatory Data Analysis) and the NSF (Astronomy and Astrophysics Grants \#1716202 and 2006517), and the Mt Cuba Astronomical Foundation.

\rev{R.L. acknowledges financial support from the Spanish Ministerio de Ciencia e Innovación, through project PID2019-109522GB-C52, and the Centre of Excellence "Severo Ochoa" award to the Instituto de Astrofísica de Andalucía (SEV-2017-0709).}

This work is partly supported by JSPS KAKENHI Grant Number P17H04574, JP18H05439, JP20K14518, JP21K13975, JST CREST Grant Number JPMJCR1761, and the Astrobiology Center of National Institutes of Natural Sciences (NINS) (Grant Numbers AB022006, AB031010, AB031014). This work is partly financed by the Spanish Ministry of Economics and Competitiveness through grant PGC2018-098153-B-C31.

The development of the MAROON-X spectrograph was funded by the David and Lucile Packard Foundation, the Heising-Simons Foundation, the Gemini Observatory, and the University of Chicago. We thank the staff of the Gemini Observatory for their assistance with the commissioning and operation of the instrument. The MAROON-X observing program is supported
by NSF grant 2108465.

This work was enabled by observations made from the Gemini North
telescope, located within the Maunakea Science Reserve and adjacent to the
summit of Maunakea. We are grateful for the privilege of observing the
Universe from a place that is unique in both its astronomical quality and
its cultural significance.

Some of the observations in the paper made use of the High-Resolution Imaging instrument Zorro obtained under Gemini LLP Proposal Number: GN/S-2021A-LP-105. Zorro was funded by the NASA Exoplanet Exploration Program and built at the NASA Ames Research Center by Steve B. Howell, Nic Scott, Elliott P. Horch, and Emmett Quigley. Zorro was mounted on the Gemini South telescope of the international Gemini Observatory, a program of NSF’s OIR Lab, which is managed by the Association of Universities for Research in Astronomy (AURA) under a cooperative agreement with the National Science Foundation. on behalf of the Gemini partnership: the National Science Foundation (United States), National Research Council (Canada), Agencia Nacional de Investigación y Desarrollo (Chile), Ministerio de Ciencia, Tecnología e Innovación (Argentina), Ministério da Ciência, Tecnologia, Inovações e Comunicações (Brazil), and Korea Astronomy and Space Science Institute (Republic of Korea).

VK gratefully acknowledges support from NASA via grant NNX17AF81G.

Based on data collected under the NGTS project at the ESO La Silla Paranal Observatory.  The NGTS facility is operated by the consortium institutes with support from the UK Science and Technology Facilities Council (STFC)  projects ST/M001962/1 and  ST/S002642/1. This work has made use of data from the European Space Agency (ESA) mission {\it Gaia} (\url{https://www.cosmos.esa.int/gaia}), processed by the {\it Gaia} Data Processing and Analysis Consortium (DPAC, \url{https://www.cosmos.esa.int/web/gaia/dpac/consortium}). Funding for the DPAC has been provided by national institutions, in particular the institutions participating in the {\it Gaia} Multilateral Agreement.

ML acknowledges support from the Swiss National Science Foundation under Grant No. PCEFP2\_194576.  The contribution of ML has been carried out within the framework of the NCCR PlanetS supported by the Swiss National Science Foundation.

CIC acknowledges support by NASA Headquarters under the NASA Earth and Space Science Fellowship Program through grant 80NSSC18K1114

NEID is funded by NASA/JPL under contract 1547612.

Observations in this paper made use of the NN-EXPLORE Exoplanet and Stellar Speckle Imager (NESSI). NESSI was funded by the NASA Exoplanet Exploration Program and the NASA Ames Research Center. NESSI was built at the Ames Research Center by Steve B. Howell, Nic Scott, Elliott P. Horch, and Emmett Quigley.

This paper is based on observations obtained from the Las Campanas Remote Observatory that is a partnership between Carnegie Observatories, The Astro-Physics Corporation, Howard Hedlund, Michael Long, Dave Jurasevich, and SSC Observatories.

We acknowledge support from NSF grants AST-190950 and 1910954.

This article is based on observations made with the MuSCAT2 instrument, developed by ABC, at Telescopio Carlos Sánchez operated on the island of Tenerife by the IAC in the Spanish Observatorio del Teide.

This material is based upon work supported by the National Science Foundation Graduate Research Fellowship under Grant No. DGE 1746045

We acknowledge the use of public TESS data from pipelines at the TESS Science Office and at the TESS Science Processing Operations Center. 

Resources supporting this work were provided by the NASA High-End Computing (HEC) Program through the NASA Advanced Supercomputing (NAS) Division at Ames Research Center for the production of the SPOC data products.

\rev{Based on data collected under the ExTrA project at the ESO La Silla Paranal Observatory. ExTrA is a project of Institut de Plan\'etologie et d'Astrophysique de Grenoble (IPAG/CNRS/UGA), funded by the European Research Council under the ERC Grant Agreement n. 337591-ExTrA. This work has been supported by a grant from Labex OSUG@2020 (Investissements d'avenir -- ANR10 LABX56).}

\end{acknowledgements}

\facilities{
NASA IRTF, Calar Alto Observatory, Gemini North, Gemini South, Fred L. Whipple Observatory, \tess, ESO La Silla Paranal Observatory, Las Cumbres Observatory, Teide Observatory, Lick Observatory, WIYN Observatory
}

\software{
Python: \texttt{pychell} \citep{Cale2019}, \texttt{EDI-Vetter Unplugged} \citep{Zink_2020}, \texttt{DAVE} \citep{Kostov_2019}, \texttt{tpfplotter} \citep{Aller_2020}, \texttt{emcee} \citep{Foreman-Mackey_2013}, \texttt{NumPy} \citep{Harris_2020}, \texttt{SciPy} \citep{Virtanen_2020}, \texttt{Matplotlib} \citep{Hunter_2007}, \texttt{AstroPy} \citep{Robitaille_2013}, \texttt{corner} \citep{Foreman-Mackey_2020}, \texttt{barycorrpy} \citep{Kanodia_2018}, \texttt{Numba} \citep{Lam_2015},  \texttt{PyTransit} \citep{Parviainen_2015}, \texttt{serval} \citep{zechmeister2018};
IDL: \texttt{EXOFASTv2} \citep{Eastman_2013}
}

\bibliography{main}{}
\bibliographystyle{aasjournal}

\clearpage
\appendix
\counterwithin{figure}{section}
\counterwithin{table}{section}

\section{Updated Methods for iSHELL Forward Modeling}
\label{sect:forward_modeling}
Once we have a full set of reduced, extracted spectra from iSHELL, the spectra must be forward-modeled to extract RVs, accounting for the stellar spectrum, gas cell, telluric absorption, fringing sources, and the line spread function of the order traces.  First, \texttt{pychell} requires an initial guess for the stellar template based on the properties of the host star.  Using the effective temperature, radius, mass, and effective gravity estimates from ExoFOP-TESS \citep{exofop} in our original radial velocity fits via \texttt{pychell}, we assume a solar metallicity and create an initial stellar template with $T = 3500$ K, $\log(g) = 4.5$, and $\mathrm{[Fe/H]} = 0.0$ using the Spanish Virtual Observatory's (SVO) theoretical spectra web server to create a BT-Settl model\footnote{\href{http://svo2.cab.inta-csic.es/theory/newov2/index.php?models=bt-settl}{http://svo2.cab.inta-csic.es/theory/newov2/index.php?models=bt-settl}}, which we further refine by Doppler broadening the spectrum to the rotational velocity of the star, which we assumed to be 2 \kms from the TRES spectra analysis (${\S}$\ref{sect:recon_spec}). Barycenter velocities are also generated as an input via the \texttt{barycorrpy} library \citep{Kanodia_2018}, based on the algorithms from \citet{2014PASP..126..838W}.

The initial fitting produced a sub-optimal stellar template with absorption lines that were too shallow and unphysical telluric velocities.  To produce better results, we performed tests by varying the stellar template temperatures from 3000 -- 4000 K, in steps of 100 K, on a subgroup of gathered spectra, and found that the 4000 K templates produced the lowest RMS flux residuals and deeper stellar absorption lines.  We used this as our initial template going forward.  The discrepancy between this initial guess and our posteriors from \texttt{EXOFASTv2} does not greatly impact our results thanks to our iterative stellar template process.  We choose to ``iterate'' the stellar template by co-adding residuals in the stellar rest frame, and repeating the forward modeling of the extracted spectra, a process known as iterative Jacobian deconvolution. We iterate for a total of 10 times to obtain final RV measurements; 10 iterations of the stellar template and repeated forward modeling is chosen because additional iterations do not yield significant further reductions in the RVs' or flux residuals' RMS. Individual radial velocity measurements for a given night are co-added using a series of statistical weighting techniques across images and orders to obtain binned nightly RV measurements and errorbars to generate our final nightly RV measurements.  \rev{After this iterative process, the final 10$^{\rm th}$ iteration stellar template is best described by an effective temperature of $\sim$3800 K rather than the starting value of 4000 K.  This demonstrates the ramifications of the iterative process and how it converges on a more accurate stellar template by using our empirical spectra, which is particularly useful when synthetic spectra are lacking in some NIR opacity sources.  This is the template that was used in our SB2 analysis (${\S}$\ref{sect:sb2_analysis}). These methods are described further in \citet{Cale2019}.}

With one co-added radial velocity measurement per observation night, we again use \texttt{pychell}, this time in combination with the co-dependent package \texttt{optimize}, which is a general-purpose Bayesian analysis tool that \texttt{pychell} expands upon with RV-specific MCMC tools and is very similar in implementation to \texttt{radvel} \citep{2018PASP..130d4504F}.  We have filtered out two individual spectra from UT 2021 February 5 and one from UT 2021 May 29, due to RV measurements that did not converge properly and were in disagreement with other spectra from the same night by 100s of km s$^{-1}$. We suspect this was due to an initially poor focus on the night of observation of $\angd{;;1.5}$ that was later improved to $\angd{;;1.1}$, giving us an overall SNR of only 87 for the night. We also removed the entire nights of UT 2020 May 17, 2020 June 14, and 2021 June 4, the latter of which is due to poor spectral fits and individual RV measurements that were also inconsistent. In this case, we suspect the cause of poor data may have been due to the high airmass of TOI 620 during observations, which reached 1.8. The first two nights do seem internally consistent and the spectral fits appear to be of the expected quality, but the final co-added RV measurements are radically different from all other data, at $>$ 150 \ms and $<$ $-70$ \ms, respectively, putting them both more than 3$\sigma$ away from the expected RV trend.  It is possible that these were caused by flare events or unfortunate slit alignments, or potentially they even may be physical if they correspond to the periastron of a highly eccentric companion.  With the data we have collected, however, we cannot say anything definitive about these outlier RV data points and remove them from our analysis.

\section{\rev{Circum-primary - Transit Times and Vetting Plots for the one-planet case, and Priors for the GP case}}
\label{sect:app_primary}

\startlongtable
\begin{deluxetable*}{lccccc}
\tablecaption{\rev{Median values and 68\% confidence interval for transit times, impact parameters, and depths}}
\tablehead{\colhead{Transit} & \colhead{Planet} & \colhead{Epoch} & \colhead{$T_T$} & \colhead{$b$} & \colhead{Depth}}
\startdata
TESS UT 2019-02-03 (TESS) & b & 0 & $2458518.00717^{+0.00050}_{-0.00051}$ & $0.886^{+0.014}_{-0.017}$ & $0.003226^{+0.00010}_{-0.000098}$\\
TESS UT 2019-02-08 (TESS) & b & 1 & $2458523.10599 \pm 0.00050$ & $0.886^{+0.014}_{-0.017}$ & $0.003226^{+0.00010}_{-0.000098}$\\
TESS UT 2019-02-13 (TESS) & b & 2 & $2458528.20481 \pm 0.00050$ & $0.886^{+0.014}_{-0.017}$ & $0.003226^{+0.00010}_{-0.000098}$\\
TESS UT 2019-02-23 (TESS) & b & 4 & $2458538.40244 \pm 0.00049$ & $0.886^{+0.014}_{-0.017}$ & $0.003226^{+0.00010}_{-0.000098}$\\
NGTS UT 2019-04-20 (R) & b & 15 & $2458594.48944 \pm 0.00045$ & $0.886^{+0.014}_{-0.017}$ & $0.00345^{+0.00026}_{-0.00027}$\\
LCO UT 2019-04-20 (z') & b & 15 & $2458594.48944 \pm 0.00045$ & $0.886^{+0.014}_{-0.017}$ & $0.00330 \pm 0.00014$\\
TMMT UT 2019-04-25 (I) & b & 16 & $2458599.58826 \pm 0.00045$ & $0.886^{+0.014}_{-0.017}$ & $0.00358 \pm 0.00022$\\
NGTS UT 2019-06-10 (R) & b & 25 & $2458645.47762 \pm 0.00042$ & $0.886^{+0.014}_{-0.017}$ & $0.00345^{+0.00026}_{-0.00027}$\\
MuSCAT2 UT 2020-01-16 (g') & b & 68 & $2458864.72679 \pm 0.00031$ & $0.886^{+0.014}_{-0.017}$ & $0.00314^{+0.00025}_{-0.00030}$\\
MuSCAT2 UT 2020-01-16 (i') & b & 68 & $2458864.72679 \pm 0.00031$ & $0.886^{+0.014}_{-0.017}$ & $0.00290 \pm 0.00018$\\
MuSCAT2 UT 2020-01-16 (r') & b & 68 & $2458864.72679 \pm 0.00031$ & $0.886^{+0.014}_{-0.017}$ & $0.00278 \pm 0.00020$\\
MuSCAT2 UT 2020-01-16 (z') & b & 68 & $2458864.72679 \pm 0.00031$ & $0.886^{+0.014}_{-0.017}$ & $0.00330 \pm 0.00014$\\
KeplerCam UT 2020-01-26 (B) & b & 70 & $2458874.92442 \pm 0.00031$ & $0.886^{+0.014}_{-0.017}$ & $0.00313^{+0.00029}_{-0.00034}$\\
MuSCAT2 UT 2020-03-02 (g') & b & 77 & $2458910.61615 \pm 0.00030$ & $0.886^{+0.014}_{-0.017}$ & $0.00314^{+0.00025}_{-0.00030}$\\
MuSCAT2 UT 2020-03-02 (i') & b & 77 & $2458910.61615 \pm 0.00030$ & $0.886^{+0.014}_{-0.017}$ & $0.00290 \pm 0.00018$\\
MuSCAT2 UT 2020-03-02 (r') & b & 77 & $2458910.61615 \pm 0.00030$ & $0.886^{+0.014}_{-0.017}$ & $0.00278 \pm 0.00020$\\
MuSCAT2 UT 2020-03-02 (z') & b & 77 & $2458910.61615 \pm 0.00030$ & $0.886^{+0.014}_{-0.017}$ & $0.00330 \pm 0.00014$\\
MuSCAT2 UT 2020-04-16 (g') & b & 86 & $2458956.50551 \pm 0.00030$ & $0.886^{+0.014}_{-0.017}$ & $0.00314^{+0.00025}_{-0.00030}$\\
MuSCAT2 UT 2020-04-16 (i') & b & 86 & $2458956.50551 \pm 0.00030$ & $0.886^{+0.014}_{-0.017}$ & $0.00290 \pm 0.00018$\\
MuSCAT2 UT 2020-04-16 (r') & b & 86 & $2458956.50551 \pm 0.00030$ & $0.886^{+0.014}_{-0.017}$ & $0.00278 \pm 0.00020$\\
MuSCAT2 UT 2020-04-16 (z') & b & 86 & $2458956.50551 \pm 0.00030$ & $0.886^{+0.014}_{-0.017}$ & $0.00330 \pm 0.00014$\\
LCRO UT 2020-11-27 (i') & b & 130 & $2459180.85350 \pm 0.00035$ & $0.886^{+0.014}_{-0.017}$ & $0.00290 \pm 0.00018$\\
MuSCAT2 UT 2021-01-07 (i') & b & 138 & $2459221.64404 \pm 0.00037$ & $0.886^{+0.014}_{-0.017}$ & $0.00290 \pm 0.00018$\\
MuSCAT2 UT 2021-01-07 (r') & b & 138 & $2459221.64404 \pm 0.00037$ & $0.886^{+0.014}_{-0.017}$ & $0.00278 \pm 0.00020$\\
MuSCAT2 UT 2021-01-07 (z') & b & 138 & $2459221.64404 \pm 0.00037$ & $0.886^{+0.014}_{-0.017}$ & $0.00330 \pm 0.00014$\\
TESS UT 2021-02-11 (TESS) & b & 145 & $2459257.33576 \pm 0.00039$ & $0.886^{+0.014}_{-0.017}$ & $0.003226^{+0.00010}_{-0.000098}$\\
TESS UT 2021-02-16 (TESS) & b & 146 & $2459262.43458 \pm 0.00039$ & $0.886^{+0.014}_{-0.017}$ & $0.003226^{+0.00010}_{-0.000098}$\\
TESS UT 2021-02-27 (TESS) & b & 148 & $2459272.63222 \pm 0.00040$ & $0.886^{+0.014}_{-0.017}$ & $0.003226^{+0.00010}_{-0.000098}$\\
ExTrA UT 2021-03-04 (J) & b & 149 & $2459277.73104 \pm 0.00040$ & $0.886^{+0.014}_{-0.017}$ & $0.00336^{+0.00013}_{-0.00012}$\\
TESS UT 2021-03-04 (TESS) & b & 149 & $2459277.73104 \pm 0.00040$ & $0.886^{+0.014}_{-0.017}$ & $0.003226^{+0.00010}_{-0.000098}$\\
ExTrA UT 2021-04-13 (J) & b & 157 & $2459318.52158^{+0.00043}_{-0.00042}$ & $0.886^{+0.014}_{-0.017}$ & $0.00336^{+0.00013}_{-0.00012}$\\
ExTrA UT 2021-04-19 (J) & b & 158 & $2459323.62040 \pm 0.00043$ & $0.886^{+0.014}_{-0.017}$ & $0.00336^{+0.00013}_{-0.00012}$\\
ExTrA UT 2021-06-03 (J) & b & 167 & $2459369.50976 \pm 0.00046$ & $0.886^{+0.014}_{-0.017}$ & $0.00336^{+0.00013}_{-0.00012}$\\
\enddata
\label{tab:transits}
\end{deluxetable*}

\begin{figure*}
    \centering
    \includegraphics[width=\textwidth]{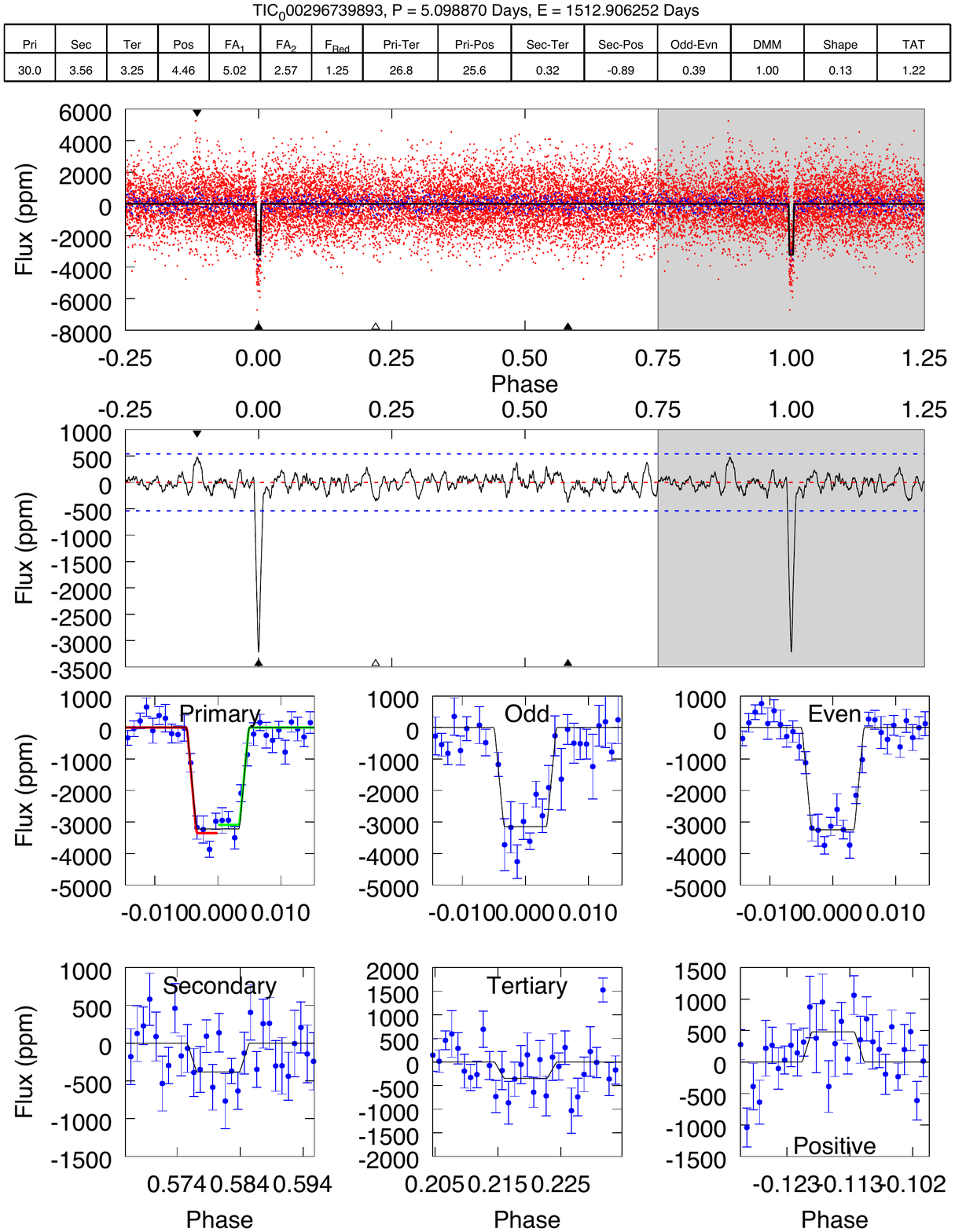}
    \caption{Sector 8 DAVE Report for TOI 620.  The top panel shows phased transit data, with \tess\ data in red, binned data in blue, and repeated data in the grey region.  The transit model is the black curve.  The middle panel shows autocorrelated flux over the same phase.  The bottom six panels depict different phased scenarios showing primary, odd, even, secondary, tertiary, and positive transits.}
    \label{fig:modshift8}
\end{figure*}

\begin{figure*}
    \centering
    \includegraphics[width=\textwidth]{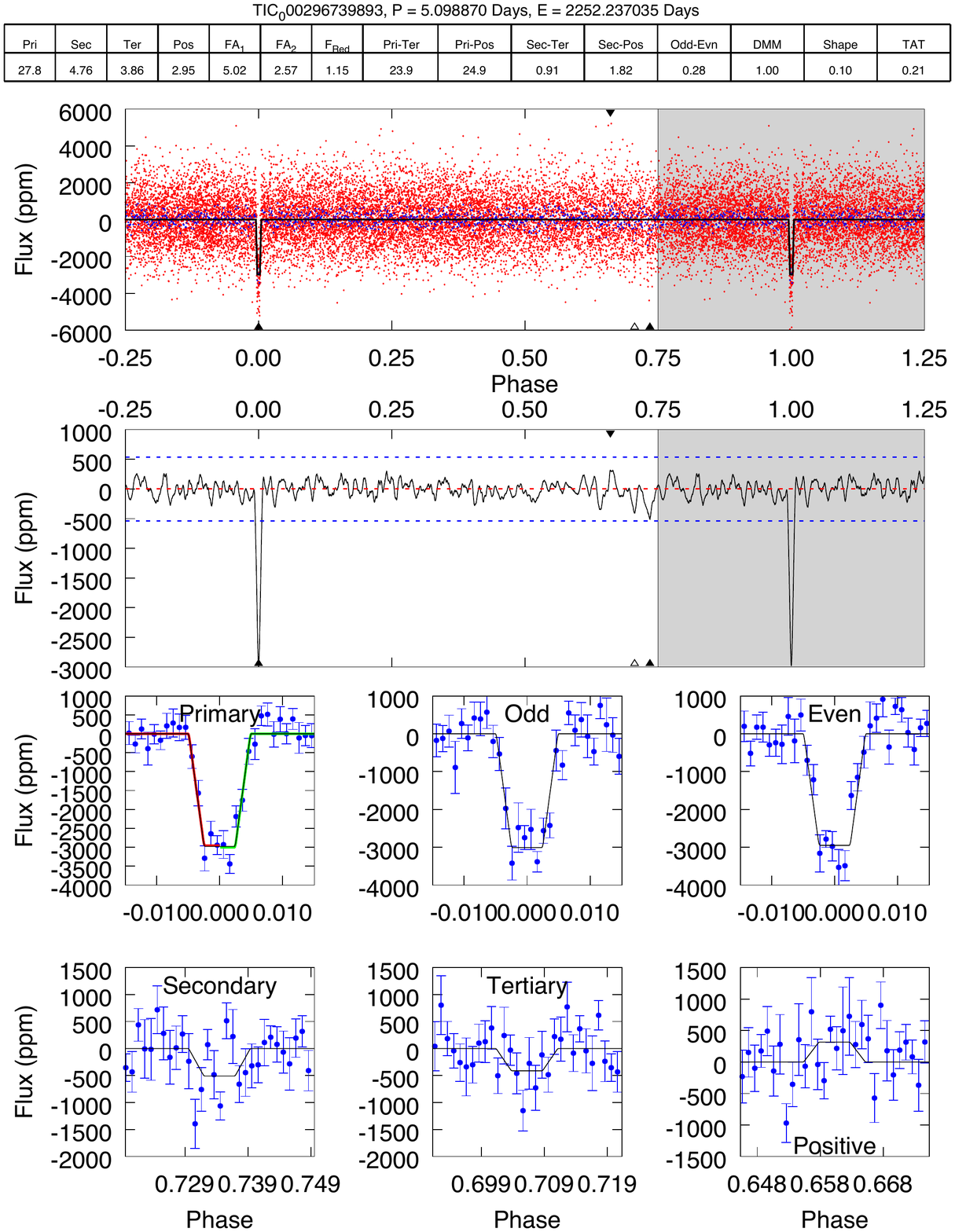}
    \caption{Sector 35 DAVE Report for TOI 620.  The top panel shows phased transit data, with \tess\ data in red, binned data in blue, and repeated data in the grey region.  The transit model is the black curve.  The middle panel shows autocorrelated flux over the same phase.  The bottom six panels depict different phased scenarios showing primary, odd, even, secondary, tertiary, and positive transits.}
    \label{fig:modshift35}
\end{figure*}

\begin{figure*}
    \centering
    \quad
    \includegraphics[width=.8\textwidth]{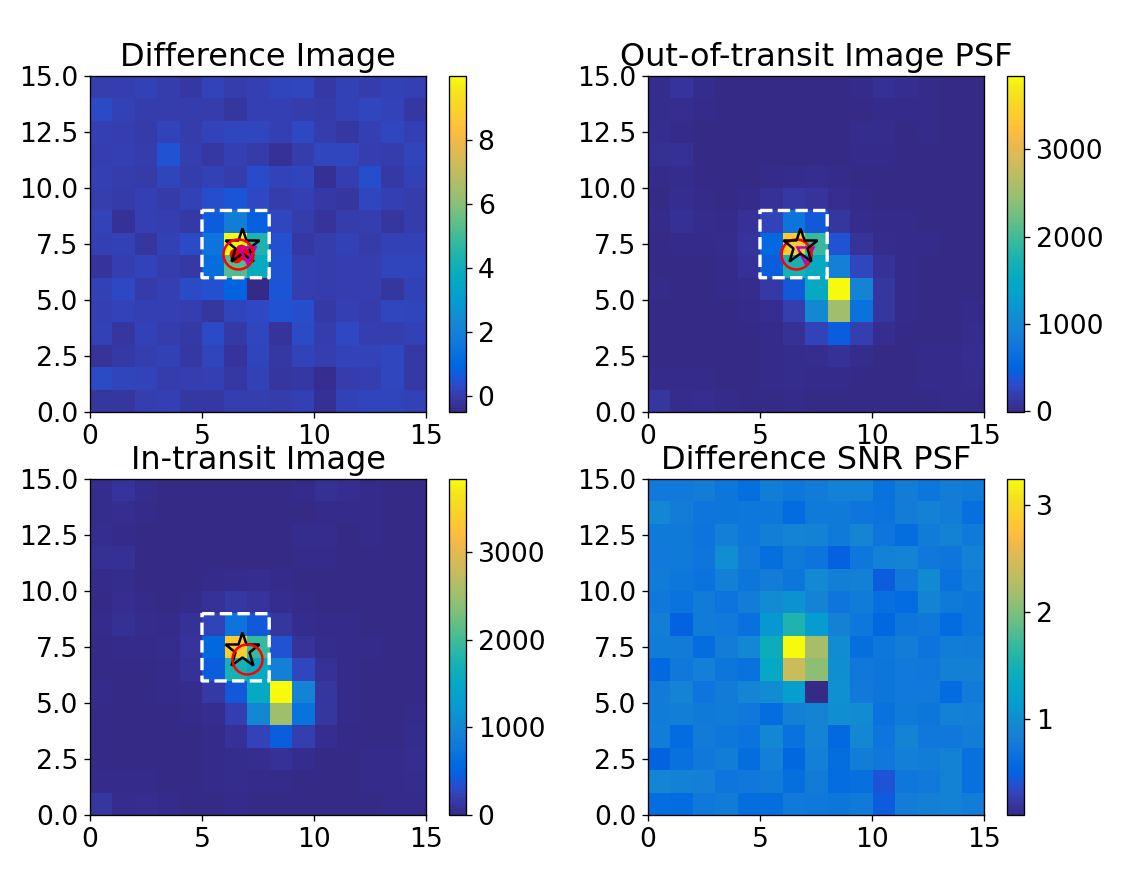}
    \includegraphics[width=.8\textwidth]{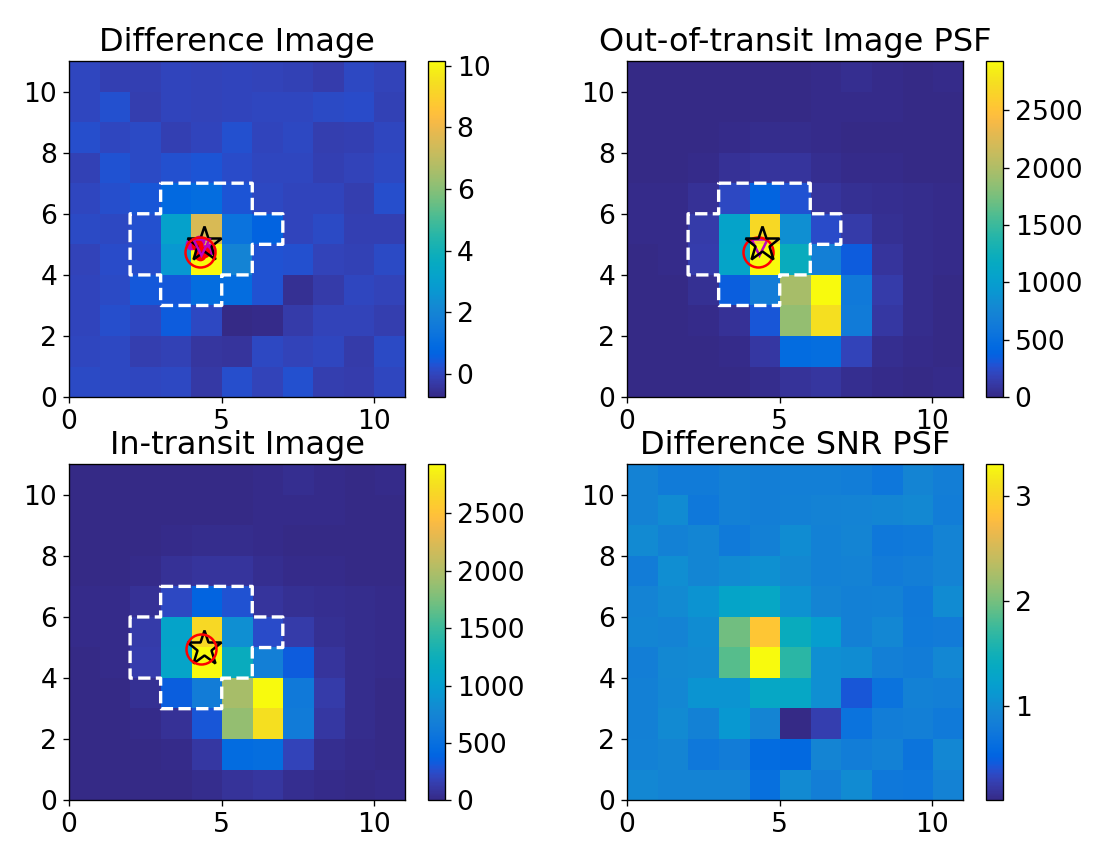}
    \caption{Photocenter difference images and PSFs for sectors 8 \rev{(top 4) and 35 (bottom 4)}.  The black-outlined star indicates the TIC position, while the red circle is the observed photocenter.  The white dashed line indicates the \tess\ target pixels used to extract the light curve, just as the orange outlines showed in the TPF plot (Figure \ref{fig:tpf}).}
    \label{fig:photocenter}
\end{figure*}

\begin{figure*}
    \centering
    \includegraphics[width=.8\textwidth]{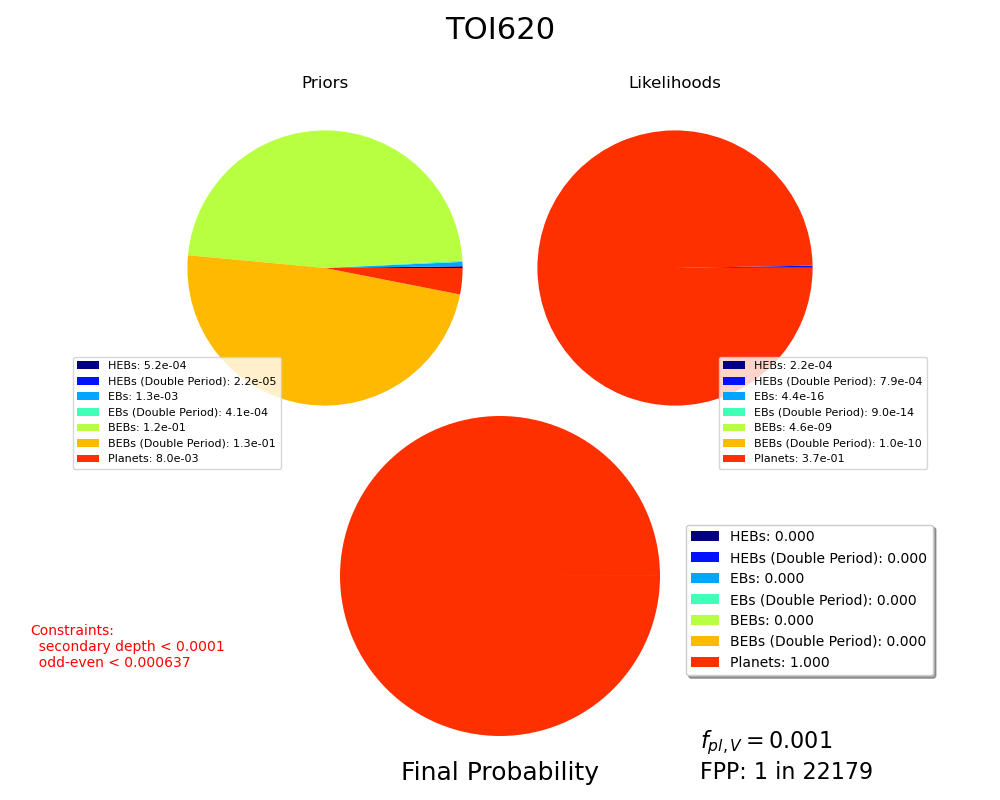}
    \includegraphics[width=.8\textwidth]{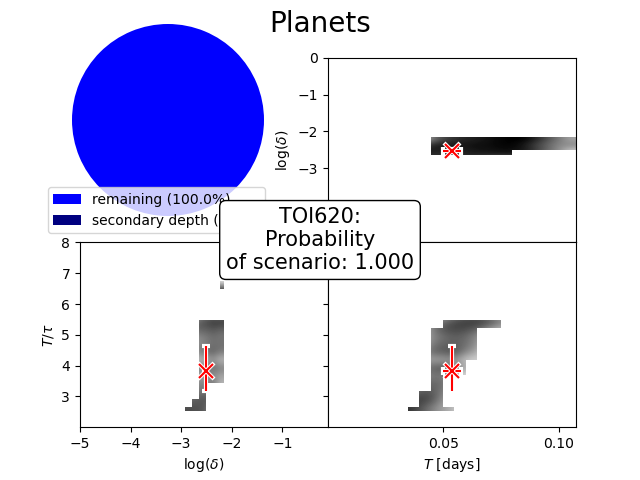}
    \caption{\rev{\textit{Top}: Pie charts of the prior probabilities, likelihoods, and posterior probabilities of TOI 620 assuming it is a planet or an EB, BEB, or HEB at 1- or 2-times the period, from \texttt{vespa}. \textit{Bottom}: The probability of the planet scenario, and the position of TOI 620 in $\log\delta$-$T$-$T/\tau$ space compared to other planets taken from galactic population statistics, also from \texttt{vespa}.}}
    \label{fig:vespa}
\end{figure*}

\begin{table*}
\centering
\begin{tabular}{cccc}
    \hline
    Parameter [units] & Initial Value ($P_{0}$) & Priors & Prior Citation \\
    \hline
    \hline
    $P_b$ [days] & 5.09881 & \faLock & this work \\
    $T_{C,b}$ [days] & 2458518.007 & \faLock & this work \\
    $e_b$ & $10^{-5}$ & $\mathcal{U}(0, 0.5)$ & this work \\
    $\omega_b$ & $\pi/2$ & $\mathcal{U}(-\pi, \pi)$ & this work \\
    $K_b$ [m s$^{-1}$] & 5 & $\mathcal{U}(0, \infty)$ & this work \\
    \hline
    $\gamma_{\rm iSHELL}$ [m s$^{-1}$] & $-8.448$  & $\mathcal{U}(P_0 \pm 100)$ & this work \\
    $\gamma_{\rm CARMENES-Vis}$ [m s$^{-1}$] & $-0.783$  & $\mathcal{U}(P_0 \pm 100)$ & this work \\
    $\gamma_{\rm CARMENES-NIR}$ [m s$^{-1}$] & $-1.510$  & $\mathcal{U}(P_0 \pm 100)$ & this work \\
    $\gamma_{\rm NEID}$ [m s$^{-1}$] & $-1.525$ & $\mathcal{U}(P_0 \pm 100)$ & this work \\
    $\gamma_{\rm MAROON-X-blue}$ [m s$^{-1}$] & $0.017$  & $\mathcal{U}(P_0 \pm 100)$ & this work \\
    $\gamma_{\rm MAROON-X-red}$ [m s$^{-1}$] & $-1.269$  & $\mathcal{U}(P_0 \pm 100)$ & this work \\
    \hline
    $\sigma_{\rm iSHELL}$ [m s$^{-1}$] & 5 & $\mathcal{N}(P_0, 2)$; $\mathcal{U}(10^{-5}, 100)$ & this work \\
    $\sigma_{\rm CARMENES-Vis}$ [m s$^{-1}$] & 5 & $\mathcal{N}(P_0, 2)$; $\mathcal{U}(10^{-5}, 100)$ & this work \\
    $\sigma_{\rm CARMENES-NIR}$ [m s$^{-1}$] & 5 & $\mathcal{N}(P_0, 2)$; $\mathcal{U}(10^{-5}, 100)$ & this work \\
    $\sigma_{\rm NEID}$ [m s$^{-1}$] & $5$ & $\mathcal{N}(P_0, 2)$; $\mathcal{U}(10^{-5}, 100)$ & this work \\
    $\sigma_{\rm MAROON-X-blue}$ [m s$^{-1}$] & 1 & $\mathcal{N}(P_0, 2)$; $\mathcal{U}(10^{-5}, 100)$ & this work \\
    $\sigma_{\rm MAROON-X-red}$ [m s$^{-1}$] & 1 & $\mathcal{N}(P_0, 2)$; $\mathcal{U}(10^{-5}, 100)$ & this work \\
    \hline
    $\dot{\gamma}$ [\msday] & $0.08$ & $\mathcal{N}(0.08, 0.01)$ & this work \\
    \hline
    $\eta_{\sigma, \rm iSHELL}$ [\ms] & 0.9 & $\mathcal{J}(0.01, 100)$ & \citet{Cale_2021} \\
    $\eta_{\sigma, \rm CARMENES-Vis}$ [\ms] & 0.9 & $\mathcal{J}(0.01, 100)$ & \citet{Cale_2021} \\
    $\eta_{\sigma, \rm CARMENES-NIR}$ [\ms] & 0.9 & $\mathcal{J}(0.01, 100)$ & \citet{Cale_2021} \\
    $\eta_{\sigma, \rm NEID}$ [\ms] & 0.9 & $\mathcal{J}(0.01, 100)$ & \citet{Cale_2021} \\
    $\eta_{\sigma, \rm MAROON-X-blue}$ [\ms] & 0.9 & $\mathcal{J}(0.01, 100)$ & \citet{Cale_2021} \\
    $\eta_{\sigma, \rm MAROON-X-red}$ [\ms] & 0.9 & $\mathcal{J}(0.01, 100)$ & \citet{Cale_2021} \\
    $\eta_\tau$ [days] & 9.41 / 11.17 & \faLock / \faLock & \citet{Cale_2021} \\
    $\eta_\ell$ & 0.15 / 0.13 & $\mathcal{N}(0.15, 0.1)$ / $\mathcal{N}(0.13, 0.1)$ & this work \\
    $\eta_P$ & 8.99 / 9.94 & \faLock / \faLock & this work \\
    \hline
\end{tabular}
\caption{The model parameters and prior distributions used in our GP RV model that considers an eccentric transiting b planet and the linear $\dot{\gamma}$ trend, \rev{as used in ${\S}$\ref{sect:ffp_analysis}}. \faLock\ indicates the parameter is fixed.  $\mathcal{N}(\mu, \sigma)$ signifies a Gaussian prior with mean $\mu$ and standard deviation $\sigma$.  $\mathcal{U}(\ell, r)$ signifies a uniform prior with left bound $\ell$ and right bound $r$.  $\mathcal{J}(\ell, r)$ signifies a Jeffrey's prior with lower bound $\ell$ and upper bound $r$.}
\label{tab:gp_rv_priors}
\end{table*}

\begin{figure*}
    \centering
    \includegraphics[width=.97\textwidth]{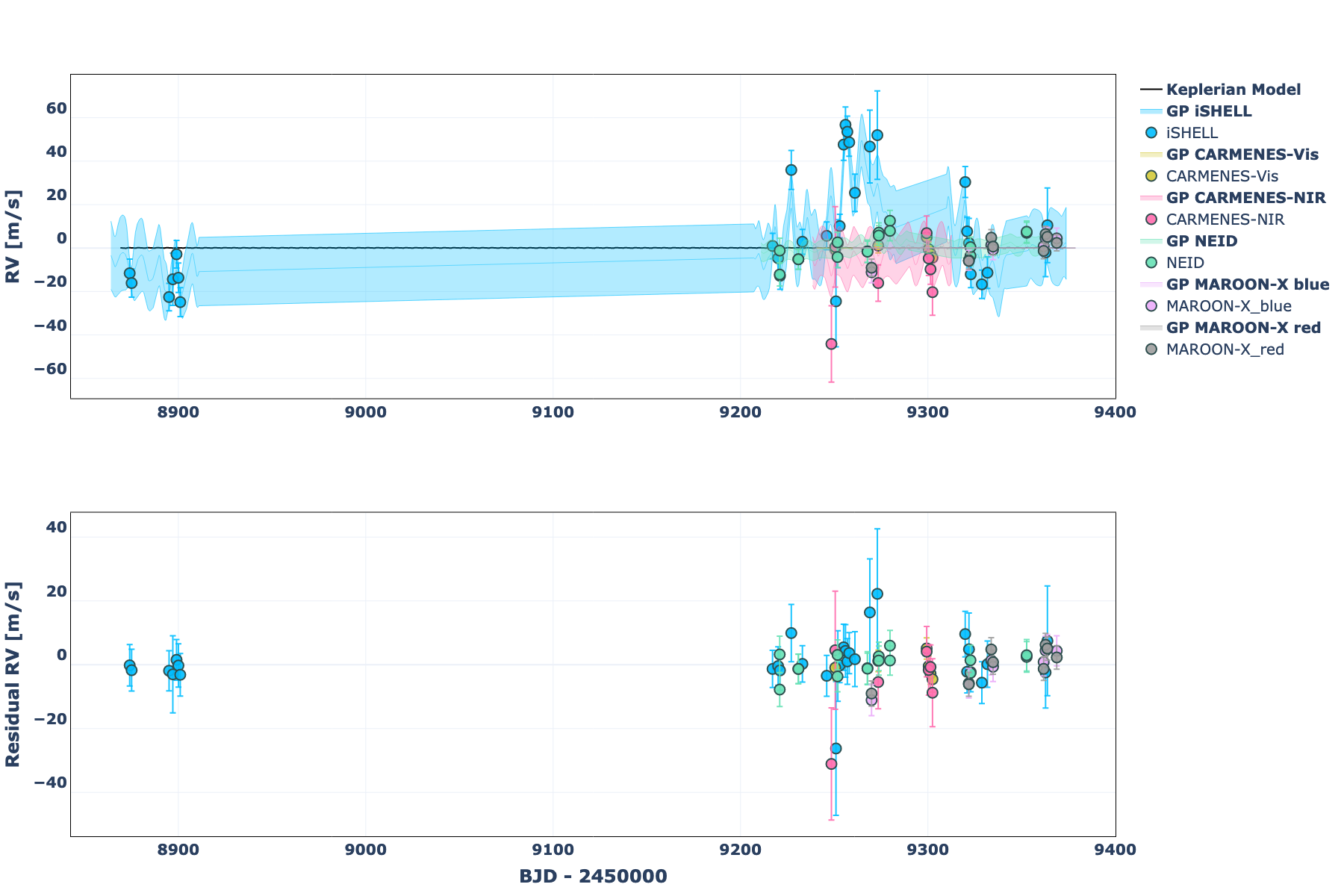}
    \includegraphics[width=.97\textwidth]{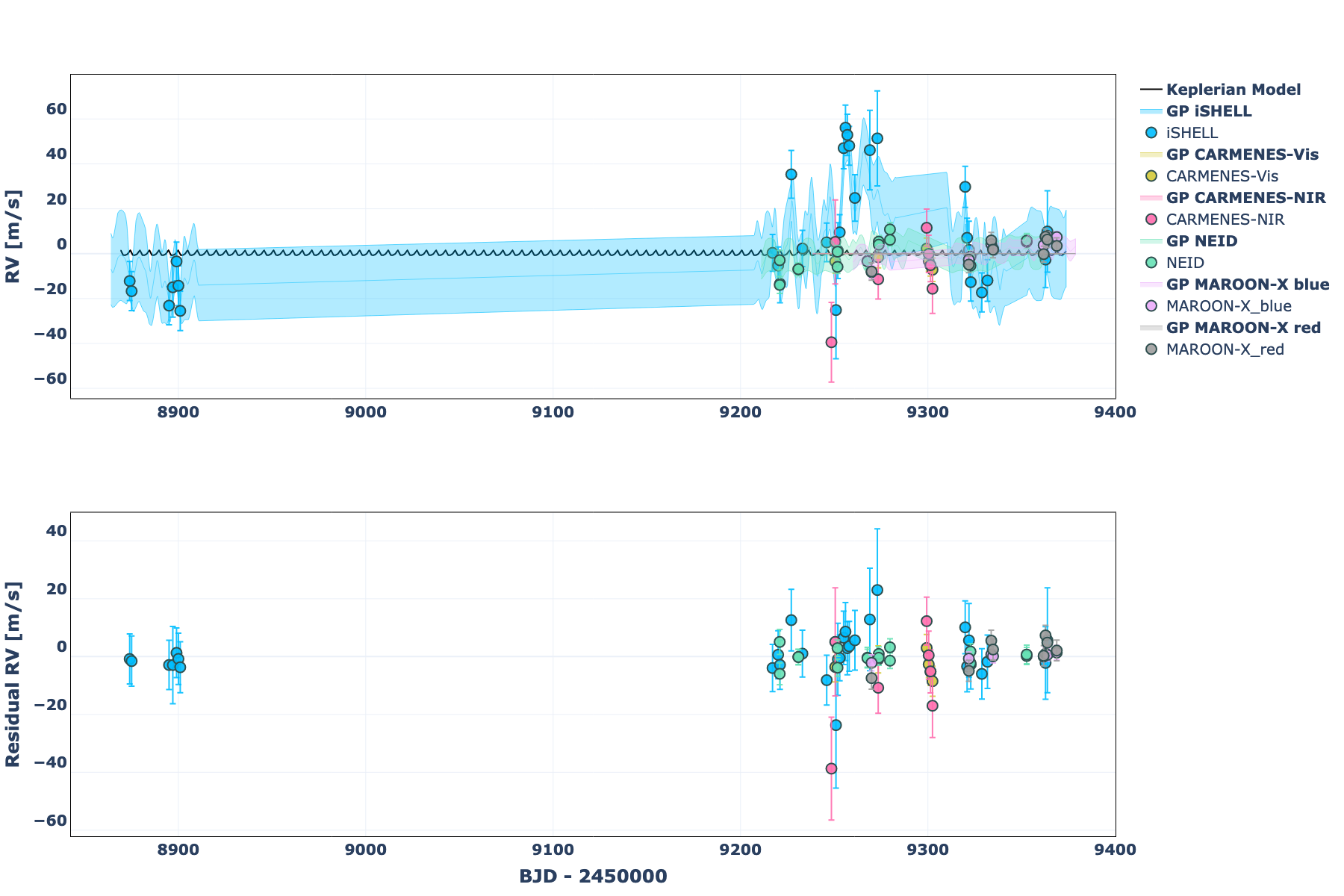}
    \caption{\rev{Full RV time-series plot for the GP model with the 8.99 day prior on $\eta_P$ (top 2), and the 9.94 day prior on $\eta_P$ (bottom 2), unphased. Residuals (data $-$ model) are shown in the lower plots.  In both cases, the GP models do not significantly improve our recovery of $K_b$.}}
    \label{fig:csec_fullrvs}
\end{figure*}

\clearpage
\onecolumngrid
\section{Circum-secondary - Analysis \& Results, Transit Times, Transit and RV Posteriors, and Full Cornerplots}
\label{sect:app_secondary}

\subsection{Transit Analysis}
\label{sect:csec_transit}

We conduct a second independent analysis of the circum-secondary scenario using the simultaneous quad-band MuSCAT2 data alone. Again, we find that the circum-secondary scenario is in contradiction with the results found in the SB2 analysis. In Figure \ref{fig:muscat_cont_csec}, we show marginal and joint posteriors for a set of parameters (effective temperature of the host and contaminant stars, impact parameter, and host stellar density) plotted against the contamination fraction. The analysis yields an upper limit to the flux from the primary of 19\% of the brightness of the secondary, which is unphysical given that we are making the assumption in the circum-secondary analysis that the secondary must be much fainter than the primary.

\begin{figure*}
    \centering
    \includegraphics[width=.5\textwidth]{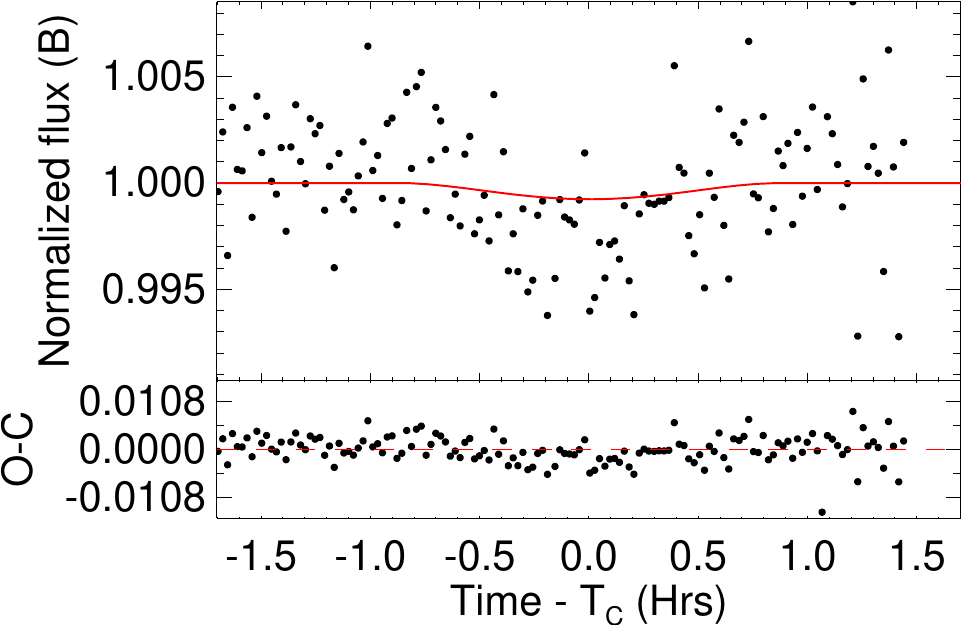}
    \caption{The MCMC transit model for TOI 620 b in the circum-secondary case from \texttt{EXOFASTv2} in the B band.  The model is the red line, and the data are the black points. It is clearly visible that the transit depth in the data is deeper than what the model predicts from the flux dilution of the secondary by the primary. \rev{The reduced $\chi^2$ for this model is 9.07.}}
    \label{fig:csec_b_transit}
\end{figure*}

\begin{figure*}
    \centering
    \includegraphics[width=\textwidth]{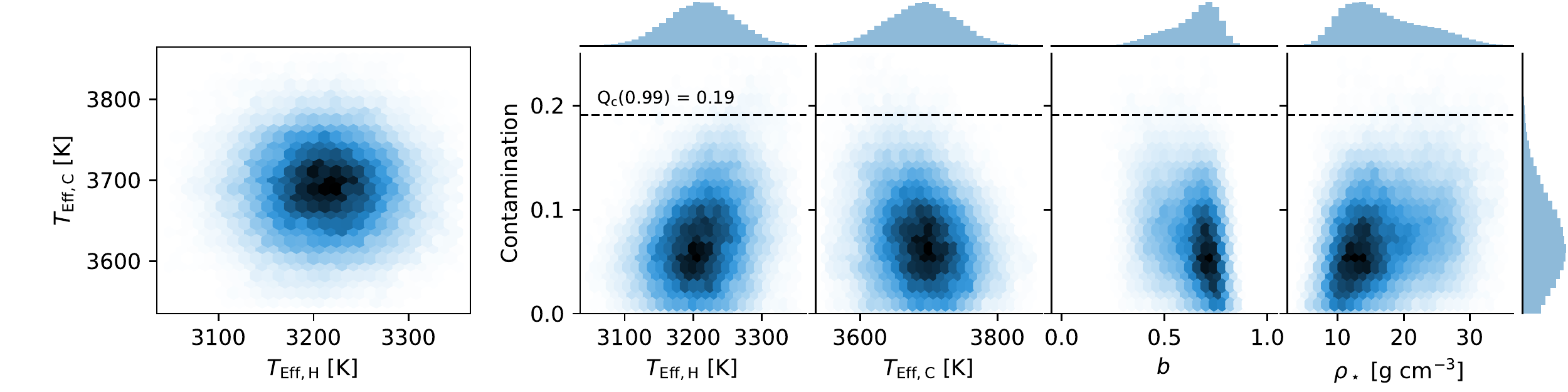}
    \caption{Posteriors and covariances for the circum-secondary MCMC of the MuSCAT2 quad-band transit light curves for a set of model parameters: effective temperature of the host [H] secondary star and contaminant [C] primary star, impact parameter, and host stellar density, plotted against the flux contamination ratio.  The upper limit to a flux ratio from the primary contributing 19\% of the light to the system rules out the circum-secondary scenario as it is not physically possible to have a bound primary companion that is both hotter and less luminous than the secondary.}
    \label{fig:muscat_cont_csec}
\end{figure*}

\begin{table*}[htp!]
    \centering
    \begin{tabular}{ccc|ccc}
        \hline
        Parameter [units] & Initial Value $(P_0)$ & Priors & Prior Citation \\
        \hline
        \hline
        $M_*$ [\msun] & 0.18 & $\mathcal{U}(0.08, 0.30)$ & this work \\
        $R_*$ [\rsun] & 0.18 & $\mathcal{U}(0.08, 0.30)$ & this work \\
        \teff~[K] & 3090 & None & this work \\
        $A_V$ [mag] & 0 & $\mathcal{U}(P_0, 0.11625)$ & \citet{Schlafly_2011} \\
        $\varpi$ [mas] & 30.28300 & $\mathcal{N}(P_0, 0.06117)$ & \citet{Gaia_Collaboration_2018} \\
        $[$Fe/H$]$ & 0 & $\mathcal{N}(0, 1)$ & this work \\
        $P$ [days] & 5.098831 & $\mathcal{U}(P_0 \pm 10\%)$ & \citet{Eastman_2019} \\
        $T_C$ [days] & 2458518.005713 & $\mathcal{U}(P_0 \pm P/3)$ & \citet{Eastman_2019} \\
        $R_p/R_*$ & 0.344 & None & this work \\
        Dilute$_{\rm B}$ & 0.991 & $\mathcal{U}(0.98, 1)$ & this work \\
        \rev{Dilute$_{\rm J}$} & 0.934 & $\mathcal{U}(0.86,1)$ & this work \\
        \rev{Dilute$_{\rm g'}$} & 0.990 & $\mathcal{U}(0.96, 1)$ & this work \\
        Dilute$_{\rm r'}$ & 0.982 & $\mathcal{U}(0.96, 1)$ & this work \\
        Dilute$_{\rm R}$ & 0.982 & $\mathcal{U}(0.96, 1)$ & this work \\
        Dilute$_{\rm i'}$ & 0.962 & $\mathcal{U}(0.92, 1)$ & this work \\
        Dilute$_{\rm TESS}$ & 0.974 & $\mathcal{U}(0.94, 1)$ & this work \\
        \rev{Dilute$_{\rm I}$} & 0.974 & $\mathcal{U}(0.94, 1)$ & this work \\
        Dilute$_{\rm z'}$ & 0.942 & $\mathcal{U}(0.88, 1)$ & this work \\
        \hline
    \end{tabular}
    \caption{Prior probability distributions for our EXOFASTv2 MCMC simulations in the circum-secondary case.  $\mathcal{N}(\mu, \sigma)$ signifies a Gaussian prior with mean $\mu$ and standard deviation $\sigma$.  $\mathcal{U}(\ell, r)$ signifies a uniform prior with left bound $\ell$ and right bound $r$.  $A_V$ is the extinction in the $V$ band, and $\varpi$ is the parallax.  Dilute is the fraction of light from from close neighboring targets.  Parameters that are missing, including orbital $e$, $\omega$ are assumed to take default values of circular.}
    \label{tab:csec_priors}
\end{table*}

\startlongtable
\begin{deluxetable*}{lccccc}
\tablecaption{\rev{Median values and 68\% confidence interval for transit times, impact parameters, and depths for the circumsecondary case}}
\tablehead{\colhead{Transit} & \colhead{Planet} & \colhead{Epoch} & \colhead{$T_T$} & \colhead{$b$} & \colhead{Depth}}
\startdata
TESS UT 2019-02-03 (TESS) & b & 0 & $2458518.00730 \pm 0.00049$ & $0.17^{+0.13}_{-0.11}$ & $0.107^{+0.017}_{-0.015}$\\
TESS UT 2019-02-08 (TESS) & b & 1 & $2458523.10612 \pm 0.00049$ & $0.17^{+0.13}_{-0.11}$ & $0.107^{+0.017}_{-0.015}$\\
TESS UT 2019-02-13 (TESS) & b & 2 & $2458528.20494 \pm 0.00048$ & $0.17^{+0.13}_{-0.11}$ & $0.107^{+0.017}_{-0.015}$\\
TESS UT 2019-02-23 (TESS) & b & 4 & $2458538.40258 \pm 0.00048$ & $0.17^{+0.13}_{-0.11}$ & $0.107^{+0.017}_{-0.015}$\\
NGTS UT 2019-04-20 (R) & b & 15 & $2458594.48958 \pm 0.00044$ & $0.17^{+0.13}_{-0.11}$ & $0.108^{+0.019}_{-0.015}$\\
LCO UT 2019-04-20 (z') & b & 15 & $2458594.48958 \pm 0.00044$ & $0.17^{+0.13}_{-0.11}$ & $0.107^{+0.018}_{-0.015}$\\
TMMT UT 2019-04-25 (I) & b & 16 & $2458599.58840^{+0.00043}_{-0.00044}$ & $0.17^{+0.13}_{-0.11}$ & $0.105^{+0.019}_{-0.015}$\\
NGTS UT 2019-06-10 (R) & b & 25 & $2458645.47777 \pm 0.00041$ & $0.17^{+0.13}_{-0.11}$ & $0.108^{+0.019}_{-0.015}$\\
MuSCAT2 UT 2020-01-16 (g') & b & 68 & $2458864.72697^{+0.00030}_{-0.00031}$ & $0.17^{+0.13}_{-0.11}$ & $0.126^{+0.025}_{-0.020}$\\
MuSCAT2 UT 2020-01-16 (i') & b & 68 & $2458864.72697^{+0.00030}_{-0.00031}$ & $0.17^{+0.13}_{-0.11}$ & $0.109^{+0.019}_{-0.015}$\\
MuSCAT2 UT 2020-01-16 (r') & b & 68 & $2458864.72697^{+0.00030}_{-0.00031}$ & $0.17^{+0.13}_{-0.11}$ & $0.106^{+0.018}_{-0.015}$\\
MuSCAT2 UT 2020-01-16 (z') & b & 68 & $2458864.72697^{+0.00030}_{-0.00031}$ & $0.17^{+0.13}_{-0.11}$ & $0.107^{+0.018}_{-0.015}$\\
KeplerCam UT 2020-01-26 (B) & b & 70 & $2458874.92461^{+0.00030}_{-0.00031}$ & $0.17^{+0.13}_{-0.11}$ & $0.126^{+0.027}_{-0.023}$\\
MuSCAT2 UT 2020-03-02 (g') & b & 77 & $2458910.61634^{+0.00029}_{-0.00030}$ & $0.17^{+0.13}_{-0.11}$ & $0.126^{+0.025}_{-0.020}$\\
MuSCAT2 UT 2020-03-02 (i') & b & 77 & $2458910.61634^{+0.00029}_{-0.00030}$ & $0.17^{+0.13}_{-0.11}$ & $0.109^{+0.019}_{-0.015}$\\
MuSCAT2 UT 2020-03-02 (r') & b & 77 & $2458910.61634^{+0.00029}_{-0.00030}$ & $0.17^{+0.13}_{-0.11}$ & $0.106^{+0.018}_{-0.015}$\\
MuSCAT2 UT 2020-03-02 (z') & b & 77 & $2458910.61634^{+0.00029}_{-0.00030}$ & $0.17^{+0.13}_{-0.11}$ & $0.107^{+0.018}_{-0.015}$\\
MuSCAT2 UT 2020-04-16 (g') & b & 86 & $2458956.50571 \pm 0.00029$ & $0.17^{+0.13}_{-0.11}$ & $0.126^{+0.025}_{-0.020}$\\
MuSCAT2 UT 2020-04-16 (i') & b & 86 & $2458956.50571 \pm 0.00029$ & $0.17^{+0.13}_{-0.11}$ & $0.109^{+0.019}_{-0.015}$\\
MuSCAT2 UT 2020-04-16 (r') & b & 86 & $2458956.50571 \pm 0.00029$ & $0.17^{+0.13}_{-0.11}$ & $0.106^{+0.018}_{-0.015}$\\
MuSCAT2 UT 2020-04-16 (z') & b & 86 & $2458956.50571 \pm 0.00029$ & $0.17^{+0.13}_{-0.11}$ & $0.107^{+0.018}_{-0.015}$\\
LCRO UT 2020-11-27 (i') & b & 130 & $2459180.85373^{+0.00034}_{-0.00035}$ & $0.17^{+0.13}_{-0.11}$ & $0.109^{+0.019}_{-0.015}$\\
MuSCAT2 UT 2021-01-07 (i') & b & 138 & $2459221.64428 \pm 0.00037$ & $0.17^{+0.13}_{-0.11}$ & $0.109^{+0.019}_{-0.015}$\\
MuSCAT2 UT 2021-01-07 (r') & b & 138 & $2459221.64428 \pm 0.00037$ & $0.17^{+0.13}_{-0.11}$ & $0.106^{+0.018}_{-0.015}$\\
MuSCAT2 UT 2021-01-07 (z') & b & 138 & $2459221.64428 \pm 0.00037$ & $0.17^{+0.13}_{-0.11}$ & $0.107^{+0.018}_{-0.015}$\\
TESS UT 2021-02-11 (TESS) & b & 145 & $2459257.33601 \pm 0.00039$ & $0.17^{+0.13}_{-0.11}$ & $0.107^{+0.017}_{-0.015}$\\
TESS UT 2021-02-16 (TESS) & b & 146 & $2459262.43483 \pm 0.00039$ & $0.17^{+0.13}_{-0.11}$ & $0.107^{+0.017}_{-0.015}$\\
TESS UT 2021-02-27 (TESS) & b & 148 & $2459272.63246^{+0.00039}_{-0.00040}$ & $0.17^{+0.13}_{-0.11}$ & $0.107^{+0.017}_{-0.015}$\\
ExTrA UT 2021-03-04 (J) & b & 149 & $2459277.73128 \pm 0.00040$ & $0.17^{+0.13}_{-0.11}$ & $0.121^{+0.021}_{-0.017}$\\
TESS UT 2021-03-04 (TESS) & b & 149 & $2459277.73128 \pm 0.00040$ & $0.17^{+0.13}_{-0.11}$ & $0.107^{+0.017}_{-0.015}$\\
ExTrA UT 2021-04-13 (J) & b & 157 & $2459318.52183 \pm 0.00042$ & $0.17^{+0.13}_{-0.11}$ & $0.121^{+0.021}_{-0.017}$\\
ExTrA UT 2021-04-19 (J) & b & 158 & $2459323.62065 \pm 0.00043$ & $0.17^{+0.13}_{-0.11}$ & $0.121^{+0.021}_{-0.017}$\\
ExTrA UT 2021-06-03 (J) & b & 167 & $2459369.51002 \pm 0.00046$ & $0.17^{+0.13}_{-0.11}$ & $0.121^{+0.021}_{-0.017}$\\
\enddata
\label{tab:transitpars_csec}
\end{deluxetable*}

\startlongtable
\begin{deluxetable*}{lcchhhhhhhhhhhhhhhhhhhhhhhhhhhhhhhh}
\tablecaption{\centering \rev{Median values and 68\% confidence interval for the circumsecondary case, created using EXOFASTv2 commit number 7971a947}}
\label{tab:TOI620.2}
\tablehead{\colhead{~~~Parameter} & \colhead{Units} & \multicolumn{28}{c}{Values}}
\startdata
\smallskip\\\multicolumn{2}{l}{Stellar Parameters:}&\smallskip\\
~~~~$M_*$\dotfill &Mass (\msun)\dotfill &$0.183^{+0.037}_{-0.035}$\\
~~~~$R_*$\dotfill &Radius (\rsun)\dotfill &$0.201\pm0.016$\\
~~~~$L_*$\dotfill &Luminosity (\lsun)\dotfill &$0.0054^{+0.0039}_{-0.0030}$\\
~~~~$\rho_*$\dotfill &Density (cgs)\dotfill &$31.8^{+2.6}_{-3.0}$\\
~~~~$\log{g}$\dotfill &Surface gravity (cgs)\dotfill &$5.091^{+0.030}_{-0.034}$\\
~~~~$T_{\rm eff}$\dotfill &Effective Temperature (K)\dotfill &$3470^{+440}_{-560}$\\
~~~~$[{\rm Fe/H}]$\dotfill &Metallicity (dex)\dotfill &$-0.40^{+0.63}_{-0.84}$\\
~~~~$[{\rm Fe/H}]_{0}$\dotfill &Initial Metallicity$^{1}$ \dotfill &$-0.43^{+0.61}_{-0.83}$\\
~~~~$Age$\dotfill &Age (Gyr)\dotfill &$6.8^{+4.8}_{-4.7}$\\
~~~~$EEP$\dotfill &Equal Evolutionary Phase$^{2}$ \dotfill &$248^{+15}_{-24}$\\
\smallskip\\\multicolumn{2}{l}{Planetary Parameters:}&b\smallskip\\
~~~~$P$\dotfill &Period (days)\dotfill &$5.0988187\pm0.0000045$\\
~~~~$R_P$\dotfill &Radius (\re)\dotfill &$6.66^{+0.67}_{-0.62}$\\
~~~~$M_P$\dotfill &Predicted Mass$^{3}$ (\me)\dotfill &$40^{+16}_{-11}$\\
~~~~$T_C$\dotfill &Time of conjunction$^{4}$ (\bjdtdb)\dotfill &$2458518.00730\pm0.00049$\\
~~~~$T_T$\dotfill &Time of minimum projected separation$^{5}$ (\bjdtdb)\dotfill &$2458518.00730\pm0.00049$\\
~~~~$T_0$\dotfill &Optimal conjunction Time$^{6}$ (\bjdtdb)\dotfill &$2458966.70335\pm0.00029$\\
~~~~$a$\dotfill &Semi-major axis (AU)\dotfill &$0.0329^{+0.0021}_{-0.0022}$\\
~~~~$i$\dotfill &Inclination (Degrees)\dotfill &$89.73^{+0.19}_{-0.23}$\\
~~~~$T_{eq}$\dotfill &Equilibrium temperature$^{7}$ (K)\dotfill &$414^{+53}_{-67}$\\
~~~~$\tau_{\rm circ}$\dotfill &Tidal circularization timescale (Gyr)\dotfill &$1.98^{+0.92}_{-0.59}$\\
~~~~$K$\dotfill &RV semi-amplitude$^{3}$ (m/s)\dotfill &$47^{+18}_{-12}$\\
~~~~$R_P/R_*$\dotfill &Radius of planet in stellar radii \dotfill &$0.304^{+0.024}_{-0.021}$\\
~~~~$a/R_*$\dotfill &Semi-major axis in stellar radii \dotfill &$35.24^{+0.92}_{-1.1}$\\
~~~~$\delta$\dotfill &$\left(R_P/R_*\right)^2$ \dotfill &$0.092^{+0.015}_{-0.012}$\\
~~~~$\delta_{\rm B}$\dotfill &Transit depth in B (fraction)\dotfill &$0.175^{+0.19}_{-0.064}$\\
~~~~$\delta_{\rm I}$\dotfill &Transit depth in I (fraction)\dotfill &$0.107^{+0.028}_{-0.018}$\\
~~~~$\delta_{\rm J}$\dotfill &Transit depth in J (fraction)\dotfill &$0.149^{+0.054}_{-0.035}$\\
~~~~$\delta_{\rm R}$\dotfill &Transit depth in R (fraction)\dotfill &$0.112^{+0.033}_{-0.019}$\\
~~~~$\delta_{\rm g'}$\dotfill &Transit depth in g' (fraction)\dotfill &$0.176^{+0.13}_{-0.057}$\\
~~~~$\delta_{\rm i'}$\dotfill &Transit depth in i' (fraction)\dotfill &$0.114^{+0.031}_{-0.020}$\\
~~~~$\delta_{\rm r'}$\dotfill &Transit depth in r' (fraction)\dotfill &$0.109^{+0.028}_{-0.019}$\\
~~~~$\delta_{\rm z'}$\dotfill &Transit depth in z' (fraction)\dotfill &$0.112^{+0.026}_{-0.019}$\\
~~~~$\delta_{\rm TESS}$\dotfill &Transit depth in TESS (fraction)\dotfill &$0.106^{+0.022}_{-0.016}$\\
~~~~$\tau$\dotfill &Ingress/egress transit duration (days)\dotfill &$0.0144^{+0.0012}_{-0.0010}$\\
~~~~$T_{14}$\dotfill &Total transit duration (days)\dotfill &$0.0595^{+0.0015}_{-0.0013}$\\
~~~~$T_{FWHM}$\dotfill &FWHM transit duration (days)\dotfill &$0.04504^{+0.00097}_{-0.00094}$\\
~~~~$b$\dotfill &Transit Impact parameter \dotfill &$0.17^{+0.13}_{-0.11}$\\
~~~~$\delta_{S,2.5\mu m}$\dotfill &Blackbody eclipse depth at 2.5$\mu$m (ppm)\dotfill &$0.36^{+1.0}_{-0.33}$\\
~~~~$\delta_{S,5.0\mu m}$\dotfill &Blackbody eclipse depth at 5.0$\mu$m (ppm)\dotfill &$114^{+100}_{-75}$\\
~~~~$\delta_{S,7.5\mu m}$\dotfill &Blackbody eclipse depth at 7.5$\mu$m (ppm)\dotfill &$670^{+330}_{-320}$\\
~~~~$\rho_P$\dotfill &Density$^{3}$ (cgs)\dotfill &$0.76^{+0.29}_{-0.19}$\\
~~~~$logg_P$\dotfill &Surface gravity$^{3}$ \dotfill &$2.95^{+0.13}_{-0.11}$\\
~~~~$\Theta$\dotfill &Safronov Number \dotfill &$0.078^{+0.029}_{-0.019}$\\
~~~~$\fave$\dotfill &Incident Flux (\fluxcgs)\dotfill &$0.0067^{+0.0042}_{-0.0034}$\\
~~~~$T_P$\dotfill &Time of Periastron (\bjdtdb)\dotfill &$2458518.00730\pm0.00049$\\
~~~~$T_S$\dotfill &Time of eclipse (\bjdtdb)\dotfill &$2458515.45789\pm0.00049$\\
~~~~$T_A$\dotfill &Time of Ascending Node (\bjdtdb)\dotfill &$2458521.83142\pm0.00049$\\
~~~~$T_D$\dotfill &Time of Descending Node (\bjdtdb)\dotfill &$2458519.28201\pm0.00049$\\
~~~~$V_c/V_e$\dotfill & \dotfill &$1.00$\\
~~~~$M_P\sin i$\dotfill &Minimum mass$^{3}$ (\me)\dotfill &$40^{+16}_{-11}$\\
~~~~$M_P/M_*$\dotfill &Mass ratio$^{3}$ \dotfill &$0.00068^{+0.00027}_{-0.00017}$\\
~~~~$d/R_*$\dotfill &Separation at mid transit \dotfill &$35.24^{+0.92}_{-1.1}$\\
~~~~$P_T$\dotfill &A priori non-grazing transit prob \dotfill &$0.01976^{+0.0010}_{-0.00094}$\\
~~~~$P_{T,G}$\dotfill &A priori transit prob \dotfill &$0.03708^{+0.0012}_{-0.00098}$\\
\smallskip\\\multicolumn{2}{l}{Wavelength Parameters:}&B&I&J&R&g'&i'&r'&z'&TESS\smallskip\\
~~~~$u_{1}$\dotfill &linear limb-darkening coeff \dotfill &$0.98^{+0.55}_{-0.62}$&$0.25^{+0.33}_{-0.18}$&$0.80^{+0.32}_{-0.36}$&$0.34^{+0.40}_{-0.24}$&$1.00^{+0.43}_{-0.50}$&$0.37^{+0.34}_{-0.25}$&$0.30^{+0.33}_{-0.21}$&$0.35^{+0.28}_{-0.23}$&$0.23^{+0.26}_{-0.17}$\\
~~~~$u_{2}$\dotfill &quadratic limb-darkening coeff \dotfill &$-0.23^{+0.54}_{-0.43}$&$0.14^{+0.36}_{-0.26}$&$-0.04^{+0.43}_{-0.36}$&$0.14^{+0.39}_{-0.32}$&$-0.23^{+0.50}_{-0.36}$&$0.15^{+0.39}_{-0.33}$&$0.10^{+0.35}_{-0.26}$&$0.13^{+0.38}_{-0.30}$&$0.33^{+0.33}_{-0.35}$\\
~~~~$A_D$\dotfill &Dilution from neighboring stars \dotfill &$0.9817^{+0.0023}_{-0.0012}$&$0.9639^{+0.0061}_{-0.0070}$&$0.9704^{+0.0042}_{-0.0048}$&$0.9673^{+0.0056}_{-0.0047}$&$0.9720^{+0.0054}_{-0.0059}$&$0.9747^{+0.0039}_{-0.0045}$&$0.9762^{+0.0039}_{-0.0044}$&$0.9693^{+0.0044}_{-0.0049}$&$0.9704^{+0.0041}_{-0.0046}$\\
\smallskip\\\multicolumn{2}{l}{Transit Parameters:}&TESS UT 2019-02-03 (TESS)&TESS UT 2019-02-08 (TESS)&TESS UT 2019-02-13 (TESS)&TESS UT 2019-02-23 (TESS)&NGTS UT 2019-04-20 (R)&LCO UT 2019-04-20 (z')&TMMT UT 2019-04-25 (I)&NGTS UT 2019-06-10 (R)&MuSCAT2 UT 2020-01-16 (g')&MuSCAT2 UT 2020-01-16 (i')&MuSCAT2 UT 2020-01-16 (r')&MuSCAT2 UT 2020-01-16 (z')&KeplerCam UT 2020-01-26 (B)&MuSCAT2 UT 2020-03-02 (g')&MuSCAT2 UT 2020-03-02 (i')&MuSCAT2 UT 2020-03-02 (r')&MuSCAT2 UT 2020-03-02 (z')&MuSCAT2 UT 2020-04-16 (g')&MuSCAT2 UT 2020-04-16 (i')&MuSCAT2 UT 2020-04-16 (r')&MuSCAT2 UT 2020-04-16 (z')&LCRO UT 2020-11-27 (i')&MuSCAT2 UT 2021-01-07 (i')&MuSCAT2 UT 2021-01-07 (r')&MuSCAT2 UT 2021-01-07 (z')&TESS UT 2021-02-11 (TESS)&TESS UT 2021-02-16 (TESS)&TESS UT 2021-02-27 (TESS)&ExTrA UT 2021-03-04 (J)&TESS UT 2021-03-04 (TESS)&ExTrA UT 2021-04-13 (J)&ExTrA UT 2021-04-19 (J)&ExTrA UT 2021-06-03 (J)\smallskip\\
~~~~$\sigma^{2}$\dotfill &Added Variance \dotfill &$-0.00000039^{+0.00000012}_{-0.00000011}$&$-0.00000019^{+0.00000012}_{-0.00000011}$&$-0.00000032^{+0.00000011}_{-0.00000010}$&$-0.00000010^{+0.00000013}_{-0.00000012}$&$-0.0000276^{+0.0000011}_{-0.0000010}$&$0.00000115^{+0.00000024}_{-0.00000021}$&$-0.00000224^{+0.00000044}_{-0.00000037}$&$-0.0000572^{+0.0000016}_{-0.0000014}$&$-0.00000172^{+0.00000034}_{-0.00000030}$&$-0.00000152^{+0.00000014}_{-0.00000013}$&$-0.00000039^{+0.00000020}_{-0.00000017}$&$-0.000001025^{+0.00000010}_{-0.000000091}$&$0.00000375^{+0.00000080}_{-0.00000071}$&$-0.00000148^{+0.00000039}_{-0.00000033}$&$-0.00000092^{+0.00000023}_{-0.00000019}$&$-0.00000158^{+0.00000026}_{-0.00000023}$&$-0.00000121^{+0.00000017}_{-0.00000015}$&$-0.00000188^{+0.00000056}_{-0.00000047}$&$-0.00000225^{+0.00000042}_{-0.00000036}$&$-0.00000098^{+0.00000037}_{-0.00000032}$&$-0.00000138^{+0.00000043}_{-0.00000036}$&$-0.0000018^{+0.0000019}_{-0.0000016}$&$-0.00000103^{+0.00000017}_{-0.00000014}$&$-0.00000141^{+0.00000019}_{-0.00000017}$&$-0.00000151^{+0.00000019}_{-0.00000017}$&$0.00000057^{+0.00000014}_{-0.00000013}$&$0.00000037^{+0.00000012}_{-0.00000011}$&$0.00000033^{+0.00000012}_{-0.00000011}$&$-0.00000023^{+0.00000033}_{-0.00000029}$&$0.00000021^{+0.00000012}_{-0.00000011}$&$-0.00000054^{+0.00000029}_{-0.00000025}$&$0.00000001^{+0.00000032}_{-0.00000027}$&$-0.00000003^{+0.00000022}_{-0.00000019}$\\
~~~~$F_0$\dotfill &Baseline flux \dotfill &$0.999887^{+0.000062}_{-0.000063}$&$0.999886\pm0.000064$&$1.000085^{+0.000062}_{-0.000061}$&$0.999680\pm0.000066$&$0.99842\pm0.00022$&$0.99970\pm0.00014$&$1.00078^{+0.00019}_{-0.00018}$&$0.99997^{+0.00039}_{-0.00040}$&$0.99996\pm0.00016$&$0.999999^{+0.000093}_{-0.000094}$&$1.00028\pm0.00011$&$1.000222\pm0.000079$&$1.00044\pm0.00020$&$1.00021\pm0.00019$&$1.00013\pm0.00013$&$0.99982\pm0.00014$&$1.00017\pm0.00011$&$0.99969\pm0.00019$&$0.99963\pm0.00016$&$0.99981\pm0.00015$&$0.99983\pm0.00016$&$1.00070\pm0.00035$&$0.99997\pm0.00011$&$0.99982\pm0.00012$&$1.00001\pm0.00011$&$1.000103\pm0.000068$&$1.000281\pm0.000064$&$1.000454\pm0.000064$&$1.00002\pm0.00013$&$0.999968^{+0.000061}_{-0.000062}$&$1.00004\pm0.00015$&$1.00003\pm0.00015$&$1.00003\pm0.00012$\\
~~~~$C_{0}$\dotfill &Additive detrending coeff \dotfill &--&--&--&--&$-0.0001\pm0.0023$&--&$0.00058\pm0.00061$&$-0.0048\pm0.0015$&$0.00045\pm0.00081$&$0.00094\pm0.00050$&$0.00057^{+0.00056}_{-0.00057}$&$0.00046^{+0.00045}_{-0.00046}$&--&$0.00032^{+0.00072}_{-0.00073}$&$0.00041\pm0.00040$&$-0.00057\pm0.00041$&$0.00062^{+0.00037}_{-0.00036}$&$-0.00152^{+0.00073}_{-0.00074}$&$-0.00207^{+0.00055}_{-0.00054}$&$-0.00087\pm0.00053$&$-0.00062\pm0.00056$&$-0.0007\pm0.0010$&$0.00001\pm0.00033$&$0.00036\pm0.00033$&$-0.00006\pm0.00027$&--&--&--&--&--&--&--&--\\
~~~~$M_{0}$\dotfill &Multiplicative detrending coeff \dotfill &--&--&--&--&$0.00013\pm0.00065$&$-0.00040^{+0.00027}_{-0.00026}$&--&$0.0037\pm0.0010$&$0.00003\pm0.00042$&$0.00032\pm0.00027$&$0.00017\pm0.00027$&$0.00023^{+0.00021}_{-0.00020}$&$0.00034^{+0.00060}_{-0.00061}$&$-0.00004\pm0.00068$&$0.00019\pm0.00051$&$0.00013\pm0.00040$&$0.00006\pm0.00029$&$-0.00025\pm0.00057$&$-0.00036\pm0.00051$&$-0.00006^{+0.00045}_{-0.00044}$&$-0.00016\pm0.00047$&--&$-0.00007\pm0.00038$&$0.00013\pm0.00044$&$-0.00008\pm0.00033$&--&--&--&--&--&--&--&--\\
~~~~$M_{1}$\dotfill &Multiplicative detrending coeff \dotfill &--&--&--&--&--&--&--&--&$-0.00059\pm0.00092$&$-0.00116\pm0.00054$&$0.00047\pm0.00063$&$0.00025^{+0.00045}_{-0.00046}$&--&$-0.00005\pm0.00070$&$-0.00016\pm0.00046$&$0.00011^{+0.00050}_{-0.00049}$&$-0.00011\pm0.00046$&$0.00073\pm0.00084$&$0.00112^{+0.00067}_{-0.00068}$&$0.00043\pm0.00065$&$0.00016\pm0.00064$&--&$-0.00010\pm0.00027$&$-0.00011\pm0.00039$&$-0.00010\pm0.00036$&--&--&--&--&--&--&--&--\\
\enddata
\tablenotetext{}{See Table 3 in \citet{Eastman_2019} for a detailed description of all parameters}
\tablenotetext{1}{The metallicity of the star at birth}
\tablenotetext{2}{Corresponds to static points in a star's evolutionary history. See \S2 in \citet{Dotter_2016}.}
\tablenotetext{3}{Uses measured radius and estimated mass from \citet{Chen_2016}}
\tablenotetext{4}{Time of conjunction is commonly reported as the "transit time"}
\tablenotetext{5}{Time of minimum projected separation is a more correct "transit time"}
\tablenotetext{6}{Optimal time of conjunction minimizes the covariance between $T_C$ and Period}
\tablenotetext{7}{Assumes no albedo and perfect redistribution}
\end{deluxetable*}

\onecolumngrid
\subsection{RV Analysis}
\label{sect:csec_rv}
\begin{figure*}[htp!]
    \centering
    \includegraphics[width=.47\textwidth]{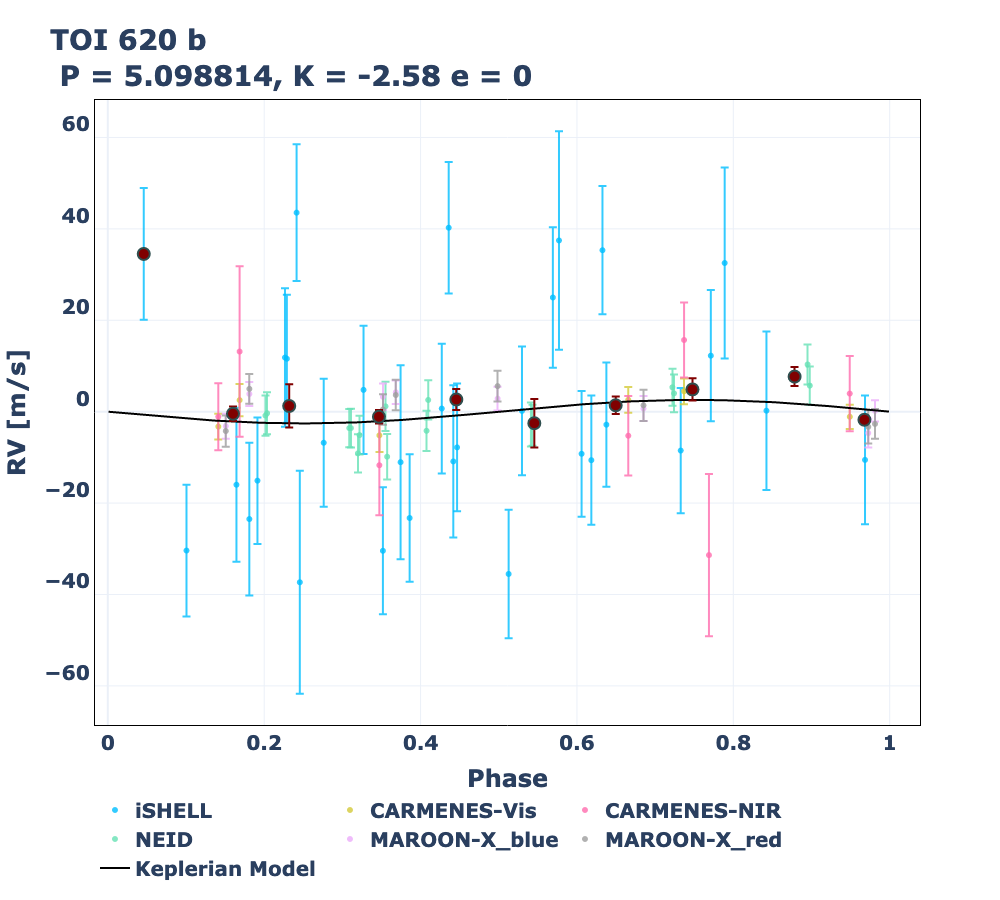}
    \includegraphics[width=.47\textwidth]{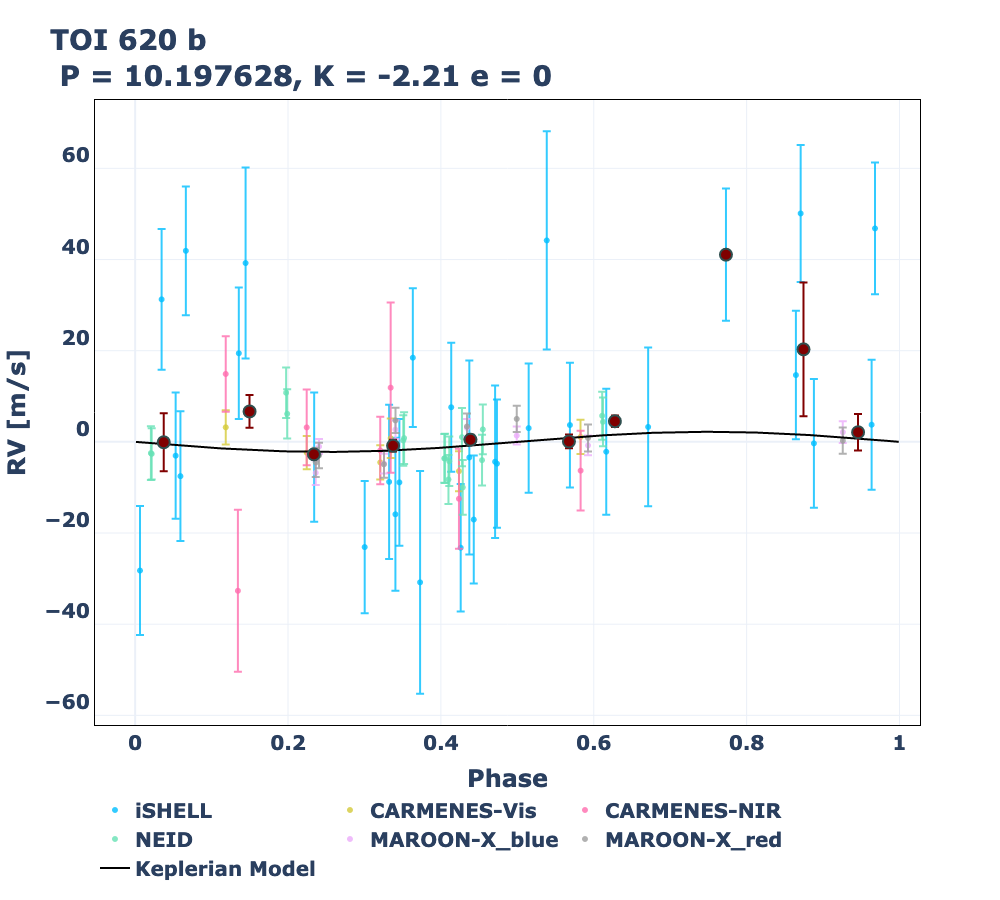}
    \caption{\rev{RV time-series plot for the circum-secondary case (left) and HEB case (right) phased to the period of b, with the black model representing the planet's MAP model fit.}}
    \label{fig:csec_phasedrvs}
\end{figure*}

Modeling the radial velocities of TOI 620 in the circum-secondary and HEB scenarios is identical to the circum-primary scenario with the exception of allowing the semi-amplitude $K$ to be negative. The physical manifestation of this result would suggest a large direct RV signal coming from a circum-secondary star or an HEB that is diluted by the primary light and thus reduced in amplitude by some (large) factor.  In this case, we recover a negative $K_b$ signal at 2.4-$\sigma$ significance (\ref{tab:csec_rv_priors}).  Each RV instrument individually supports a negative $K$ at varying statistical significance except MAROON-X. The RV plots for this scenario are presented in Figures \ref{fig:csec_phasedrvs} and \ref{fig:csec_fullrvs}. All of the posteriors are well behaved and uncorrelated.

\begin{table*}[htp!]
\centering
\begin{tabular}{cccccc}
    \hline
    Parameter [units] & MAP Value & MCMC Posterior \\
    \hline
    \hline
    $P_b$ [days] & 5.09881 (Locked) & -- \\
    $T_{C,b}$ [days] & 2458518.007 (Locked) & -- \\
    $e_b$ & 0 (Locked) & -- \\
    $\omega_b$ & $0$ (Locked) & -- \\
    $K_b$ [m s$^{-1}$] & -2.58 & $-2.62^{+0.96}_{-0.93}$ \\
    \hline
    $\gamma_{\rm iSHELL}$ [m s$^{-1}$] & 4.11 & $4.01^{+2.96}_{-2.94}$ \\
    $\gamma_{\rm CARMENES-Vis}$ [m s$^{-1}$] & -1.63 & $-1.65^{+1.58}_{-1.56}$ \\
    $\gamma_{\rm CARMENES-NIR}$ [m s$^{-1}$] & -2.56 & $-2.42^{+3.61}_{-3.71}$ \\
    $\gamma_{\rm NEID}$ [m s$^{-1}$] & 0.59 & $0.64^{+1.16}_{-1.23}$ \\
    $\gamma_{\rm MAROON-X-blue}$ [m s$^{-1}$] & -6.50 & $-6.41^{+1.36}_{-1.32}$ \\
    $\gamma_{\rm MAROON-X-red}$ [m s$^{-1}$] & -7.59 & $-7.50^{+1.58}_{-1.60}$ \\
    \hline
    $\sigma_{\rm iSHELL}$ [m s$^{-1}$] & 13.14  & $13.31^{+1.14}_{-1.05}$ \\
    $\sigma_{\rm CARMENES-Vis}$ [m s$^{-1}$] & 2.26 & $3.23^{+1.75}_{-1.48}$ \\
    $\sigma_{\rm CARMENES-NIR}$ [m s$^{-1}$] & 4.88  & $5.10^{+1.84}_{-1.94}$ \\
    $\sigma_{\rm NEID}$ [m s$^{-1}$] & 3.69 & $4.08^{+1.09}_{-0.94}$\\
    $\sigma_{\rm MAROON-X-blue}$ [m s$^{-1}$] & 1.93  & $2.26^{+1.02}_{-0.87}$ \\
    $\sigma_{\rm MAROON-X-red}$ [m s$^{-1}$] & 3.17  & $3.52^{+0.91}_{-0.71}$ \\
    \hline
    $\dot{\gamma}$ [\msday] & 0.08 & $0.08^{+0.01}_{-0.01}$ \\  
    \hline
\end{tabular}
\caption{The model parameters and posterior distributions used in our RV model that considers the transiting b planet as circum-secondary and the linear $\dot{\gamma}$ trend.  The priors are identical to the circum-primary run except the $K_b$ value is allowed to be negative.}
\label{tab:csec_rv_priors}
\end{table*}

\begin{figure*}[htp!]
    \centering
    \includegraphics[width=\textwidth]{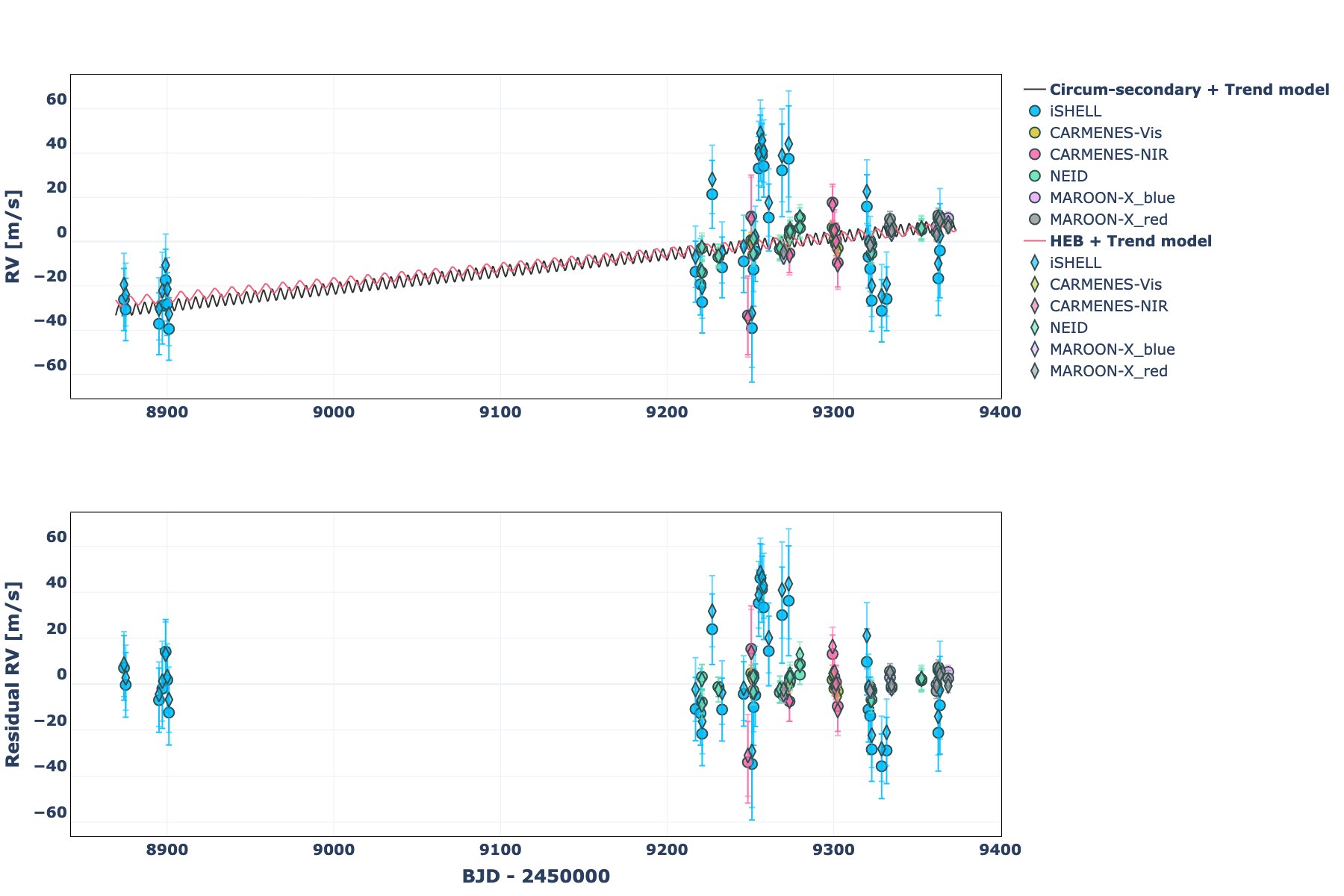}
    \caption{\rev{Full RV time-series plot for the circum-secondary and HEB cases, with the black line representing the circum-secondary model and the red line representing the HEB model.  The top plot shows the RVs for each instrument and errorbars over the full time baseline of observations, while the bottom plot shows the residuals (data $-$ model).}}
    \label{fig:csec_fullrvs}
\end{figure*}

\clearpage
\onecolumngrid
\section{HEB - Analysis \& Results, Transit Times, Transit and RV Posteriors, and Full Cornerplots}
\label{sect:app_heb}

\subsection{Transit Analysis}
\label{sect:heb_transit}

\begin{figure*}
    \centering
    \includegraphics[width=.5\textwidth]{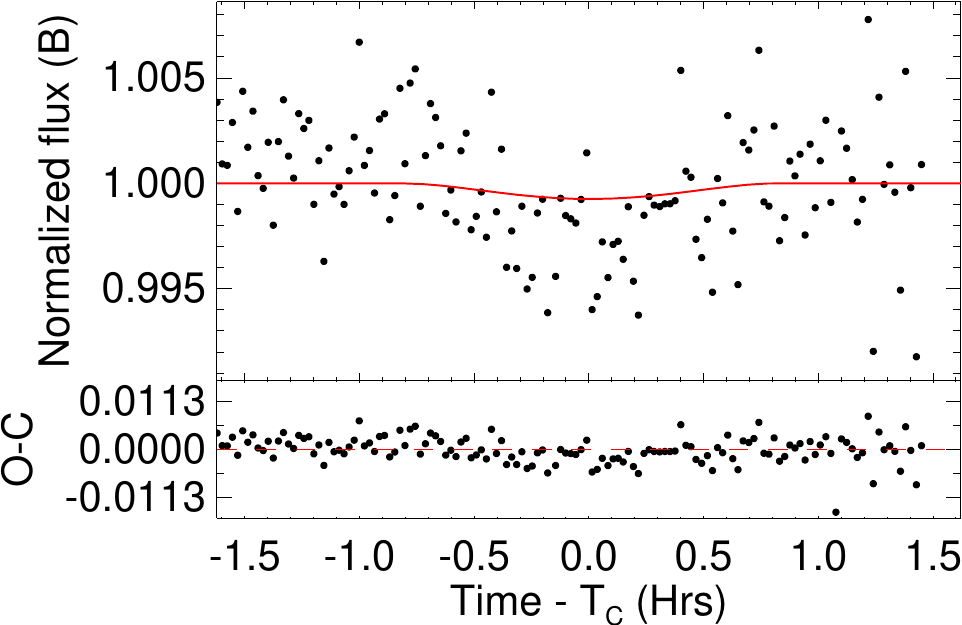}
    \caption{The MCMC transit model for TOI 620 b in the HEB case from \texttt{EXOFASTv2} in the B band.  The model is the red line, and the data are the black points. It is clearly visible that the transit depth in the data is deeper than what the model predicts from the flux dilution of the secondary by the primary. \rev{The reduced $\chi^2$ for this model is 9.28.}}
    \label{fig:heb_b_transit}
\end{figure*}

\clearpage
\begin{table*}[htp!]
    \centering
    \begin{tabular}{ccccc|c}
        \hline
        Parameter [units] & $(P_0)$, 1 & $(P_0)$, 2 & $(P_0)$, 3 & Priors & Prior Citation \\
        \hline
        \hline
        $M_*$ [\msun] & 0.18 & 0.18 & 0.18 & $\mathcal{U}(0.08, 0.30)$ & this work \\
        $R_*$ [\rsun] & 0.18 & 0.18 & 0.18 & $\mathcal{U}(0.08, 0.30)$ & this work \\
        \teff~[K] & 3090 & 3090 & 3090 & None & this work \\
        $A_V$ [mag] & 0 & 0 & 0 & $\mathcal{U}(P_0, 0.11625)$ & \citet{Schlafly_2011} \\
        $\varpi$ [mas] & 30.283 & 30.283 & 30.283 & $\mathcal{N}(P_0, 0.06117)$ & \citet{Gaia_Collaboration_2018} \\
        $[$Fe/H$]$ & 0 & 0 & 0 & $\mathcal{N}(0, 1)$ & this work \\
        $P$ [days] & 5.098831 & 10.197662 & 10.197662 & $\mathcal{U}(P_0 \pm 10\%)$ & \citet{Eastman_2019} \\
        $T_C$ [days] & 2458518.005713 & 2458518.005713 & 2458523.105 & $\mathcal{U}(P_0 \pm P/3)$ & \citet{Eastman_2019} \\
        $R_p/R_*$ & 0.88 & 0.88 & 0.88 & $\mathcal{U}(0.5, 2)$ & this work \\
        $M_p$ [\msun] & 0.16 & 0.16 & 0.16 &$\mathcal{N}(P_0, 0.05)$ & this work \\
        Dilute$_{\rm B}$ & 0.991 & 0.991 & 0.991 & $\mathcal{U}(0.98, 1)$ & this work \\
        \rev{Dilute$_{\rm J}$} & 0.934 & 0.934 & 0.934 & $\mathcal{U}(0.86,1)$ & this work \\
        \rev{Dilute$_{\rm g'}$} & 0.990 & 0.990 & 0.990 & $\mathcal{U}(0.96, 1)$ & this work \\
        Dilute$_{\rm r'}$ & 0.982 & 0.982 & 0.982 & $\mathcal{U}(0.96, 1)$ & this work \\
        Dilute$_{\rm R}$ & 0.982 & 0.982 & 0.982 & $\mathcal{U}(0.96, 1)$ & this work \\
        Dilute$_{\rm i'}$ & 0.962 & 0.962 & 0.962 & $\mathcal{U}(0.92, 1)$ & this work \\
        Dilute$_{\rm TESS}$ & 0.974 & 0.974 & 0.974 & $\mathcal{U}(0.94, 1)$ & this work \\
        \rev{Dilute$_{\rm I}$} & 0.974 & 0.974 & 0.974 & $\mathcal{U}(0.94, 1)$ & this work \\
        Dilute$_{\rm z'}$ & 0.942 & 0.942 & 0.942 & $\mathcal{U}(0.88, 1)$ & this work \\
        \hline
    \end{tabular}
    \caption{Prior probability distributions for our EXOFASTv2 MCMC simulations in the HEB case.  The starting value $P_0$ has 3 columns corresponding to the 3 cases that were run assuming a 5.09d period and a 10.19d period with even and odd transits.  $\mathcal{N}(\mu, \sigma)$ signifies a Gaussian prior with mean $\mu$ and standard deviation $\sigma$.  $\mathcal{U}(\ell, r)$ signifies a uniform prior with left bound $\ell$ and right bound $r$.  $A_V$ is the extinction in the $V$ band, and $\varpi$ is the parallax.  Dilute is the fraction of light from from close neighboring targets.  Parameters that are missing, including orbital $e$, $\omega$ are assumed to take default values of circular.}
    \label{tab:heb_priors}
\end{table*}

\startlongtable
\begin{deluxetable*}{lccccc}
\tablecaption{\rev{Median values and 68\% confidence interval for transit times, impact parameters, and depths for the HEB case}}
\tablehead{\colhead{Transit} & \colhead{Planet} & \colhead{Epoch} & \colhead{$T_T$} & \colhead{$b$} & \colhead{Depth}}
\startdata
TESS UT 2019-02-03 (TESS) & b & 0 & $2458518.00675^{+0.00068}_{-0.00091}$ & $1.80^{+0.23}_{-0.13}$ & $0.0914^{+0.0020}_{-0.0037}$\\
TESS UT 2019-02-08 (TESS) & b & 1 & $2458523.10557^{+0.00068}_{-0.00091}$ & $1.80^{+0.23}_{-0.13}$ & $0.0914^{+0.0020}_{-0.0037}$\\
TESS UT 2019-02-13 (TESS) & b & 2 & $2458528.20438^{+0.00068}_{-0.00089}$ & $1.80^{+0.23}_{-0.13}$ & $0.0914^{+0.0020}_{-0.0037}$\\
TESS UT 2019-02-23 (TESS) & b & 4 & $2458538.40203^{+0.00068}_{-0.00088}$ & $1.80^{+0.23}_{-0.13}$ & $0.0914^{+0.0020}_{-0.0037}$\\
NGTS UT 2019-04-20 (R) & b & 15 & $2458594.48908^{+0.00064}_{-0.00082}$ & $1.80^{+0.23}_{-0.13}$ & $0.0892^{+0.0042}_{-0.0069}$\\
LCO UT 2019-04-20 (z') & b & 15 & $2458594.48908^{+0.00064}_{-0.00082}$ & $1.80^{+0.23}_{-0.13}$ & $0.0663^{+0.0035}_{-0.0028}$\\
TMMT UT 2019-04-25 (I) & b & 16 & $2458599.58791^{+0.00064}_{-0.00082}$ & $1.80^{+0.23}_{-0.13}$ & $0.0859^{+0.0061}_{-0.0050}$\\
NGTS UT 2019-06-10 (R) & b & 25 & $2458645.47730^{+0.00062}_{-0.00076}$ & $1.80^{+0.23}_{-0.13}$ & $0.0892^{+0.0042}_{-0.0069}$\\
MuSCAT2 UT 2020-01-16 (g') & b & 68 & $2458864.72658^{+0.00047}_{-0.00040}$ & $1.80^{+0.23}_{-0.13}$ & $0.0824^{+0.011}_{-0.0074}$\\
MuSCAT2 UT 2020-01-16 (i') & b & 68 & $2458864.72658^{+0.00047}_{-0.00040}$ & $1.80^{+0.23}_{-0.13}$ & $0.0821^{+0.0071}_{-0.0073}$\\
MuSCAT2 UT 2020-01-16 (r') & b & 68 & $2458864.72658^{+0.00047}_{-0.00040}$ & $1.80^{+0.23}_{-0.13}$ & $0.0913^{+0.0066}_{-0.010}$\\
MuSCAT2 UT 2020-01-16 (z') & b & 68 & $2458864.72658^{+0.00047}_{-0.00040}$ & $1.80^{+0.23}_{-0.13}$ & $0.0663^{+0.0035}_{-0.0028}$\\
KeplerCam UT 2020-01-26 (B) & b & 70 & $2458874.92422^{+0.00046}_{-0.00039}$ & $1.80^{+0.23}_{-0.13}$ & $0.0831^{+0.0052}_{-0.0074}$\\
MuSCAT2 UT 2020-03-02 (g') & b & 77 & $2458910.61595^{+0.00048}_{-0.00036}$ & $1.80^{+0.23}_{-0.13}$ & $0.0824^{+0.011}_{-0.0074}$\\
MuSCAT2 UT 2020-03-02 (i') & b & 77 & $2458910.61595^{+0.00048}_{-0.00036}$ & $1.80^{+0.23}_{-0.13}$ & $0.0821^{+0.0071}_{-0.0073}$\\
MuSCAT2 UT 2020-03-02 (r') & b & 77 & $2458910.61595^{+0.00048}_{-0.00036}$ & $1.80^{+0.23}_{-0.13}$ & $0.0913^{+0.0066}_{-0.010}$\\
MuSCAT2 UT 2020-03-02 (z') & b & 77 & $2458910.61595^{+0.00048}_{-0.00036}$ & $1.80^{+0.23}_{-0.13}$ & $0.0663^{+0.0035}_{-0.0028}$\\
MuSCAT2 UT 2020-04-16 (g') & b & 86 & $2458956.50536^{+0.00044}_{-0.00034}$ & $1.80^{+0.23}_{-0.13}$ & $0.0824^{+0.011}_{-0.0074}$\\
MuSCAT2 UT 2020-04-16 (i') & b & 86 & $2458956.50536^{+0.00044}_{-0.00034}$ & $1.80^{+0.23}_{-0.13}$ & $0.0821^{+0.0071}_{-0.0073}$\\
MuSCAT2 UT 2020-04-16 (r') & b & 86 & $2458956.50536^{+0.00044}_{-0.00034}$ & $1.80^{+0.23}_{-0.13}$ & $0.0913^{+0.0066}_{-0.010}$\\
MuSCAT2 UT 2020-04-16 (z') & b & 86 & $2458956.50536^{+0.00044}_{-0.00034}$ & $1.80^{+0.23}_{-0.13}$ & $0.0663^{+0.0035}_{-0.0028}$\\
LCRO UT 2020-11-27 (i') & b & 130 & $2459180.85360^{+0.00036}_{-0.00044}$ & $1.80^{+0.23}_{-0.13}$ & $0.0821^{+0.0071}_{-0.0073}$\\
MuSCAT2 UT 2021-01-07 (i') & b & 138 & $2459221.64421^{+0.00031}_{-0.00048}$ & $1.80^{+0.23}_{-0.13}$ & $0.0821^{+0.0071}_{-0.0073}$\\
MuSCAT2 UT 2021-01-07 (r') & b & 138 & $2459221.64421^{+0.00031}_{-0.00048}$ & $1.80^{+0.23}_{-0.13}$ & $0.0913^{+0.0066}_{-0.010}$\\
MuSCAT2 UT 2021-01-07 (z') & b & 138 & $2459221.64421^{+0.00031}_{-0.00048}$ & $1.80^{+0.23}_{-0.13}$ & $0.0663^{+0.0035}_{-0.0028}$\\
TESS UT 2021-02-11 (TESS) & b & 145 & $2459257.33598^{+0.00027}_{-0.00049}$ & $1.80^{+0.23}_{-0.13}$ & $0.0914^{+0.0020}_{-0.0037}$\\
TESS UT 2021-02-16 (TESS) & b & 146 & $2459262.43481^{+0.00025}_{-0.00050}$ & $1.80^{+0.23}_{-0.13}$ & $0.0914^{+0.0020}_{-0.0037}$\\
TESS UT 2021-02-27 (TESS) & b & 148 & $2459272.63244^{+0.00026}_{-0.00049}$ & $1.80^{+0.23}_{-0.13}$ & $0.0914^{+0.0020}_{-0.0037}$\\
ExTrA UT 2021-03-04 (J) & b & 149 & $2459277.73127^{+0.00027}_{-0.00048}$ & $1.80^{+0.23}_{-0.13}$ & $0.0604^{+0.0033}_{-0.0039}$\\
TESS UT 2021-03-04 (TESS) & b & 149 & $2459277.73127^{+0.00027}_{-0.00048}$ & $1.80^{+0.23}_{-0.13}$ & $0.0914^{+0.0020}_{-0.0037}$\\
ExTrA UT 2021-04-13 (J) & b & 157 & $2459318.52183^{+0.00032}_{-0.00048}$ & $1.80^{+0.23}_{-0.13}$ & $0.0604^{+0.0033}_{-0.0039}$\\
ExTrA UT 2021-04-19 (J) & b & 158 & $2459323.62066^{+0.00031}_{-0.00048}$ & $1.80^{+0.23}_{-0.13}$ & $0.0604^{+0.0033}_{-0.0039}$\\
ExTrA UT 2021-06-03 (J) & b & 167 & $2459369.51006^{+0.00031}_{-0.00049}$ & $1.80^{+0.23}_{-0.13}$ & $0.0604^{+0.0033}_{-0.0039}$\\
\enddata
\label{tab:transitpars_heb}
\end{deluxetable*}

\startlongtable
\begin{deluxetable*}{lcchhhhhhhhhhhhhhhhhhhhhhhhhhhhhhhh}
\tablecaption{\centering \rev{Median values and 68\% confidence interval for the HEB case, created using EXOFASTv2 commit number 7971a947}}
\label{tab:TOI620.3}
\tablehead{\colhead{~~~Parameter} & \colhead{Units} & \multicolumn{28}{c}{Values}}
\startdata
\smallskip\\\multicolumn{2}{l}{Stellar Parameters:}&\smallskip\\
~~~~$M_*$\dotfill &Mass (\msun)\dotfill &$0.151^{+0.076}_{-0.021}$\\
~~~~$R_*$\dotfill &Radius (\rsun)\dotfill &$0.278\pm0.012$\\
~~~~$L_*$\dotfill &Luminosity (\lsun)\dotfill &$0.0077^{+0.0072}_{-0.0026}$\\
~~~~$\rho_*$\dotfill &Density (cgs)\dotfill &$10.0^{+3.3}_{-1.6}$\\
~~~~$\log{g}$\dotfill &Surface gravity (cgs)\dotfill &$4.726^{+0.14}_{-0.056}$\\
~~~~$T_{\rm eff}$\dotfill &Effective Temperature (K)\dotfill &$3260^{+490}_{-360}$\\
~~~~$[{\rm Fe/H}]$\dotfill &Metallicity (dex)\dotfill &$-0.21^{+0.38}_{-0.49}$\\
~~~~$[{\rm Fe/H}]_{0}$\dotfill &Initial Metallicity$^{1}$ \dotfill &$-0.30^{+0.44}_{-0.39}$\\
~~~~$Age$\dotfill &Age (Gyr)\dotfill &$0.058^{+0.057}_{-0.013}$\\
~~~~$EEP$\dotfill &Equal Evolutionary Phase$^{2}$ \dotfill &$152^{+18}_{-10.}$\\
\smallskip\\\multicolumn{2}{l}{Planetary Parameters:}&b\smallskip\\
~~~~$P$\dotfill &Period (days)\dotfill &$5.0988228^{+0.0000061}_{-0.0000068}$\\
~~~~$R_P$\dotfill &Radius (\re)\dotfill &$35.0^{+8.0}_{-3.4}$\\
~~~~$M_P$\dotfill &Predicted Mass (\me)\dotfill &$41100^{+14000}_{-9400}$\\
~~~~$T_C$\dotfill &Time of conjunction$^{3}$ (\bjdtdb)\dotfill &$2458518.00675^{+0.00068}_{-0.00091}$\\
~~~~$T_T$\dotfill &Time of minimum projected separation$^{4}$ (\bjdtdb)\dotfill &$2458518.00675^{+0.00068}_{-0.00091}$\\
~~~~$T_0$\dotfill &Optimal conjunction Time$^{5}$ (\bjdtdb)\dotfill &$2459032.98773^{+0.00035}_{-0.00037}$\\
~~~~$a$\dotfill &Semi-major axis (AU)\dotfill &$0.0380^{+0.0035}_{-0.0023}$\\
~~~~$i$\dotfill &Inclination (Degrees)\dotfill &$86.45^{+0.28}_{-0.27}$\\
~~~~$T_{eq}$\dotfill &Equilibrium temperature$^{6}$ (K)\dotfill &$423^{+61}_{-47}$\\
~~~~$\tau_{\rm circ}$\dotfill &Tidal circularization timescale (Gyr)\dotfill &$1.61^{+0.71}_{-0.99}$\\
~~~~$K$\dotfill &RV semi-amplitude (m/s)\dotfill &$35900^{+7000}_{-7200}$\\
~~~~$R_P/R_*$\dotfill &Radius of planet in stellar radii \dotfill &$1.18^{+0.22}_{-0.11}$\\
~~~~$a/R_*$\dotfill &Semi-major axis in stellar radii \dotfill &$29.6^{+1.7}_{-1.2}$\\
~~~~$\delta$\dotfill &$\left(R_P/R_*\right)^2$ \dotfill &$1.39^{+0.57}_{-0.25}$\\
~~~~$\delta_{\rm B}$\dotfill &Transit depth in B (fraction)\dotfill &$0.066^{+0.028}_{-0.026}$\\
~~~~$\delta_{\rm I}$\dotfill &Transit depth in I (fraction)\dotfill &$0.083^{+0.011}_{-0.021}$\\
~~~~$\delta_{\rm J}$\dotfill &Transit depth in J (fraction)\dotfill &$-0.39^{+0.17}_{-0.46}$\\
~~~~$\delta_{\rm R}$\dotfill &Transit depth in R (fraction)\dotfill &$0.082^{+0.012}_{-0.014}$\\
~~~~$\delta_{\rm g'}$\dotfill &Transit depth in g' (fraction)\dotfill &$0.072^{+0.017}_{-0.038}$\\
~~~~$\delta_{\rm i'}$\dotfill &Transit depth in i' (fraction)\dotfill &$0.071^{+0.021}_{-0.045}$\\
~~~~$\delta_{\rm r'}$\dotfill &Transit depth in r' (fraction)\dotfill &$0.0921^{+0.0075}_{-0.018}$\\
~~~~$\delta_{\rm z'}$\dotfill &Transit depth in z' (fraction)\dotfill &$-0.112^{+0.067}_{-0.27}$\\
~~~~$\delta_{\rm TESS}$\dotfill &Transit depth in TESS (fraction)\dotfill &$0.0933^{+0.0050}_{-0.0089}$\\
~~~~$\tau$\dotfill &Ingress/egress transit duration (days)\dotfill &$0.03393^{+0.00097}_{-0.0010}$\\
~~~~$T_{14}$\dotfill &Total transit duration (days)\dotfill &$0.0679^{+0.0019}_{-0.0021}$\\
~~~~$T_{FWHM}$\dotfill &FWHM transit duration (days)\dotfill &$0.03393^{+0.00097}_{-0.0010}$\\
~~~~$b$\dotfill &Transit Impact parameter \dotfill &$1.80^{+0.23}_{-0.13}$\\
~~~~$\delta_{S,2.5\mu m}$\dotfill &Blackbody eclipse depth at 2.5$\mu$m (ppm)\dotfill &$8.2^{+35}_{-5.9}$\\
~~~~$\delta_{S,5.0\mu m}$\dotfill &Blackbody eclipse depth at 5.0$\mu$m (ppm)\dotfill &$2100^{+3500}_{-990}$\\
~~~~$\delta_{S,7.5\mu m}$\dotfill &Blackbody eclipse depth at 7.5$\mu$m (ppm)\dotfill &$11600^{+12000}_{-3400}$\\
~~~~$\rho_P$\dotfill &Density (cgs)\dotfill &$5.1^{+1.5}_{-1.9}$\\
~~~~$logg_P$\dotfill &Surface gravity \dotfill &$4.505^{+0.064}_{-0.12}$\\
~~~~$\Theta$\dotfill &Safronov Number \dotfill &$19.9^{+6.9}_{-5.6}$\\
~~~~$\fave$\dotfill &Incident Flux (\fluxcgs)\dotfill &$0.0073^{+0.0052}_{-0.0027}$\\
~~~~$T_P$\dotfill &Time of Periastron (\bjdtdb)\dotfill &$2458518.00675^{+0.00068}_{-0.00091}$\\
~~~~$T_S$\dotfill &Time of eclipse (\bjdtdb)\dotfill &$2458515.45733^{+0.00068}_{-0.00091}$\\
~~~~$T_A$\dotfill &Time of Ascending Node (\bjdtdb)\dotfill &$2458521.83086^{+0.00068}_{-0.00091}$\\
~~~~$T_D$\dotfill &Time of Descending Node (\bjdtdb)\dotfill &$2458519.28145^{+0.00068}_{-0.00091}$\\
~~~~$V_c/V_e$\dotfill & \dotfill &$1.00$\\
~~~~$M_P\sin i$\dotfill &Minimum mass (\me)\dotfill &$41000^{+14000}_{-9400}$\\
~~~~$M_P/M_*$\dotfill &Mass ratio \dotfill &$0.76^{+0.35}_{-0.19}$\\
~~~~$d/R_*$\dotfill &Separation at mid transit \dotfill &$29.6^{+1.7}_{-1.2}$\\
~~~~$P_T$\dotfill &A priori non-grazing transit prob \dotfill &$-0.0061^{+0.0039}_{-0.0070}$\\
~~~~$P_{T,G}$\dotfill &A priori transit prob \dotfill &$0.0747^{+0.0041}_{-0.0040}$\\
\smallskip\\\multicolumn{2}{l}{Wavelength Parameters:}&B&I&J&R&g'&i'&r'&z'&TESS\smallskip\\
~~~~$u_{1}$\dotfill &linear limb-darkening coeff \dotfill &$0.70^{+0.25}_{-0.50}$&$0.41^{+0.35}_{-0.21}$&$1.770^{+0.099}_{-0.12}$&$0.42^{+0.25}_{-0.28}$&$0.60^{+0.38}_{-0.30}$&$0.64^{+0.40}_{-0.43}$&$0.22^{+0.30}_{-0.15}$&$1.53^{+0.23}_{-0.15}$&$0.19^{+0.19}_{-0.13}$\\
~~~~$u_{2}$\dotfill &quadratic limb-darkening coeff \dotfill &$0.04^{+0.55}_{-0.34}$&$0.23^{+0.27}_{-0.34}$&$-0.826^{+0.12}_{-0.079}$&$0.11^{+0.21}_{-0.27}$&$-0.01^{+0.36}_{-0.34}$&$0.11^{+0.40}_{-0.46}$&$0.25^{+0.20}_{-0.29}$&$-0.68^{+0.17}_{-0.15}$&$0.27^{+0.33}_{-0.29}$\\
\smallskip\\\multicolumn{2}{l}{Transit Parameters:}&TESS UT 2019-02-03 (TESS)&TESS UT 2019-02-08 (TESS)&TESS UT 2019-02-13 (TESS)&TESS UT 2019-02-23 (TESS)&NGTS UT 2019-04-20 (R)&LCO UT 2019-04-20 (z')&TMMT UT 2019-04-25 (I)&NGTS UT 2019-06-10 (R)&MuSCAT2 UT 2020-01-16 (g')&MuSCAT2 UT 2020-01-16 (i')&MuSCAT2 UT 2020-01-16 (r')&MuSCAT2 UT 2020-01-16 (z')&KeplerCam UT 2020-01-26 (B)&MuSCAT2 UT 2020-03-02 (g')&MuSCAT2 UT 2020-03-02 (i')&MuSCAT2 UT 2020-03-02 (r')&MuSCAT2 UT 2020-03-02 (z')&MuSCAT2 UT 2020-04-16 (g')&MuSCAT2 UT 2020-04-16 (i')&MuSCAT2 UT 2020-04-16 (r')&MuSCAT2 UT 2020-04-16 (z')&LCRO UT 2020-11-27 (i')&MuSCAT2 UT 2021-01-07 (i')&MuSCAT2 UT 2021-01-07 (r')&MuSCAT2 UT 2021-01-07 (z')&TESS UT 2021-02-11 (TESS)&TESS UT 2021-02-16 (TESS)&TESS UT 2021-02-27 (TESS)&ExTrA UT 2021-03-04 (J)&TESS UT 2021-03-04 (TESS)&ExTrA UT 2021-04-13 (J)&ExTrA UT 2021-04-19 (J)&ExTrA UT 2021-06-03 (J)\smallskip\\
~~~~$\sigma^{2}$\dotfill &Added Variance \dotfill &$-0.000000386^{+0.00000010}_{-0.000000073}$&$-0.000000118^{+0.00000011}_{-0.000000099}$&$-0.00000030^{+0.00000015}_{-0.00000012}$&$-0.00000013^{+0.00000013}_{-0.00000012}$&$-0.0000272^{+0.0000020}_{-0.0000012}$&$0.00000125^{+0.00000033}_{-0.00000022}$&$-0.00000217^{+0.0000011}_{-0.00000037}$&$-0.0000565^{+0.0000018}_{-0.0000028}$&$-0.00000152^{+0.00000061}_{-0.00000029}$&$-0.00000145^{+0.00000014}_{-0.00000016}$&$-0.00000031^{+0.00000014}_{-0.00000015}$&$-0.000000958^{+0.000000075}_{-0.000000076}$&$0.00000465^{+0.00000085}_{-0.00000090}$&$-0.00000139^{+0.00000051}_{-0.00000026}$&$-0.00000093^{+0.00000051}_{-0.00000024}$&$-0.00000141\pm0.00000029$&$-0.00000103^{+0.00000025}_{-0.00000030}$&$-0.00000173^{+0.00000091}_{-0.00000054}$&$-0.00000215^{+0.00000041}_{-0.00000040}$&$-0.00000108^{+0.00000055}_{-0.00000031}$&$-0.00000116^{+0.00000043}_{-0.00000038}$&$-0.0000003^{+0.0000030}_{-0.0000019}$&$-0.00000098^{+0.00000013}_{-0.00000016}$&$-0.00000122^{+0.00000025}_{-0.00000019}$&$-0.00000143^{+0.00000022}_{-0.00000010}$&$0.00000058^{+0.00000013}_{-0.00000017}$&$0.000000386^{+0.000000093}_{-0.00000015}$&$0.00000038^{+0.00000014}_{-0.00000017}$&$-0.00000021^{+0.00000046}_{-0.00000039}$&$0.000000120^{+0.000000076}_{-0.000000083}$&$-0.00000048^{+0.00000050}_{-0.00000021}$&$0.00000005^{+0.00000026}_{-0.00000037}$&$-0.00000001^{+0.00000028}_{-0.00000022}$\\
~~~~$F_0$\dotfill &Baseline flux \dotfill &$0.999914^{+0.000042}_{-0.000087}$&$0.999858^{+0.000041}_{-0.000047}$&$1.000140^{+0.000057}_{-0.000085}$&$0.999681^{+0.000087}_{-0.000072}$&$0.99787^{+0.00027}_{-0.00013}$&$0.99971^{+0.00014}_{-0.00011}$&$1.00056^{+0.00011}_{-0.00013}$&$0.99890^{+0.00027}_{-0.00029}$&$0.999440^{+0.00012}_{-0.000100}$&$1.000035^{+0.000065}_{-0.00011}$&$1.000052^{+0.000086}_{-0.00010}$&$1.000249^{+0.00013}_{-0.000090}$&$1.00012^{+0.00022}_{-0.00011}$&$0.99947^{+0.00022}_{-0.00010}$&$1.00019^{+0.00017}_{-0.00013}$&$0.99954^{+0.00014}_{-0.00013}$&$1.000255^{+0.000094}_{-0.00020}$&$0.99930^{+0.00023}_{-0.00018}$&$0.99974^{+0.00017}_{-0.00022}$&$0.999627^{+0.000078}_{-0.00015}$&$0.99989^{+0.00014}_{-0.00022}$&$1.00049^{+0.00064}_{-0.00042}$&$1.000035^{+0.000052}_{-0.00010}$&$0.99953^{+0.00013}_{-0.00015}$&$1.00010^{+0.00010}_{-0.00013}$&$1.000103^{+0.00010}_{-0.000073}$&$1.000277^{+0.000075}_{-0.00010}$&$1.000499^{+0.000091}_{-0.00011}$&$0.99995^{+0.00022}_{-0.00016}$&$0.999948\pm0.000060$&$1.00015^{+0.00012}_{-0.00024}$&$1.00003^{+0.00014}_{-0.00012}$&$1.00006^{+0.00014}_{-0.00018}$\\
~~~~$C_{0}$\dotfill &Additive detrending coeff \dotfill &--&--&--&--&$0.0060^{+0.0013}_{-0.0016}$&--&$0.00045^{+0.00054}_{-0.0011}$&$-0.0042^{+0.0020}_{-0.0013}$&$0.00207^{+0.00077}_{-0.0012}$&$0.00106^{+0.00038}_{-0.00063}$&$0.00077^{+0.00058}_{-0.00049}$&$0.00014^{+0.00060}_{-0.00051}$&--&$-0.00245^{+0.00068}_{-0.00052}$&$0.00065^{+0.00039}_{-0.00045}$&$-0.00170^{+0.00055}_{-0.00032}$&$0.00114^{+0.00034}_{-0.00036}$&$-0.00339^{+0.0015}_{-0.00094}$&$-0.00242^{+0.00074}_{-0.00097}$&$-0.00185^{+0.0010}_{-0.00075}$&$-0.00066^{+0.00037}_{-0.00049}$&$-0.00002^{+0.00088}_{-0.0030}$&$0.00000^{+0.00053}_{-0.00028}$&$0.00114^{+0.00024}_{-0.00022}$&$-0.00016^{+0.00026}_{-0.00046}$&--&--&--&--&--&--&--&--\\
~~~~$M_{0}$\dotfill &Multiplicative detrending coeff \dotfill &--&--&--&--&$0.00009^{+0.00061}_{-0.00043}$&$-0.00054^{+0.00045}_{-0.00025}$&--&$0.00485^{+0.00062}_{-0.0012}$&$0.00010^{+0.00052}_{-0.00032}$&$0.00047^{+0.00026}_{-0.00027}$&$0.00027^{+0.00036}_{-0.00026}$&$0.00041^{+0.00017}_{-0.00025}$&$0.00045^{+0.00056}_{-0.0013}$&$-0.00020^{+0.00075}_{-0.00042}$&$0.00007^{+0.00035}_{-0.00064}$&$0.00006^{+0.00032}_{-0.00033}$&$0.00004\pm0.00047$&$-0.00068^{+0.00071}_{-0.00056}$&$-0.00039^{+0.00046}_{-0.00045}$&$-0.00007^{+0.00058}_{-0.00038}$&$-0.00047^{+0.00059}_{-0.00057}$&--&$-0.00016^{+0.00050}_{-0.00034}$&$0.00029^{+0.00048}_{-0.00063}$&$-0.00026^{+0.00038}_{-0.00028}$&--&--&--&--&--&--&--&--\\
~~~~$M_{1}$\dotfill &Multiplicative detrending coeff \dotfill &--&--&--&--&--&--&--&--&$-0.00399^{+0.00081}_{-0.00066}$&$-0.00122^{+0.00062}_{-0.00043}$&$-0.00069^{+0.00048}_{-0.00070}$&$0.00053^{+0.00046}_{-0.00041}$&--&$0.00076^{+0.00068}_{-0.00062}$&$-0.00025^{+0.00068}_{-0.00061}$&$0.00044^{+0.00069}_{-0.00082}$&$-0.00049^{+0.00048}_{-0.00042}$&$0.00138^{+0.0012}_{-0.00082}$&$0.00168^{+0.00068}_{-0.0012}$&$0.00096^{+0.00095}_{-0.0014}$&$0.00042^{+0.00042}_{-0.00063}$&--&$0.00006^{+0.00037}_{-0.00033}$&$0.00015^{+0.00034}_{-0.00043}$&$-0.00004^{+0.00052}_{-0.00039}$&--&--&--&--&--&--&--&--\\
\enddata
\tablenotetext{}{See Table 3 in \citet{Eastman_2019} for a detailed description of all parameters}
\tablenotetext{1}{The metallicity of the star at birth}
\tablenotetext{2}{Corresponds to static points in a star's evolutionary history. See \S2 in \citet{Dotter_2016}.}
\tablenotetext{3}{Time of conjunction is commonly reported as the "transit time"}
\tablenotetext{4}{Time of minimum projected separation is a more correct "transit time"}
\tablenotetext{5}{Optimal time of conjunction minimizes the covariance between $T_C$ and Period}
\tablenotetext{6}{Assumes no albedo and perfect redistribution}
\end{deluxetable*}

\onecolumngrid
\subsection{RV Analysis}
\label{sect:heb_rv}
The analysis of TOI 620's RVs in the circum-secondary case, by allowing a negative $K$, works equally well for an HEB scenario, and it is difficult to distinguish between the two from RVs alone.  However, unique to the HEB scenario is the possibility of the true period being 10.20 days. Our priors are identical to the circum-secondary analysis with the exception of the period being doubled. We obtain nearly identical results, with a recovered negative $K_b$ at 2.4-$\sigma$ significance (\ref{tab:heb_rv_priors}). The RV plots for this scenario are presented in Figures \ref{fig:csec_phasedrvs} and \ref{fig:csec_fullrvs}. All posteriors are well behaved and uncorrelated.



\begin{table*}[htp!]
\centering
\begin{tabular}{cccccc}
    \hline
    Parameter [units] & MAP Value & MCMC Posterior \\
    \hline
    \hline
    $P_b$ [days] & 10.19762 (Locked) & -- \\
    $T_{C,b}$ [days] & 2458518.007 (Locked) & -- \\
    $e_b$ & 0 (Locked) & -- \\
    $\omega_b$ & $0$ (Locked) & -- \\
    $K_b$ [m s$^{-1}$] & -2.21 & $-2.48^{+0.96}_{-0.95}$ \\
    \hline
    $\gamma_{\rm iSHELL}$ [m s$^{-1}$] & -2.62 & $4.12^{+2.90}_{-3.01}$ \\
    $\gamma_{\rm CARMENES-Vis}$ [m s$^{-1}$] & -0.11 & $-0.06^{+1.89}_{-2.02}$ \\
    $\gamma_{\rm CARMENES-NIR}$ [m s$^{-1}$] & -1.51 & $-0.61^{+3.86}_{-3.81}$ \\
    $\gamma_{\rm NEID}$ [m s$^{-1}$] & 0.20 & $0.88^{+1.41}_{-1.48}$ \\
    $\gamma_{\rm MAROON-X-blue}$ [m s$^{-1}$] & -4.43 & $-5.09^{+1.21}_{-1.21}$ \\
    $\gamma_{\rm MAROON-X-red}$ [m s$^{-1}$] & -6.43 & $-7.00^{+1.50}_{-1.50}$ \\
    \hline
    $\sigma_{\rm iSHELL}$ [m s$^{-1}$] & 13.24 & $13.17^{+1.14}_{-1.08}$ \\
    $\sigma_{\rm CARMENES-Vis}$ [m s$^{-1}$] & 3.36 & $4.07^{+1.54}_{-1.30}$ \\
    $\sigma_{\rm CARMENES-NIR}$ [m s$^{-1}$] & 5.00  & $5.48^{+1.08}_{-0.93}$ \\
    $\sigma_{\rm NEID}$ [m s$^{-1}$] & 5.00 & $5.14^{+1.08}_{-0.93}$\\
    $\sigma_{\rm MAROON-X-blue}$ [m s$^{-1}$] & 1.00  & $1.21^{+1.03}_{-0.80}$ \\
    $\sigma_{\rm MAROON-X-red}$ [m s$^{-1}$] & 2.68 & $3.11^{+0.86}_{-0.66}$ \\
    \hline
    $\dot{\gamma}$ [\msday] & 0.07 & $0.08^{+0.01}_{-0.01}$ \\  
    \hline
\end{tabular}
\caption{The model parameters and posterior distributions used in our RV model that considers the transiting b planet as an HEB and the linear $\dot{\gamma}$ trend.  The priors are identical to the circum-secondary run except the $P_b$ value is locked at twice the \tess period.}
\label{tab:heb_rv_priors}
\end{table*}

\onecolumngrid
\section{RV Table}

\startlongtable
\begin{deluxetable*}{lcccccc}
\label{tab:RVs}
\tablecaption{A full list of the radial velocity values, times, and errors used in this paper.}
\tablehead{\colhead{~~~\bjdtdb [days]} & \colhead{RV [\ms]} & \colhead{Error [\ms]} & \colhead{Instrument} & \colhead{Offset$^\dagger$ [\ms]} & \colhead{Offset Error$^\dagger$ [\ms]}}
\startdata
2458874.042603 \dotfill & -22.03 \dotfill & 4.96 \dotfill & iSHELL & -- & -- \\
2458875.075875 \dotfill & -26.54 \dotfill & 5.03 \dotfill & iSHELL & -- & -- \\ 
2458895.047871 \dotfill & -32.99 \dotfill & 4.71 \dotfill & iSHELL & -- & -- \\
2458897.065268 \dotfill & -24.80 \dotfill & 11.31 \dotfill & iSHELL & -- & -- \\
2458899.036716 \dotfill & -13.29 \dotfill & 4.83 \dotfill & iSHELL & -- & -- \\
2458900.046926 \dotfill & -24.11 \dotfill & 5.34 \dotfill & iSHELL & -- & -- \\ 
2458901.022179 \dotfill & -35.33 \dotfill & 5.21 \dotfill & iSHELL & -- & -- \\
2459217.085069 \dotfill & -9.48 \dotfill & 4.07 \dotfill & iSHELL & -- & -- \\
2459220.070212 \dotfill & -15.13 \dotfill & 4.32 \dotfill & iSHELL & -- & -- \\
2459220.905235 \dotfill & -2.32 \dotfill & 3.95 \dotfill & NEID & -- & -- \\
2459220.915972 \dotfill & -13.34 \dotfill & 3.31 \dotfill & NEID & -- & -- \\
2459221.062629 \dotfill & -23.22 \dotfill & 4.63 \dotfill & iSHELL & -- & -- \\
2459227.095393 \dotfill & 25.50 \dotfill & 7.97 \dotfill & iSHELL & -- & -- \\
2459230.866622 \dotfill & -6.31 \dotfill & 1.94 \dotfill & NEID & -- & -- \\
2459230.877358 \dotfill & -6.31 \dotfill & 2.06 \dotfill & NEID & -- & -- \\
2459233.027990 \dotfill & -7.54 \dotfill & 3.89 \dotfill & iSHELL & -- & -- \\
2459245.996343 \dotfill & -4.82 \dotfill & 4.87 \dotfill & iSHELL & -- & -- \\
2459248.508540 \dotfill & -35.88 \dotfill & 17.06 \dotfill & CARMENES-NIR & -- & -- \\
2459250.546740 \dotfill & 8.82 \dotfill & 18.00 \dotfill & CARMENES-NIR & -- & -- \\
2459250.546800 \dotfill & -0.89 \dotfill & 2.73 \dotfill & CARMENES-Vis & -- & -- \\
2459250.939338 \dotfill & -34.97 \dotfill & 20.55 \dotfill & iSHELL & -- & -- \\
2459251.765798 \dotfill & -5.28 \dotfill & 2.41 \dotfill & NEID & -- & -- \\
2459251.776534 \dotfill & 1.46 \dotfill & 2.20 \dotfill & NEID & -- & -- \\
2459251.940203 \dotfill & -8.45 \dotfill & 10.23 \dotfill & iSHELL & -- & -- \\
2459252.938448 \dotfill & -0.33 \dotfill & 3.46 \dotfill & iSHELL & -- & -- \\
2459255.020459 \dotfill & 37.18 \dotfill & 5.89 \dotfill & iSHELL & -- & -- \\
2459256.017137 \dotfill & 46.30 \dotfill & 7.11 \dotfill & iSHELL & -- & -- \\ 
2459257.009042 \dotfill & 43.08 \dotfill & 5.83 \dotfill & iSHELL & -- & -- \\
2459258.011730 \dotfill & 38.23 \dotfill & 4.90 \dotfill & iSHELL & -- & -- \\
2459261.040202 \dotfill & 14.99 \dotfill & 7.52 \dotfill & iSHELL & -- & -- \\
2459267.742876 \dotfill & -2.67 \dotfill & 3.09 \dotfill & NEID & -- & -- \\
2459267.753612 \dotfill & -2.82 \dotfill & 2.54 \dotfill & NEID & -- & -- \\
2459269.006339 \dotfill & 36.29 \dotfill & 16.24 \dotfill & iSHELL & -- & -- \\
2459269.942464 \dotfill & -11.47 \dotfill & 2.51 \dotfill & MAROON-X  blue & -1.5 \dotfill & 0.5 \dotfill \\
2459269.942464 \dotfill & -11.20 \dotfill & 1.77 \dotfill & MAROON-X  red & -2.0 \dotfill & 0.5 \dotfill \\
2459273.024693 \dotfill & 41.53 \dotfill & 19.96 \dotfill & iSHELL & -- & -- \\
2459273.476340 \dotfill & 0.95 \dotfill & 1.70 \dotfill & CARMENES-Vis & -- & -- \\
2459273.476430 \dotfill & -7.83 \dotfill & 7.20 \dotfill & CARMENES-NIR & -- & -- \\
2459273.764119 \dotfill & 5.94 \dotfill & 1.68 \dotfill & NEID & -- & -- \\
2459273.774855 \dotfill & 4.61 \dotfill & 1.90 \dotfill & NEID & -- & -- \\
2459279.744662 \dotfill & 11.40 \dotfill & 2.38 \dotfill & NEID & -- & -- \\
2459279.758946 \dotfill & 6.76 \dotfill & 1.99 \dotfill & NEID & -- & -- \\
2459299.334560 \dotfill & 4.82 \dotfill & 1.67 \dotfill & CARMENES-Vis & -- & -- \\
2459299.334360 \dotfill & 15.12 \dotfill & 6.62 \dotfill & CARMENES-NIR & -- & -- \\
2459300.414680 \dotfill & 3.48 \dotfill & 6.62 \dotfill & CARMENES-NIR & -- & -- \\
2459300.415170 \dotfill & -0.68 \dotfill & 1.38 \dotfill & CARMENES-Vis & -- & -- \\
2459301.395920 \dotfill & -1.51 \dotfill & 5.45 \dotfill & CARMENES-NIR & -- & -- \\
2459301.396090 \dotfill & -2.75 \dotfill & 1.66 \dotfill & CARMENES-Vis & -- & -- \\
2459302.446160 \dotfill & -12.04 \dotfill & 9.77 \dotfill & CARMENES-NIR & -- & -- \\
2459302.446330 \dotfill & -4.59 \dotfill & 2.81 \dotfill & CARMENES-Vis & -- & -- \\
2459319.904094 \dotfill & 19.94 \dotfill & 5.76 \dotfill & iSHELL & -- & -- \\
2459320.909564 \dotfill & -2.78 \dotfill & 5.12 \dotfill & iSHELL & -- & -- \\
2459321.840076 \dotfill & -5.94 \dotfill & 1.93 \dotfill & MAROON-X  blue & 0.0\dotfill & 0.0\dotfill \\
2459321.840076 \dotfill & -8.10 \dotfill & 1.26 \dotfill & MAROON-X  red & 0.0\dotfill & 0.0\dotfill \\
2459321.909542 \dotfill & -8.14 \dotfill & 10.57 \dotfill & iSHELL & -- & -- \\
2459322.701506 \dotfill & -4.73 \dotfill & 1.99 \dotfill & NEID & -- & -- \\
2459322.712242 \dotfill & -0.73 \dotfill & 2.12 \dotfill & NEID & -- & -- \\
2459322.864615 \dotfill & -22.52 \dotfill & 4.54 \dotfill & iSHELL & -- & -- \\
2459328.783324 \dotfill & -27.13 \dotfill & 5.02 \dotfill & iSHELL & -- & -- \\
2459331.781398 \dotfill & -21.77 \dotfill & 5.93 \dotfill & iSHELL & -- & -- \\
2459333.810183 \dotfill & 1.01 \dotfill & 1.71 \dotfill & MAROON-X  blue & 0.0\dotfill & 0.0\dotfill \\
2459333.810183 \dotfill & 2.66 \dotfill & 1.09 \dotfill & MAROON-X  red & 0.0\dotfill & 0.0\dotfill \\
2459334.761529 \dotfill & -1.08 \dotfill & 1.93 \dotfill & MAROON-X  blue & 0.0\dotfill & 0.0\dotfill \\
2459334.761529 \dotfill & -1.46 \dotfill & 1.28 \dotfill & MAROON-X  red & 0.0\dotfill & 0.0\dotfill \\
2459352.691482 \dotfill & 5.86 \dotfill & 2.41 \dotfill & NEID & -- & -- \\
2459352.702216 \dotfill & 6.37 \dotfill & 2.71 \dotfill & NEID & -- & -- \\
2459361.765208 \dotfill & 0.49 \dotfill & 1.29 \dotfill & MAROON-X  blue & 2.0\dotfill & 1.0\dotfill \\
2459361.765208 \dotfill & -3.45 \dotfill & 0.80 \dotfill & MAROON-X  red & 2.5\dotfill & 1.0\dotfill \\
2459362.784009 \dotfill & -12.48 \dotfill & 10.33 \dotfill & iSHELL & -- & -- \\
2459362.784227 \dotfill & 4.28 \dotfill & 1.80 \dotfill & MAROON-X  blue & 2.0\dotfill & 1.0\dotfill \\
2459362.784227 \dotfill & 4.29 \dotfill & 0.74 \dotfill & MAROON-X  red & 2.5\dotfill & 1.0\dotfill \\
2459363.739140 \dotfill & 4.71 \dotfill & 1.70 \dotfill & MAROON-X  blue & 2.0\dotfill & 1.0\dotfill \\
2459363.739140 \dotfill & 3.03 \dotfill & 1.00 \dotfill & MAROON-X  red & 2.5\dotfill & 1.0\dotfill \\
2459363.770313 \dotfill & 0.04 \dotfill & 16.68 \dotfill & iSHELL & -- & -- \\
2459368.754018 \dotfill & 4.11 \dotfill & 2.17 \dotfill & MAROON-X  blue & 2.0\dotfill & 1.0\dotfill \\
2459368.754018 \dotfill & 0.24 \dotfill & 1.05 \dotfill & MAROON-X  red & 2.5\dotfill & 1.0\dotfill \\
\enddata
\tablenotetext{\dagger}{See explanation for the offset and offset error in \rev{section \ref{sec:maroonx}.}}

\end{deluxetable*}
~

\twocolumngrid

\end{document}